\documentclass[showpacs,showkeys,aps]{revtex4}
\usepackage{amssymb}
\usepackage{graphicx}% Include figure files
\usepackage{dcolumn}% Align table columns on decimal point
\usepackage{bm}% bold math
\usepackage{graphicx}
\usepackage{latexsym}
\usepackage{hyperref}
\usepackage{ifthen}
\usepackage{amsmath,amssymb,amsthm,mathrsfs,amsfonts,dsfont}
\usepackage{lscape}
\usepackage{booktabs}
\usepackage{tabularx}
\usepackage{amssymb}
\usepackage{color}
\begin{document}
\newcommand{\ti}[1]{\mbox{\tiny{#1}}}
\def\be{\begin{equation}}
\def\ee{\end{equation}}
\def\bea{\begin{eqnarray}}
\def\eea{\end{eqnarray}}
\newcommand{\bt}[1]{\mathbf{\mathtt{#1}}}
\newcommand{\tb}[1]{\textbf{\texttt{#1}}}
\newcommand{\diff}{\mathrm{d}}
\newcommand{\btb}[1]{\textcolor[rgb]{0.00,0.00,1.00}{\tb{#1}}}
\newcommand{\il}{~}
\newcommand{\rtb}[1]{\textcolor[rgb]{1.00,0.00,0.00}{\tb{#1}}}
\newcommand{\ptb}[1]{\textcolor[rgb]{0.50,0.00,0.50}{\tb{#1}}}
\title{Perfect fluid tori orbiting Kehagias-Sfetsos naked singularities}
\author{Zden\v{e}k Stuchl\'{\i}k, Daniela Pugliese, Jan Schee, and  Hana Ku\v{c}\'{a}kov\'{a}}
\email{zdenek.stuchlik@physics.cz; daniela.pugliese@fpf.slu.cz; jan.schee@fpf.slu.cz; hana.kucakova@centrum.cz}
\affiliation{Institute of Physics, Faculty of Philosophy \& Science, Silesian University in Opava, Bezru\v{c}ovo n\'{a}m\v{e}st\'{i} 13, CZ-74601 Opava, Czech Republic}
%\date{\today}
\begin{abstract}
We construct perfect fluid tori in the field of the Kehagias-Sfetsos (K-S) naked singularity  representing spherically symmetric vacuum solution of the modified Ho\v{r}ava quantum gravity that is characterized by a dimensionless parameter $\omega M^2$, combining the gravitational mass parameter $M$ of the spacetime with the Ho\v{r}ava parameter $\omega$ reflecting the role of the quantum corrections. In dependence on the value of $\omega M^2$, the K-S naked singularities demonstrate a variety of qualitatively different behavior of their circular geodesics that is fully reflected in the properties of the toroidal structures. In all of the K-S naked singularity spacetimes the tori are located above an ``antigravity'' sphere where matter can stay in stable equilibrium position,  that is  relevant for   the stability of  the orbiting   fluid toroidal  accretion structures.
\end{abstract}
%Quantum gravity, 04.60.-m
%Relativistic stars, 04.40.Dg
%97.10.Gz   Accretion and accretion discs stellar  97.10.Gz
%%% ----------------------------------------------------------------------
\maketitle
\section{Introduction}
The Ho\v{r}ava quantum gravity \cite{Hor:2009:PHYSR4} is a  recent promising approach to the quantum gravity  attracting   great attention, being a field-theory based on the ideas of the solid state physics that uses an anisotropic scaling of space and time \cite{Lif:1941:ZhEK-SpTheorPhys} where the  time dimension is scaled as $t \rightarrow b^{z}t$,  with an integer $z$, while  the space dimensions are scaled as $x \rightarrow bx$,  and, as a consequence, its Lagrangian is Lorentz invariant at low energies, but the Lorentz invariance is broken at high energies.  In fact,  the parameter $z$ changes from $z=3$, in the UV limit, to $z=1$ in the IR limit where the Lorentz invariance is  recovered \cite{Hor:2009:PHYSR4,Hor:2009:PHYSRL,Hen-Kle-Gom:2009:PRD,Hor-MelT:2010:PHYSR4,And-etal:2012:PHYSR4,Gri-Hor-MelT:2012:JHEP,Gri-Hor-MelT:2013:PHYSRL,Ver-Sot:2012:PHYSR4,Lib-Mac-Sot:2012:PHYSRL}.
The solutions of the Ho\v{r}ava-Lifshitz  effective gravitational equations have been discovered by \cite{Bar-Sot:2012:PHYSRL,Bar-Sot:2013:PHYSRL,Bar-Sot:2013:PHYSR4}, while a spherically symmetric solution  of the modified Ho\v{r}ava gravity,   known as Kehagias-Sfetsos (K-S) metric,  having asymptotically Schwarzschildian character and compatible with the Minkowski vacuum  was found in \cite{Keh-Sfe:2009:PhysLetB,Park Mu-In:2009:JHEP}. The character of the K-S geometries  is governed by the dimensionless product $\omega M^2$ where   the parameter $M$ is the gravitational mass determining the distance scales in the K-S  metric, and   $\omega$ is  the Ho\v{r}ava parameter that reflects the influence of the quantum effects.
If we assume that the Ho\v{r}ava parameter is an universal constant then  the character of the spacetime is only regulated  by the mass  $M$: for $\omega M^2 \geq 1/2$, the K-S metric describes a black hole solution, while for $\omega M^2 < 1/2$, there are no event horizons and the metric describes a naked singularity. The current observational limits on $\omega$ \cite{Ior-Rug:2010:IJMPA,Liu-etal:2011:GenRelGrav,Ior-Rug:2011:IJMPD} do not exclude the existence of stellar-mass K-S naked singularities. The Solar system test puts limit $\omega > 3.2 \times 10^{-20} cm^{-2}$ \cite{Ior-Rug:2011:IJMPD} implying that  the  gravitational mass of K-S naked singularities cannot exceed $2.6 \times 10^{4}M_{\odot}$.
Thus, it is of key interest to search for any signatures of K-S naked singularity spacetimes, for example in  the emergence of accretion  phenomena, which allow  to be distinguished clearly from the black hole attractors.
In this regard, the K-S black hole spacetimes has been investigated in a series of works on  the particle motion \cite{Hak-etal:2010:MPLA,Ali-Sen:2010:PHYSR4,Abd-Ahm-Hak:2011:PHYSR4,Ior-Rug:2011:IJMPD,Eno-etal:2011:PHYSR4,Hor-Ger-Ker:2011:PHYSR4,Hak-Abd-Ahm:2013:PHYSR4} and optical phenomena \cite{Har-Kov-Lob:2011:CLAQG,Ama-Eir:2012:PHYSR4,Eir-Sen:2012:PHYSR4,Ata-Abd-Ahm:2013:AstrSpaScie}.
While the K-S naked singularity spacetimes were also  extensively   studied: in particular  the effective potential of the test particle motion and the circular geodesics   were studied in \cite{Vie-etal:2014:PHYSR4}, where close similarity with properties of the circular geodesics of the Reissner-Nordstr\"om (RN) naked singularity spacetimes \cite{Stu-Hle:2002:ActaPhysSlov,Pug-etal:2011:PHYSR4} has been demonstrated. The ultra-high-energy particle collisions in the deepest parts of the gravitational potential of K-S naked singularity spacetimes have been studied in \cite{Stu-Sche-Abd:2014:PHYSR4}. Wide range of optical phenomena related to innermost parts of Keplerian discs (direct and indirect images, spectral continuum of thermal radiation, and profiled spectral lines) had been exposed in \cite{Stuchlik:2014yaa}. For all of the considered phenomena, related to the K-S naked singularities, an ``antigravity'' effect \cite{Vie-etal:2014:PHYSR4,Stuchlik:2014yaa} plays a fundamental role, a  similar situation has been also   discovered in some braneworld naked singularity spacetimes \cite{Kot-Stu-Tor:2008:CLAQG,Stu-Kot:2009:GenRelGrav,Sche-Stu:2009:IJMPD,Stu-Kol:2012:JCAP,Ali-etal:2013:CLAQG}, but it seems to be different in comparison to the rotating Kerr naked singularity spacetimes \cite{Stu:1980:BAC,Stu-Sche:2010:CLAQG,Stu-Hle-Tru:2011:CLAQG,Pat-Jos:2011:CLAQG,Stu-Sche:2012a:CLAQG,Stu-Sche:2013:CLAQG,Sche-Stu:2013:JCAP,Kol-Stu:2013:PHYSR4},
(for a detailed analysis of the magnetized  thick discs   see \cite{Adamek:2013dza}).
However in all of these studies, the motion of photons, particles and fluid has been assumed to be governed by the General Relativity laws, i.e., the IR (low energy) end of the Ho\v{r}ava theory, and possible effects of the Lorentz invariance violation, expected at the UV (high energy) end of the theory were not considered \cite{Hor:2009:PHYSR4,Hor:2009:PHYSRL,Keh-Sfe:2009:PhysLetB,Park Mu-In:2009:JHEP}. Thus, the Lorentz invariance violation remains still an open issue of the Ho\v{r}ava gravity, although the extremely complex character of the particle motion in a situation where this violation is relevant has been exposed in \cite{Cap-Pol:2010:JEP}. Here we keep the assumption of the general relativistic approximation and the standard geodesic motion that were  used in the previous research.

All the K-S naked singularity spacetimes contain a ``static'' sphere (or  also ``antigravity'' sphere),   where test particles, as subjected to an  ``antigravity'' effect,  are in a stable equilibrium position and the sphere radius    locates  also   the innermost limit on  the existence of circular geodesics, corresponding  to orbits with zero angular momentum and zero angular frequency \cite{Vie-etal:2014:PHYSR4,Stuchlik:2014yaa}. Nevertheless, the properties of the circular geodesics, in the K-S naked singularity spacetimes, are strongly determined  by the dimensionless parameter $\omega M^2$ \cite{Vie-etal:2014:PHYSR4,Stu-Sche-Abd:2014:PHYSR4,Stuchlik:2014yaa}: as it is  $0 < \omega M^2 < (\omega M^2)_{mso} = 0.2811$, then  stable circular orbits exist between the static sphere at radius $r_{stat}$ and infinity, while the  spacetimes  $(\omega M^2)_{mso} < \omega M^2 < (\omega M^2)_{\gamma} = 0.3849$ are characterized by two separated regions of stable circular geodesics, the outer region extends between the inner marginally stable orbit at $r_{isco}$ and infinity  and the inner one extends between the static radius $r_{stat}$ and the outer marginally stable orbit at $r_{osco}$, unstable circular geodesics are located at $r_{osco} < r < r_{isco}$. In the K-S spacetime,  with  $(\omega M^2)_{\gamma} < \omega M^2 < 1/2$, the inner region of stable circular orbits is bounded by a (stable) photon circular orbit located  $r_{\gamma}^o$, while the unstable circular geodesics located at $r_{\gamma}^i < r < r_{isco}$  are bounded by the (unstable) photon circular orbit in  $r_{\gamma}^i$,  located under the outer region of stable circular geodesics: then   no circular geodesics are possible   in the region  $r_{\gamma}^o < r < r_{\gamma}^i$ \cite{Vie-etal:2014:PHYSR4}.
As discussed in \cite{Stuchlik:2014yaa}, the standard Keplerian accretion disc can exist  in the region of the outer stable circular geodesics that is the  region of the stable circular geodesics in  the K-S geometries with  $0 < \omega M^2 < (\omega M^2)_{mso} = 0.2811$, while in the other K-S naked singularity spacetimes, the Keplerian disc has to be limited by $r_{isco}$ from below. The  gradient of the angular frequency $\Omega_K$ of the circular geodesic motion vanishes at some radius $r_{\Omega}^{Max}$ inside the Keplerian disc in all the K-S naked singularity spacetimes, and the condition for Keplerian accretion $d\Omega_K/dr < 0$ is then violated \cite{Stuchlik:2014yaa}. We expect that near the radius $r_{\Omega}^{Max}$, where the Keplerian accretion stops to work, matter has to be accumulated and a fluid toroidal structure has to be created successively. Therefore, it seems that in the K-S naked singularity spacetimes the fluid toroidal structures have to be related even to accretion processes that were of the Keplerian character at the beginning. Of course, the toroidal perfect fluid equilibrium configurations are related also to the inner regions of the circular geodesics, and could be thus doubled, as in the case of the RN naked singularity spacetimes \cite{Kuc-Sla-Stu:2011:JCAP}.

Here, we study the structure and shape of perfect fluid toroidal equilibrium tori orbiting the K-S naked singularities, considering the tori for whole range of the dimensionless parameter $\omega M^2 < 1/2$ corresponding to naked singularity spacetimes. All the various kinds of the character of the circular geodesic structure will be considered and their impact on the structure of the toroidal configurations will be reflected.

The present article is structured as follows: in Sec.\il(\ref{Sec:K-Sgeo}) we introduce the Kehagias-Sfetsos (K-S) geometry.  Section \il(\ref{Sec:cic-ge}) is devoted to the analysis of circular geodesics in Kehagias-Sfetsos spacetimes.  In Sec.\il(\ref{Sec:strucut}) we discuss the structure of toroidal  perfect fluid  orbiting a  Kehagias-Sfetsos  attractor and the equilibrium tori in the  K-S naked singularity spacetimes are considered in Sec.\il(\ref{Sec:Eq}). A classification of toroidal configurations in K-S spacetimes is introduced and detailed in Sec.\il(\ref{Sec:class}). Finally  the article closes with the conclusions  in Sec.\il(\ref{Sec:concl}).
\section{Kehagias-Sfetsos geometry}\label{Sec:K-Sgeo}
The Kehagias-Sfetsos solution of the modified Ho\v{r}ava gravity in the standard Schwarzschild coordinates and the geometric units ($c=G=1$) reads \cite{Keh-Sfe:2009:PhysLetB}
\bea\label{Eq:metric}
ds^{2}=-f(r)dt^{2}+\frac{1}{f(r)}dr^{2}+r^{2}d\theta^{2}+r^{2}\sin^{2}\theta d\phi^{2},
\\
f(r)\equiv1+r^{2}\omega\left(1-\sqrt{1+{4M}/{\omega r^{3}}}\right),
\eea
where $\omega$ is a  free parameter  governing  the role of the modified Ho\v{r}ava quantum gravity, the parameter $M$ gives the gravitational mass and the dimensional scale of the solution. Although the K-S solution of the modified Ho\v{r}ava gravity is a vacuum solution, it is not a Ricci flat solution. In the Schwarzschild limit, $\omega \rightarrow \infty$, the Ricci scalar has the asymptotic behavior
$
                       R \sim {1}/{\omega}
$.
More information can be found in \cite{Stu-Sche-Abd:2014:PHYSR4}. The physical singularity of the K-S spacetimes located at $r=0$ is of a special character, as the metric coefficients $g_{tt}(r=0)=g_{rr}(r=0)=1$ are finite there, while their radial derivatives are divergent. This implies that particles freely radially falling from infinity can reach the physical singularity, if they do not loose energy during their fall; however, any particle with non-zero angular momentum is repulsed by the centrifugal repulsive barrier, acting in accord with the ``antigravity'' effect, and cannot reach the physical singularity \cite{Vie-etal:2014:PHYSR4}. The ``antigravity'' effect can be illustrated by the behavior of the lapse function $f(r)$ at vicinity of the physical singularity. For $r \to 0$ it is
$
f(r) \sim 1 - 2\sqrt{\omega M r}
$,
indicating gravitation repulsion of the quintessential-field type \cite{Kis:2003:CLAQG}.
Properties of the K-S spacetimes are governed by the dimensionless parameter $\omega M^2$ \cite{Vie-etal:2014:PHYSR4,Stu-Sche-Abd:2014:PHYSR4,Stuchlik:2014yaa}. If the Ho\v{r}ava parameter $\omega$ is assumed  fixed, then the gravitational mass parameter $M$ governs the character of the K-S spacetimes.
%\begin{equation}
%1+r^{2}\omega\left(1-\sqrt{1+\frac{4M}{\omega r^{3}}}\right)=0
%\end{equation}
%and they

Two horizons of the K-S black hole spacetimes exist at radii
\begin{equation}
 r_{\pm}\equiv M\pm\sqrt{M^{2}-\frac{1}{2\omega}}\quad\mbox{if}\quad
\omega M^2 \geq \omega M^{2}_{h}\equiv\frac{1}{2}.
\end{equation}
When the equality holds, the horizons coincide, giving an extreme black hole (BH) K-S spacetime. The K-S naked singularity  (NS) spacetimes occur for
$
\omega M^2 < \omega M^{2}_{h}$.
Therefore, assuming the parameter $\omega$ fixed, the gravitational mass $M$ small enough guarantees the existence of a K-S naked singularity. Recent works on the influence of the dimensionless Ho\v{r}ava parameter in astrophysical systems \cite{Ior-Rug:2010:OAJ,Har-Kov-Lob:2011:CLAQG,Dwo-Hor-Ger:2013:AstrNach} put very weak limit
$
         \omega M^2 > 8 \times 10^{-10}
$
that is very far from the limit corresponding to the K-S black hole spacetimes, and the K-S naked singularity spacetimes have to be considered in astrophysical phenomena related to compact objects having mass $M < 10^4 M_{\odot}$, if the modified Ho\v{r}ava gravity is considered relevant.
\section{Circular geodesics}\label{Sec:cic-ge}
The motion of test particles and photons is governed by the spacetime geodesics. Due to the axial symmetry and stationarity of the K-S spacetimes two constants of the geodesic motion arise being related to the covariant components of the particle (photon) 4-momentum
\(
 P_{\phi}=L,  P_{t}=-E
\).
The radial component of the 4-momentum of test particles reads
\begin{equation}
\left[P^{r}\right]^{2}=E^{2}-V_{eff},\quad V_{eff}\equiv f(r)\left(m^2 + \frac{L^{2}}{r^{2}}\right) .
\end{equation}
where an effective potential $ V_{eff}$ is introduced,
$m$ being the particle mass.
Circular geodesics were studied for the K-S black hole spacetimes in \cite{Chen-Wang:2010:IJMPA,Abd-Ahm-Hak:2011:PHYSR4} and for the K-S naked singularity spacetimes in \cite{Vie-etal:2014:PHYSR4,Stu-Sche-Abd:2014:PHYSR4,Stuchlik:2014yaa}. We give a short summary of the properties of the circular geodesics in dependence on the parameter $\omega M^2$.

\subsection{Photon circular geodesics}
We consider  the equatorial motion ($\theta = \pi/2$),
the photon orbits exist  for
\begin{equation}
 \omega M^2 > \omega M^{2}_{\gamma}\equiv\frac{2}{3\sqrt{3}}.
\end{equation}
In the K-S black hole spacetimes, $\omega M^2 \geq \omega M^2_{h}=1/2$, one photon circular geodesics always exists, and its radius is given by
\begin{equation}
r_{\gamma}^i\equiv2\sqrt{3}M\cos\left(\frac{1}{3}\cos^{-1}\left(-\frac{2}{3\sqrt{3}\omega M^2}\right)-\frac{2\pi}{3}\right).\label{eq_rph1}
\end{equation}
This orbit is unstable relative to radial perturbations and represents an inner boundary for existence of circular geodesics.
For the K-S naked singularity spacetimes with $\omega M^2_{h}  > \omega M^2 > \omega M^2_{\gamma}$, two photon circular orbits exist - the outer (unstable) one at the radius given by formula (\ref{eq_rph1}) and the inner (stable) one at
\begin{equation}
r_{\gamma}^o\equiv2\sqrt{3}M\cos\left(\frac{1}{3}\cos^{-1}\left(-\frac{2}{3\sqrt{3}\omega M^2}\right)\right).
\end{equation}
Between the stable and unstable circular orbits no circular geodesics are possible. This stable circular photon orbit is the outer boundary of the inner region of the circular geodetical orbits. As it  is $r_{\gamma}^o(\omega) < r_{\gamma}^i(\omega)$, hereafter we adopt the notation $r_{\gamma}^-\equiv r_{\gamma}^o$ and $r_{\gamma}^+\equiv r_{\gamma}^i$, so that $r_{\gamma}^-<r_{\gamma}^+$. For $\omega M^2 = \omega M^2_{\gamma}$, only one photon circular orbit at the radius $r_{\gamma}=\sqrt{3}M$ remains, while for $\omega M^2 < \omega M^2_{\gamma}$, no photon circular orbit exists.
%\end{enumerate}
In the K-S naked singularity spacetimes with $\omega M^2_{h} > \omega M^2 > \omega M^2_{\gamma}$, allowing existence of the stable and unstable photon circular geodesics, a region of trapped photon orbits exists in the vicinity of the stable photon circular geodesic. This situation is similar to those discovered in the field of Kerr naked singularity spacetimes \cite{Stu-Sche:2010:CLAQG}, where the trapped photons can strongly influence the accretion phenomena, especially in the near-extreme Kerr naked singularity states \cite{Stu:1980:BAC,Stu-Hle-Tru:2011:CLAQG,Stu-Sche:2012a:CLAQG,Stu-Sche:2012b:CLAQG}. We expect a similar strong influence in the K-S naked singularity spacetimes too \cite{Stuchlik:2014yaa,Vie-etal:2014:PHYSR4}.
In the following, we simplify calculations putting $M=1$ making both the radial coordinate and the Ho\v{r}ava parameter dimensionless; of course, also the time coordinate becomes dimensionless. The resulting formulae can be easily transformed to expressions containing the mass parameter $M$ by the transformations $r/M \rightarrow r$ and $\omega M^2\rightarrow \omega$.
\subsection{Keplerian circular orbits}
The equatorial circular orbits of test particles are determined by the condition ${dV_{eff}}/{dr}=0$. The radial profile of the specific angular momentum $L_{K}$(related to the unit rest mass of the particle)  and energy $E_{K}$  take the form
\bea
\frac{L_{K}^{2}(r;\omega)}{m^2}\equiv\frac{r^{2}}{r A(r;\omega)-3}\left[r^{3}\omega  A(r;\omega)-(1+r^{3}\omega)\right] ,\quad A(r;\omega)\equiv\sqrt{1+\frac{4}{r^{3}\omega}}, \label{eq_L2circ1}
\\
\frac{E_{K}^{2}(r;\omega)}{m^2}\equiv\left[1+r^{2}\omega\left(1-\sqrt{1+\frac{4}{r^{3}\omega}}\right)\right]\left(1+\frac{1}{rA-3}\left[r^{3}\omega A-(1+r^{3}\omega)\right]\right).
\eea
In the following, we use the simplified notation $L_{K}/m \to L_{K}$ and $E_{K}/m \to E_{K}$.
The radial profile of the angular frequency of the test particle motion on the Keplerian (geodesic) circular orbits takes the form
\begin{equation}
\Omega_{K}(r;\omega)\equiv\frac{U^{\phi}}{U^{t}}=\frac{f(r)}{r^{2}}\frac{L_{K}}{E_{K}}=\sqrt{\frac{r^3\omega [A(r;\omega)-1]-1}{r^3A(r;\omega)}}.\label{eq30}
\end{equation}
\begin{figure}[h]
%\begin{center}
\includegraphics[scale=0.33]{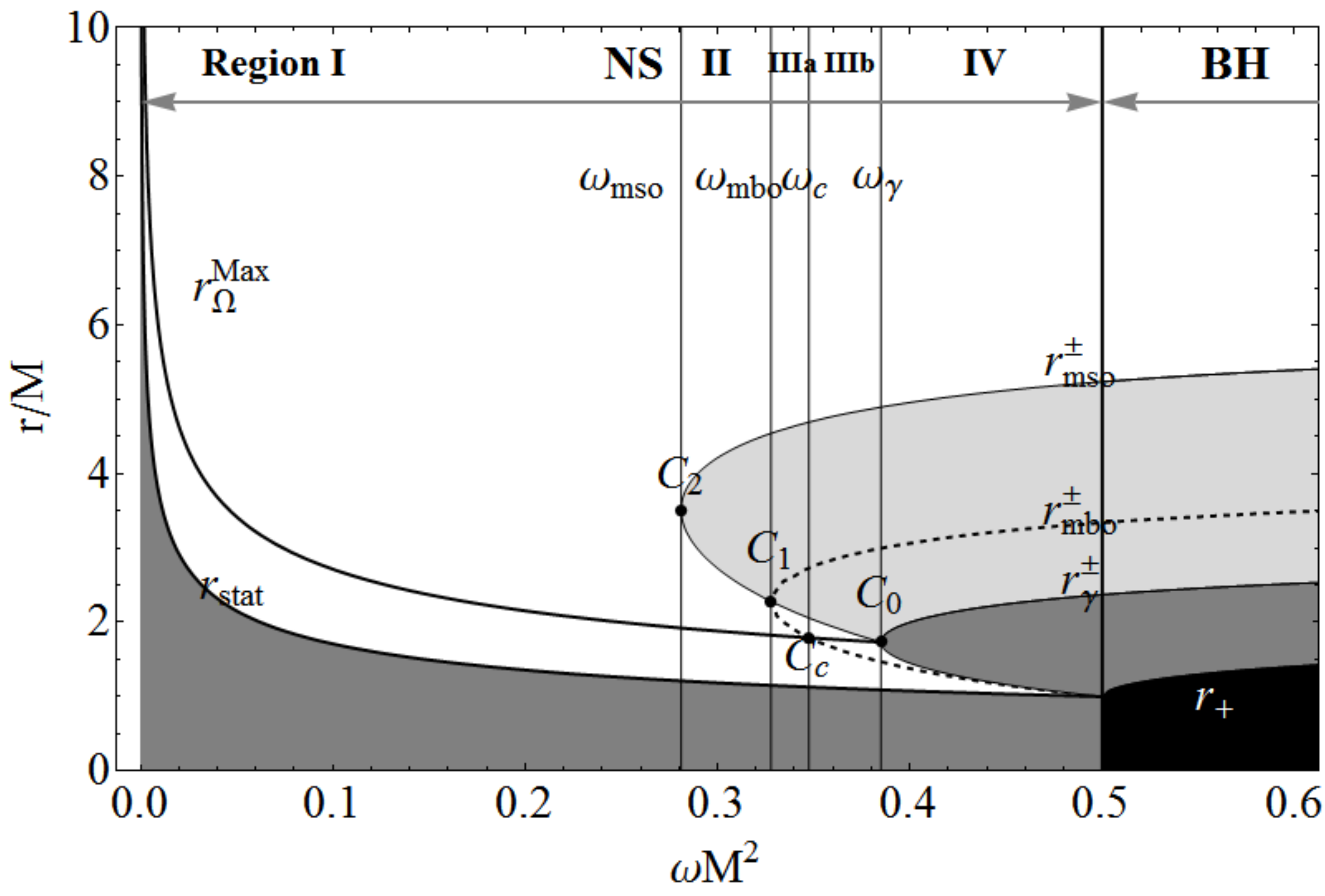}%{fig3}
%%CPlotoiscom
%\includegraphics[width=.61\textwidth]{stright.pdf}
\includegraphics[scale=.3]{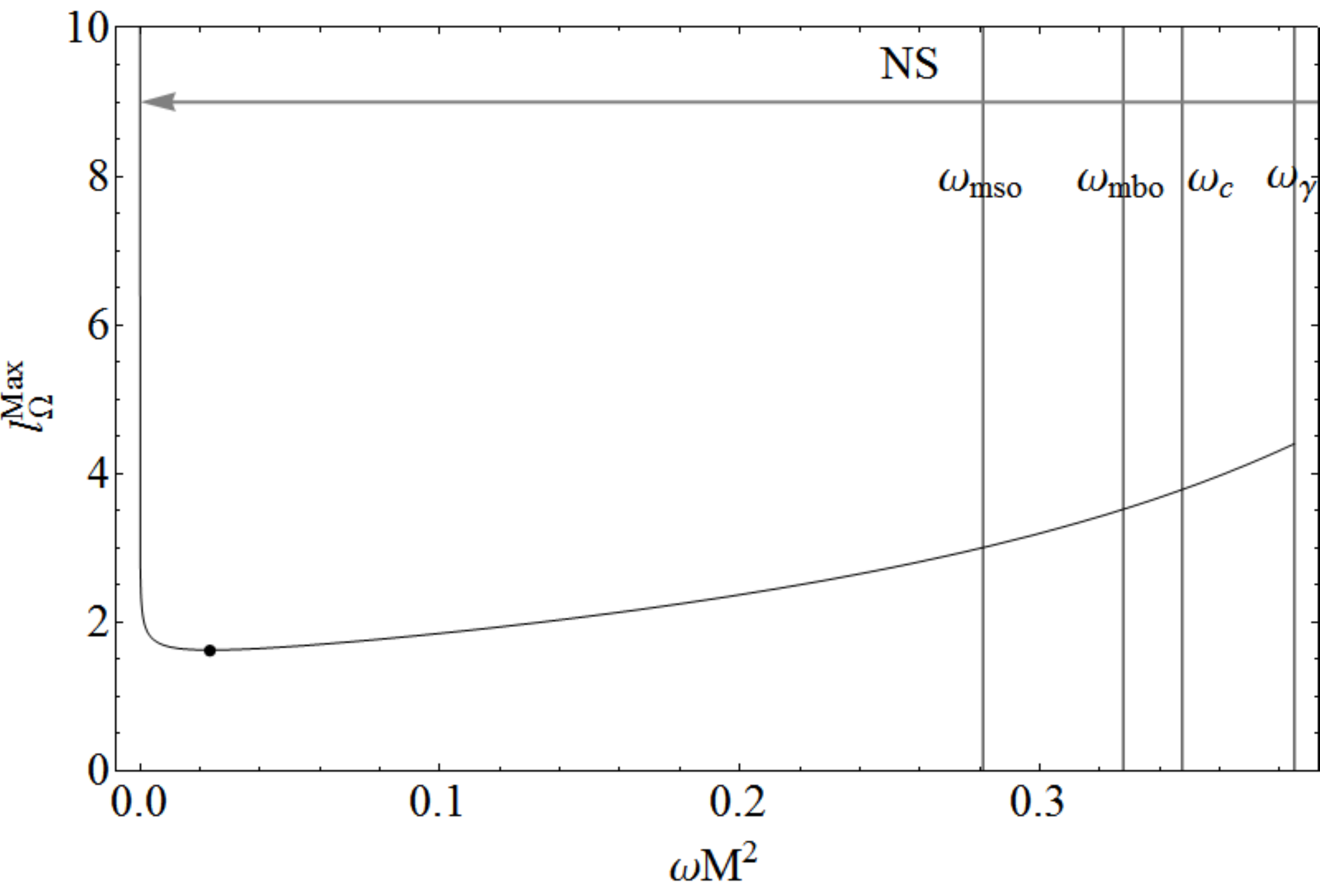}
\caption{Left panel: Regions of the stable and unstable circular geodetical orbits are given in dependence on the Ho\v{r}ava parameter $\omega$ along with the critical radii of the geodesic motion. Naked singularity  (NS) case ($\omega M^2\in]0,0,5[$) and black hole case (BH) ($\omega M^2>0,5$). Radii of  the set $\mathfrak{R}\equiv\{r_{mso}^{\pm}, r_{mbo}^{\pm},r_{\gamma}^{\pm}, r_{\Omega}^{Max},r_{stat}\}$ as function of the Ho\v{r}ava parameter $\omega$ in the plane $(r/M,\omega M^2)$. Black region is $r<r_+$, where $r_+$ is the black hole  horizon, gray regions  are forbidden, no circular motion is possible, regions of unstable circular orbits are in light-gray. White regions mark the stability regions. The marginally bounded orbits are $r_{mbo}^+>r_{mbo}^-$,   and  $r_{mso}^+>r_{mso}^-$ are marginally stable circular orbits, $r_{\gamma}^+>r_{\gamma}^-$  are limiting  circular orbits (photon-like orbits). The radius is in units of mass $M$ as it is  $r/M \rightarrow r$ and  also $\omega M^2\rightarrow \omega$.
The static radius $r_{stat}$  and the radius $r_{\Omega}^{Max}$ of the local maximum of the angular velocity are also plotted. The standard MRI viscosity mechanism can be at work in the region of stability up to the radius $r_{\Omega}^{Max}$ where the angular velocity gradient governing the viscosity vanishes.   The critical points $C_0, C_1, C_2, C_3$ and $C_c$ are defined in the text. Regions \textbf{I-IV} of naked singularities  sources  are also signed. Right panel: the fluid angular momentum $l_{\Omega}^{Max}\equiv l(r_{\Omega}^{Max})$, evaluated on the critical point $r_{\Omega}^{Max}$ for the frequency $\Omega$, as function of the Ho\v{r}ava parameter $\omega$, the point on the curve signs a minimum of  $l_{\Omega}^{Max}$ respect to $\omega$, see Eq.\il(\ref{Eq:ang-lo}).}\label{Fig:complete}
%\end{center}
\end{figure}
The radial profile of the specific angular momentum demonstrates a zero point  ($L^2_{K}(r;\omega)=0$)  giving the static, or ``antigravity'', radius where a stable equilibrium position of the test particles occurs giving the lower limit for existence of circular orbits of test particles in the field of all K-S naked singularity spacetimes. The static radius is given by \cite{Vie-etal:2014:PHYSR4}
\begin{equation}
   r_{stat}(\omega) \equiv \frac{1}{(2\omega)^{1/3}} .
\end{equation}
No zero points of the specific energy radial profile exist.
 At the static radius, the specific energy of the Keplerian circular orbits reaches its minimum
\(
                    E_{K}(r=r_{stat},\omega) \equiv E_{stat} = 1 - (2\omega)^{1/3} .
\)
Since $L_{K}=0$ at the static radius, there is naturally also $\Omega_{K}(r_{stat},\omega)=0$ in all K-S naked singularity spacetimes. This implies that a turning point (maximum) of the radial profile of the Keplerian angular velocity has to exist. The condition ${d\Omega_{K}}/{dr}=0$ implies the location of the local maximum at
\begin{equation}
                   r_{\Omega}^{Max}(\omega) \equiv \left(\frac{2}{\omega}\right)^{1/3} = 4^{1/3} r_{stat}(\omega),
\end{equation}
i.e., close to the static radius. The function $r_{\Omega}^{Max}(\omega)$ terminates at $
                   r_{\Omega}^{Max}(\omega = \omega_{\gamma}) = \sqrt{3} = r_{\gamma}(\omega = \omega_{\gamma})
$.
The circular geodesics with zero gradient of the angular velocity exist only in the K-S naked singularity spacetimes with $\omega < \omega_{\gamma}$. In the spacetimes with $\omega_{h} > \omega > \omega_{\gamma}$, the radius is located in the region forbidden for the circular geodesics -- the angular velocity increases with decreasing radius in the external region of the circular geodesics located above the $r_{\gamma}^+$, and it decreases with decreasing radius in the internal region of the circular geodesics, located between $r_{stat}$ and $r_{\gamma}^-$ - see \cite{Stuchlik:2014yaa,Vie-etal:2014:PHYSR4}.
The marginally stable circular geodesics satisfy simultaneously the conditions
$
{dV_{eff}}/{dr}=0$ and ${d^{2}V_{eff}}/{dr^{2}}=0
$,
these lead to  the orbital angular momentum
\begin{equation}
L_{mso}^{2}(r;\omega)\equiv\frac{r^{2}\left[2-4r^{3}\omega(2-A(r))-r^{6}\omega^{2}(1-A(r))\right]}{30-3A(r)r(4+r^{3}\omega)+12r^{3}\omega}.\label{eq_L2circ2}
\end{equation}
The critical value of the Ho\v{r}ava parameter, $\omega_{mso}$, separating spacetimes admitting two marginally stable radii $r_{osco}$ and $r_{isco}$ and no marginally stable orbits reads
\(
              \omega_{mso} \equiv 0.281100\), as it is $r_{osco}\leq r_{isco}$ hereafter we adopt the notation $r_{mso}^-\equiv r_{osco}$ and  $r_{mso}^+\equiv r_{isco}$ so that $r_{mso}^-\leq r_{mso}^+$.
The marginally bound orbits, i.e., the unstable circular geodesics having the specific energy $E_{K} = 1$ are given by the condition
\(
                  E_{mbo} \equiv E_{K}(r_{mbo},L_{mbo}) = 1
\)
that governs both the radius $r_{mbo}$ and the angular momentum $L_{mbo}$ of the Keplerian orbit. Note that this condition determines both the unstable outer orbits with $E_{K}=1$ and $L_{K}=L_{mbo}$ and the stable inner orbits with the same energy $E_{K}=1$, but different angular momentum $L > L_{mbo}$, if they exist. Clearly, these orbits can exist only in the spacetimes allowing for existence of two stable circular geodesics for a given angular momentum, i.e. the spacetimes with $\omega > \omega_{mso}$.
There exists a critical value of the Ho\v{r}ava parameter
\(
              \omega_{mbo} \equiv 0.327764
\)
separating spacetimes where the marginally bound unstable circular orbits are allowed, and where they do not exist, since the inner (outermost) marginally stable orbits have $E_{mso}^- < 1$ in such spacetimes.
The marginally stable orbits radii $r_{mso}\in (r_{mso}^-$, $r_{mso}^+$), and the marginally bound orbit radii $r_{mbo}$ are given as functions of the parameter $\omega$ in Figure 3, along with the radii of the photon circular geodesics $r_{\gamma}$ ($r_{\gamma}^-, r_{\gamma}^+$), the static radius, $r_{stat}$, and the radius where the gradient of the angular velocity of the circular geodesics vanishes, $r_{\Omega}^{Max}$. The critical points where the curves $r_{\gamma}(\omega)$, $r_{mbo}(\omega)$ and $r_{mso}(\omega)$ have the local extremum are determined by the points $
C_i$ with $i\in\{0,..,3\}$   of the plane $(r-\omega)$:
\begin{eqnarray}\nonumber
                C_{0}& \equiv& (r_{\gamma}, \omega_{\gamma}) = (1.7321, 0.3849),\quad          C_{1}\equiv(r_{mbo}, \omega_{mbo}) = (2.2818, 0.327764),
                \\\label{C-SET}
                C_{2}& \equiv& (r_{mso}, \omega_{mso}) = (3.5140, 0.2811),\quad C_{3}\equiv(r_{\Omega}^{Max},\omega_{mso})=(1.921,0.2811).
\end{eqnarray}
The point $C_{3}$ determining the radius $r_{\Omega}^{Max}(\omega)$ for the critical value $\omega_{mso}$ is important for the Keplerian disc structure during the transition accross the critical point in the evolution of a K-S naked singularity due to the accreting matter.
Vanishing of the angular velocity gradient is relevant for Keplerian accretion in the K-S spacetimes with $\omega < \omega_{mso}$, since $r_{\Omega}^{Max}(\omega) < r_{mso}(\omega)$ in the spacetimes with $\omega > \omega_{mso}$. For  $\omega > 1/2$ corresponding to the K-S black holes, there are no circular orbits under the inner horizon, contrary to the case of the standard general relativistic RN or Kerr black holes. The points $C_0$ and $C_2$ (or $C_1$) give the classification of the K-S naked singularity spacetimes due to the properties of circular geodesics governing the Keplerian accretion discs. The points $C_0$, $C_1$ and $C_2$ also give the maximal extension of the inner Keplerian discs in the corresponding classes of the K-S naked singularity spacetimes -- for details see discussion in \cite{Stuchlik:2014yaa}. Here we shortly present the classification of the K-S spacetimes reflecting the properties of circular geodesics.
\subsection{On the Keplerian discs   and classification of the K-S spacetimes}\label{Sec:Classes-Keplerian}
In dependence on the (dimensionless) parameter $\omega M^2$, four classes of the K-S spacetimes were introduced in \cite{Stuchlik:2014yaa}. For each class of the K-S spacetimes, typical radial profiles of the specific energy, specific angular momentum, and angular velocity of the circular geodesics governing behavior of the Keplerian discs was presented in Figure 5 in \cite{Stuchlik:2014yaa}; the corresponding behavior of the effective potential is reflected in \cite{Vie-etal:2014:PHYSR4}. In all the classes of the K-S naked singularity spacetimes a sphere of matter can be created at the static radius. Such a sphere could be a final state of the accreted matter, if dissipative mechanisms (viscosity, gravitational radiation, etc.) could work at time scales long enough.
The four classes of the K-S spacetimes are given in dependence on the magnitude of the Ho\v{r}ava dimensionless parameter $\omega$ (using simplification of $M=1$).
\subsubsection{Class I - Naked singularities with $\omega < \omega_{mso}$}\label{Subsec:ClassI}
The K-S naked singularity spacetimes containing no photon circular orbits and no unstable circular orbits. Only stable circular orbits are allowed for the specific angular momentum from the interval $L^2 \in]0+ \infty[$, Fig.\il(\ref{Fig:complete}). Since all the circular geodesic orbits are stable, with energy and angular momentum decreasing with decreasing radius, we can consider the Keplerian discs in the whole range of $r\in]r_{stat}, +\infty[$. A critical point of the Keplerian discs in these K-S naked singularity spacetimes occurs at the radius $r_{\Omega}^{Max}$ where the angular-velocity gradient vanishes. At this radius and its vicinity, the standard viscosity mechanism based on the non-zero gradient of the angular velocity of the accreting matter ceases its relevance. It can be substituted there by the mechanism based on the gravitational radiation of the orbiting matter, as corresponding looses of energy and angular momentum can transmit the accreting matter to the region where the viscosity mechanism causing heat production and thermally radiating Keplerian discs could continue its work. A more realistic expectation is related to the possibility of cumulation of accreting matter near the radius $r_{\Omega}^{Max}$ where a fluid toroid with pressure gradients governing its shape and structure could naturally arise
\footnote{We can assume that the standard Keplerian accretion discs are limited by the condition $\frac{d\Omega_{K}}{dr} < 0$ that is required by the Magneto-Rotational Instability (MRI) viscosity mechanism \cite{Bal-Haw:1998:RevModPhys}. If MRI is the only mechanism generating the viscosity effects that enable accretion of matter in the Keplerian discs, their inner edge has to be located at the critical radius $r_{\Omega}^{Max}$  \cite{Vie-etal:2014:PHYSR4}. If some other viscosity mechanism related to the oppositely oriented gradient of the angular velocity (${d\Omega_{K}}/{dr} > 0$) works, the internal parts of the Keplerian discs could also heat up and radiate thermally.}
\subsubsection{Class II - Naked singularities with $ \omega\in]  \omega_{mso},\omega_{\gamma}[$}\label{Subsec:ClassII}
These K-S naked singularity spacetimes containing no photon circular orbits, but two marginally stable circular orbits at $r_{mso}^-$ and $r_{mso}^+$  Fig.\il(\ref{Fig:complete}). The stable circular orbits, potentially corresponding to Keplerian discs, are located in the region $r\in]r_{stat}, r_{mso}^-[$ (the inner Keplerian disc) and $r\in]r_{mso}^+ , \infty[$ (the outer Keplerian disc). For the marginally stable circular orbits, the energy and angular momentum satisfy the conditions $E_{mso}^+ < E_{mso}^-$ and $L_{mso}^+ < L_{mso}^-$. The unstable circular geodesics are located at $r\in]r_{mso}^-, r_{mso}^+[$, and their energy and angular momentum belong to the intervals $E\in]E_{mso}^+ , E_{mso}^-[$ and $L\in]L_{mso}^+, L_{mso}^-[$.
Since the critical radius $r_{\Omega}^{Max} < r_{mso}^-$, we can conclude that the complete outer Keplerian disc is quite regular relative to the MRI viscosity mechanism; the critical point of the vanishing angular-velocity gradient is related to the inner Keplerian disc only. Therefore, we can conclude that the accretion Keplerian discs work quite well down to the isco circular orbit and the complexities related to the vanishing of the angular-velocity gradient are relevant only in the inner Keplerian discs, if such discs will be created -- for details see \cite{Stuchlik:2014yaa}.
In the K-S naked singularity spacetimes allowing for existence of the unstable circular geodesics, the marginally bound (unstable) circular geodesics with specific energy given by the condition $E_{mbo} = 1$ play a crucial role. Such orbits can be approached approximately by particles starting at rest at infinity with specific energy $E=1$ and angular momentum $L \sim L_{mbo}$ and are significant in theory of toroidal configurations of perfect fluid orbiting black holes or naked singularities, giving the upper limit on the existence of marginally stable toroidal configurations \cite{Koz-Abr-Jar:1978:ASTRA,Stu-Sla-Hle:2000:ASTRA,Stu:2005:MPLA}. The marginally bound orbits can exist only in the K-S naked singularity spacetimes having the Ho\v{r}ava parameter
\(
                 \omega \in] \omega_{mbo}, \omega_h [\).
In the K-S spacetimes with $\omega < \omega_{b}$ no unstable circular orbits with $E>1$ exist - then even particles following unstable circular orbits can remain on a bounded orbit, if perturbed radially, and the equilibrium toroidal configurations of perfect fluid cannot extend up to infinity in such spacetimes \cite{Koz-Abr-Jar:1978:ASTRA,Stu-Sla-Hle:2000:ASTRA} and as we shall discuss below.
In the inner Keplerian disc, the critical point of the vanishing angular velocity gradient and its relation to the standard MRI viscosity mechanism is relevant, influencing thus the properties of the disc along the lines discussed in the previous case of the K-S naked singularity spacetimes. The creation of some toroidal structures can be thus relevant also for this reason.
\subsubsection{Class III - Naked singularities with $\omega\in]\omega_h ,\omega_{\gamma}[$}\label{Sub:ClassIII}
The K-S naked singularity spacetimes containing two photon circular orbits, the inner stable one at $r_{\gamma}^-$, the outer unstable one at $r_{\gamma}^+>r_{\gamma}^-$, and one marginally stable orbit at $r_{mso}^+ > r_{\gamma}^+$. The stable circular orbits are located at $r\in]r_{stat}, r_{\gamma}^-[$, and at $r > r_{mso}^+$. Unstable circular orbits are located at $r\in]r_{\gamma}^+ , r_{mso}^+[$. At the region $r\in]r_{\gamma}^-,r_{\gamma}^+[$, no circular geodesics are allowed. There is no critical point related to the vanishing of the angular-velocity gradient. The angular velocity increases (decreases) with decreasing radius at the outer (inner) Keplerian disc.
In the outer Keplerian disc the viscosity MRI mechanism can work quite well and we can expect the standard Keplerian thermally radiating disc in the whole region above $r_{mso}^+$. In the inner Keplerian disc some other viscosity mechanism could be working, however, its structure is quite non-standard, as the energy and angular momentum diverge and change extremely fastly near the stable  photon circular orbit, especially in the case of near-extreme K-S naked singularity spacetimes. Creation of such a Keplerian structure by natural astrophysical processes seems to be very improbable -- for details see \cite{Stuchlik:2014yaa}.
\subsubsection{Class IV - Black holes with $\omega > \omega_h$}\label{Sub: ClassIV}
The K-S black hole spacetimes have the circular geodesics above the outer horizon only. The inner limit is given by the photon circular orbit $r_{\gamma}$. The unstable circular orbits are located between the photon circular orbit and the marginally stable orbit, $r_{mso}^+$, above which the stable circular orbits are located, similarly to the case of the Schwarzschild spacetimes. Of course, the standard Keplerian thermally radiating disc with the MRI viscosity mechanism extends in whole the region above the isco.
\section{Structure of perfect fluid tori}\label{Sec:strucut}
The structure and shape of the so called test perfect fluid tori has been extensively studied for a variety of general relativistic spacetimes \cite{Koz-Abr-Jar:1978:ASTRA,Abr-Cal-Nob:1980:APJ,Stu-Sla-Hle:2000:ASTRA,Sla-Stu:2005:CLAQG,Stu:2005:MPLA,Stu-Sla-Kov:2009:CLAQG,Kuc-Sla-Stu:2011:JCAP,
Adamek:2013dza,Pug-Mon-Bern:2012:MONRAS}. Here, the toroidal structures will be investigated in the K-S naked singularity spacetimes of the modified Ho\v{r}ava quantum gravity, but the standard general relativistic methods of treating the test perfect fluid configurations will be applied, as they correspond to the IR end of the Ho\v{r}ava gravity.

The perfect fluid energy-momentum tensor components $T_{\mu\nu}$ relative to coordinate basis read
\begin{equation}
	T_{\mu\nu}=(p+\rho)U_\mu U_\nu - p g_{\mu\nu}
\end{equation}
where $p$ ($\rho$) is the perfect fluid pressure (energy density) and $g_{\mu\nu}$ are components of the metric tensor. The elements of the perfect fluid follow circular trajectories, i.e., their four-velocity reads
\(
	U^{\mu}=\left(U^t,0,0,U^\phi\right)
\).
The Euler equation governing the structure and the shape of the perfect fluid configurations in a given spacetime can be cast in the form
\begin{equation}
	\frac{\nabla_\mu p}{p+\rho}=-\nabla_\mu\ln(U_t)+\frac{\Omega \nabla_\mu l}{1-\Omega l} \label{euler_eq}
\end{equation}
where $\Omega$ ($l$) is the angular velocity (specific angular momentum) of the fluid element, defined by the relations
\begin{equation}
	\Omega=\frac{U^\phi}{U^t},\, l=-\frac{U_\phi}{U_t}.
\end{equation}
For barotropic fluid assumed here, ($p=p(\rho)$), it follows from Eq. (\ref{euler_eq}) that there exists an invariant function $\Omega=\Omega(l)$ and surfaces of constant pressure are given by Boyer's condition \cite{Koz-Abr-Jar:1978:ASTRA}
\begin{equation}
	\int^p_0\frac{\diff p}{p+\rho}=W(p)-W(0)=-\ln\frac{U_t}{(U_t)_{in}}+\int^l_{l_in}\frac{\Omega\diff l}{1-\Omega l}.
\end{equation}
To obtain a particular structure, we have to specify the functions $\Omega=\Omega(l)$ and $l=l(r,\theta)$. Usually, the simplest case of the marginally stable toroidal structures is studied, as it gives whole the substantial information on the structure of more complex toroidal configurations \cite{Koz-Abr-Jar:1978:ASTRA,Abr-Cal-Nob:1980:APJ}. For a marginally stable torus, the specific angular momentum of the fluid element remains constant accross the toroid, $l(r,\theta)=l_0=const$. The angular velocity of the fluid then reads
\begin{equation}
	\Omega=-\frac{g_{tt}}{g_{\phi\phi}}l_0=\frac{f(r)}{r^2\sin^2\theta}l_0.
\end{equation}
Under these assumptions, the function $W=W(r,\theta)$ takes the simple form
\begin{equation}
	W(r,\theta)=\frac{1}{2}\ln U_t^2=\frac{1}{2}\ln\left[\frac{f(r)r^2\sin^2\theta}{r^2\sin^2\theta-f(r)l_0^2}\right].
 \label{potential}
\end{equation}
where $U_t$ follows from normalization of the four-velocity $U^\mu$ %and reads
%\begin{equation}
Structure and shape of the perfect fluid equilibrium configurations is governed by the geodesic circular motion, as the pressure gradient vanishes at their centre that is governed by stable circular geodesics, while the cusps determining their edge where outflow of matter is possible from the equilibrium configurations are determined by unstable circular geodesics \cite{Koz-Abr-Jar:1978:ASTRA,Stu-Sla-Hle:2000:ASTRA}. For this reason, the specific angular momentum of the geodesic motion, i.e. angular momentum related to the energy,
\(
         l_K \equiv \frac{L_K}{E_K}
\)
plays an important role in constructing the equilibrium toroidal configurations. The shape of the equilibrium toroidal structures is given by the relation
\begin{equation}
         \frac{d\theta}{dr} = - \frac{\partial U_t/\partial r}{\partial U_t/\partial \theta}.
\end{equation}
\section{Equilibrium tori in the K-S naked singularity spacetimes}\label{Sec:Eq}
\begin{figure}[H]
\begin{center}
\begin{tabular}{ccc}
 \includegraphics[scale=0.4]{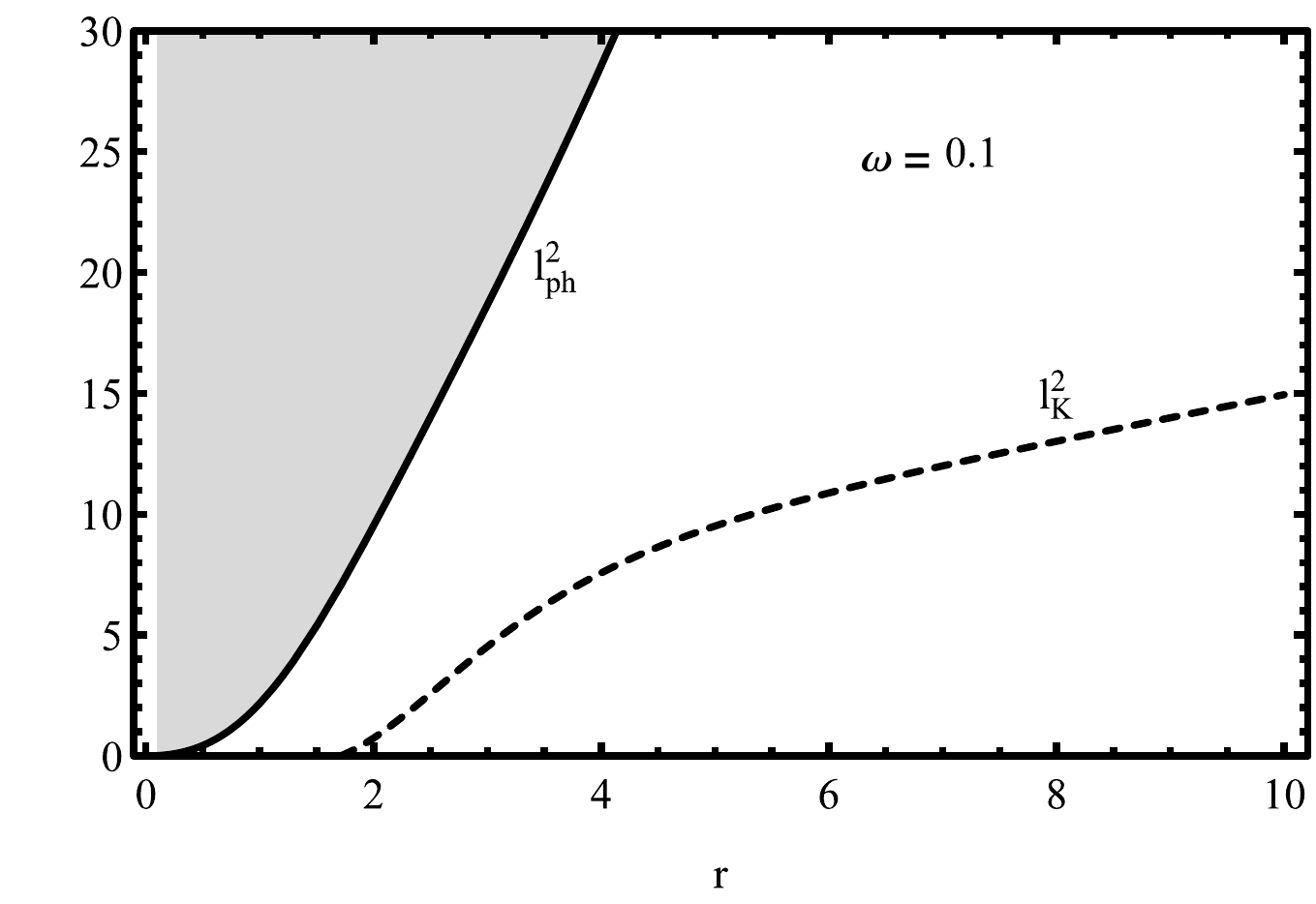}
 \includegraphics[scale=0.4]{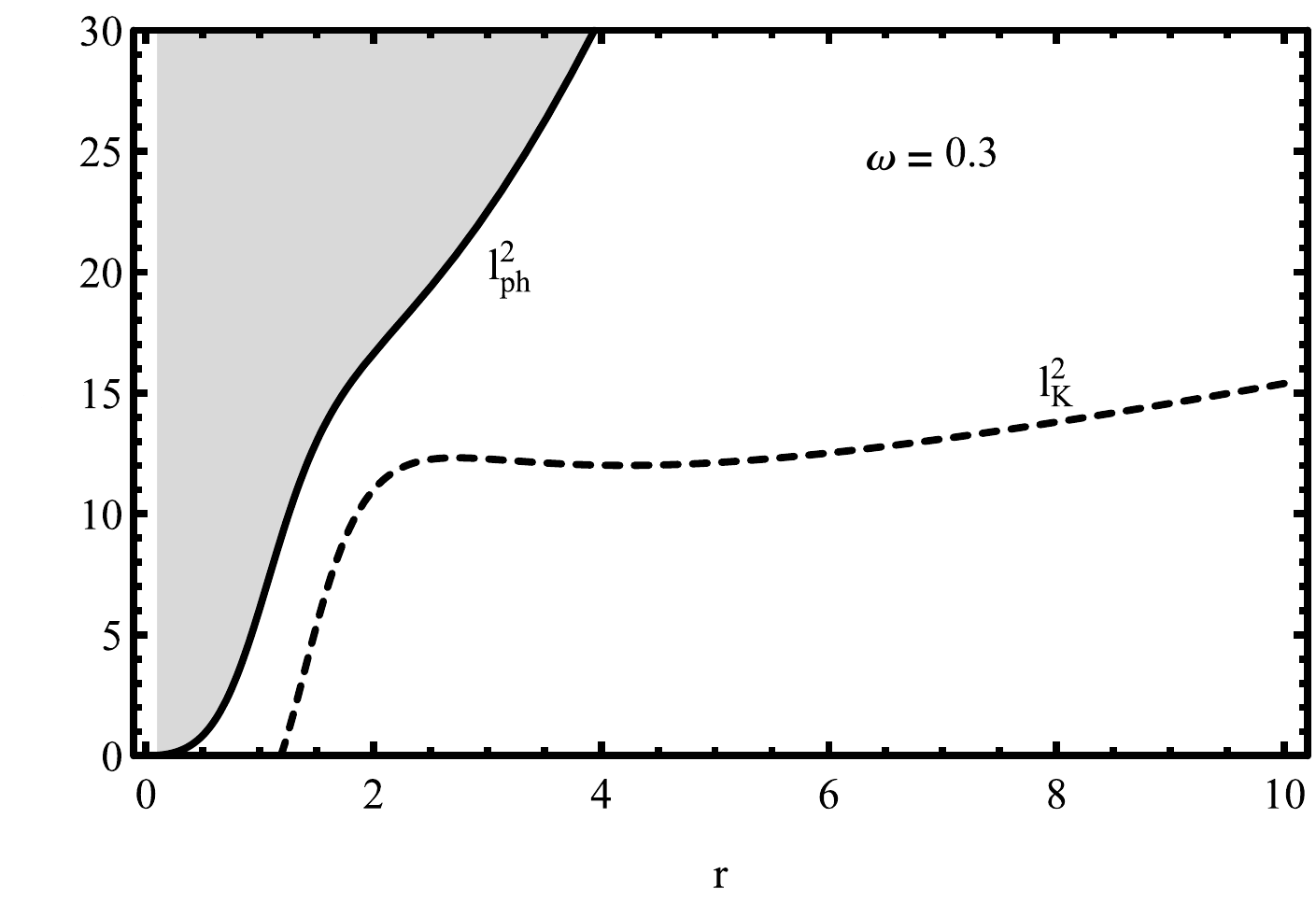}
	 \includegraphics[scale=0.45]{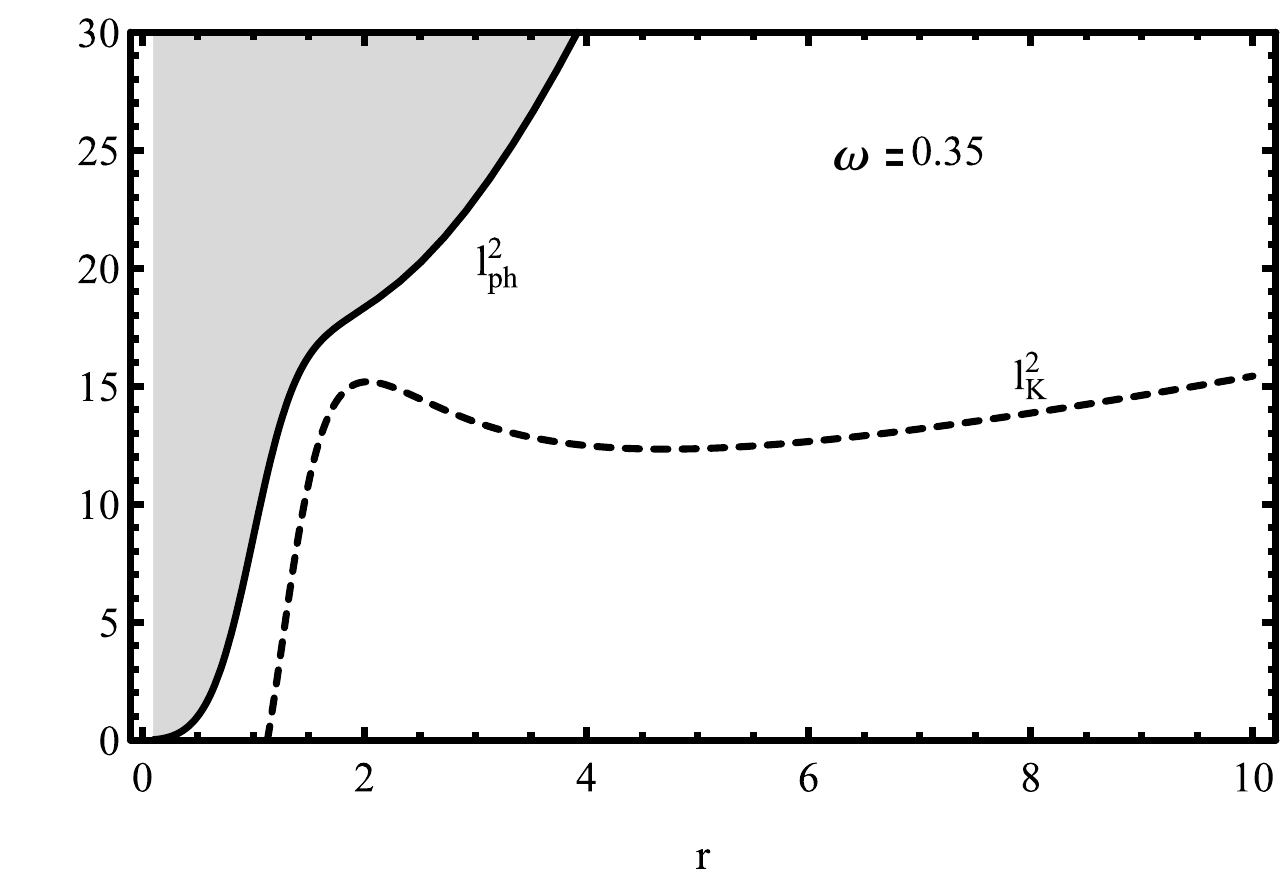}
	 \\
	 \includegraphics[scale=0.44]{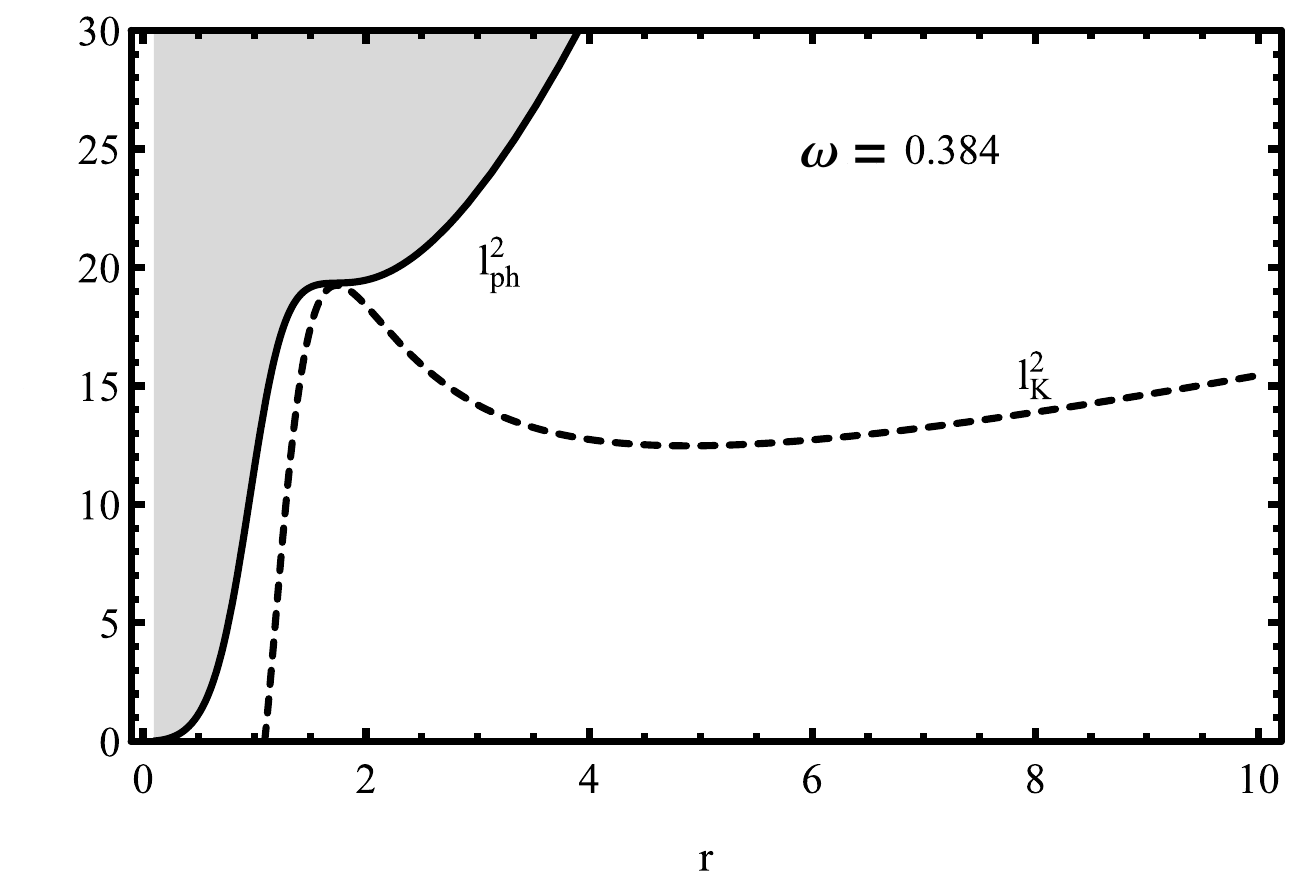}
	 \includegraphics[scale=0.44]{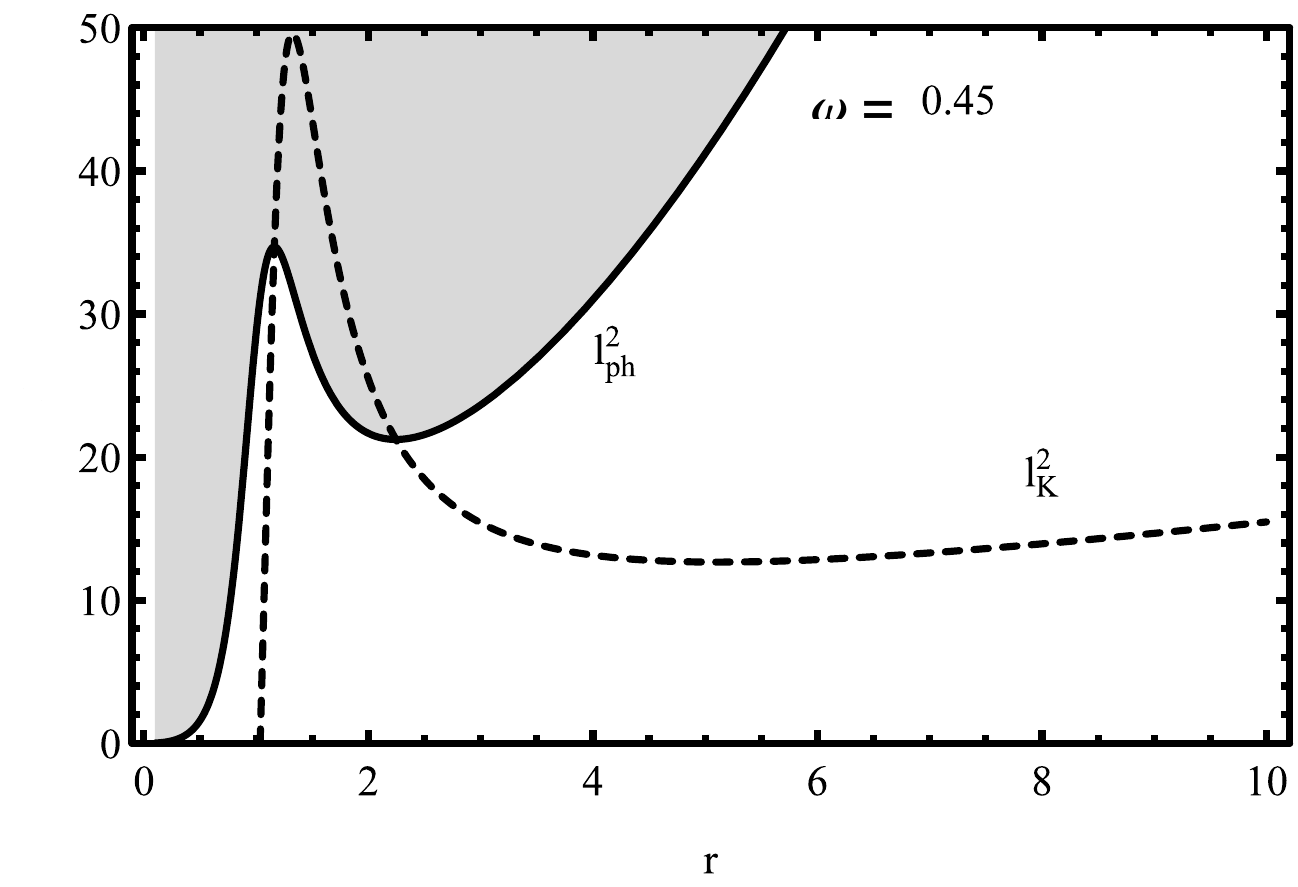}
\includegraphics[scale=0.44]{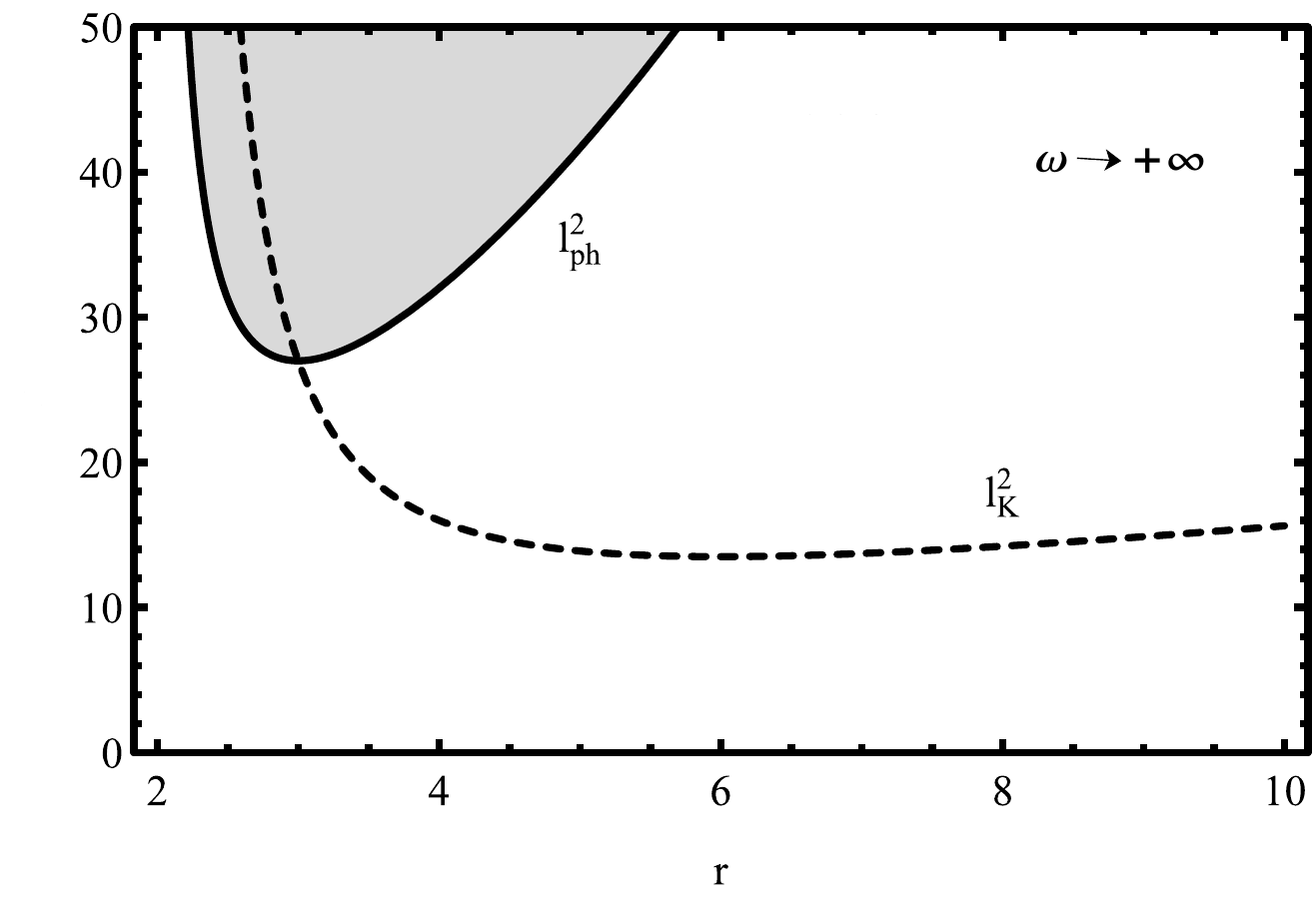}
\end{tabular}
\caption{Behavior of functions $l^2_{ph}$ and $l^2_{K}$, governing the equilibrium configurations with uniform distribution of the
specific angular momentum, in five qualitatively different situations compared with  the Schwarzschild black hole case for values of the parameter $\omega$ characterizing all the classes of the K-S spacetimes.  The radius is in units of mass $M$ as it is  $r/M \rightarrow r$ and  also $\omega M^2\rightarrow \omega$.}\label{Figs:lphlsch}
\end{center}
\end{figure}

In the K-S spacetimes, the potential $W$ in Eq.\il(\ref{potential}) governing the equilibrium tori with uniform distribution of the specific angular momentum reads
\begin{equation}
       W(r,\theta;l) = \frac{1}{2}\ln U_t^2=\frac{1}{2}\ln\frac{\left[1 + r^2 \omega\left(1 - \sqrt{1 + \frac{4}{\omega r^3}}\right)\right]r^2\sin^2\theta}{r^2\sin^2\theta-l^2 \left[1 + r^2 \omega\left(1 - \sqrt{1 + \frac{4}{\omega r^3}}\right)\right]}.
\end{equation}
Its behavior in the equatorial plane is crucial, as it determines all the relevant properties of the equilibrium tori. In the equatorial plane, two conditions have to be satisfied for reality of the potential $W$, namely
\begin{equation}
	f(r;\omega)=1 + r^2 \omega\left(1 - \sqrt{1 + \frac{4}{\omega r^3}}\right)\geq 0,\quad
	r^2-l^2(1 + r^2 \omega\left(1 - \sqrt{1 + \frac{4}{\omega r^3}}\right)\geq 0
\end{equation}
The first condition means that the tori could be located in the stationary regions of the spacetime, while the second condition is clearly related to the photon geodesic motion, implying the condition
\begin{equation}
      l^2 < {l_{ph}^{2}}\equiv \frac{r^{2}}{1+r^{2}\omega\left(1-\sqrt{1+\frac{4}{r^{3}\omega}}\right)}.
\end{equation}
The photon motion thus gives the reality condition of the equilibrium toroidal structures.
The local extrema of the potential $W(r,\theta=\pi/2)$ are given by the condition
\begin{equation}
   l^2 = l^{2}_{K}(r,\omega) \equiv \frac{L_{K}^{2}(r;\omega)}{E_{K}^{2}(r;\omega)}=\frac{r^{2}\left[r^{3}\omega A(r,\omega)-(1+r^{3}\omega)\right]}{rA(r,\omega)f^{2}(r,\omega)} , \label{eq_l2circ1}
\end{equation}
i.e., they are determined by the radial profile of the Keplerian specific angular momentum. Properties of the equilibrium tori are related to the properties of the circular geodesic motion. Therefore, classification of the K-S spacetimes according to the properties of the equilibrium tori have to be analogue to the classification according to the circular geodesics summarized in Sec.\il(\ref{Sec:Classes-Keplerian}). Of course, we have to distinguish also the K-S spacetimes allowing for existence of the marginally bound circular orbits with $E=1$. We give the relevant behavior of the functions $l^2_{K}(r,\omega)$ and $l^2_{ph}(r,\omega)$ in Figures\il(\ref{Figs:lphlsch}) for values of the parameter $\omega$ characterizing all the classes of the K-S spacetimes.
\section{Classification of toroidal configurations in K-S spacetimes}\label{Sec:class}
In the case of $l$=const tori, we can  classify  all the possible structures of equipotential
surfaces $W =$const, and, thus, possible toroidal configurations of perfect fluid, according
to the values of $l$. Therefore we  present behavior of the potential in the equatorial plane, i.e.
the function $W (r, \theta = \pi/2)$, and meridional sections through the structure of equipotential
surfaces.
We consider  the square of the particle  angular momentum $L^2$ and the fluid angular momentum $l^2$ as the the potentials  $V_{eff}(L)$ and $V_{eff}(l)$ are  function of $L^2$ or $l^2$ respectively, and  the dynamics is  independent from the orbital angular momentum sign.
We concentrate our attention on the set $\mathfrak{R}\equiv\{r_{mso}^{\pm}, r_{mbo}^{\pm},r_{\gamma}^{\pm}, r_{\Omega}^{Max},r_{stat}\}$ and the regions  with boundaries in  $\mathfrak{R}$,  the angular momenta $l_i\in\mathfrak{L}\equiv\{l_{mso}^{\pm}, l_{mbo}^{\pm},l_{\gamma}^{\pm}, l_{\Omega}^{Max},l_{stat}\}$ where $l_i\equiv l(r_i)$ and in general we adopt an analogue notation for any function $\mathbf{Q}(r):\;\mathbf{Q}_i\equiv\mathbf{Q}(r_i)$. %and the spacetime parameter $\omega$ is normalized in $\omega ^2$.
%
%\textbf{Introducing a further point}
%
It is convenient to introduce with   $
C_i$  defined  in Eq.\il(\ref{C-SET})
 the  couple:
\be
\mathbf{C}_c\equiv\{\omega_c, r_{c}\}, \quad \omega_cM^2\equiv\frac{1}{256} \left(37+30 \sqrt{3}\right)\approx0.3475059,\quad r_c\equiv\frac{8}{11} \left(2 \sqrt{3}-1\right) M\approx1.79207M
\ee
 where
 $r_{mbo}^-(\omega_c)=r_{\Omega}^{Max}(\omega_c)=r_c$, and
$r_{stat}<r_{\Omega}^{Max}=r_{mbo}^-<r_{mso}^-<r_{mbo}^+<r_{mso}^+$, see Fig.\il(\ref{Fig:complete}).
We note however that the critical points of $\Omega(l)$ are related to the photon orbits for $\omega M^2 \geq{2}/{3 \sqrt{3}}$
as
\be\label{Eq:ang-lo}
\Omega=\frac{f(r)}{r^2 \sin^2\theta}l,\quad
l^{Max}_{\Omega}\equiv l(r_{\Omega}^{Max}) =\sqrt{\frac{2^{2/3} \left(\sqrt{3}-2\right)}{4\left[ \left(\sqrt{3}-1\right) \omega ^{2/3}+\ 2^{2/3} \left(\sqrt{3}-2\right) \omega\right]-2^{1/3} \omega ^{1/3} }},
\ee
the angular momentum $l^{Max}_{\Omega}$ has a critical point in $\omega M^2=\frac{1}{432} \left(5+3 \sqrt{3}\right)= 0.0236022$ see Fig.\il(\ref{Fig:complete}).
We shall consider separately the
naked singularity  case $(\omega M^2\in]0,1/2[)$ in Sec.\il(\ref{Sec:K-S-NS}) and the black hole  case  $(\omega M^2>1/2)$ in  Sec.\il(\ref{Sec:K-S-BH}). The extreme BH-case $(\omega M^2=1/2)$ is addressed in Sec.\il(\ref{Sec:K-S-ExBH}).
\subsection{Naked singularity case: $\omega M^2\in]0,0.5[$}\label{Sec:K-S-NS}
In the K-S naked singularity spacetimes it is convenient  to consider the analysis in four   regions of $\omega$ and correspondingly for  four classes of naked singularity sources, namely:
{\textbf{Region I}: $\omega\in]0,\omega_{mso}[$},  explored in Sec.\il(\ref{Sec:NS-RegionI}),    corresponds to the naked singularities in  Class III in Sec.\il(\ref{Subsec:ClassI}),
{\textbf{Region II}: $\omega\in]\omega_{mso},\omega_{mbo}[$} in Sec.\il(\ref{Sec:NSRegionII}),
{\textbf{Region III-a}: $\omega\in]\omega_{mbo},\omega_c[$} in Sec.\il(\ref{Sec:NSRegionIII}),
{\textbf{Region III-b}: $\omega\in]\omega_c,\omega_{\gamma}[$} in Sec.\il(\ref{Sec:NSRegionIIIb}). The set of geometries of \textbf{Region II-IIIb} corresponds  to the spacetimes  Class II  discussed in Sec.\il(\ref{Subsec:ClassII}).
 Finally {\textbf{Region IV}: $\omega\in]\omega_{\gamma},0.5[$}, considered in Sec.\il(\ref{Sec:NSregionIV}), corresponds to naked singularities of Class III discussed for the Keplerian orbits profile in Sec.\il(\ref{Sub:ClassIII}).
The particular source
 $\omega=\omega_{mso}$ is considered in Sec.\il(\ref{Sec:NS1par}),  $\omega=\omega_{mbo}$ in Sec.\il(\ref{Sec:NSpar2}),
$\omega=\omega_{c}$ in Sec.\il(\ref{NSparc}) and
$\omega=\omega_{\gamma}$ in Sec.\il(\ref{Sec:NSpar4}).
%-----------------------------------------
%
\subsubsection{Region I: $\omega\in]0,\omega_{mso}[$}\label{Sec:NS-RegionI}
\begin{figure}[h]
\includegraphics[width=.481\textwidth]{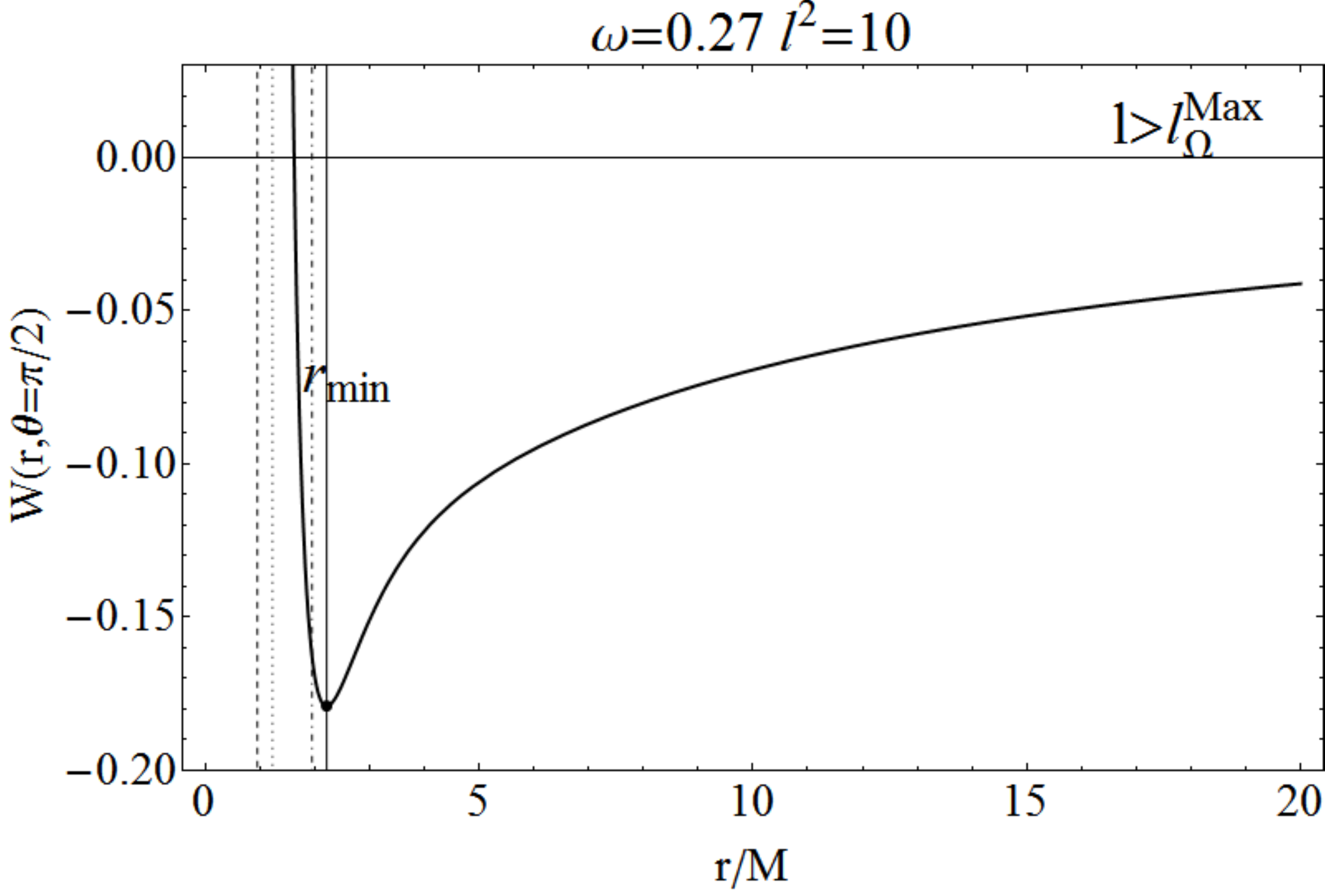}
\includegraphics[width=.481\textwidth]{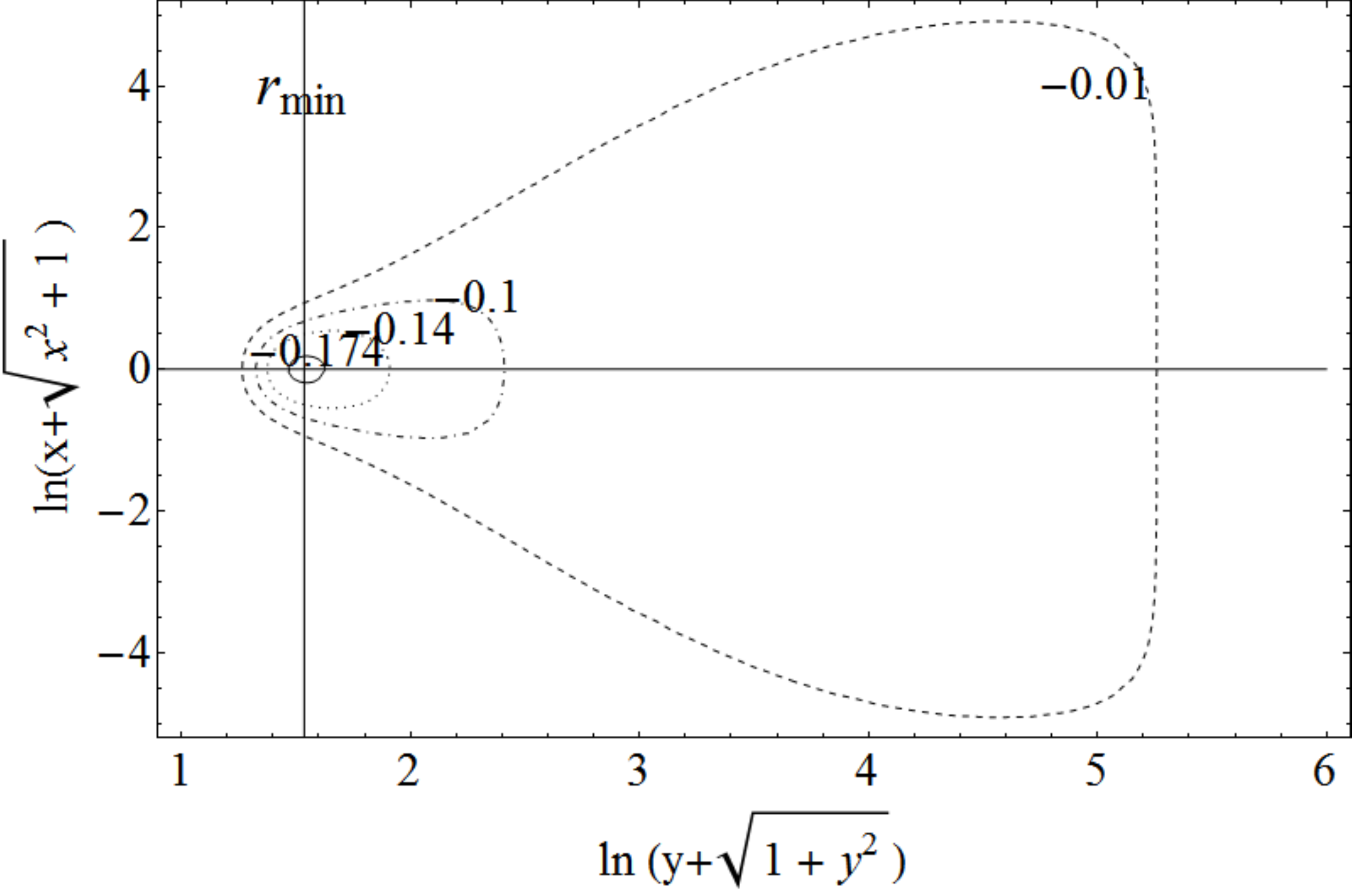}
%\\
\\
\includegraphics[width=.481\textwidth]{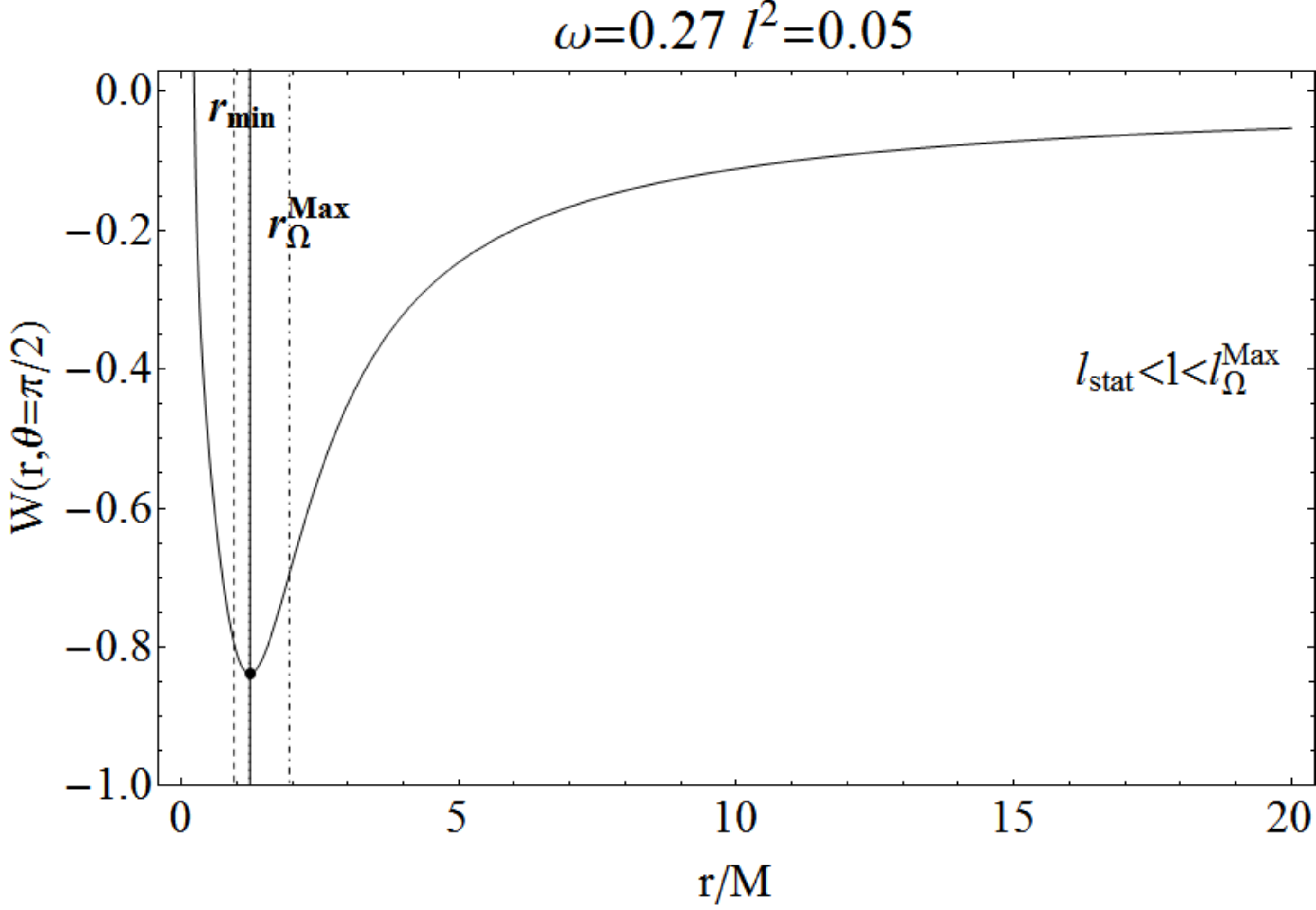}
\includegraphics[width=.481\textwidth]{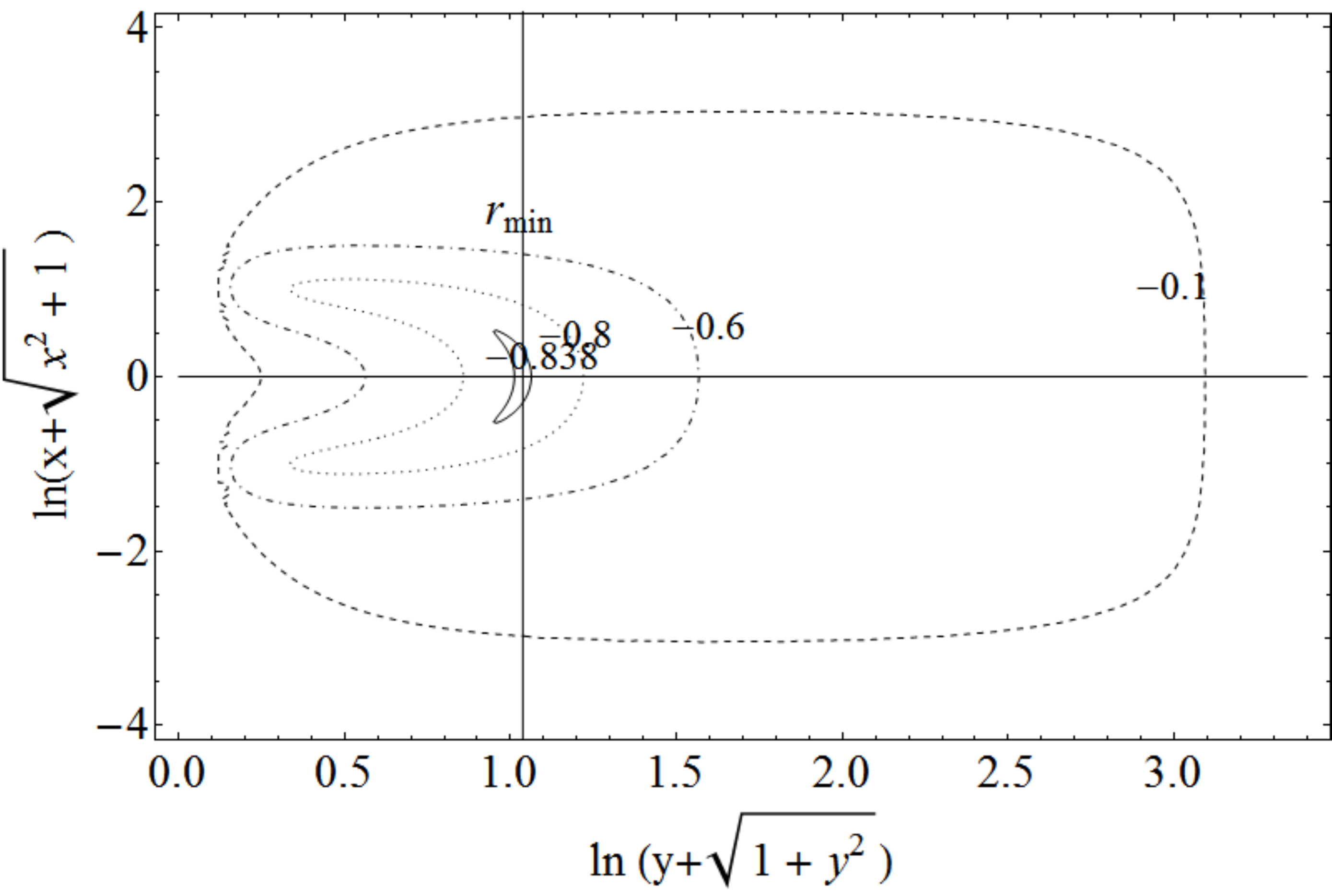}
\caption{Region I: $\omega\in]0,\omega_{mso}[$, naked singularity $\omega M^2=0.27$.
 It  is $l_{\Omega}^{Max}>l_{stat} $, and $r_{stat}<r_{\Omega}^{Max}$.  The radius is in units of mass $M$ as it is  $r/M \rightarrow r$ and   $\omega M^2\rightarrow \omega$. It is $r/M=\sqrt{x^2+y^2}$ where $(x,y)$ are Cartesian coordinates. Vertical lines in right panels set the $r_i\in\mathfrak{R}$ and  the effective potential critical points.}
\label{Fig:Sxxipdffirt}
\end{figure}

We focus of the naked singularity sources at $\omega\in]0,\omega_{mso}[$ where $\omega_{mso}M^2=0.2810998$ $r_{mso}^{\pm}(\omega_{mso})= 3.51399M$. The properties of the test particle circular motion have been detailed in Sec.\il(\ref{Subsec:ClassI}). There is then no last stable circular orbit,  or photon circular orbit and
no unstable circular orbit. In this region
it  is $l_{\Omega}^{Max}>l_{stat} $, and $r_{stat}<r_{\Omega}^{Max}$.
Only stable circular orbits for $V_{eff}(L)$ are allowed.  Since all the circular geodesic orbits are stable, it is possible to consider the Keplerian discs centered  in $r\in]r_{stat},+\infty[$. In  \textbf{Region I} no unstable modes  are possible.
We summarize the situation as follows:
\begin{description}
\item[-) $l>l_{\Omega}^{Max}$]
The radius $r_{\Omega}^{Max}$ locates the critical points of the   angular frequency $ \Omega(r, L)$. For fluid angular momentum $l>l_{\Omega}^{Max}$ there are always stable orbits  with $V_{eff}(l, r)<1$. This minimum of the effective potential locates the  disc center, see Fig.\il(\ref{Fig:Sxxipdffirt})-a.
\item[-) $l=l_{\Omega}^{Max}$]
In this particular case the center of the disc (maximum of the hydrostatic pressure) is located on $r_{\Omega}^{Max}$ and the inner edge of the disc is in the region $r<r_{\Omega}^{Max}$, the discs morphology is analogue to the situation in Fig.\il(\ref{Fig:Sxxipdffirt})-a for  $l>l_{\Omega}^{Max}$.
An  analogue situation happens for the following case:
 \item[-) ${l\in]l_{stat},l_{\Omega}^{Max}]}$] See  Fig.\il(\ref{Fig:Sxxipdffirt})-b. As $l$ decreases approaching  the limit  $l\approx0$, the closed surfaces thicken approaching to the sources.
\end{description}
%
%\clearpage
%
\subsubsection{Naked singularity: $\omega=\omega_{mso}$}\label{Sec:NS1par}
In this section we analyse  the  naked singularity  geometry with $\omega=\omega_{mso}$, separating geometries  with two marginally stable radii $r_{mso}^{\pm}$ (\textbf{Region II-IV}) and the class of naked singularity spacetime where  no marginally stable
orbits are (\textbf{Region I}). It is then   $r_{stat}<r_{\Omega}^{Max}<r_{mso}^-=r_{mso}^+$ and  $l_{mso}^-= l_{mso}^+>l_{\Omega}^{Max}>l_{stat} $.
\begin{figure}[h]
%%Plot0ms23
\includegraphics[width=.481\textwidth]{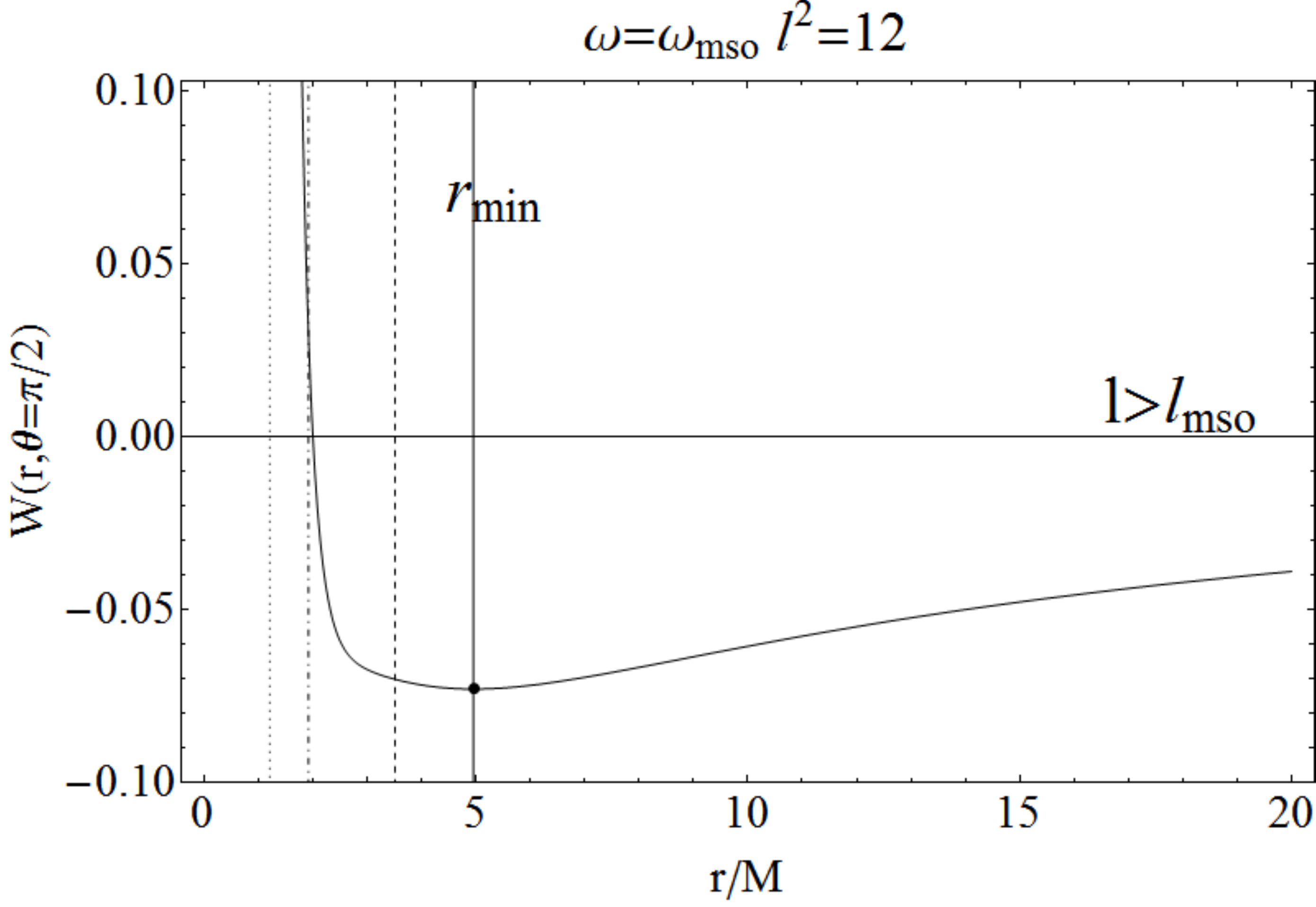}
\includegraphics[width=.481\textwidth]{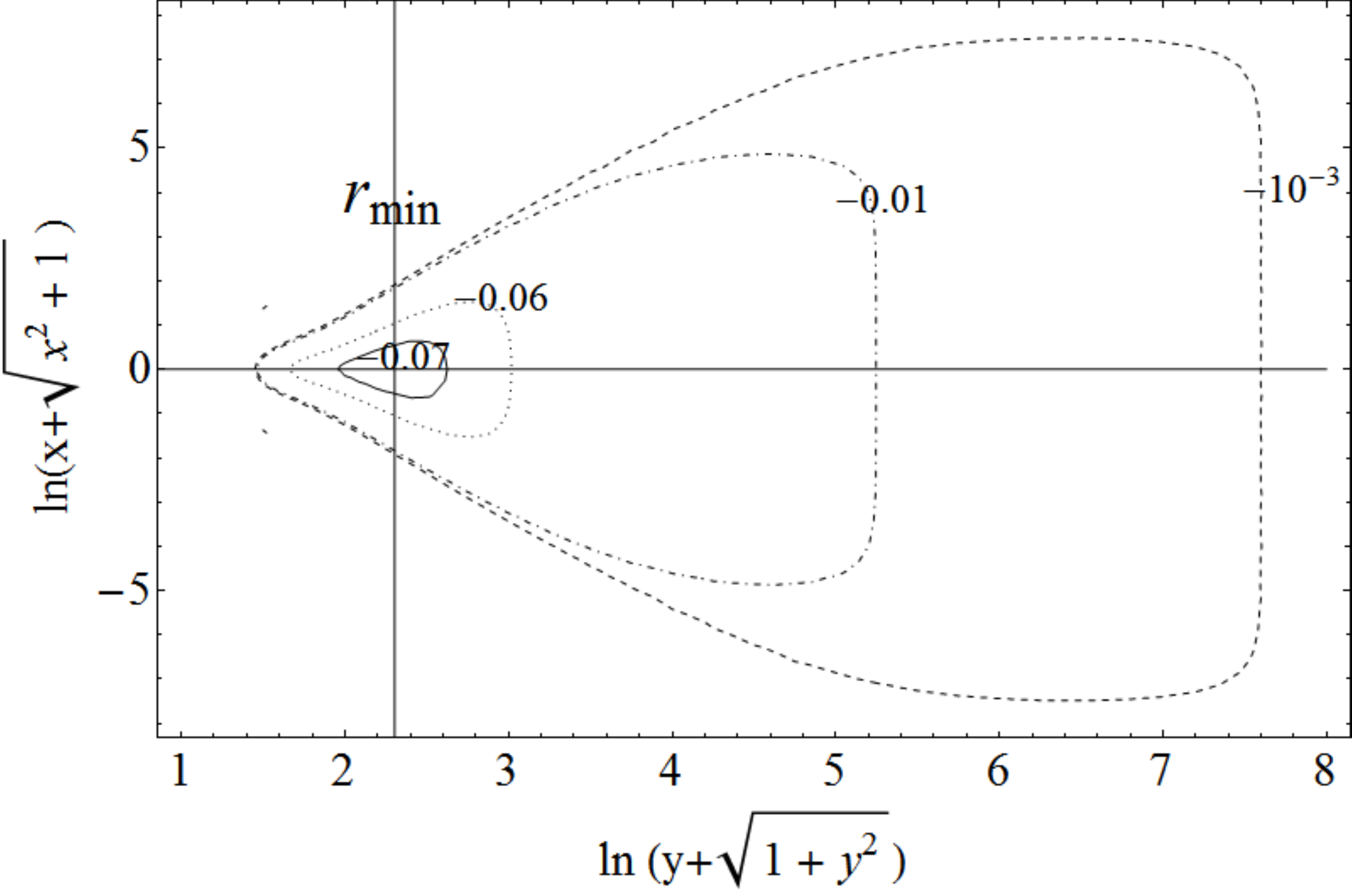}
%\\
\includegraphics[width=.481\textwidth]{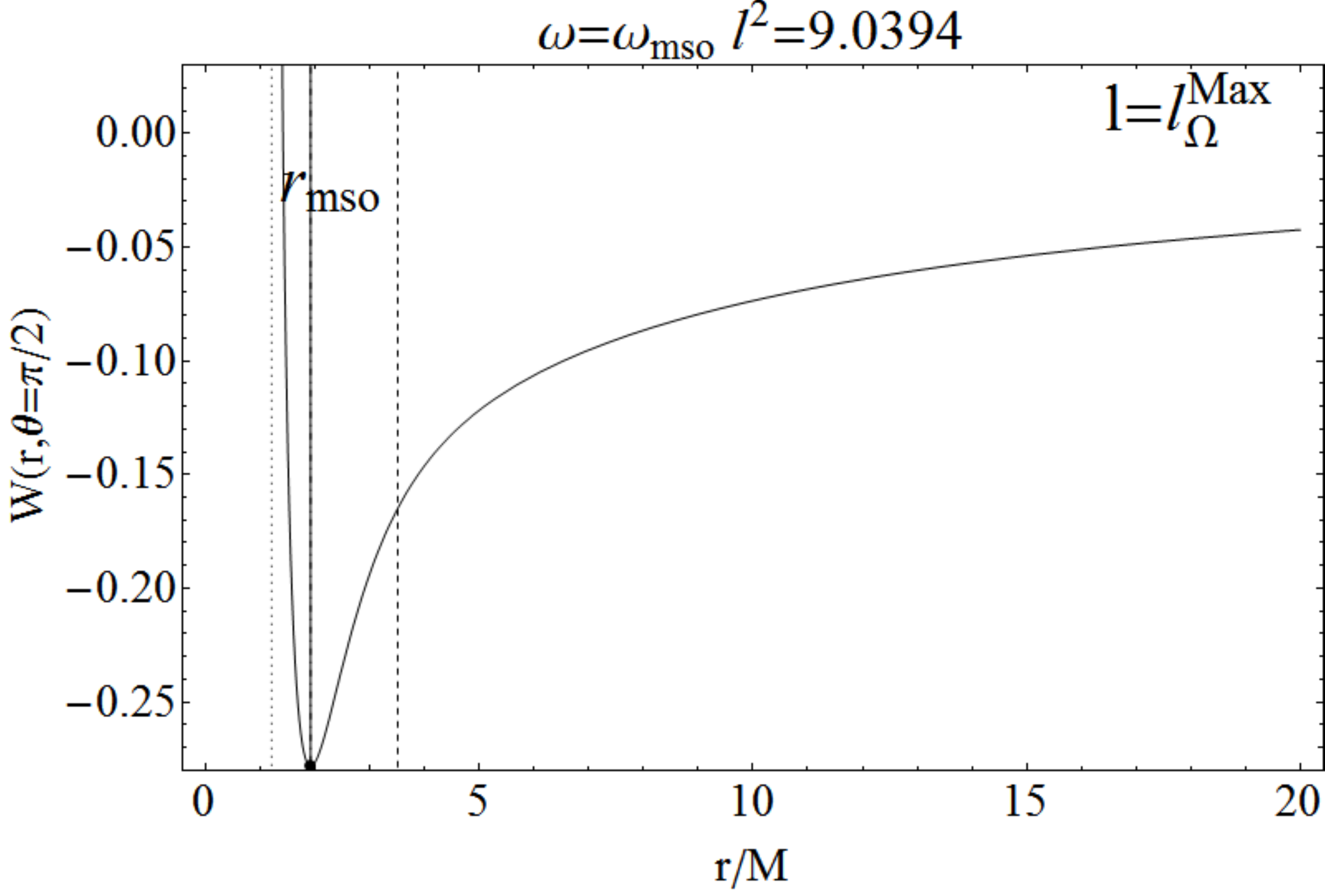}
\includegraphics[width=.481\textwidth]{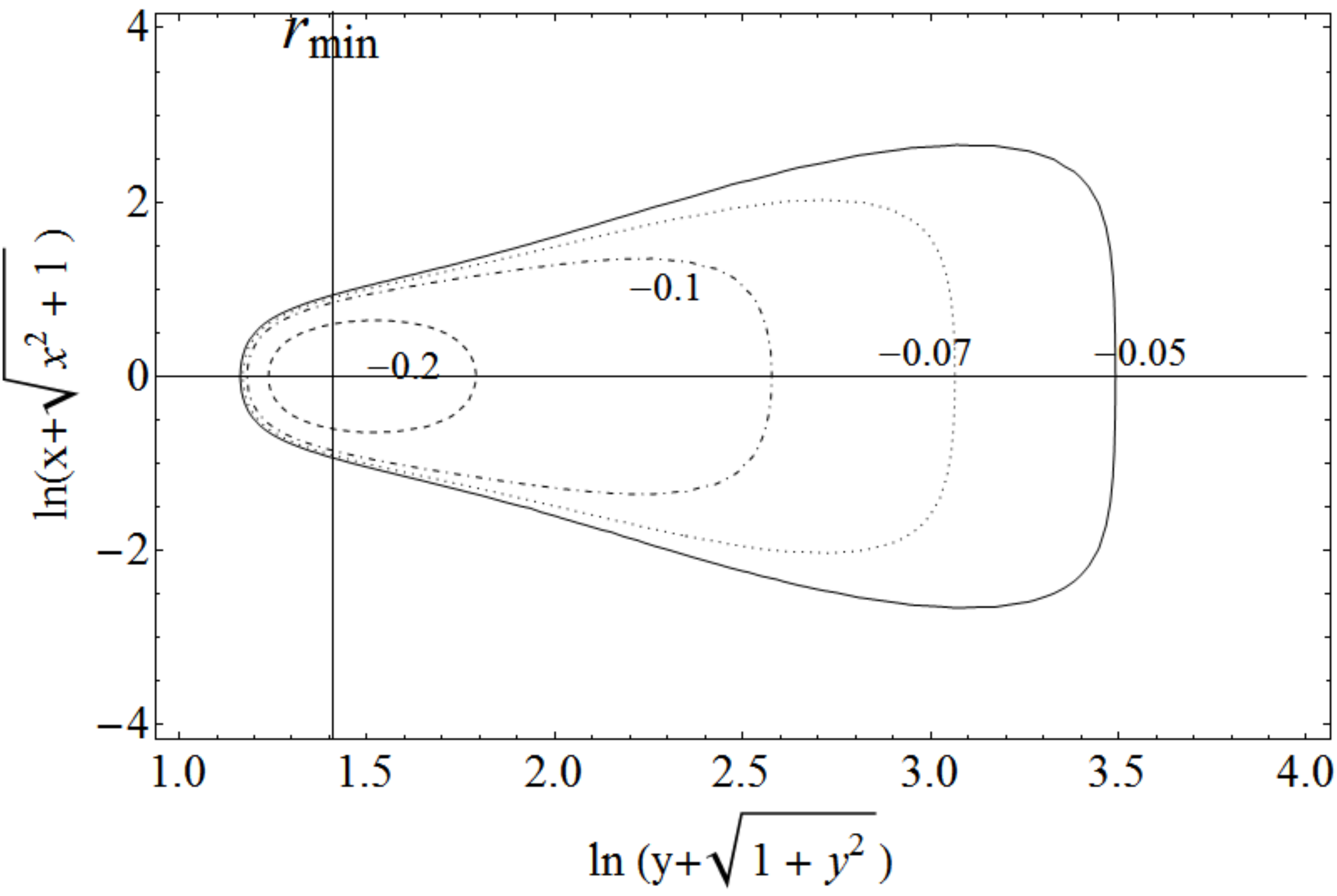}
%\\
\includegraphics[width=.481\textwidth]{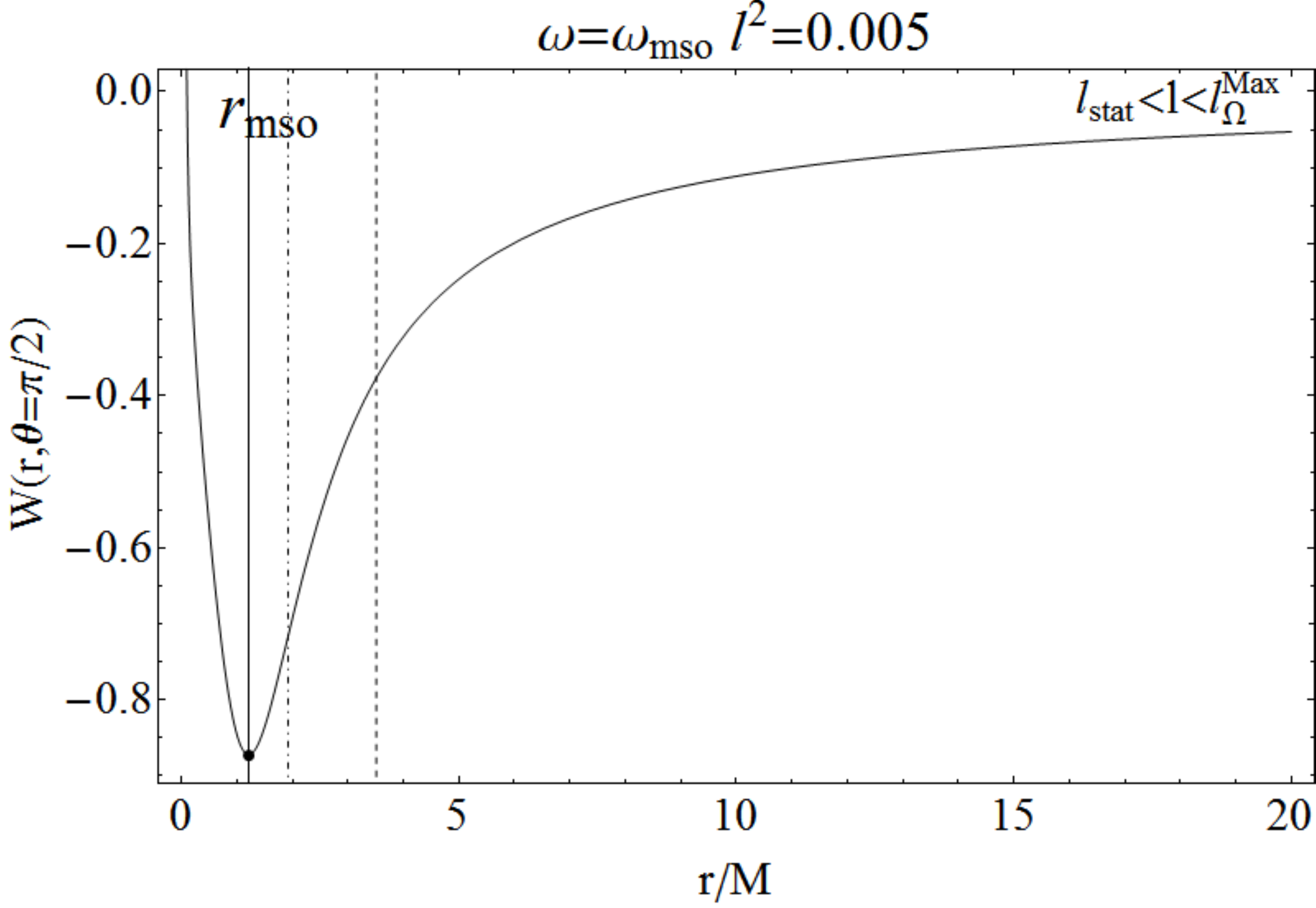}
\includegraphics[width=.481\textwidth]{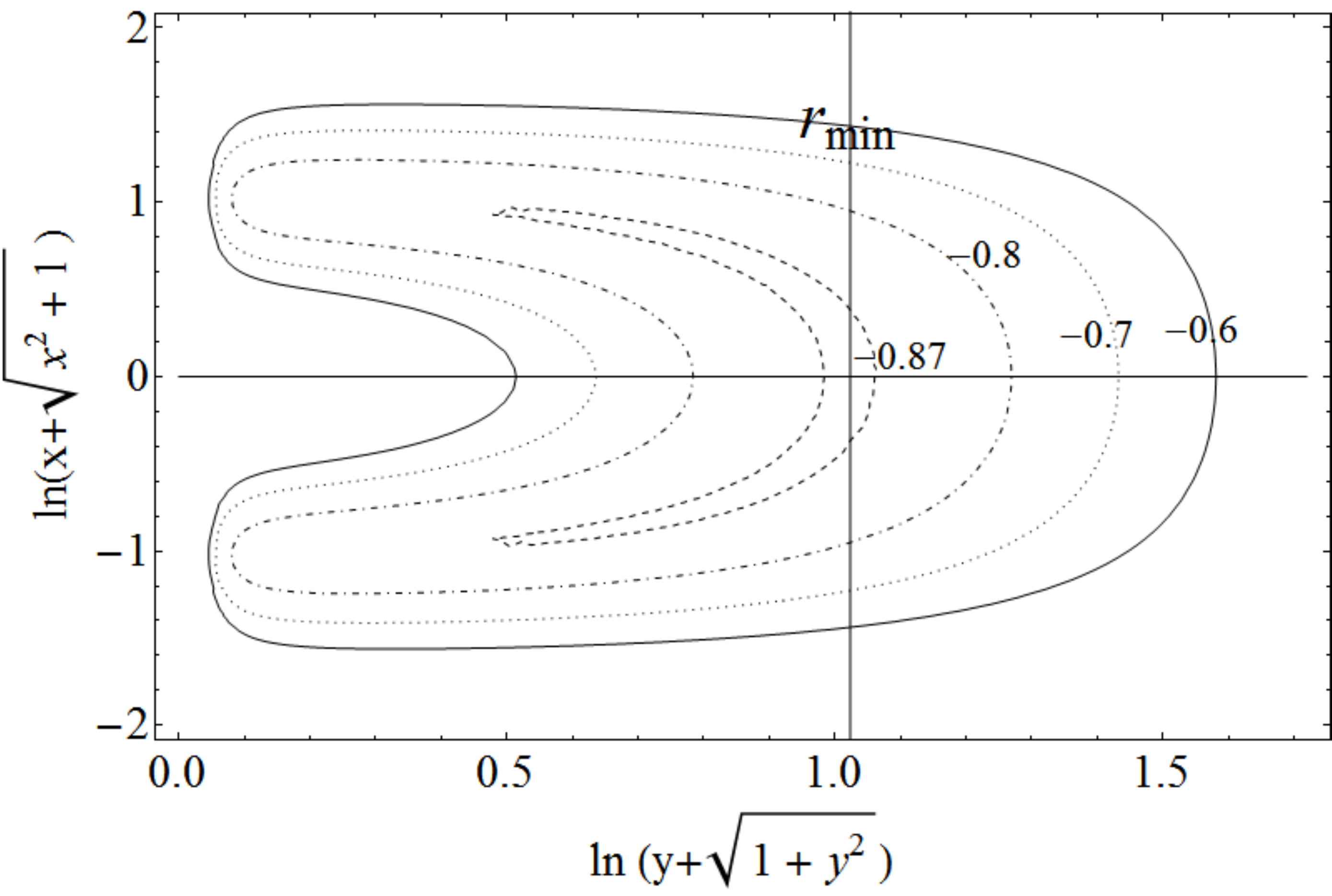}
\caption{Naked singularity $\omega=\omega_{mso}$.
 It  is $l_{mso}^-= l_{mso}^+>l_{\Omega}^{Max}>l_{stat} $, and $r_{stat}<r_{\Omega}^{Max}<r_{mso}^-=r_{mso}^+$.  Vertical lines in right panels set the $r_i\in\mathfrak{R}$ and  the effective potential critical points. It is  $r/M \rightarrow r$ and  also $\omega M^2\rightarrow \omega$, where $r/M=\sqrt{x^2+y^2}$ and  $(x,y)$ are Cartesian coordinates.}
\label{Fig:Sisecon1}
\end{figure}
\begin{description}
\item[-) $l>l_{mso}$] Only minimum points of the effective potential are possible and closed fluid surfaces centered on the minimum points Fig.\il(\ref{Fig:Sisecon1})-a.
\item[-) $l=l_{mso}$] The critical point of the effective potential is  located on $r_{mso}$, and for $(r=r_{mso},l=l_{mso})$ it is $V=0.925491$, $\partial_rV=0$ and  $\partial^{(2)}_rV=0$. The configuration for $l\approx l_{mso}$ is analogue to the case of  $l>l_{mso}$ in Fig.\il(\ref{Fig:Sisecon1})-a.
\item[-) ${l\in]l_{stat},l_{mso}[}$]  This case includes the ranges  $l\in ]l_{stat},l_{\Omega}^{Max}[ \cup [l_{\Omega}^{Max},l_{mso}[$. There are only  closed surfaces as in  Fig.\il(\ref{Fig:Sisecon1})-(b,c).
\end{description}
%%
%%
%\clearpage
\subsubsection{Region II: $\omega\in]\omega_{mso},\omega_{mbo}[$}\label{Sec:NSRegionII}
\begin{figure}[h]
%%CPlotoiscomPlotoocl14Plotoocl14
\includegraphics[width=.471\textwidth]{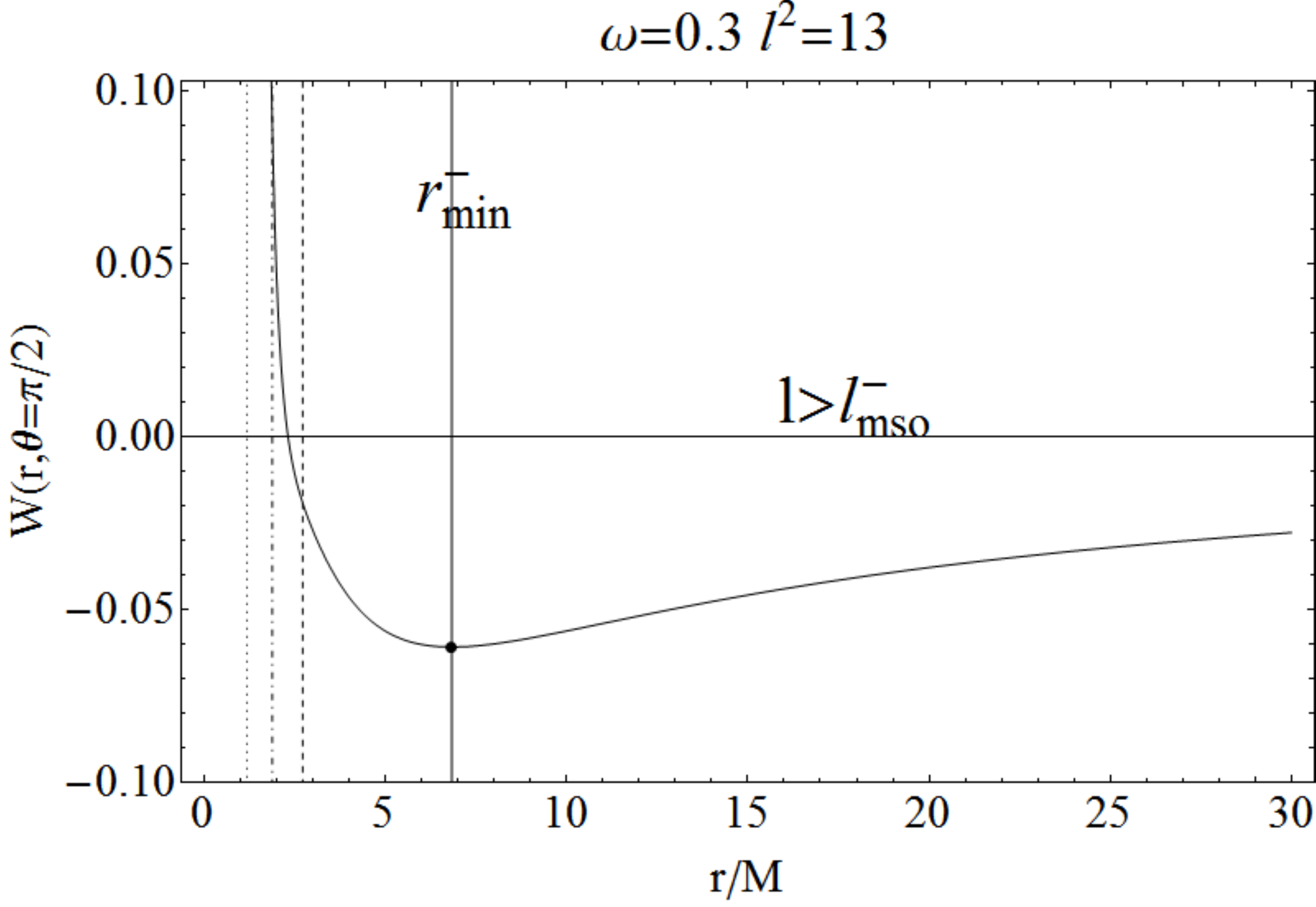}
\includegraphics[width=.41\textwidth]{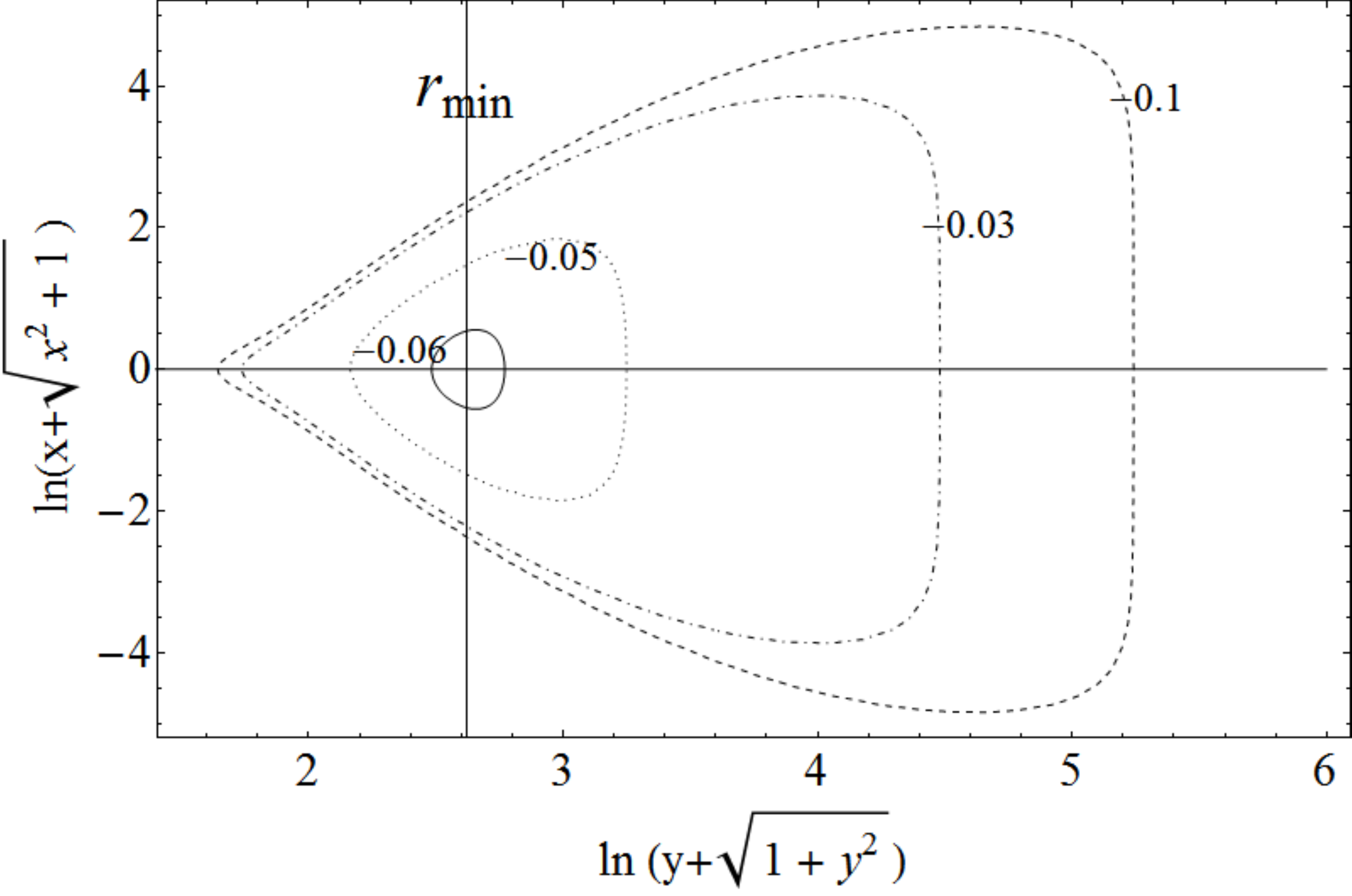}
\\
\includegraphics[width=.481\textwidth]{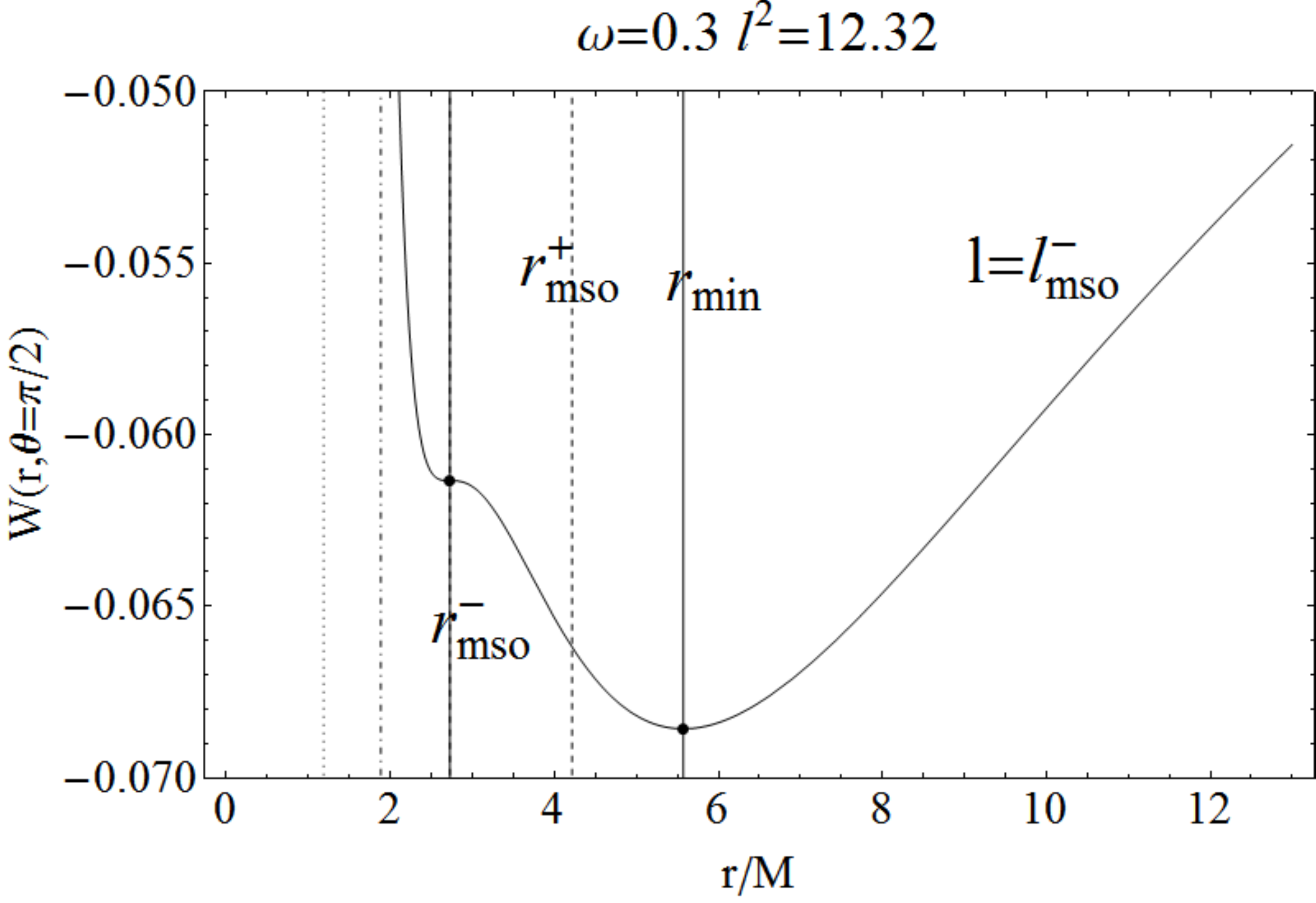}
\includegraphics[width=.31\textwidth]{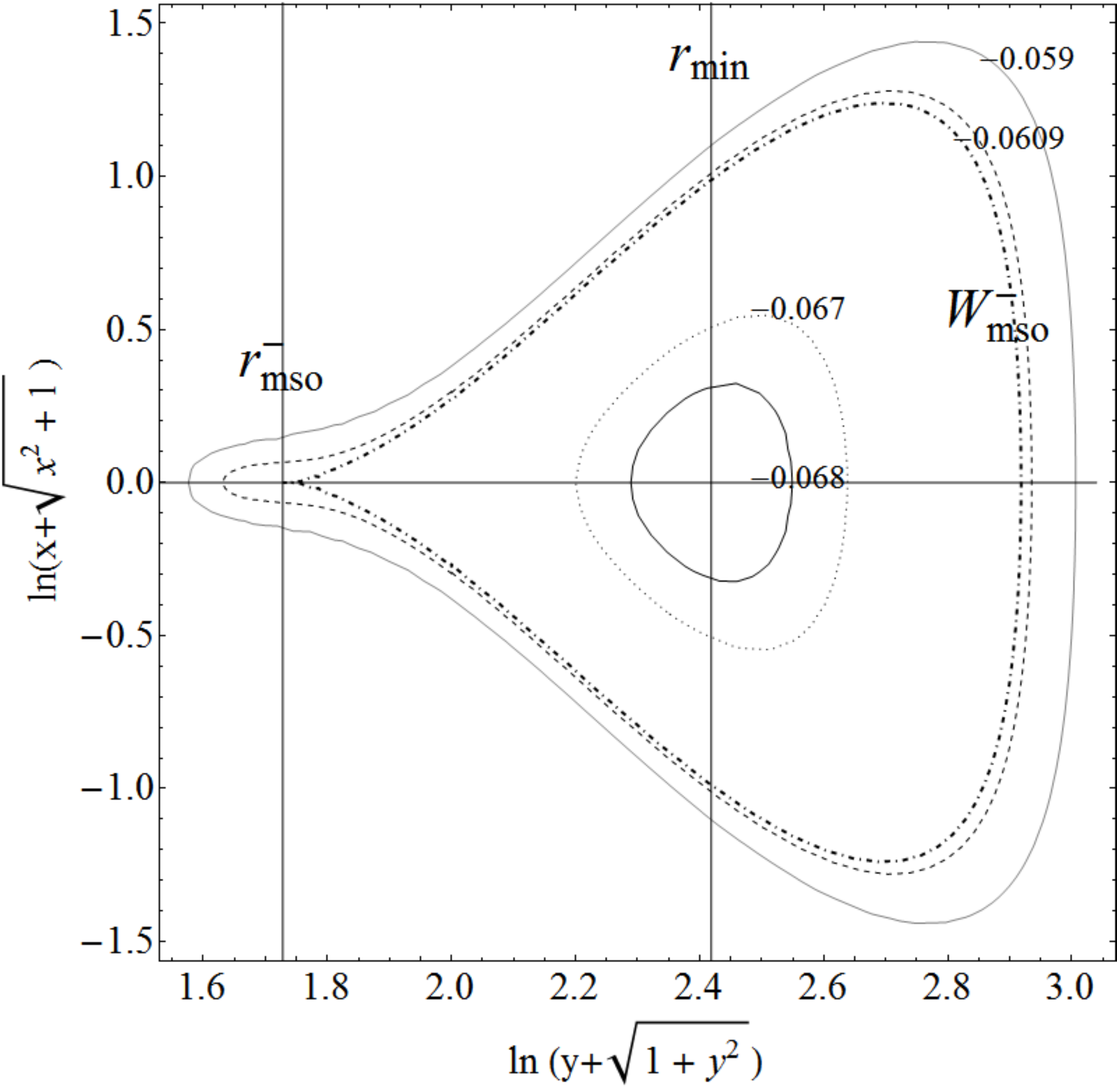}
\\
\includegraphics[width=.481\textwidth]{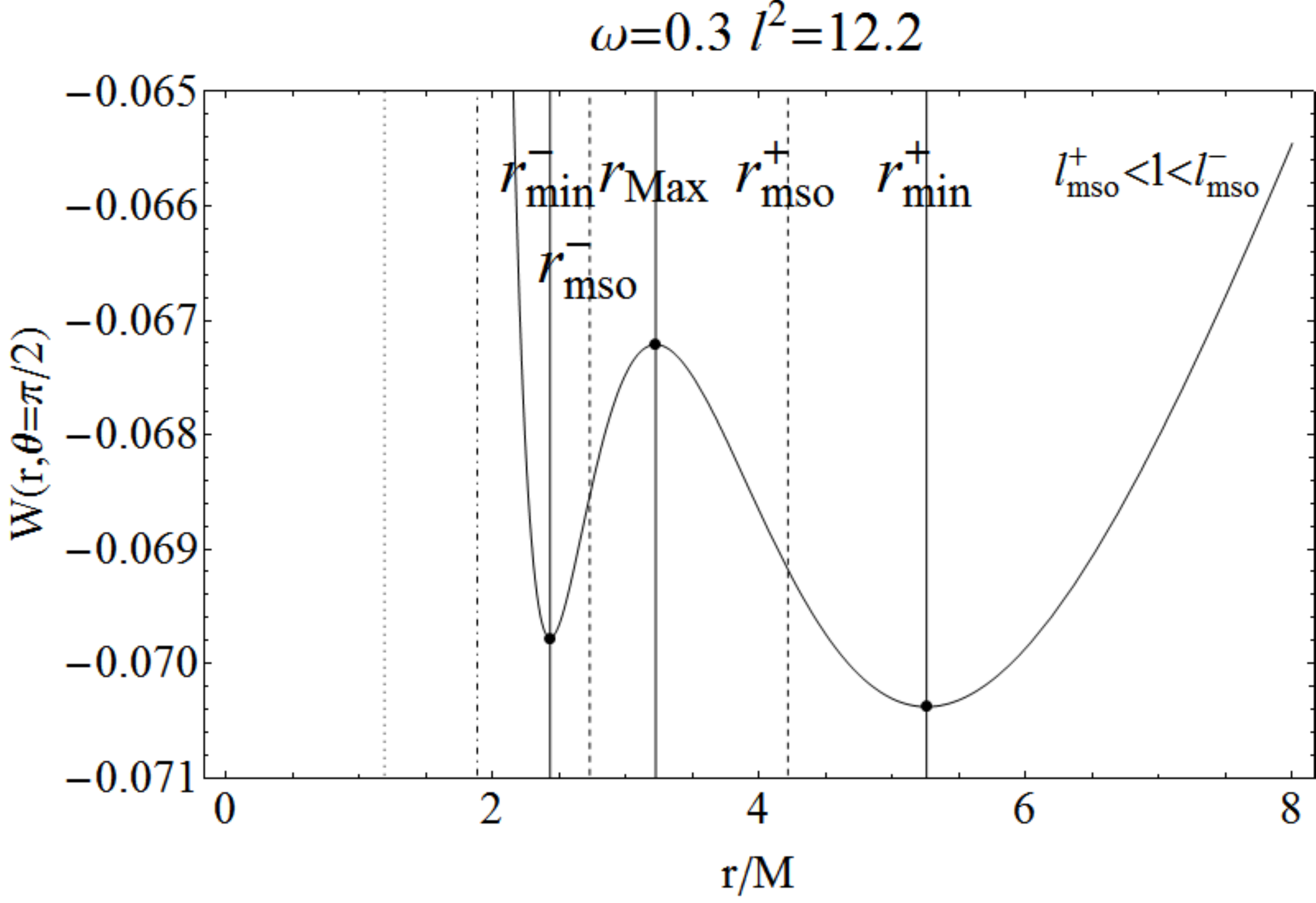}
\includegraphics[width=.41\textwidth]{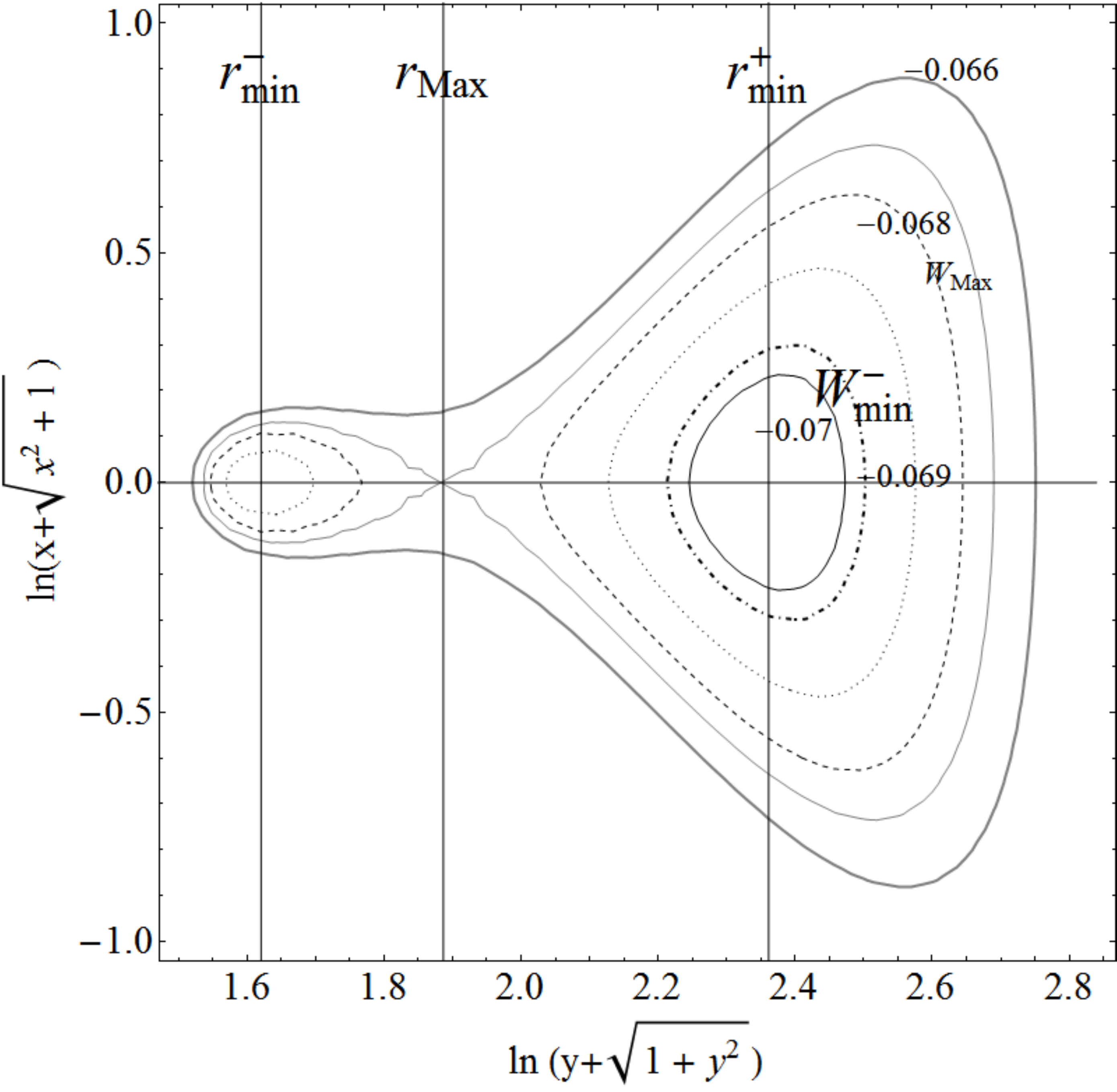}
\caption{  Region II: $\omega\in]\omega_{mso},\omega_{mbo}[$, naked singularity $\omega M^2=0.3$.  Where $\omega M^2\rightarrow \omega$.  Vertical lines in right panels set the $r_i\in\mathfrak{R}$ and  the effective potential critical points.
 It  is $l_{mso}^-> l_{mso}^+>l_{\Omega}^{Max}>l_{stat} $, and $r_{stat}<r_{\Omega}^{Max}<r_{mso}^-<r_{mso}^+$, where $r/M=\sqrt{x^2+y^2}$ and  $(x,y)$ are Cartesian coordinates.}
\label{Fig:newfigokey}
\end{figure}

\begin{figure}[h]
%\\
\includegraphics[width=.481\textwidth]{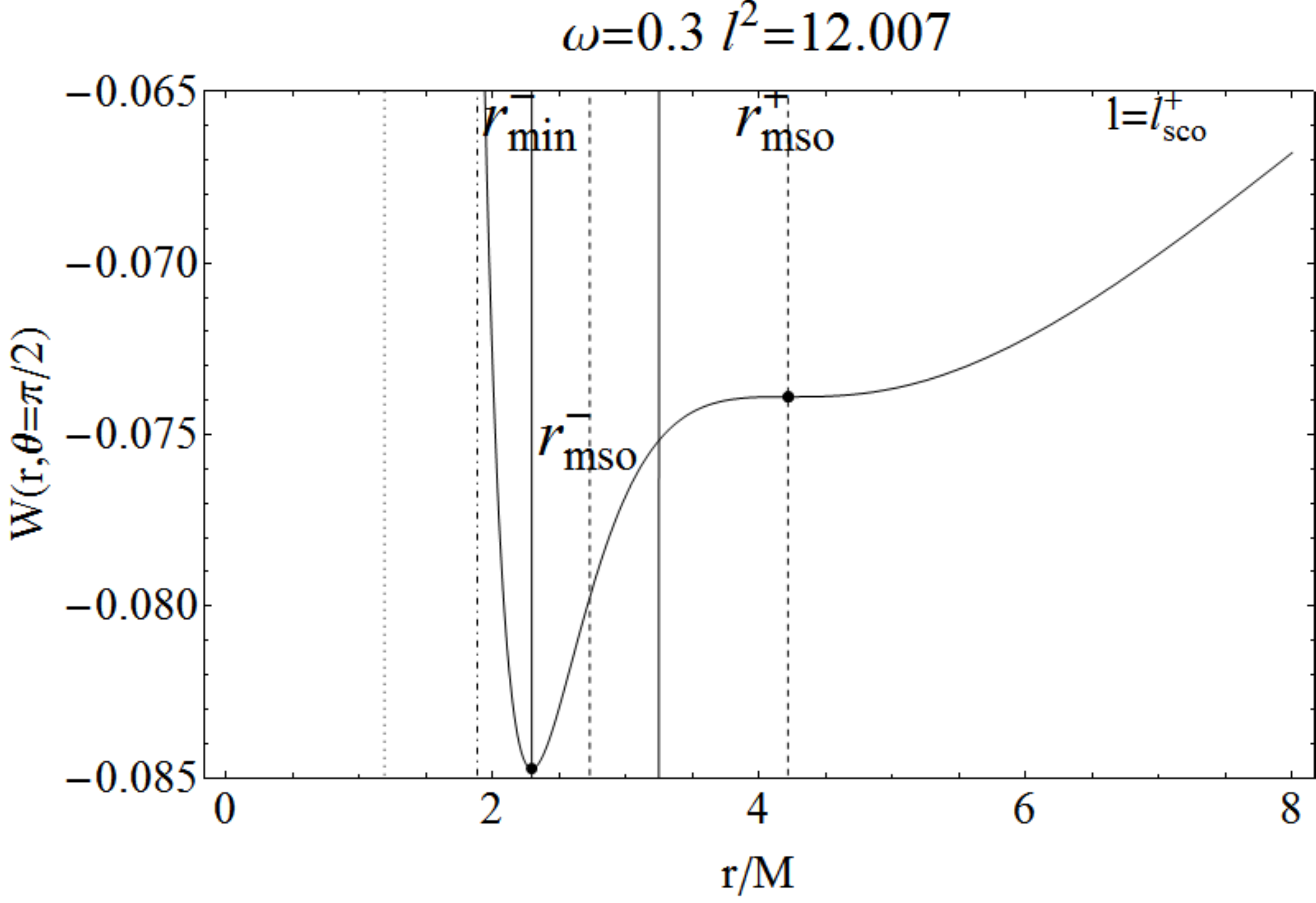}
\includegraphics[width=.41\textwidth]{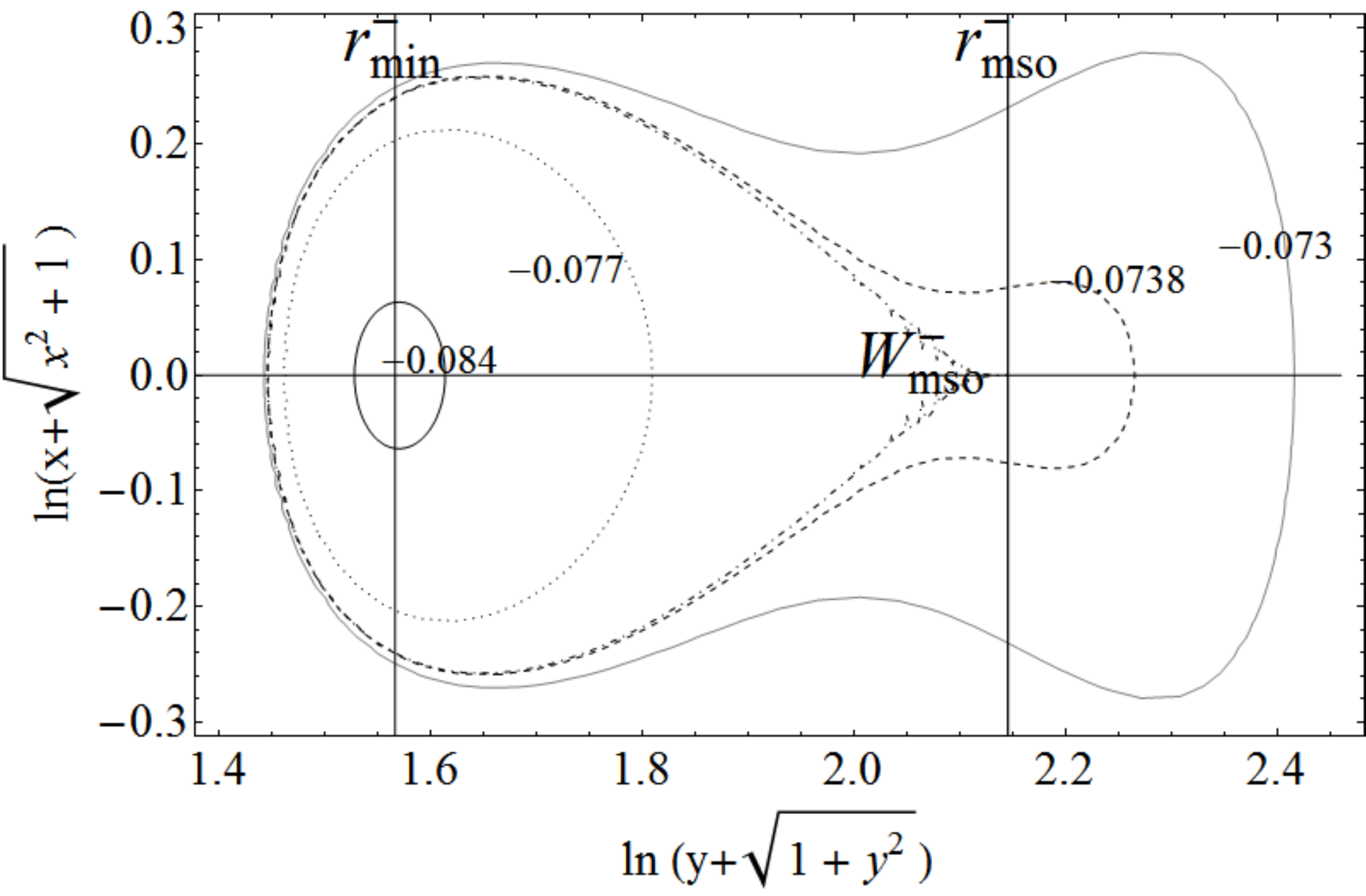}
\\
\includegraphics[width=.481\textwidth]{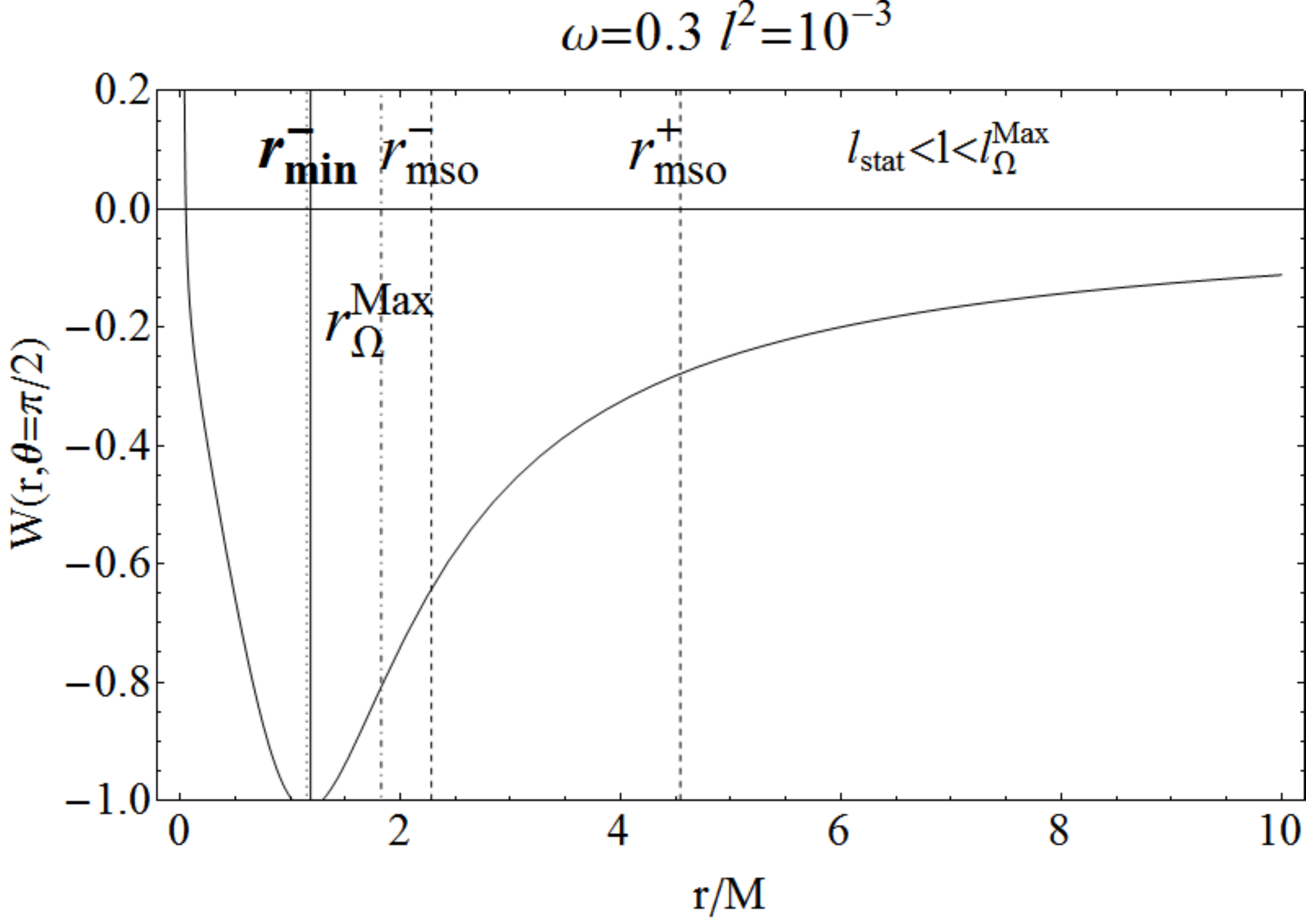}
\includegraphics[width=.3\textwidth]{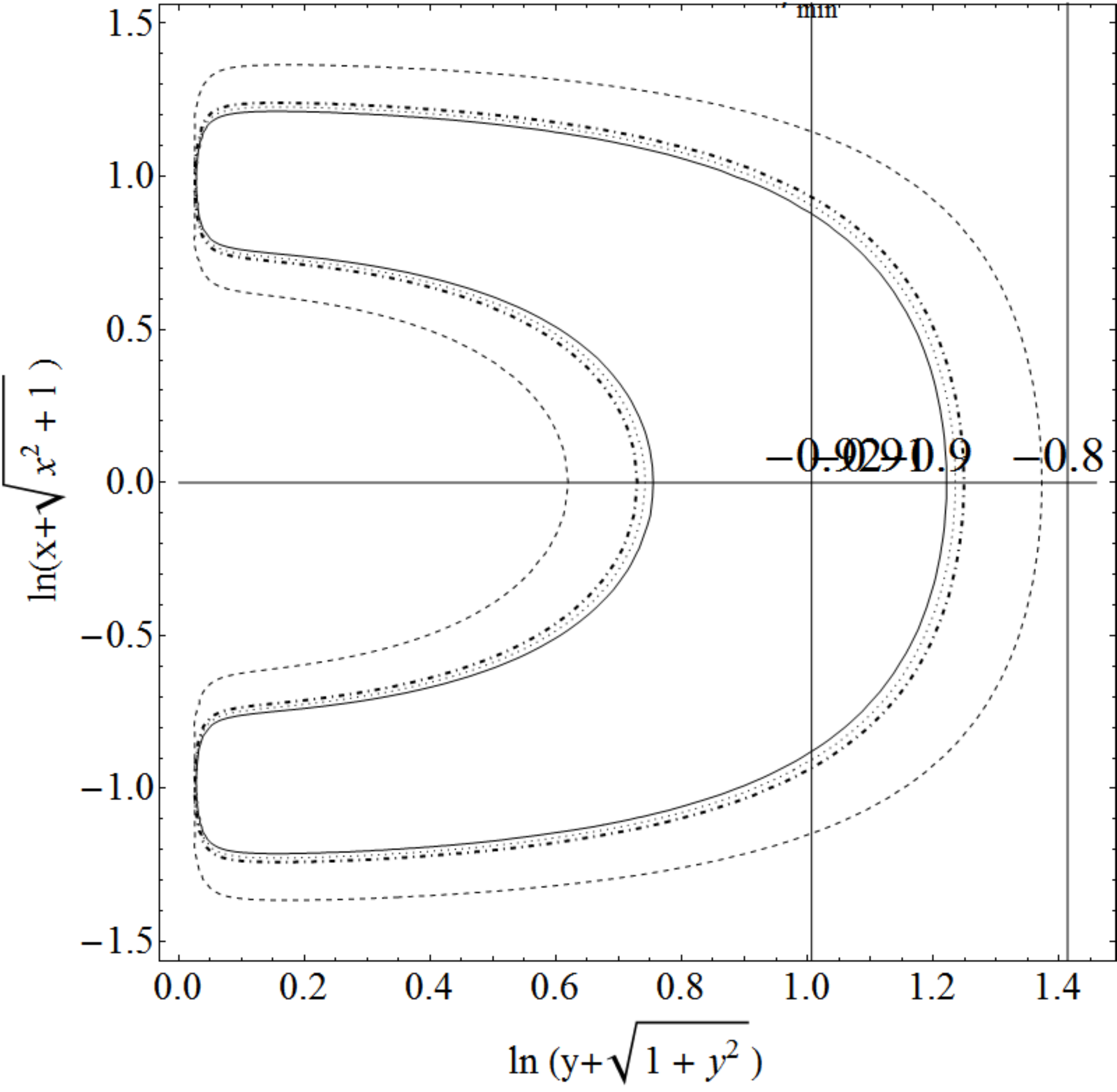}
\caption{ Region II: $\omega\in]\omega_{mso},\omega_{mbo}[$,  naked singularity $\omega M^2=0.3$. Where $\omega M^2\rightarrow \omega$.
 It  is $l_{mso}^-> l_{mso}^+>l_{\Omega}^{Max}>l_{stat} $, and $r_{stat}<r_{\Omega}^{Max}<r_{mso}^-<r_{mso}^+$, where $r/M=\sqrt{x^2+y^2}$ and  $(x,y)$ are Cartesian coordinates.  Vertical lines in right panels set the $r_i\in\mathfrak{R}$ and  the effective potential critical points.}
\label{Fig:reportthis}
\end{figure}

The K-S naked singularity spacetimes of   \textbf{Region II}  contain no photon circular orbits or marginally bounded orbits, but
two marginally stable circular orbits at $r_{mso}^{\pm}:\;r_{mso}^{-}<r_{mso}^{+}$, respectively.  It  is $r_{stat}<r_{\Omega}^{Max}<r_{mso}^-<r_{mso}^+$ and $l_{mso}^-> l_{mso}^+>l_{\Omega}^{Max}>l_{stat} $. As detailed in Sec.\il(\ref{Subsec:ClassII}) there are  stable circular orbits  in $r_{min}^+>r_{mso}^+$  (which corresponds to the outer $C^+$   disc) and $r_{min}^-\in]r_{stat},r_{mso}^-[$ (which corresponds to the center of the inner $C^-$  disc). Unstable circular orbits are in $r_{Max}\in]r_{mso}^{-},r_{mso}^{+}[$.  The energy of circular orbits is  $V_{eff}(r,l)<1$. In $r_{stat}$ it is $l_{stat}=0$, $V_{stat}\in]0,1[$,
$\partial_rV_{stat}=0$ and
$\partial^{2}_rV_{stat}>0$.
As two disconnected regions of orbital stability characterize these spacetimes,  it is foreseen the existence of a  double family of closed configurations, an outer $C^+$ with center in $r_{min}^+$ and an inner one $C^-$ centered in $r_{min}^-$, as it is  $r_{min}^-<r_{Max}<r_{min}^+$, an instability point, cross for the disc surface, occurs in between the two families of closed surfaces,  as $V_{Max}<1$, the crossed surface is closed as well, having two lobes. No accretion into the singularity is possible.  This is a characteristic of all the K-S-naked singularity solutions. Matter is somehow prevented to accrete the singularity, thus one can say that the naked singularity  is  embedded in a
region of  ``fluid-impermeable'' vacuum. For $\omega>\omega_{mbo}$, an excretion of matter can occur, this phenomenon is interpreted as a repulsive gravity effect, and it characterizes in general geometries with naked singularities.
The    cusp
enables outflow of fluid from one disc to the other, by violation of hydrostatic equilibrium and  giving rise properly to a feeding between the
discs:
the outer edge of the inner torus is  located above the (outer) marginally stable circular geodesic, i.e. $r_{mso}^-$, just as in a binary system one disc is fed by the outcome of matter from the companion.
The absence of an accretion point for this class of K-S naked singularity could be compared with the analogue cases where  the gravitational attraction is  balanced for example
by the cosmic repulsion due  to a cosmological constant, for instance  in relativistic thick discs in the Kerr-de Sitter backgrounds \cite{Sla-Stu:2005:CLAQG} or  in the Reissner-Nordstr\"om-anti-de Sitter \cite{Kuc-Sla-Stu:2011:JCAP}.
It should be noted that these last cases can be associated to  effective potential functions admitting two cups  due to  double unstable points separated by a stability region where  the center of the disc is located.
So that the crossed configuration  allows accretion through the
inner cusp and/or excretion through the outer cusp.
In the present case a  double configuration is  possible only in a specific set of   K-S-naked singularities,  namely those with  critical parameter $\omega$ sufficiently close to the extreme black hole case, i.e. the  sources belonging to \textbf{Region II-IV}, the repulsive effect is due to  combination of  the ``antigravity'' effect  typical  of the NS-cases  and  the
centrifugal repulsion.
The main difference however with the aforementioned  cosmological spacetimes, is that for a K-S-naked singularity there are no double cusp configurations, but only  double centered ones.
Two main aspects of the effective potential function  regulate  the  excretion of matter  and  the feeding from  one  disc to the other     as well as other essential characteristics of the  torus morphology such as the thickness, namely: the presence of (also stable) orbits  at $ E> 1 $ (characterizing  spacetimes in \textbf{Region III-IV}) and the gap between the $W$-values  as in the two minimum points i.e.   three cases  should be distinguished;  \textbf{I} for $W_{min}^->W_{min}^+$,  \textbf{II}  for $W_{min}^-=W_{min}^+$ and  \textbf{III} for  $W_{min}^-<W_{min}^+$,  correspondingly three types of configurations  belonging to the \textbf{I-II-III} cases   should be considered respectively.
As the outer minimum is always negative (see the location of the orbits $r_{mbo}^{\pm}$, compared to $r_{mso}^{\pm}$ in Fig.\il(\ref{Fig:complete}), the case in which the interior minimum  is positive or zero is easily located in the \textbf{I}-class (for  $W> 0$ in  $r_{mso}<r_ {min} <r_{mbo}$).%\rtb{!}.
%It is therefore necessary to study first the limiting case \textbf{III} \rtb{!}.
\begin{description}
\item[-){$l>l_{mso}^-$}] There are closed stable configurations. The effective potential has only one minimum point and no maxima, Fig.\il(\ref{Fig:newfigokey}-a).
\item[-)$l=l_{mso}^-$] This case is illustrated in Fig.\il(\ref{Fig:newfigokey}-b): at $r_{mso}^-$ there is a saddle point of the effective potential, a minimum is located at $r_{min}>r_{mso}^+$. There are only closed configuration. The value $W_{mso}^+$ is actually a critical point, correspondent to the  inner edge of the disc centered in $r_{min}$. As in Fig.\il(\ref{Fig:newfigokey}-b) it is $|W_{mso}^-|<|W_{mso}^+|$
the maximum value of $W$ for the appearance of  the thick torus  is $|W_{min}|$, as $W\gtrapprox W_{min}^+$ there starts a sequence of centered configurations, that thickens increasing $W$ until the values $W_{mso}^-$, where a cusped-inner edge is formed, and then other closed configuration filled the region.
\item[-) ${ l\in ]l_{mso}^+,l_{mso}^-[ }$ ] There are two minimum and one maximum point:
$r_{stat}<r_{min}^-<r_{mso}^-<r_{max}<r_{mso}^+<r_{min}^+$, as a consequence of this  there is a double closed configuration of one inner torus  $C^-$ centred in $r_{min}^-$  and $C^+$ in $r_{min}^+$ (the outer one). As $W$ increases towards $W_{Max}$ the  outer edge of $C^-$ approaches the inner edges of  $C^+$ until a double centered, closed crossed configuration appears for $W_{Max}$. As $W$ approaches asymptotically $W=0$, for $W\in]W_{Max},0[$
only one closed configuration appears with two maxima of thickness in $r_{min}$.
In Fig.\il(\ref{Fig:newfigokey})-c the case \textbf{I}  is shows, as it is the maximum values of  $|W|$ for the formation of a thick configuration is $|W_{min}^+|$: the outer configuration $C^+$ is formed  and as  $W$ fills the gap $\Delta=|W_{min}^+-W_{min}^-|$ a second disc $C^-$ appears, thickens increasing $W$ until $W=W_{max}$.
\item[-) ${l=l_{mso}^+}$] This case is discussed in Fig.\il(\ref{Fig:reportthis})-a. In $r_{mso}^+$  is a saddle point of the potential $W$. A minimum is located in $r_{min}^-<r_{mso}^-$, that corresponds to the center of the closed configuration. A closed (stable) disc is formed as $W\gtrapprox W_{min}^-$, as $W= W_{mso}^+$ a configuration similar to that in Fig.\il(\ref{Fig:newfigokey}-d) that is $l=l_{mso}^-$, where the center of the disc was in $r_{min}^+$, we note that the location of the ``cusp'' in the two cases is reversed. Increasing again $W>W_{mso}^+$ one closed, stable configuration with two maximum points of thickness appears. Finally we could read this peculiar case shown in Fig.\il(\ref{Fig:newfigokey}-d) where only one minimum and a saddle point are as \textbf{III}
$W_{min}^-<W_{mso}^+$.
\item[-)${l\in]l_{stat},l_{mso}^+[}$]  This range includes $]l_{stat},l_{mso}^+[=]l_{stat},l_{\Omega}^{Max}[\cup[l_{\Omega}^{Max},l_{mso}^+[  $.  There are  only closed, stable configurations with center located in $r=r_{min}^+$. This case is similar to the analogue situation for attractors in 	 \textbf{Region I}, see   Fig.\il(\ref{Fig:reportthis}).
\end{description}
%%
%\end{figure}
%HERE I AM.
%\clearpage
\subsubsection{Naked singularity: $\omega=\omega_{mbo}$}\label{Sec:NSpar2}
\begin{figure}[h]
\includegraphics[width=.481\textwidth]{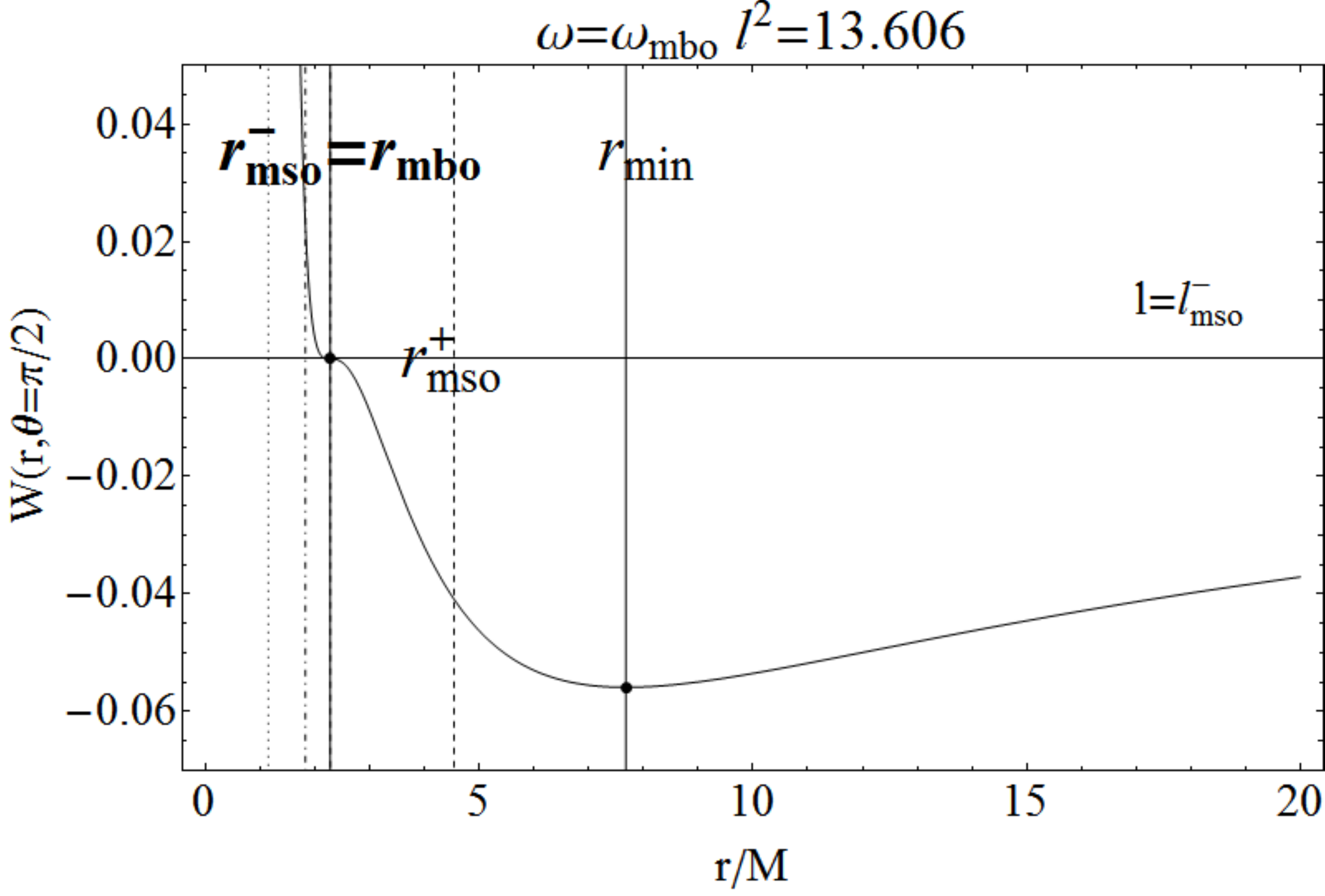}
\includegraphics[width=.481\textwidth]{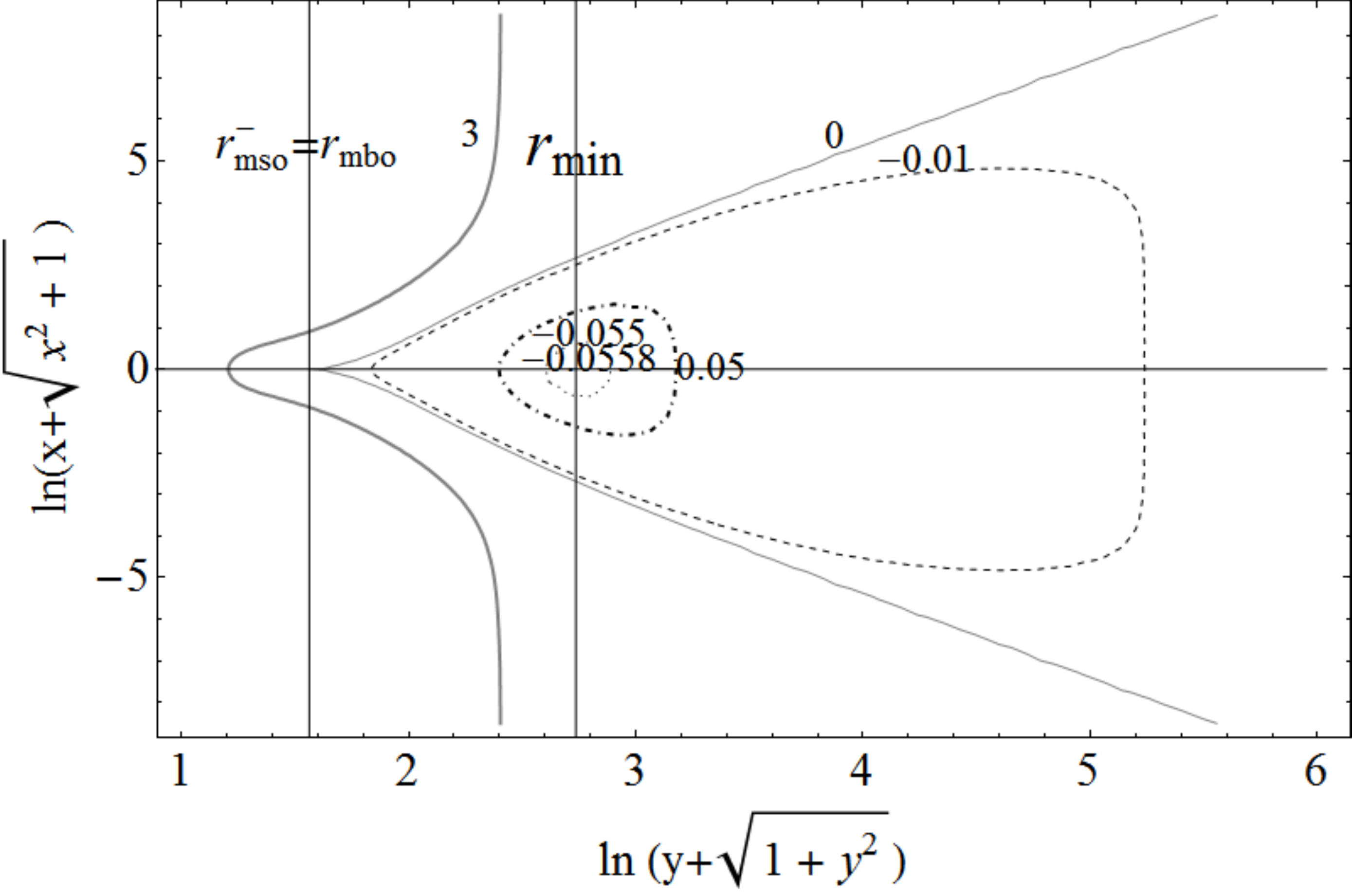}
\\
%\begin{figure}[h]
\includegraphics[width=.481\textwidth]{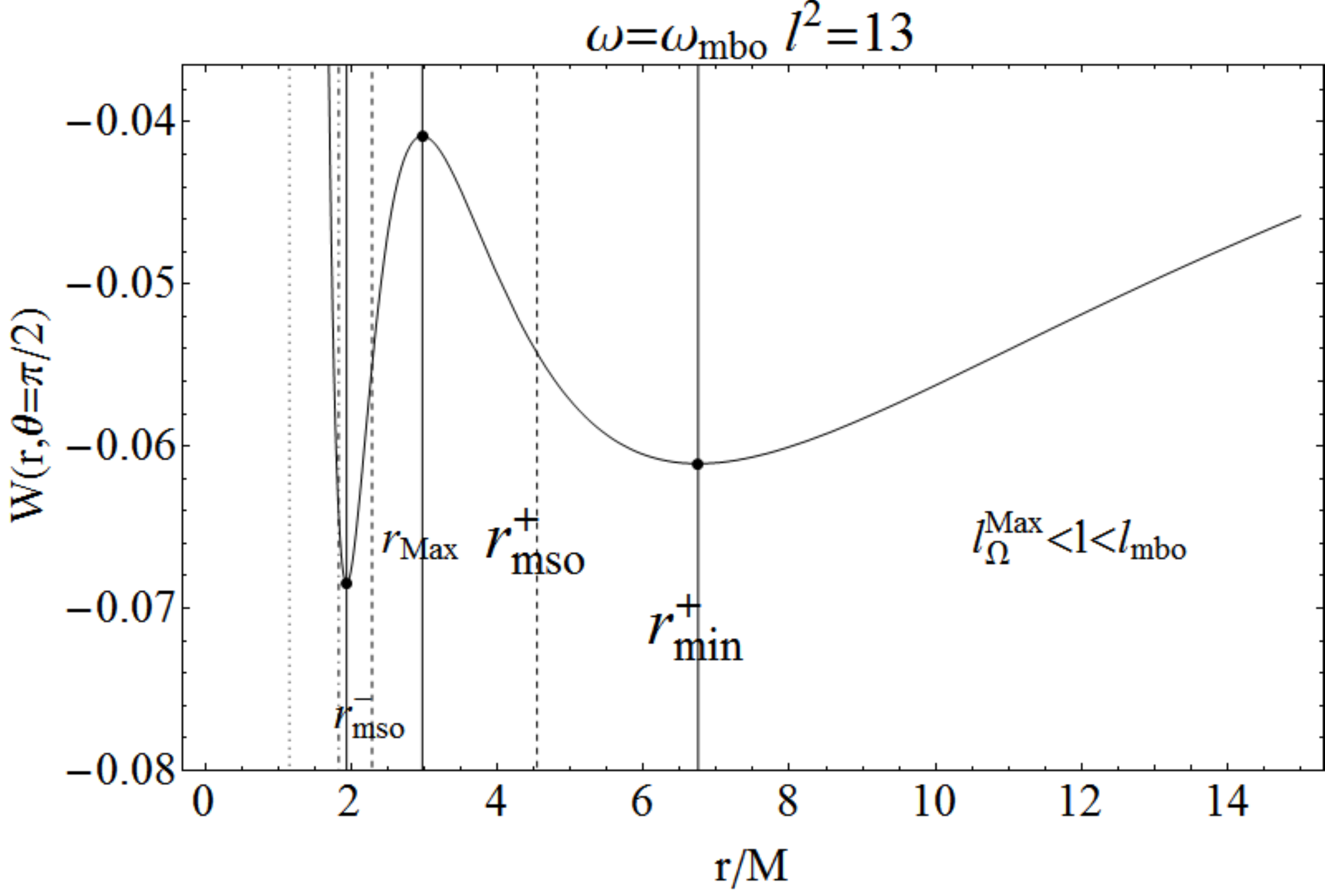}
\includegraphics[width=.41\textwidth]{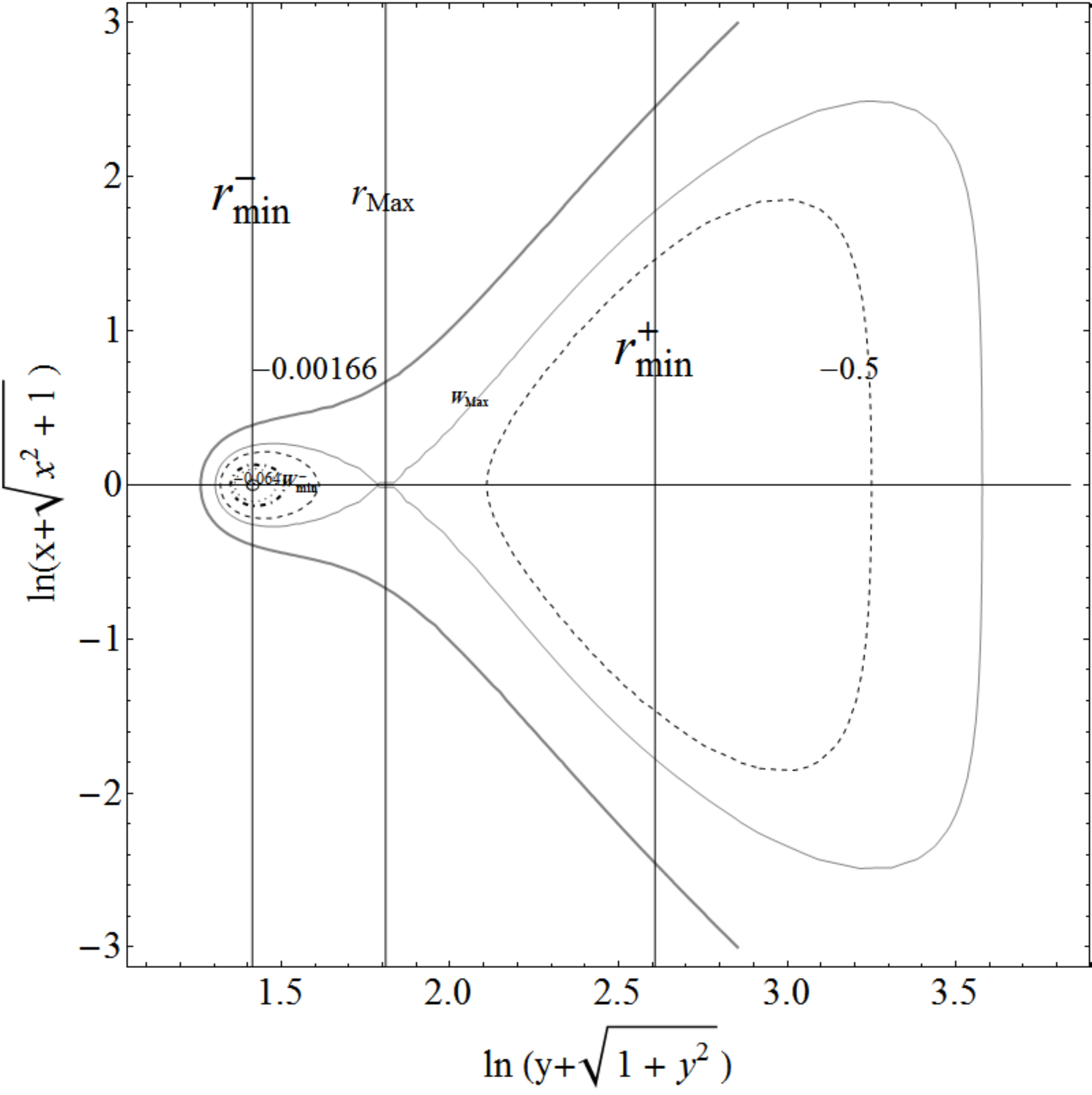}
\\
\includegraphics[width=.481\textwidth]{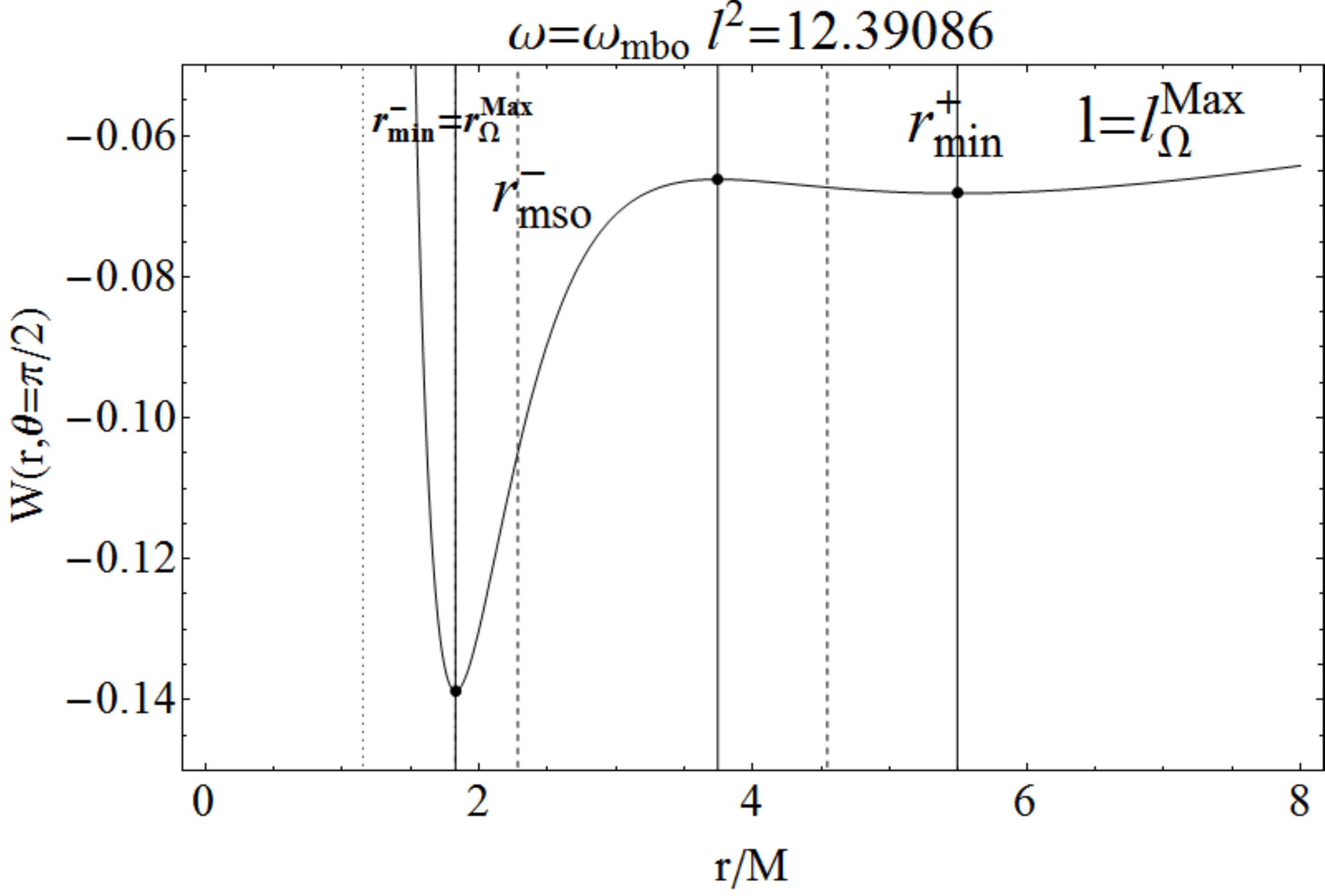}
\includegraphics[width=.381\textwidth]{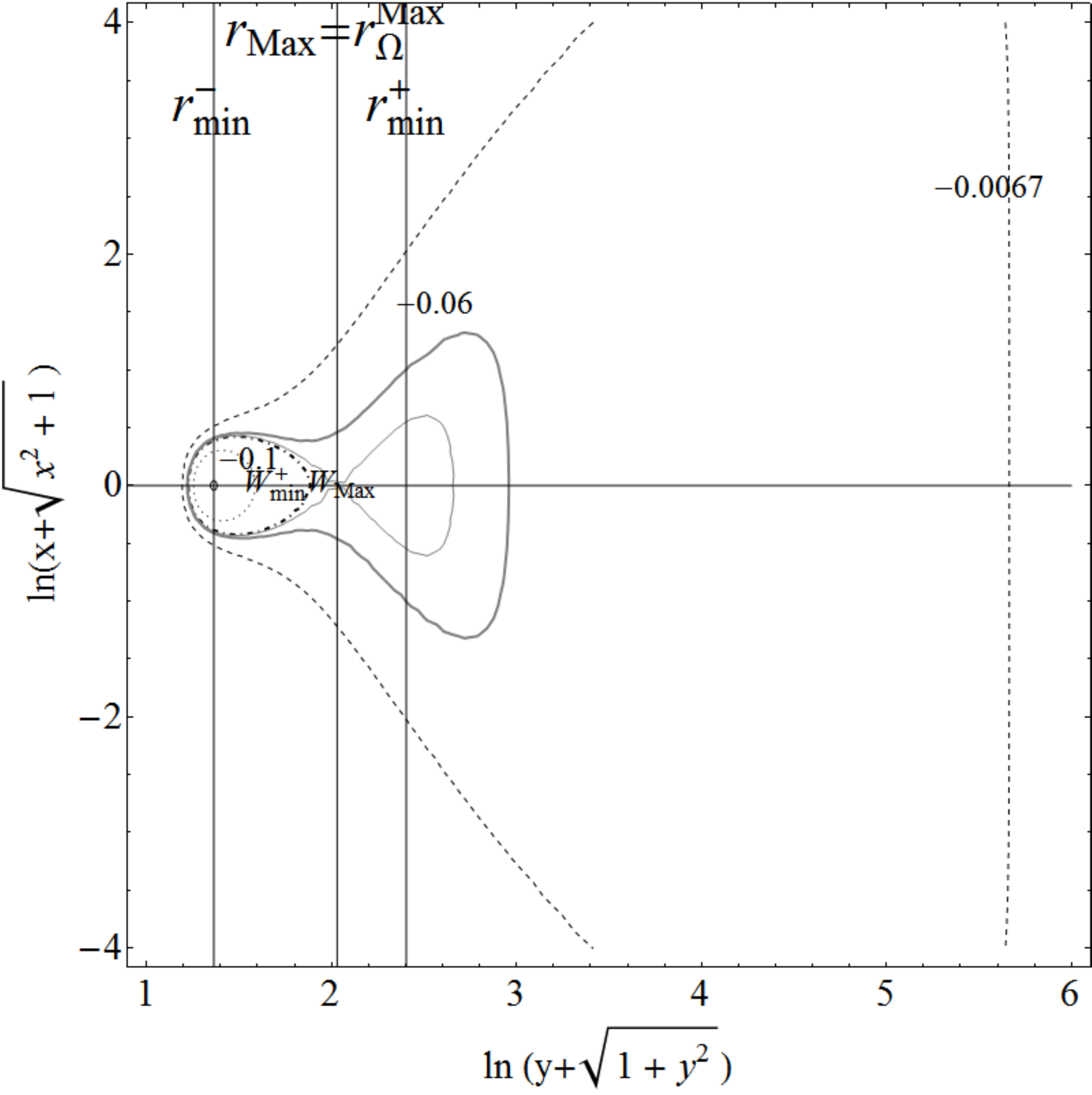}
\caption{ Naked singularity $\omega=\omega_{mbo}$.
 It  $l_{mso}^-=l^-_{mbo}= l^+_{mbo}>l_{\Omega}^{Max}> l_{mso}^+$, and $r_{stat}<r_{\Omega}^{Max}<r_{mbo}^-=r_{mso}^-=r_{mbo}^+<r_{mso}^+$.  Vertical lines in right panels set the $r_i\in\mathfrak{R}$ and  the effective potential critical points. It is  $\omega M^2\rightarrow \omega$, where $r/M=\sqrt{x^2+y^2}$ and  $(x,y)$ are Cartesian coordinates.}
\label{Fig:ban-or}
\end{figure}

\begin{figure}[h]
%%CPlotoiscomPlotoocl14Plotoocl14
\includegraphics[width=.481\textwidth]{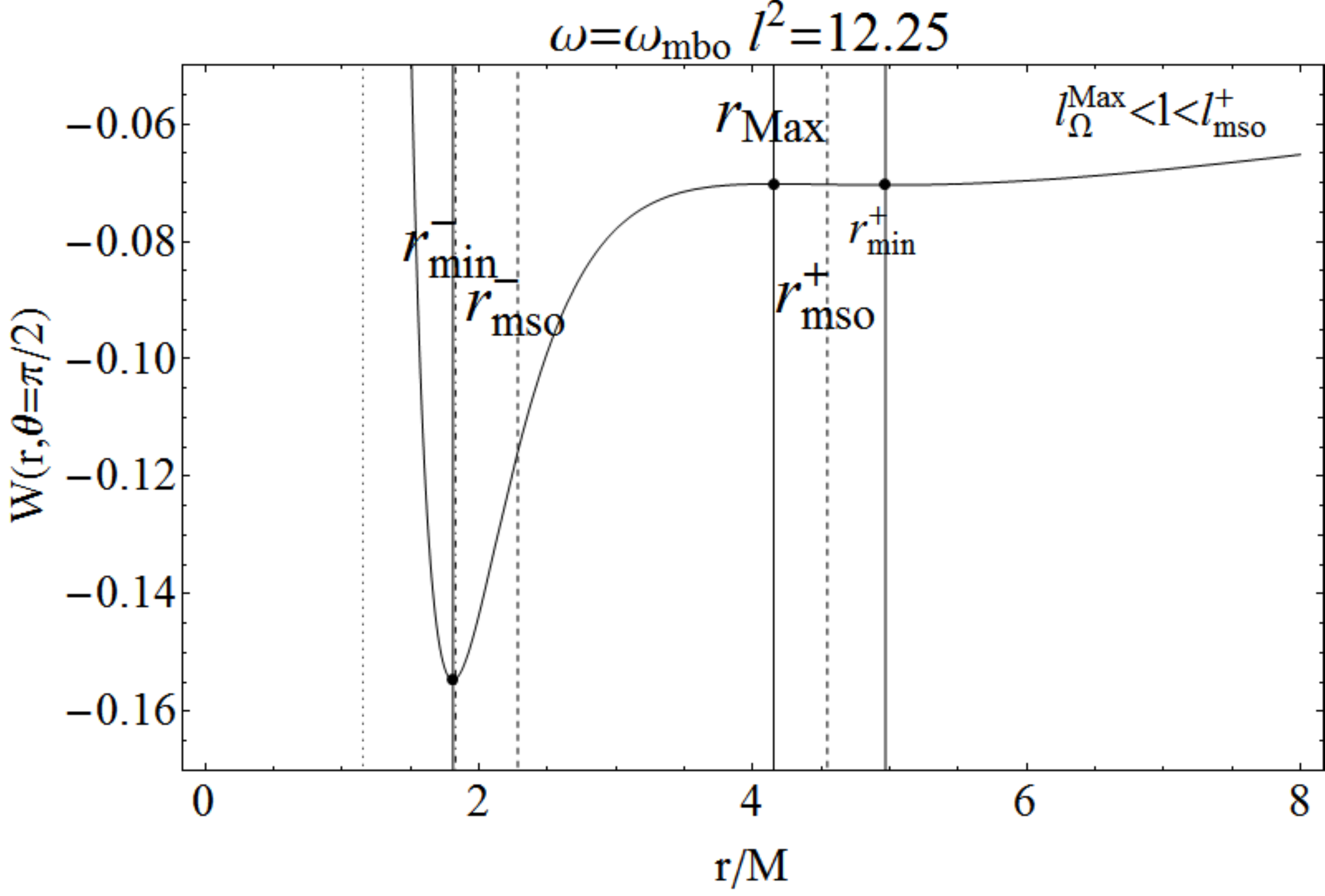}
\includegraphics[width=.41\textwidth]{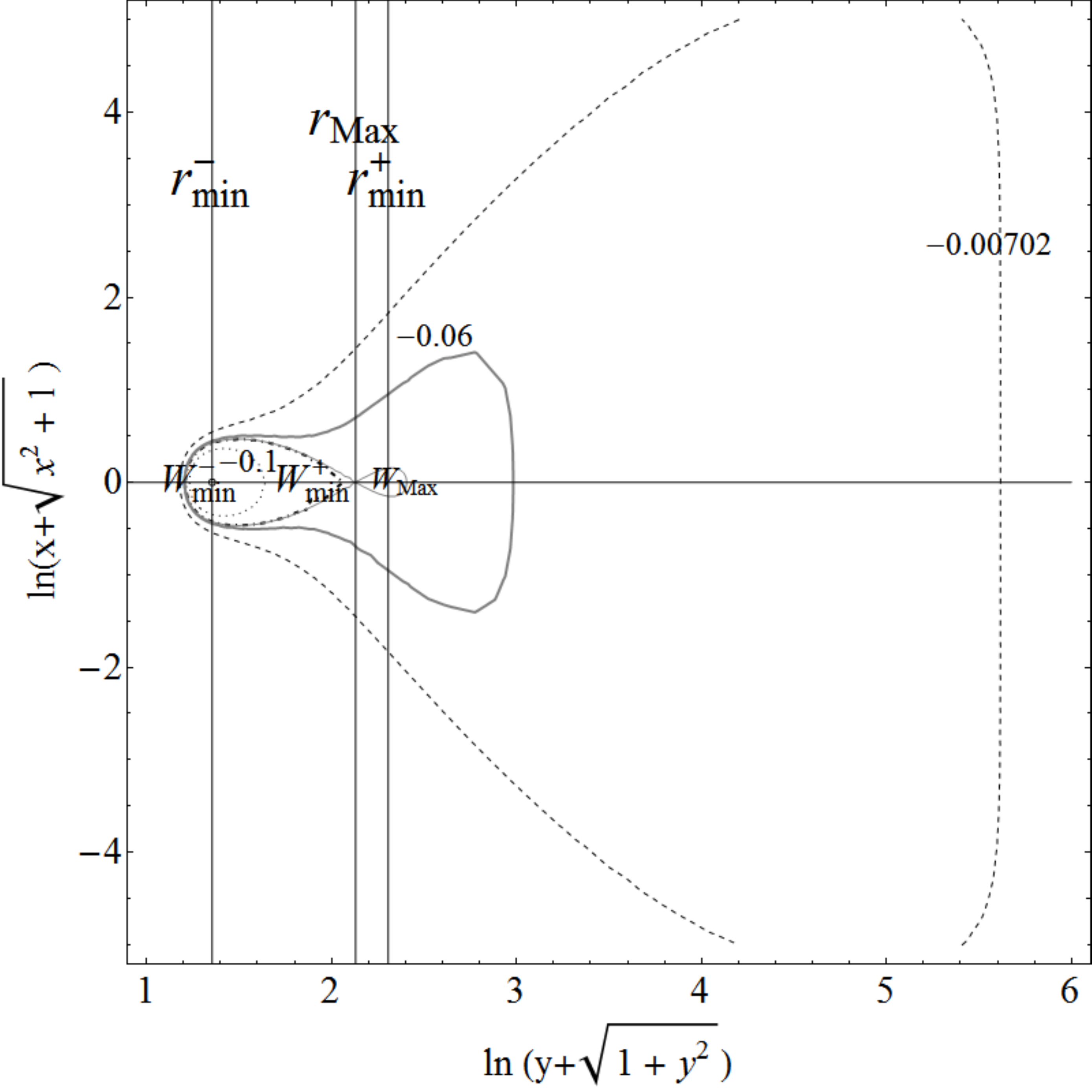}
\\
\includegraphics[width=.481\textwidth]{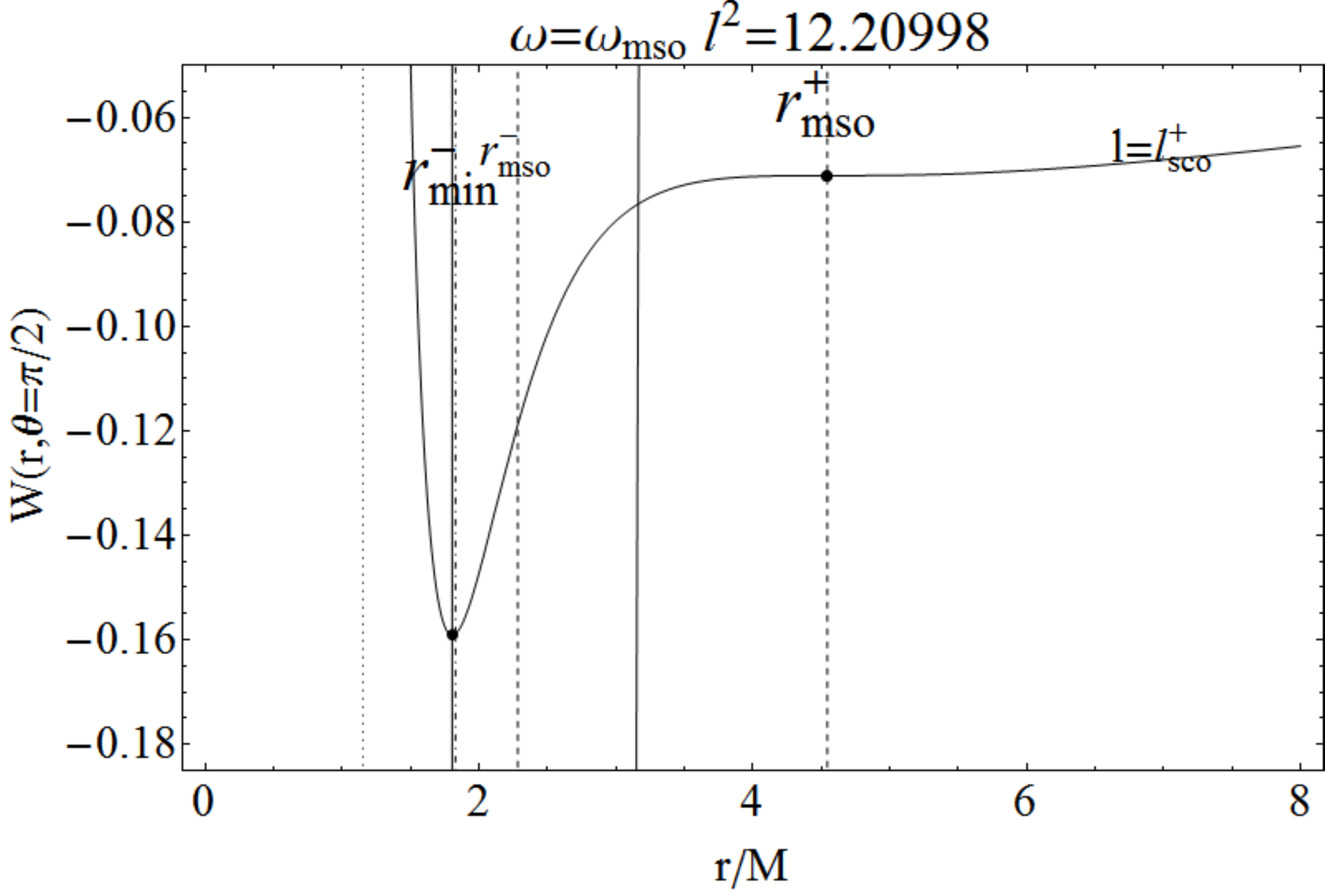}
\includegraphics[width=.481\textwidth]{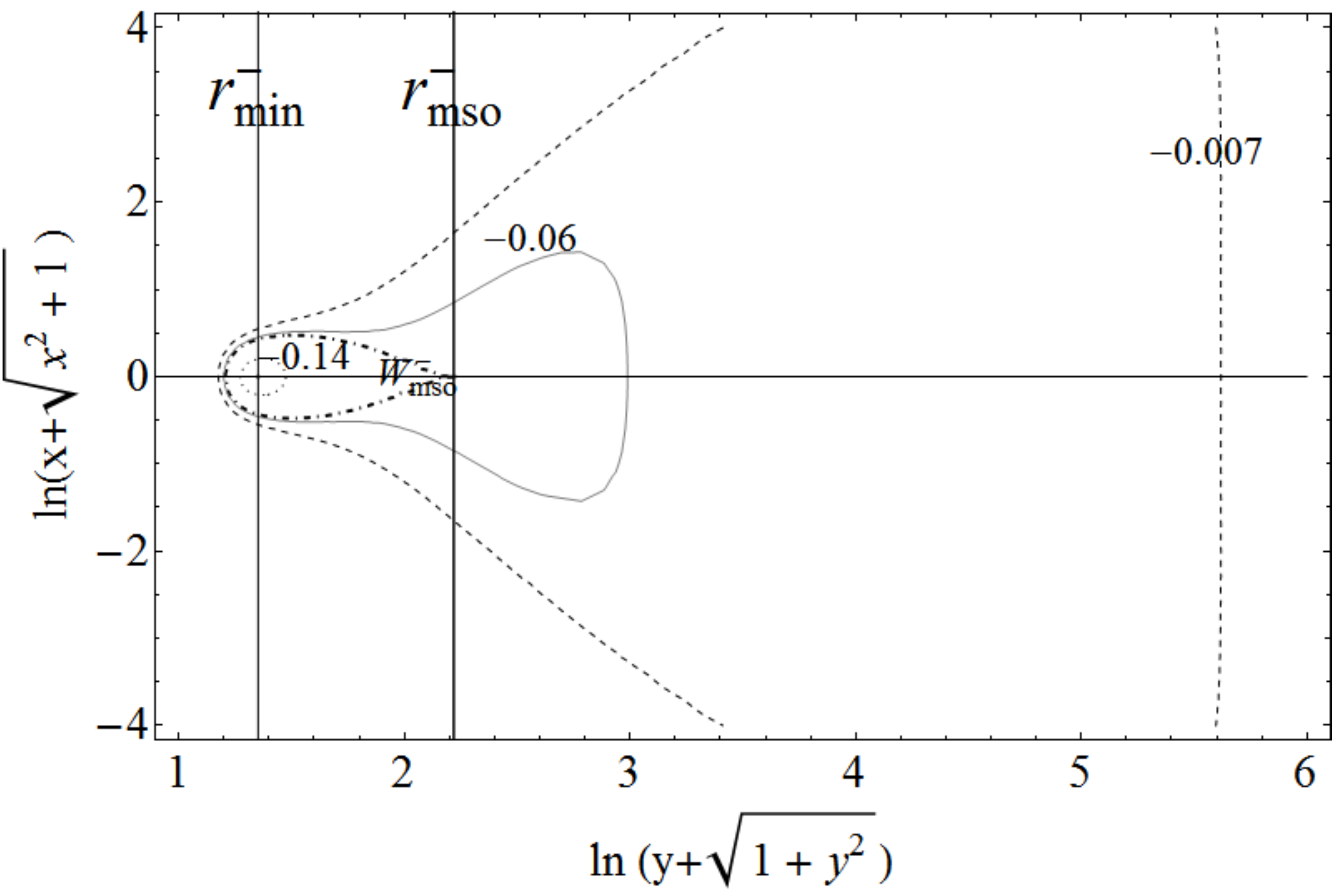}
%\\
\\
\includegraphics[width=.481\textwidth]{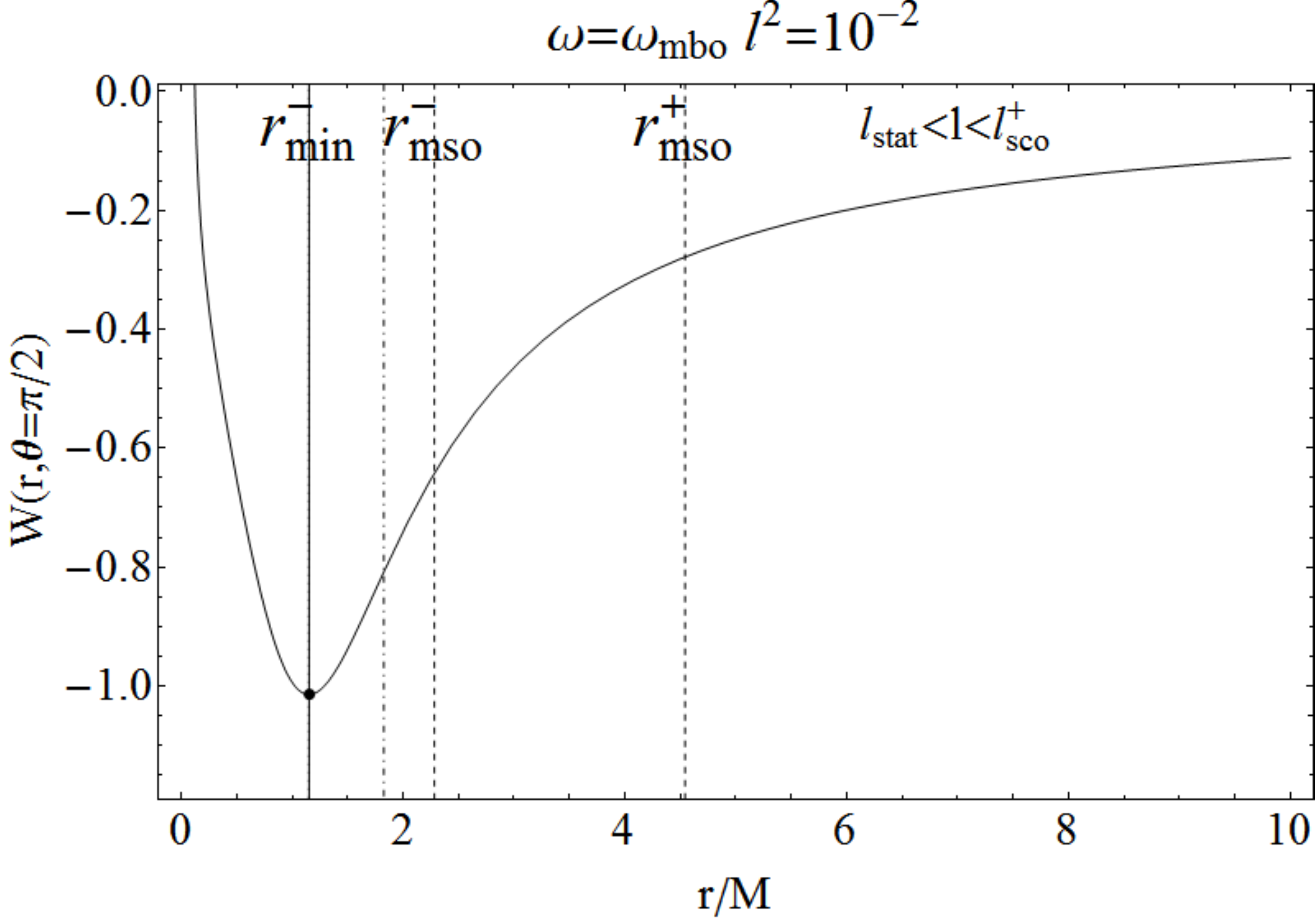}
\includegraphics[width=.481\textwidth]{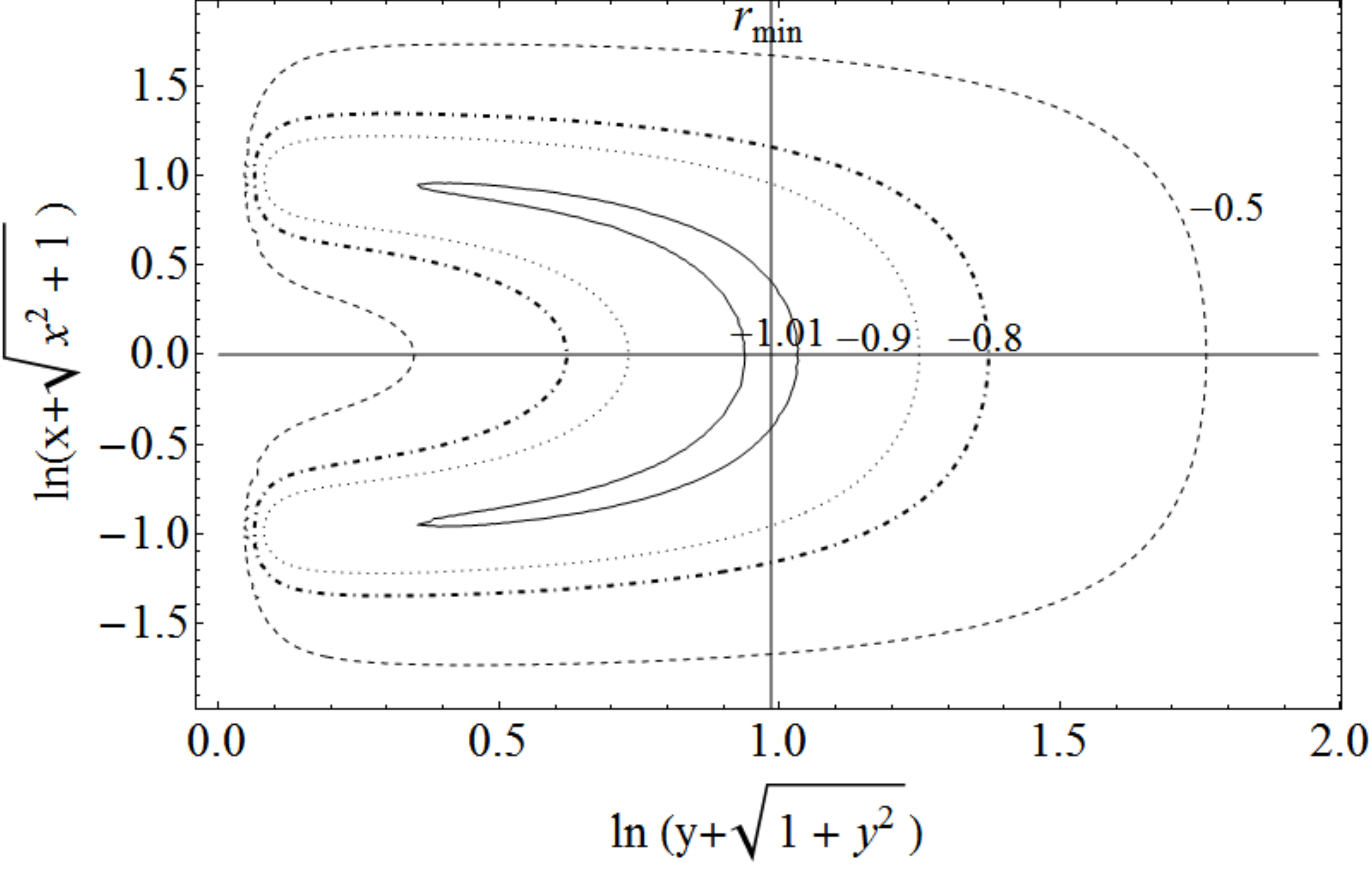}
\caption{ Naked singularity $\omega=\omega_{mbo}$. With  $\omega M^2\rightarrow \omega$.
 It  $l_{mso}^-=l^-_{mbo}= l^+_{mbo}>l_{\Omega}^{Max}> l_{mso}^+$, and $r_{stat}<r_{\Omega}^{Max}<r_{mbo}^-=r_{mso}^-=r_{mbo}^+<r_{mso}^+$.  Vertical lines in right panels set the $r_i\in\mathfrak{R}$ and  the effective potential critical points.}
\label{Fig:tis-bre}
\end{figure}

In this section we consider the particular  case of a naked singularity spacetime at $\omega=\omega_{mbo}$, where $r_{stat}<r_{\Omega}^{Max}<r_{mbo}^-=r_{mso}^-=r_{mbo}^+<r_{mso}^+$ and
  $l_{mso}^-=l^-_{mbo}= l^+_{mbo}>l_{\Omega}^{Max}> l_{mso}^+$. In this limiting spacetime the inner last stable circular orbits  is located to the marginally bounded orbits, and it is a critical parameter as for the Ho\v{r}ava parameter $\omega\in]\omega_{mbo},1/2[$ two branches $r_{mbo}^{\pm}$ appear.
   The marginally bound  circular geodesics  is thus an unstable orbit with
specific energy $ E_{mbo} = 1$.
We summarize this case  as follows:
\begin{description}
\item[-) $l>l_{mso}^-$] The effective potential has a minimum $r_{min}>r_{mso}^-$, center of  closed surfaces. The situation is not different from the cases $l>l_{mso}^-$ in the \textbf{Region I-II},  see Fig.\il(\ref{Fig:newfigokey}-a).
\item[-) $l=l_{mso}^-$] This case is illustrated in Fig.\il(\ref{Fig:ban-or})-a. The effective potential has  a minimum $W_{min}<0$ in $r_{min}>r_{mso}^+$. At $r=r_{mso}^-=r_{mbo}$ the effective potential has a saddle point where $W_{mso}^-=0$, correspondingly a set of closed configurations  cetered in $r_{min}$ are as $W\in]W_{min},0[$, more and more stretched on the equatorial plane as  $W$ approaches the limit $W=W_{mso}^-$ where an open surface appears.
\item[-)${l\in]l_{\Omega}^{Max},l_{mso}^-[}$] There are three critical points of the hydrostatic pressure for $W_{crit}<0$. Correspondingly there are generally  two closed configuration $C^{\pm}$ centered in $r_{min}^-<r_{min}^+$ respectively, the closed-cross surface is for $W=W_{Max}<0$. This case is illustrated in Fig.\il(\ref{Fig:ban-or})-b, considerations analogue to the case in Fig.\il(\ref{Fig:newfigokey}-a)  for singularities in \textbf{Region II}  with $l\in]l_{mso}^+,l_{mso}^-[$. This is a double $\mathbf{III}$ configuration as $W_{min}^-<W_{min}^+$
\item[-) $l=l_{\Omega}^{Max}$] There is a minimum in $r_{\Omega}^{Max}<r_{mso}^-$, a minimum $r_{min}^+>r_{mso}^+$ and a maximum point $r_{Max}$. The situation is similar to the case $l\in]l_{\Omega}^{Max},l_{mbo}[$, see Fig.\il(\ref{Fig:ban-or})-b and Fig.\il(\ref{Fig:ban-or})-c. The situation is analogue to the case ${l\in]l_{\Omega}^{Max},l_{mso}^-[}$ and leads to a double $\mathbf{III}$ configuration.
The torus thickness, the spatial distances between the inner edge of the outer torus and the outer edge in the inner configuration are regulated by the magnitude of the $W$ and the gap $|W^+_{min}-W_{min}^-|$.
\item[-) ${l\in]l_{mso}^+,l_{\Omega}^{Max} [}$] This case is illustrated in Fig.\il(\ref{Fig:tis-bre})-a. There are closed $C^{\pm}$,  $\mathbf{III}$ type toroidal surfaces and  a critical surface. The behavior of equipotential surfaces is
analogue to the situation in Fig.\il(\ref{Fig:ban-or})-b and Fig.\il(\ref{Fig:ban-or})-c.
\item[-) $l=l_{mso}^+$] There is a sequence of \textbf{III}-type of closed $C^-$ toroidal discs centered in $r_{min}^-$ and a critical closed (but not crossed) surface correspondent to the saddle point of the effective potential. As this critical point is located in $r_{mso}^+$, the ``cusp'' is towards the outer space, and the situation, sketched  in Fig.\il(\ref{Fig:tis-bre})-b is similar to  other cases where $l=l_{mso}^+$,  for example in Fig.\il(\ref{Fig:newfigokey})-b.
\item[-) ${l\in]l_{stat},l_{mso}^+[}$] There are only closed $C^-$ (stable) \textbf{III}-type configurations centered in $r_{min}<r_{mso}^-$, see  Fig.\il(\ref{Fig:tis-bre})-c.
\end{description}
\clearpage
\subsubsection{Region III-a: $\omega\in]\omega_{mbo},\omega_c[$}\label{Sec:NSRegionIII}
\begin{figure}[h]
%\\
\includegraphics[width=.481\textwidth]{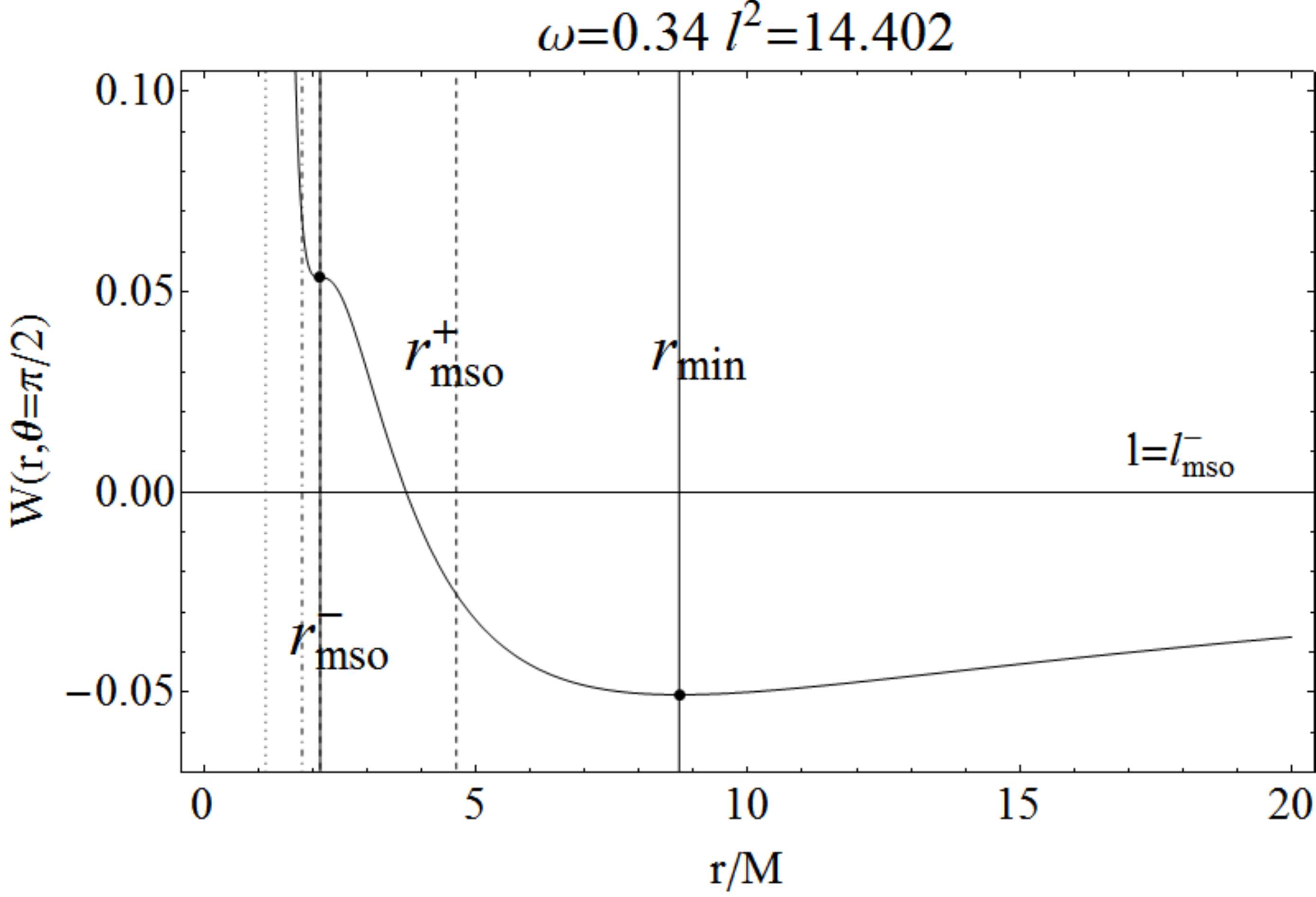}
\includegraphics[width=.481\textwidth]{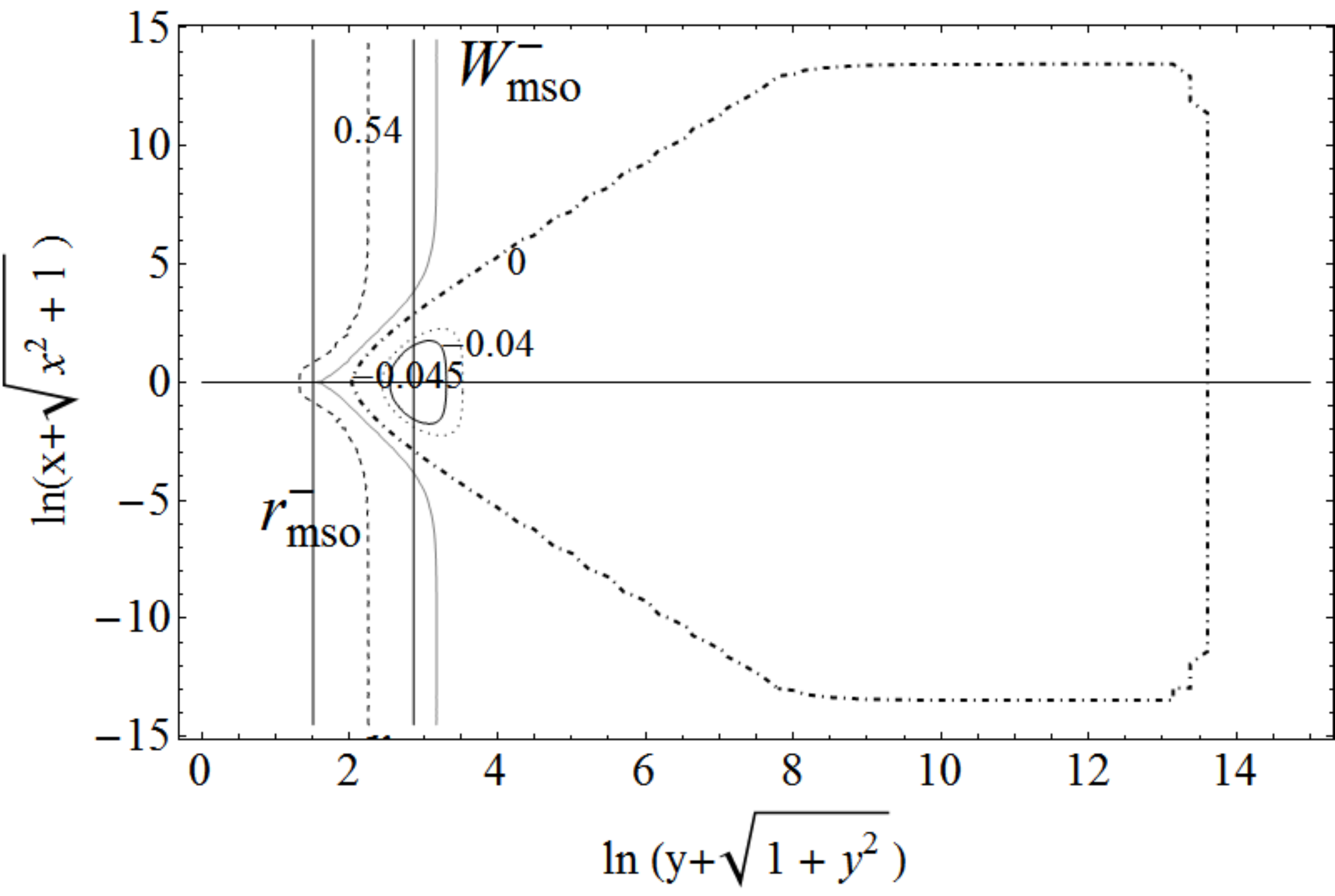}
\\
\includegraphics[width=.481\textwidth]{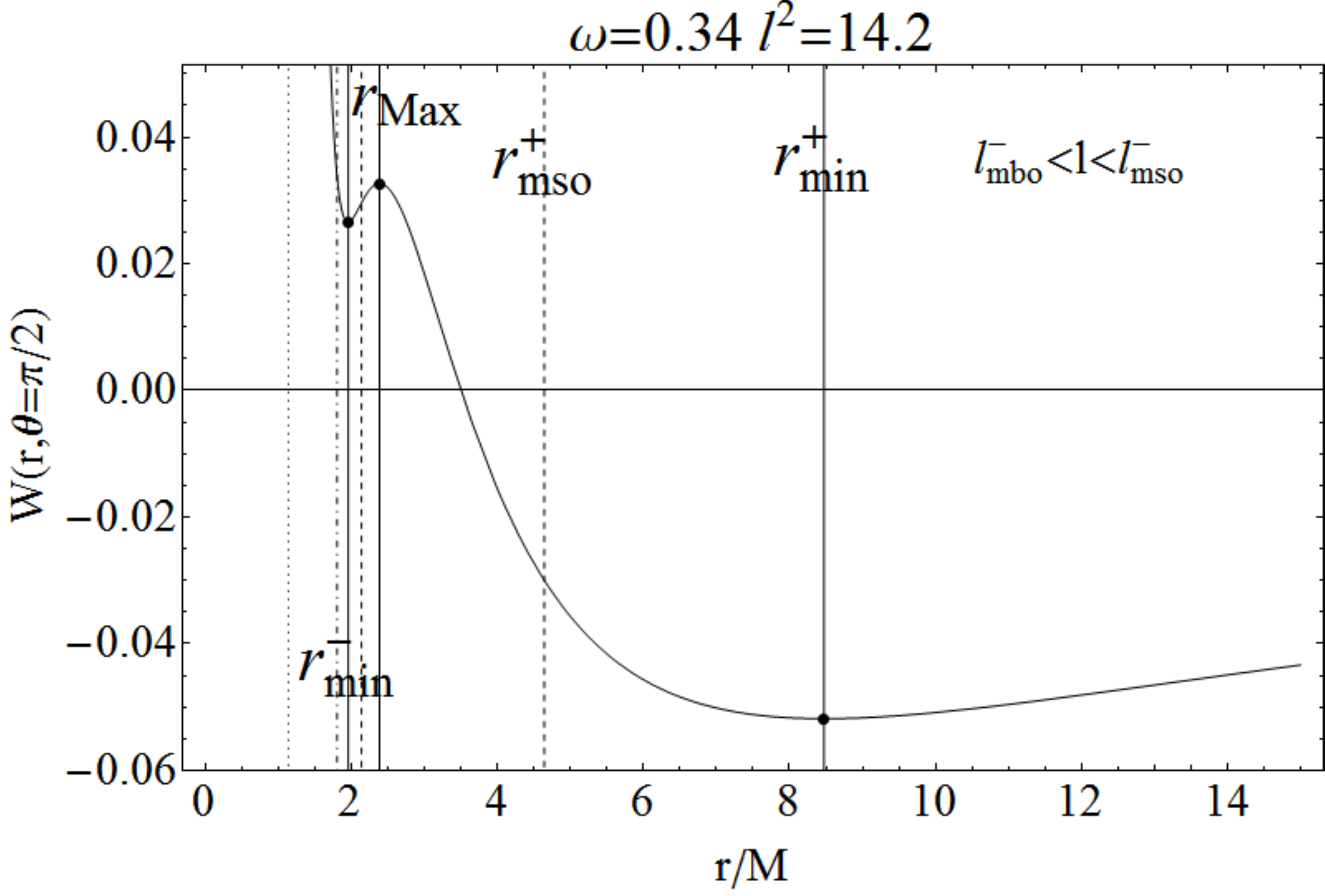}
\includegraphics[width=.41\textwidth]{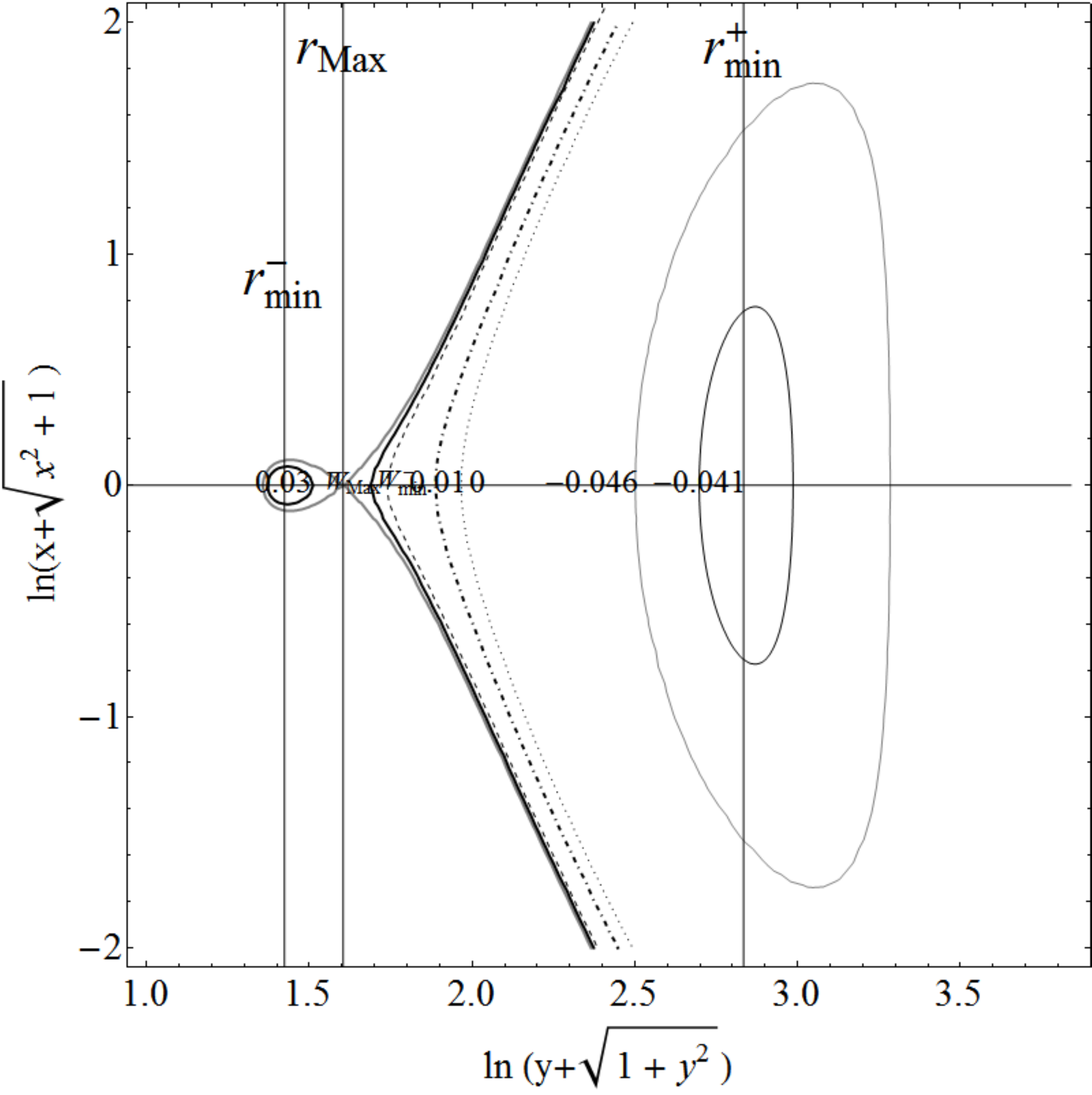}
\\
\includegraphics[width=.481\textwidth]{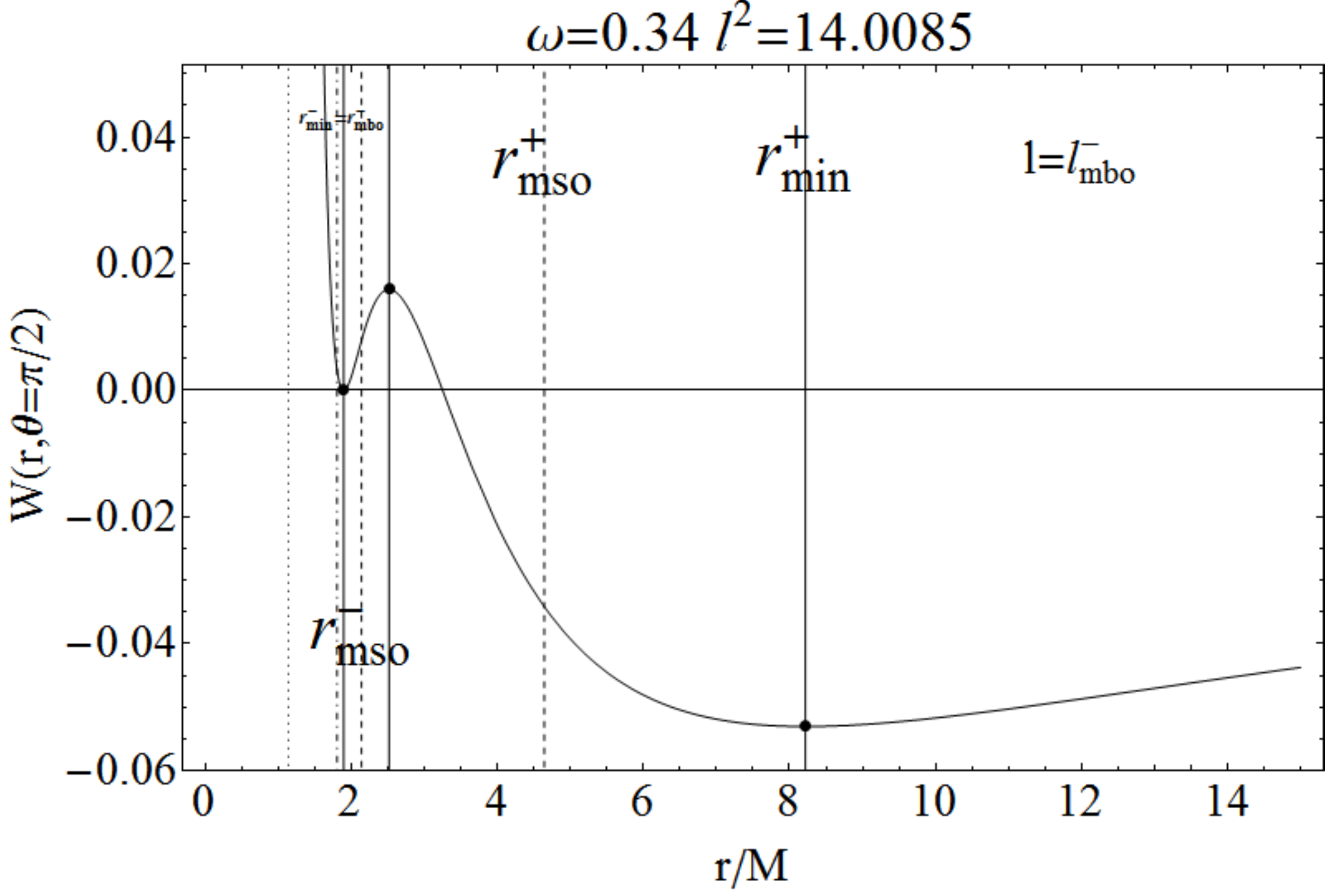}
\includegraphics[width=.481\textwidth]{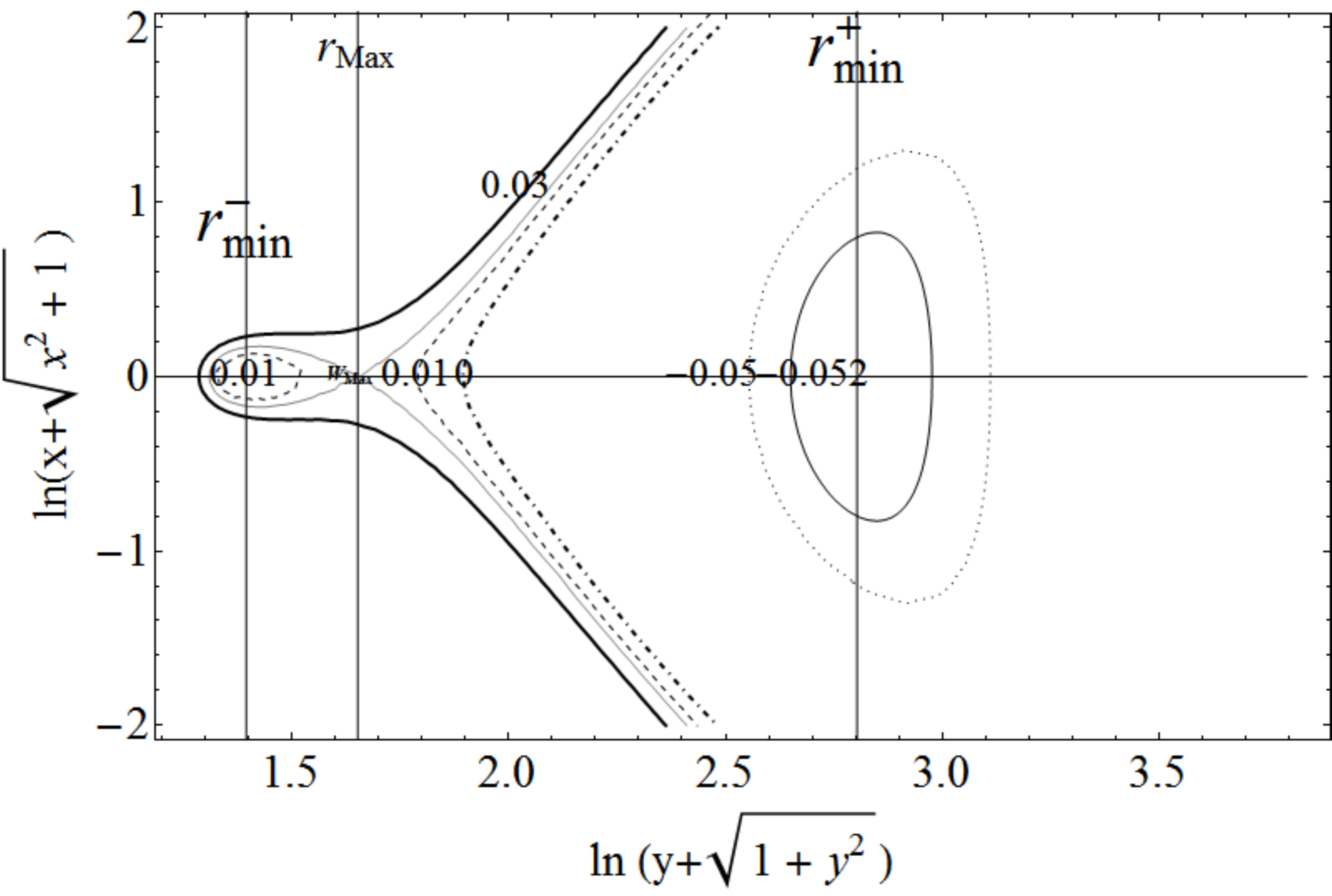}
\caption{Region III-a: $\omega\in]\omega_{mbo},\omega_c[$, naked singularity  $\omega=0.34$. Where $\omega M^2\rightarrow \omega$. It  $l_{mso}^->l^-_{mbo}> l^+_{mbo}>l_{\Omega}^{Max}>l_{mso}^+$, and $r_{stat}<r_{\Omega}^{Max}<r_{mbo}^-<r_{mso}^-<r_{mbo}^+<r_{mso}^+$, where $r/M=\sqrt{x^2+y^2}$ and  $(x,y)$ are Cartesian coordinates.  Vertical lines in right panels set the $r_i\in\mathfrak{R}$ and  the effective potential critical points.}
\label{Fig:lun-1}
\end{figure}

\begin{figure}[h]
%\\
\includegraphics[width=.481\textwidth]{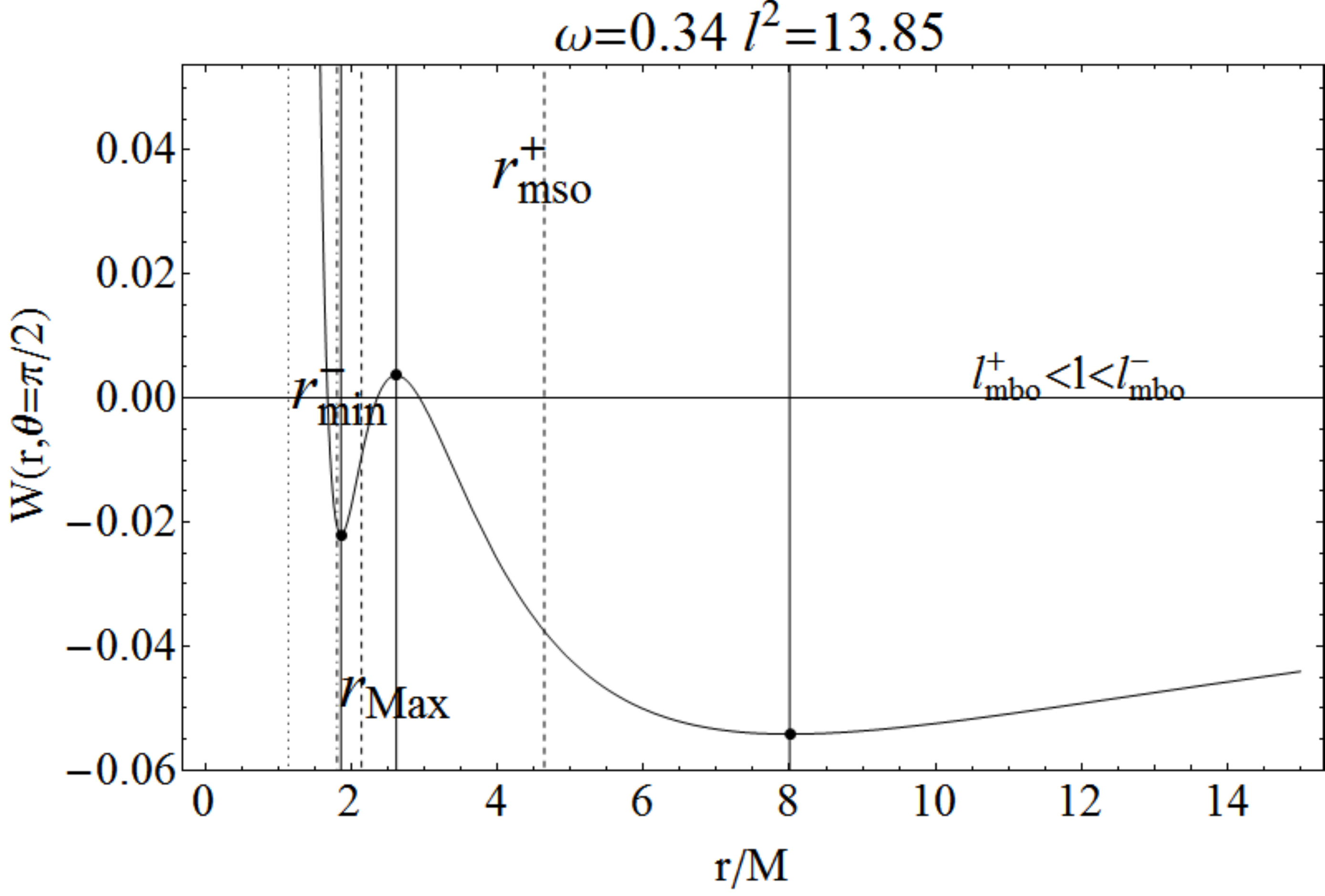}
\includegraphics[width=.481\textwidth]{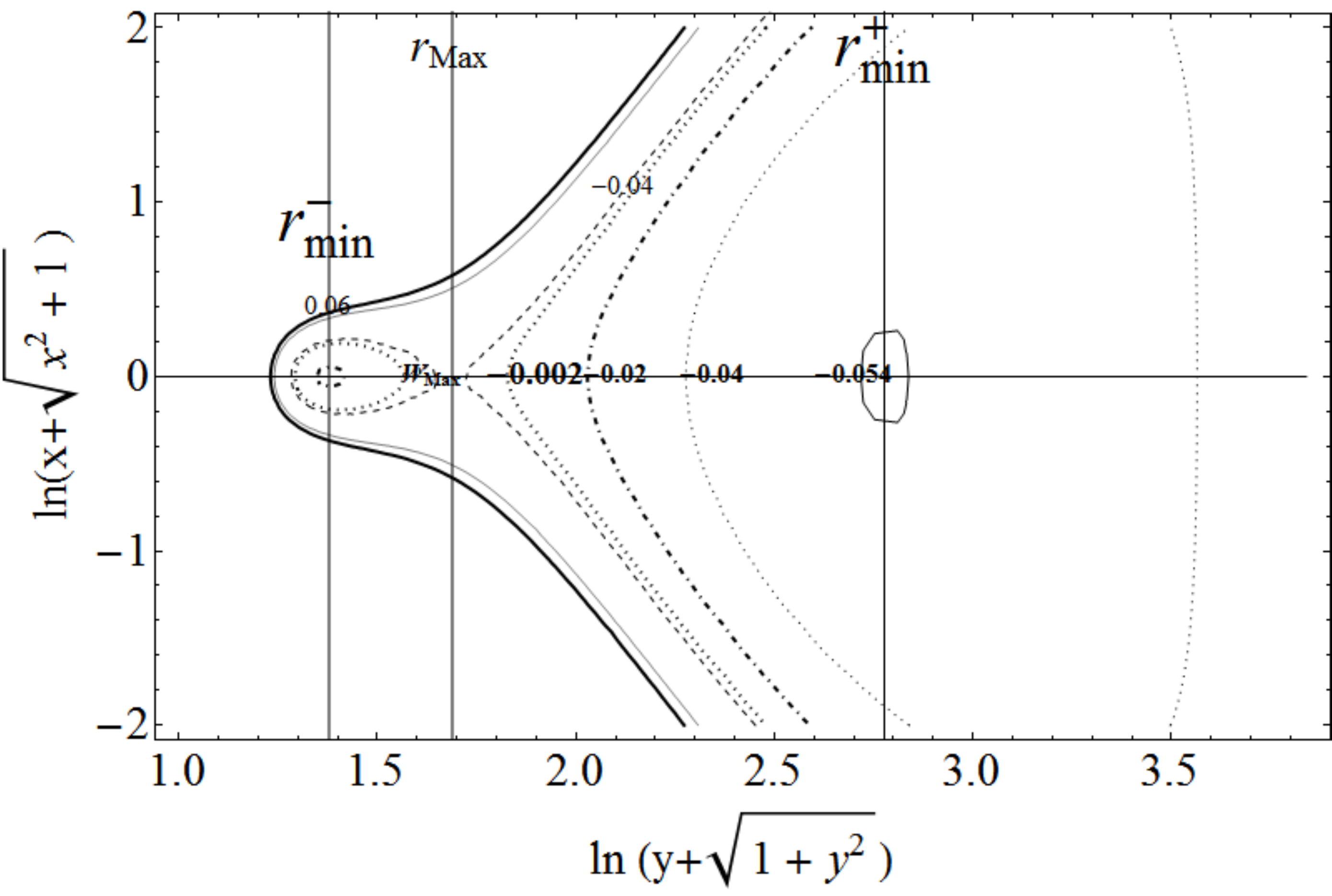}
\\
\includegraphics[width=.481\textwidth]{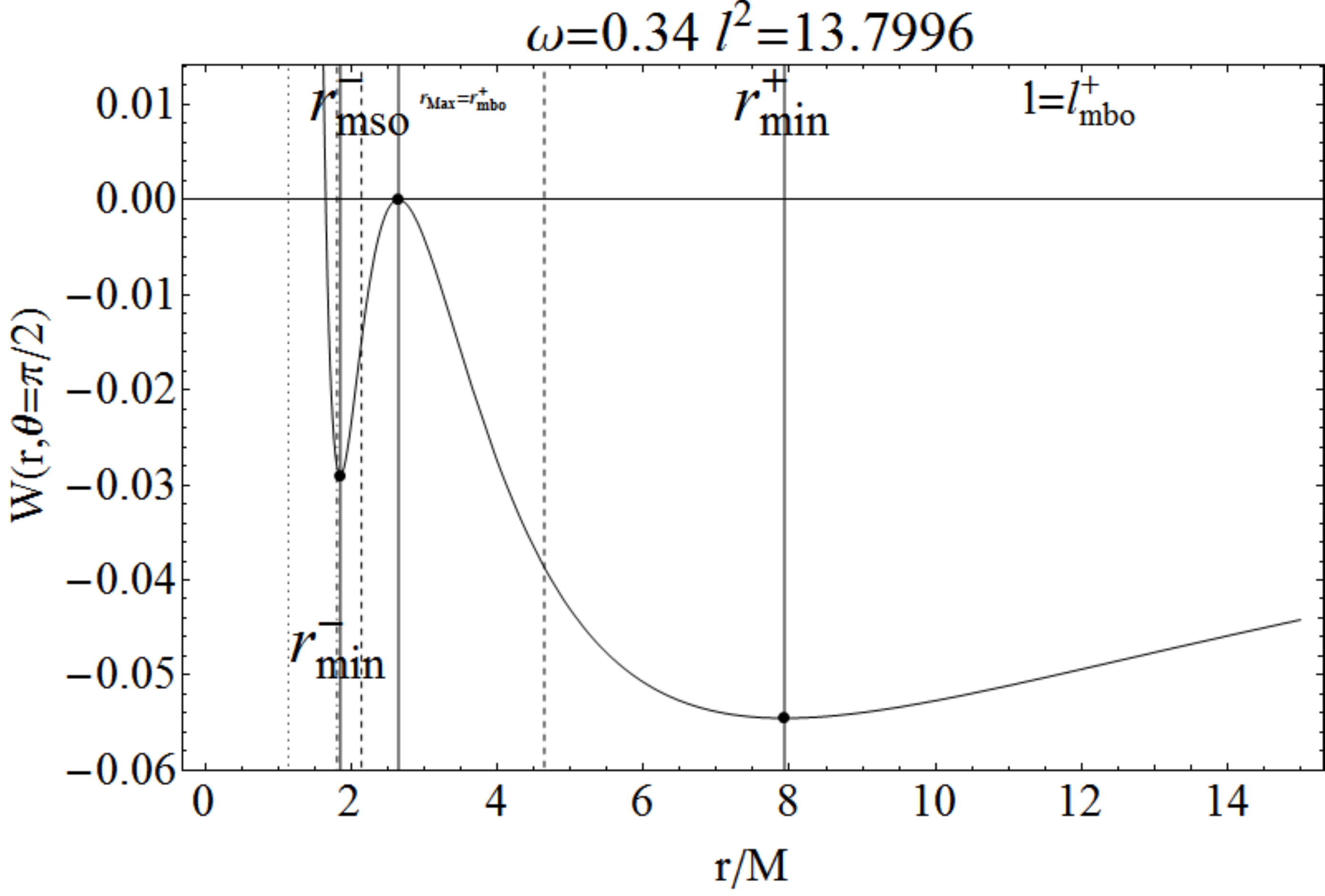}
\includegraphics[width=.481\textwidth]{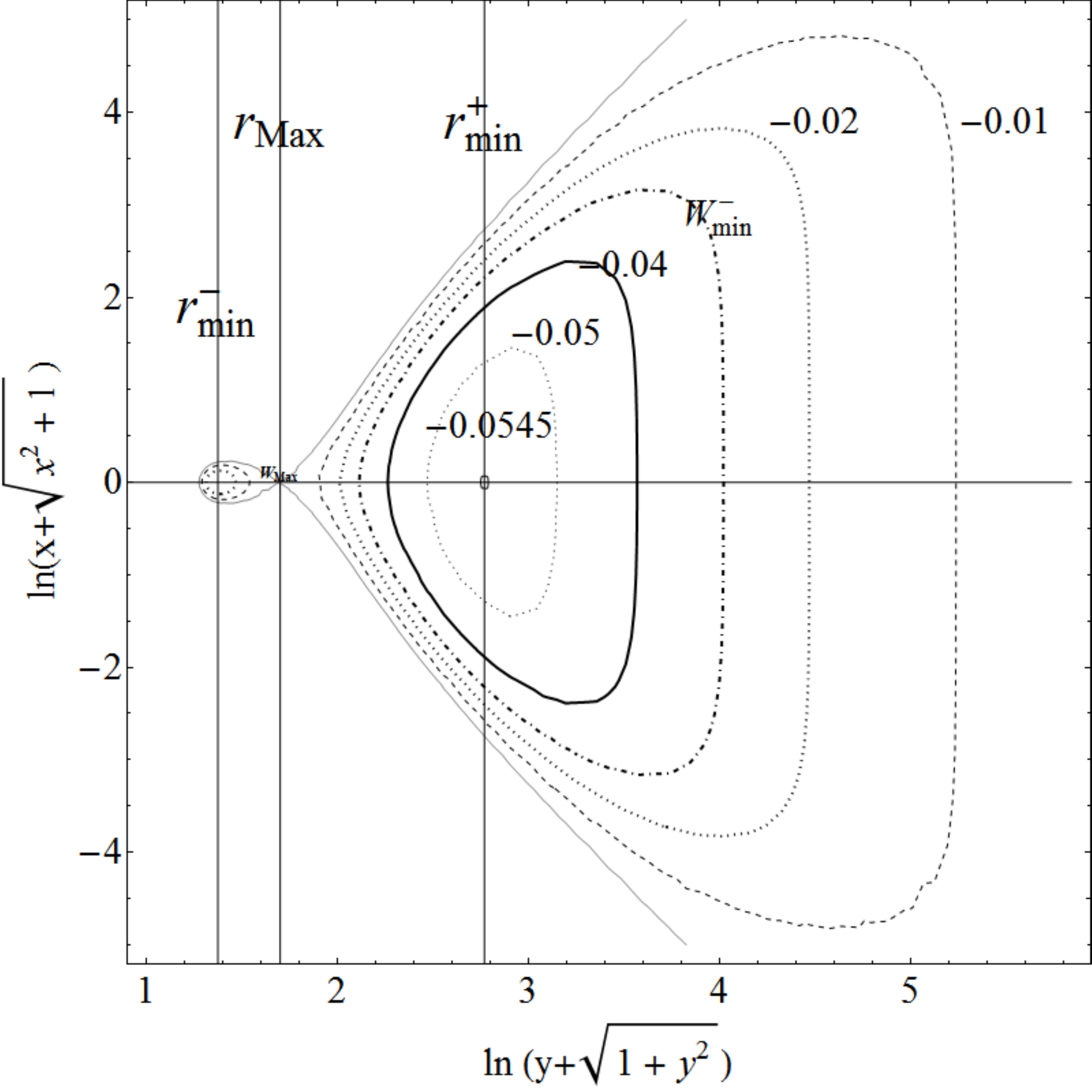}
\\
\includegraphics[width=.481\textwidth]{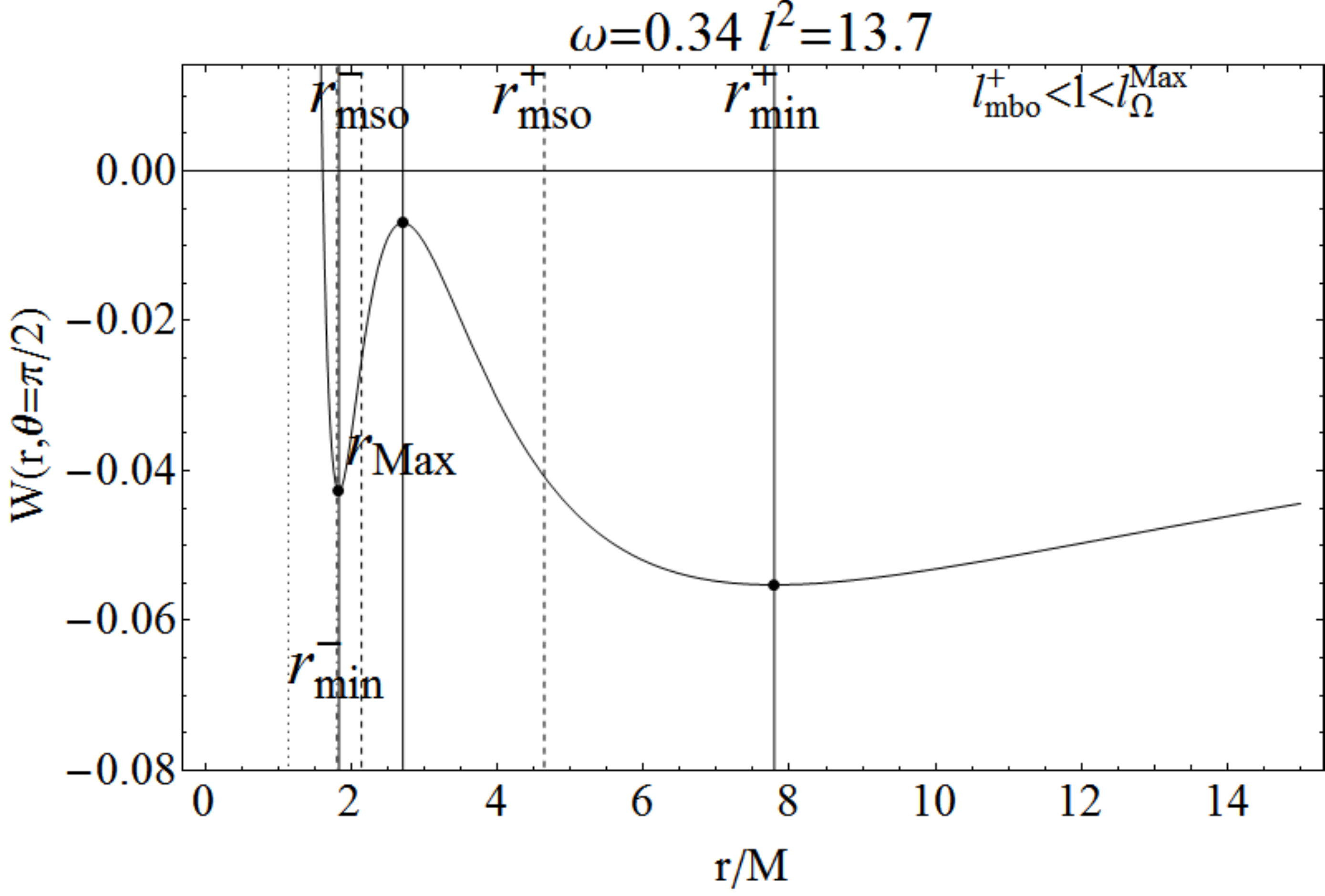}
\includegraphics[width=.481\textwidth]{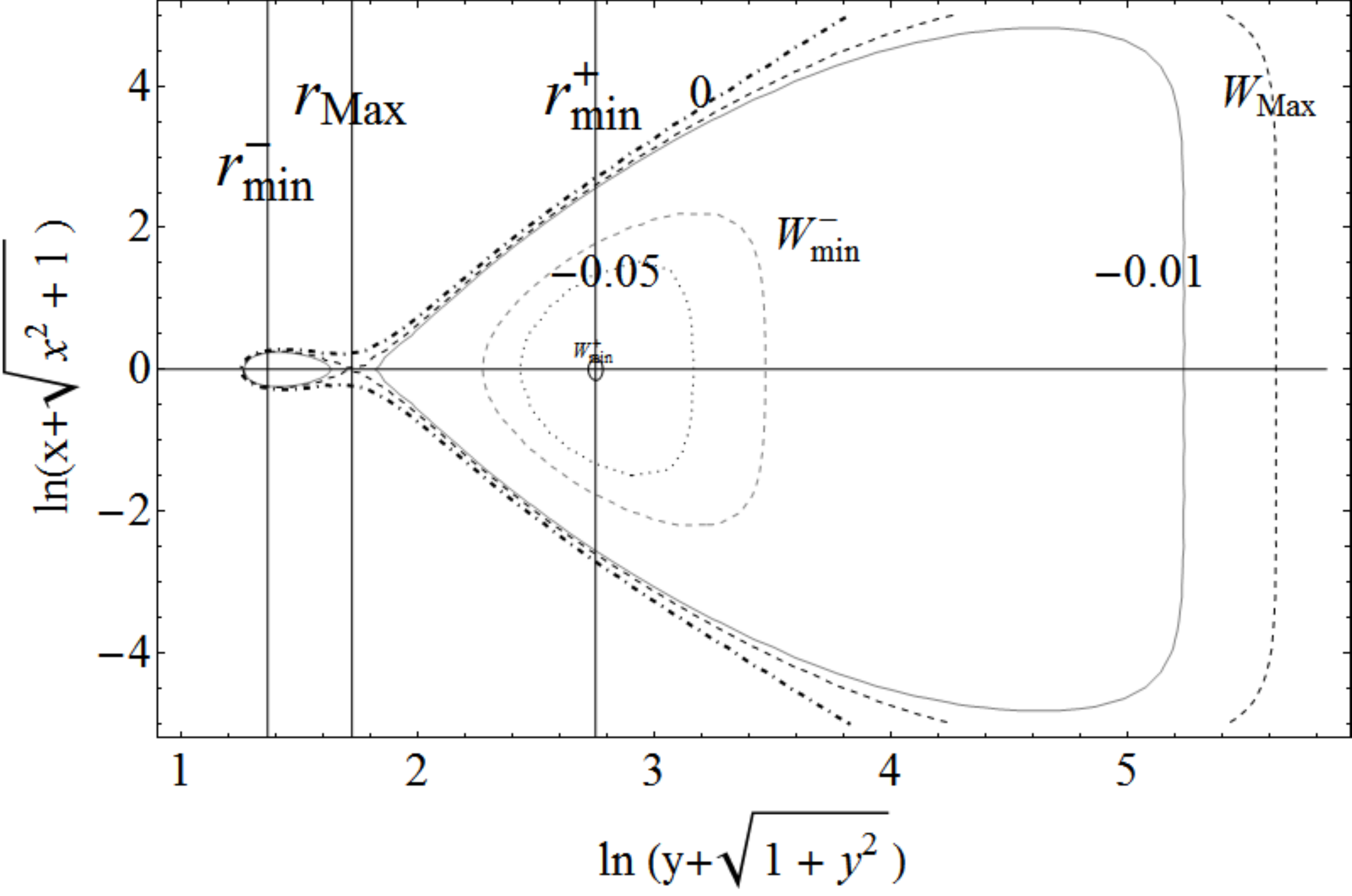}
\caption{ Region III-a: $\omega\in]\omega_{mbo},\omega_c[$ naked singularity  $\omega M^2=0.34$. Where $\omega M^2\rightarrow \omega$.  Vertical lines in right panels set the $r_i\in\mathfrak{R}$ and  the effective potential critical points. It  $l_{mso}^->l^-_{mbo}> l^+_{mbo}>l_{\Omega}^{Max}> l_{mso}^+$, and $r_{stat}<r_{\Omega}^{Max}<r_{mbo}^-<r_{mso}^-<r_{mbo}^+<r_{mso}^+$, where $r/M=\sqrt{x^2+y^2}$ and  $(x,y)$ are Cartesian coordinates.}
\label{Fig:Rip-second}
\end{figure}

\begin{figure}[h]
%\\
\includegraphics[width=.481\textwidth]{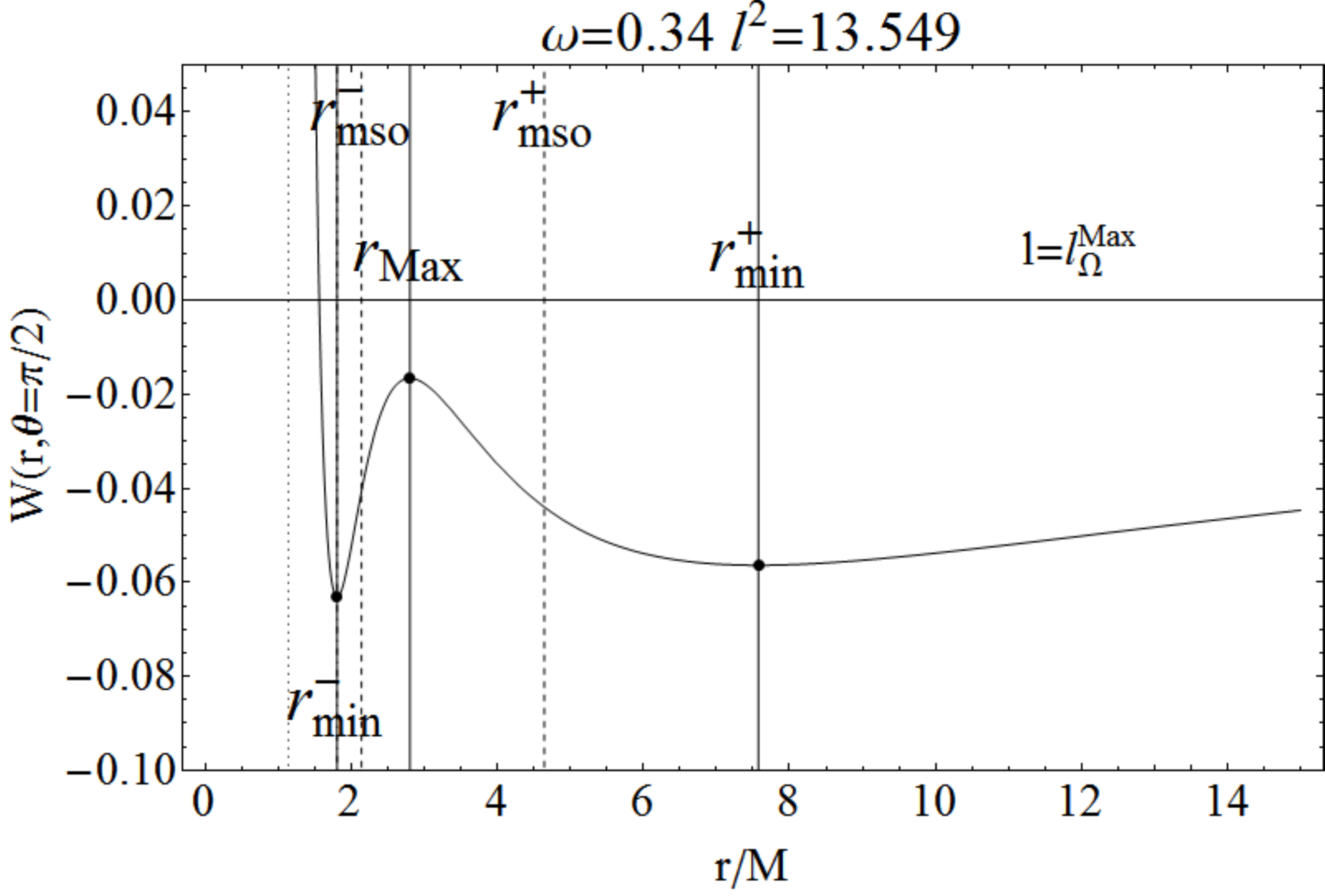}
\includegraphics[width=.481\textwidth]{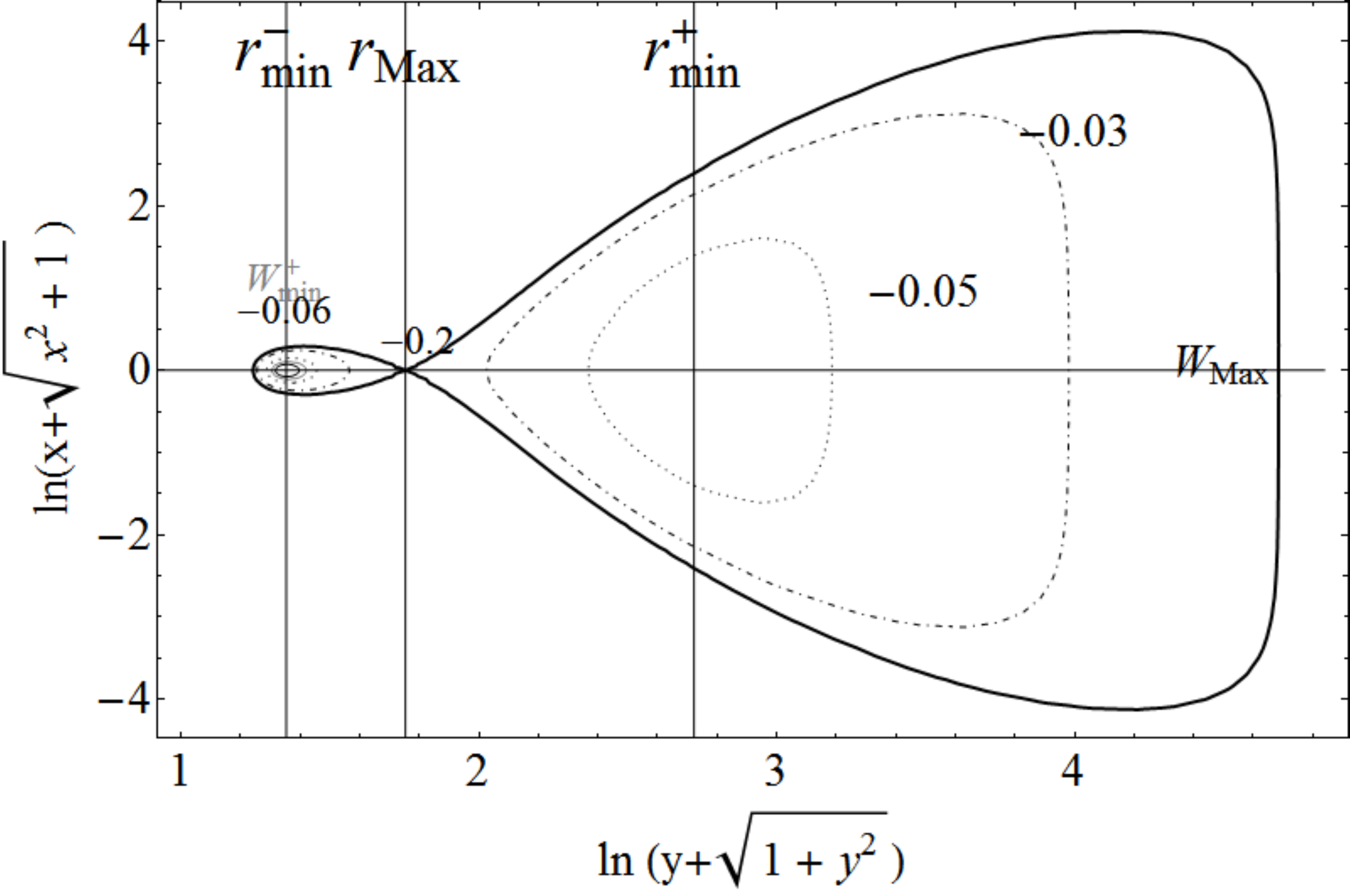}
\\
\includegraphics[width=.481\textwidth]{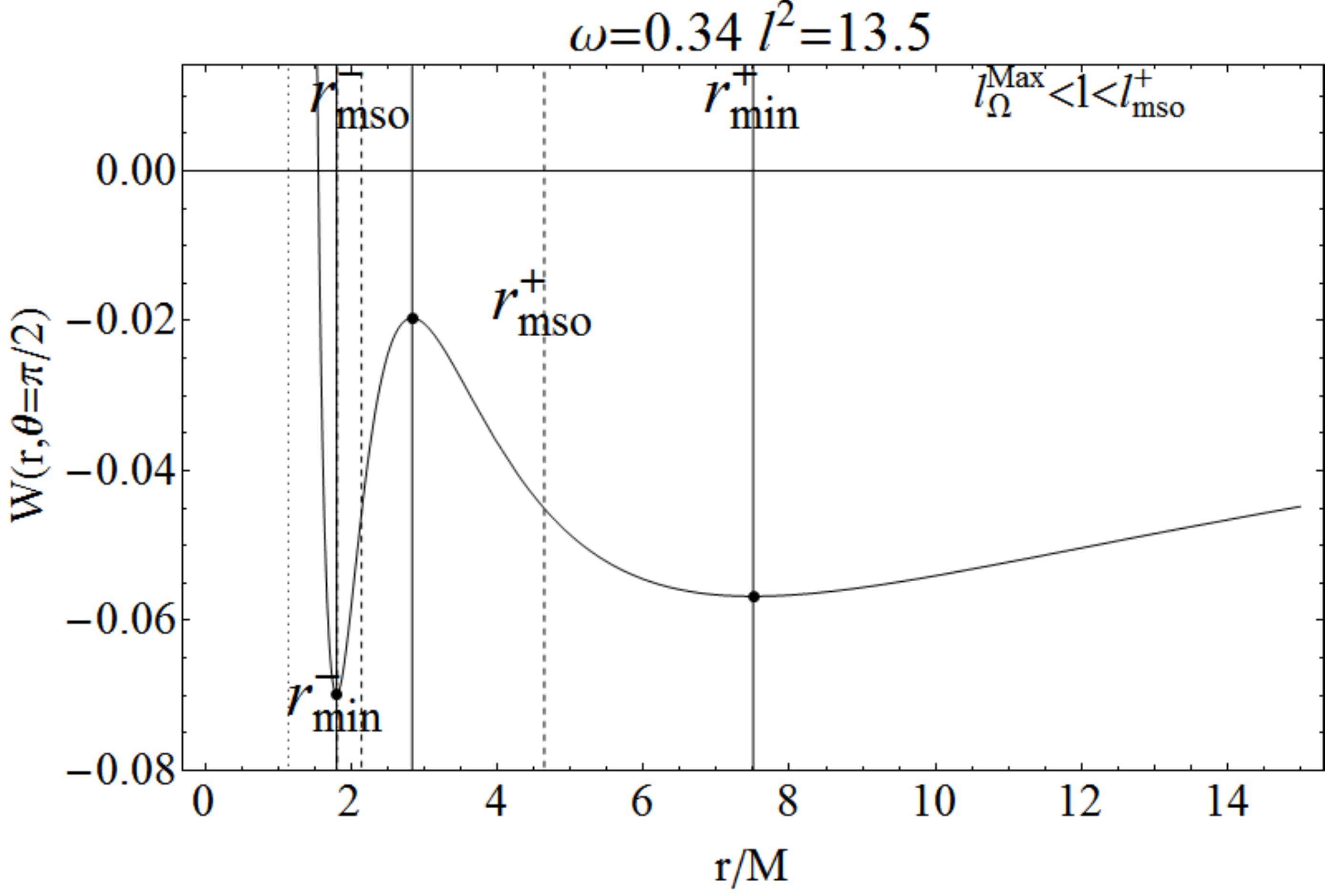}
\includegraphics[width=.481\textwidth]{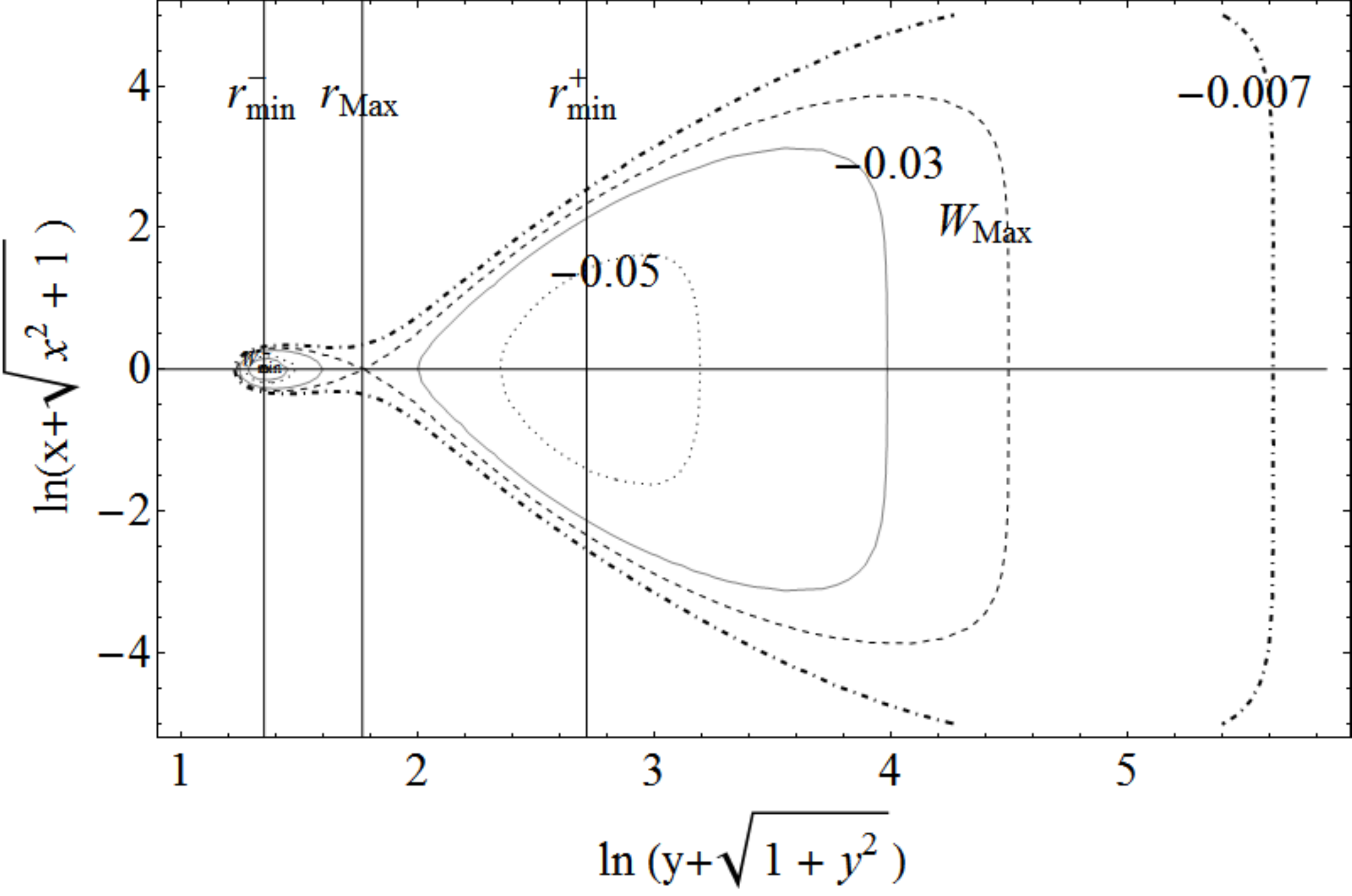}
\\
\includegraphics[width=.481\textwidth]{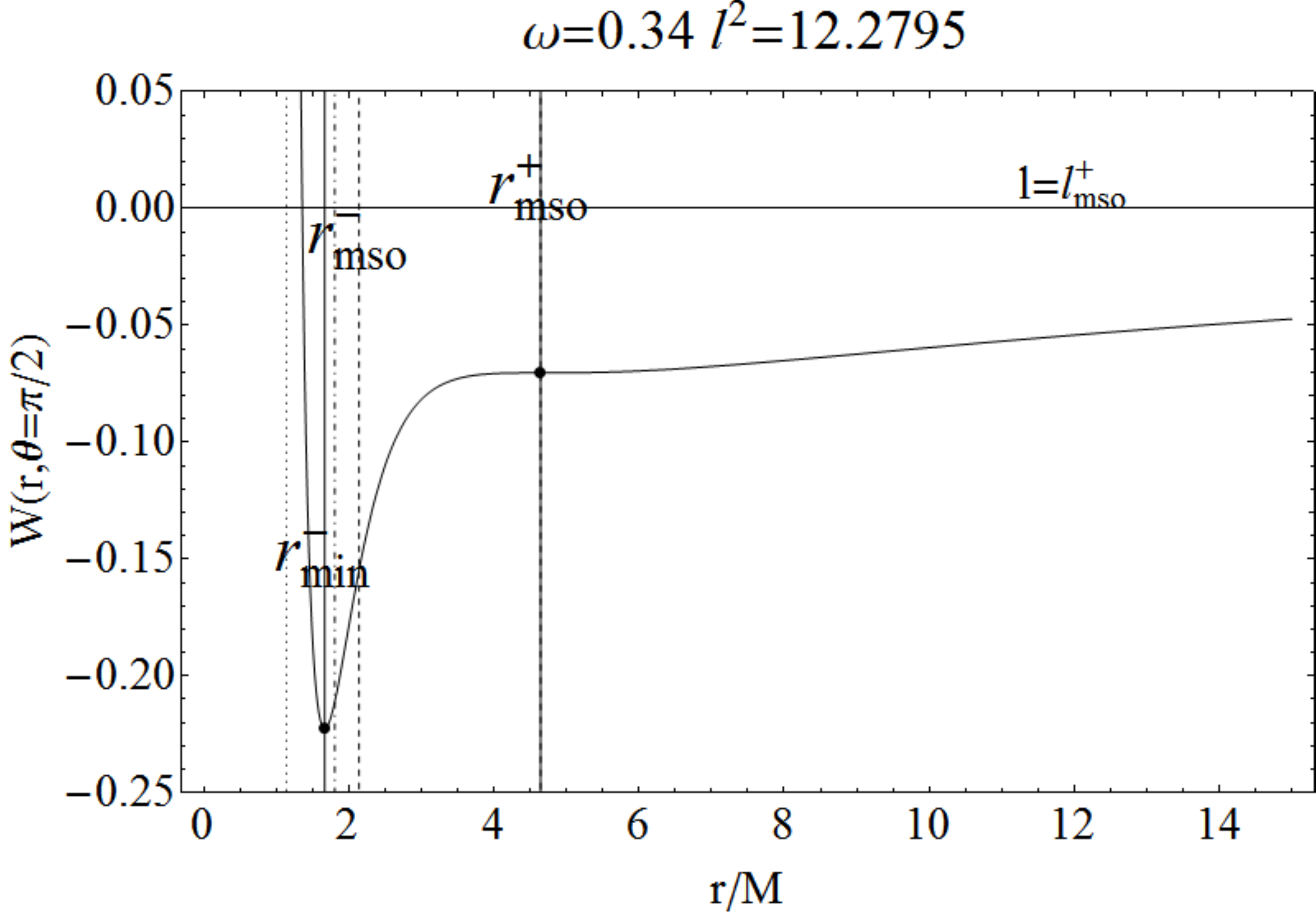}
\includegraphics[width=.481\textwidth]{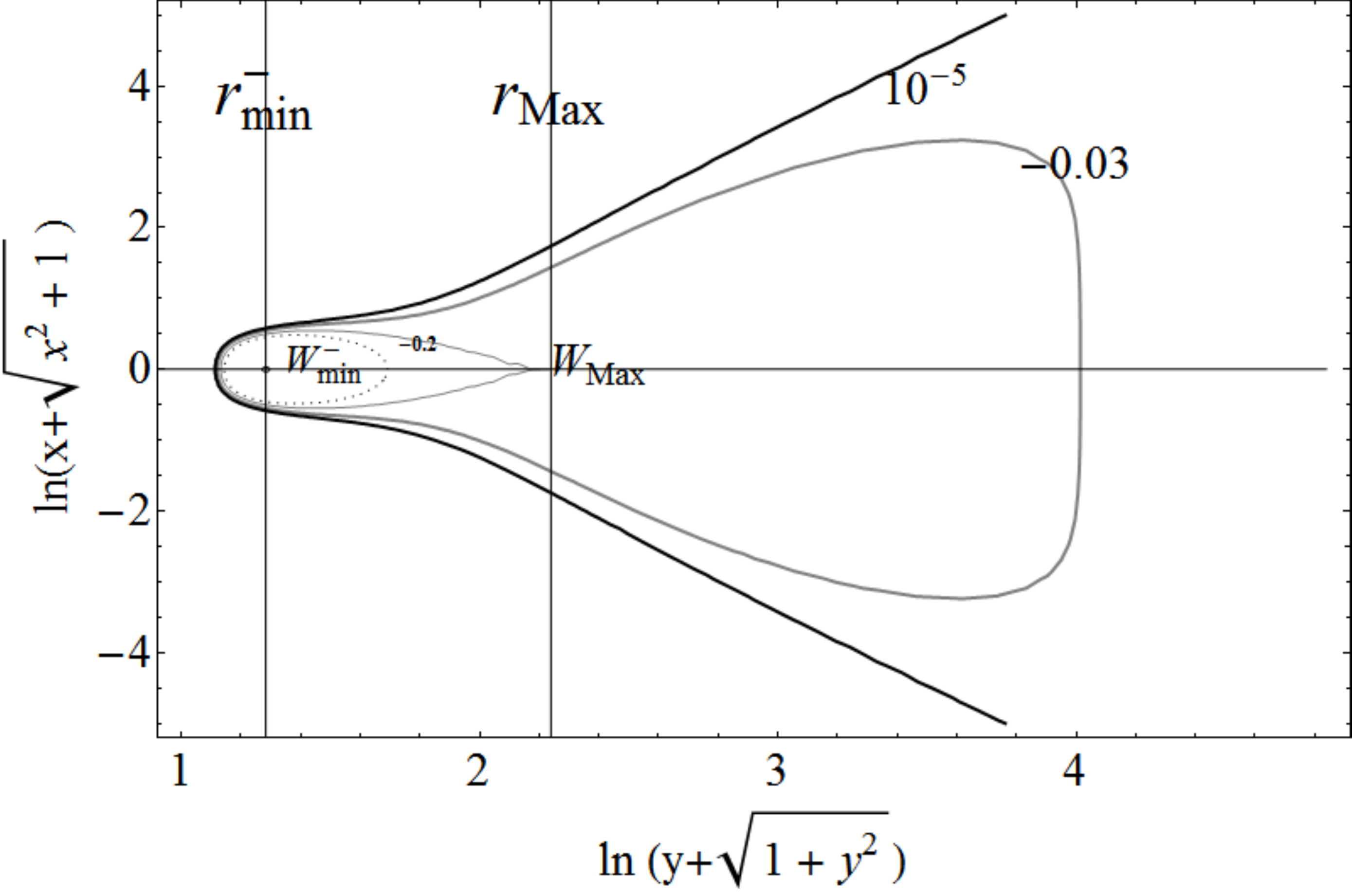}
\\
\includegraphics[width=.481\textwidth]{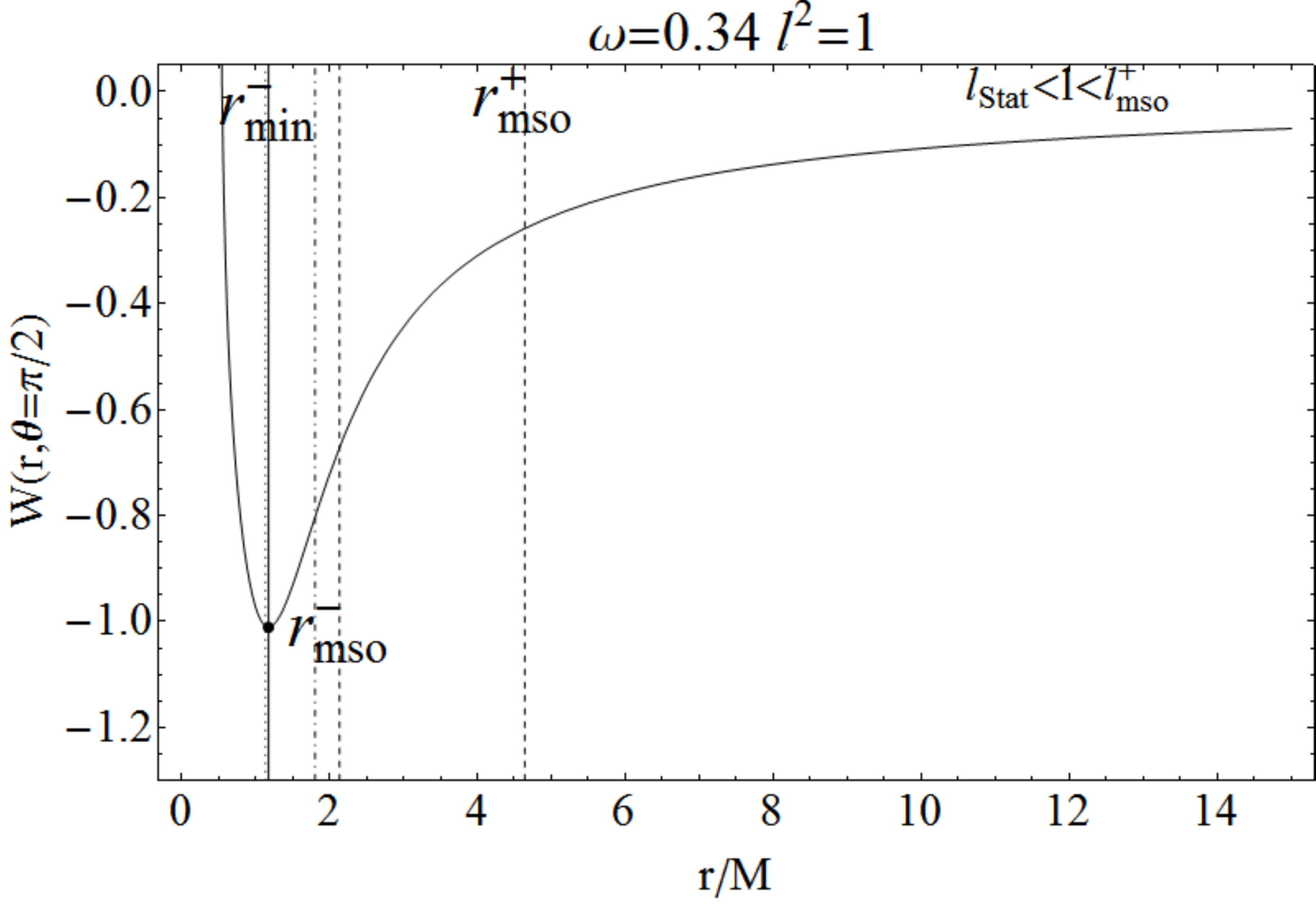}
\includegraphics[width=.481\textwidth]{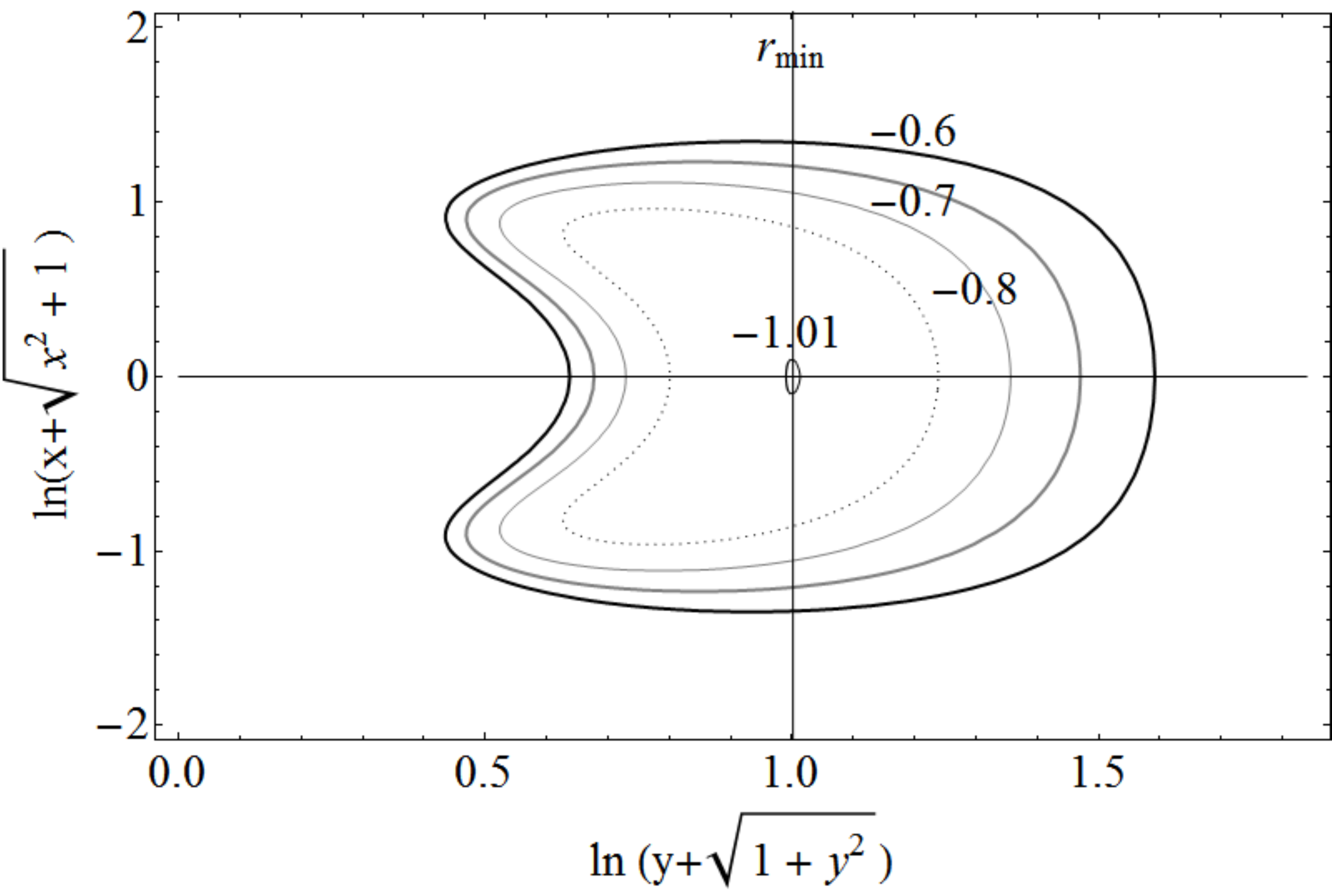}
\caption{ Region III-a: $\omega\in]\omega_{mbo},\omega_c[$, naked singularity $\omega M^2=0.34$. Where $\omega M^2\rightarrow \omega$. It  $l_{mso}^->l^-_{mbo}> l^+_{mbo}>l_{\Omega}^{Max}> l_{mso}^+$, and $r_{stat}<r_{\Omega}^{Max}<r_{mbo}^-<r_{mso}^-<r_{mbo}^+<r_{mso}^+$.  Vertical lines in right panels set the $r_i\in\mathfrak{R}$ and  the effective potential critical points. It is $r/M=\sqrt{x^2+y^2}$ and  $(x,y)$ are Cartesian coordinates.}
\label{Fig:ira-n-b}
\end{figure}

In Kehagias-Sfetsos naked singularity spacetimes  with Ho\v{r}ava parameter in \textbf{Region III-a}   the situation is the following:
it  is $r_{stat}<r_{\Omega}^{Max}<r_{mbo}^-<r_{mso}^-<r_{mbo}^+<r_{mso}^+$ where $l_{mso}^->l^-_{mbo}> l^+_{mbo}>l_{\Omega}^{Max}>l_{mso}^+$ and $W_{mbo}^{\pm}=0$. Naked singularities  in  \textbf{Region III-a} belong to the attractors of Class II, and the structure of the test particular circular motion has been therefore discussed in Sec.\il(\ref{Subsec:ClassII}).  Toroidal configurations in these naked singularity spacetimes are characterized by the {presence of excretion points}. In details:
\begin{description}
\item[-)$l>l_{mso}^-$] There are only stable  toroidal surfaces as in Fig.\il(\ref{Fig:newfigokey}-a).
\item[-) $l=l_{mso}^-$] There is a saddle point located in $r_{mso}^-$, see Fig.\il(\ref{Fig:lun-1})-a, where potential $W_{mso}>0$, a minimum is  at $r_{min}^+>r_{mso}^+$: toroidal stable configurations are as $W\in]W_{min}^+,0[$, only one ``critical'' open surface is located in $r_{mso}^-$, that corresponds therefore to an ``inner cusp''. We could compare this situation with the cases $l=l_{mso}^-$  at $\omega=\omega_{mso}$ in Fig.\il(\ref{Fig:Sisecon1})-b where $W_{mso}^-=0$. No accretion into the singularity is in any case allowed.
\item[-) ${l\in]l_{mbo}^-, l_{mso}^-[}$]  The inner minimum  is located at $r_{min}^-:\; W_{min}^->0$, there is a maximum $r_{Max:\; }W_{Max}>0$ and the outer minimum  is in $r_{min}^+:\; W_{min}^+<0$.
A toroidal  set of closed and  stable configurations is  at $W\in]W_{min}^+, 0[$: as  the potential increases the surface stretches along the equatorial plane and thickens so far it approaches the asymptotic superior limit of $W=0$.
A further set of inner closed $C^-$ configurations is  centered at $r_{min}^-$: as it is  $(W_{min}^->0,W_{min}^+<0)$  there are no  double closed configuration at equal $W$ and $l$. According to the tori classification proposed for configurations with  ${ l\in ] l_{mso}^+,l_{mso}^-[ }$  in  \textbf{Region II} (Fig.\il(\ref{Fig:newfigokey})-a ), the torus model illustrated  in Fig.\il(\ref{Fig:lun-1})-b is  considered to be of the type
\textbf{I}  as it is $W_{min}^->W_{min}^+$.
As long as $W$ increases towards its maximum an outer cusp appears, leading  to a crossed  point, opened towards the exterior and therefore to an excretion of matter.
Finally we note that the outer (stable) configuration can be  larger and  thicker  than the inner (unstable and excreting) one. The thickness as well as other geometrical torii  features are regulated  by the gaps $\Delta_i\equiv |W_{min}^--W_{Max}|<\Delta_o\equiv |W_{min}^+-0|$ and $\Delta^r_i\equiv|r_{min}^--r_{Max}|$ and the location of the crossing points $r_{x}^o:\;\left.W(r^o_x)\right|_{\theta=\pi/2}=0$, and  $r_{x}^i:\;\left.W(r^i_x)\right|_{\theta=\pi/2}=W_{min}^-$.
\item[-)$l=l_{mbo}^-$] This case is illustrated in Fig.\il(\ref{Fig:lun-1})-c is a limiting case respect to Fig.\il(\ref{Fig:lun-1})-b as  it is $W_{min}^-=0$ and  the situation is analogue. We could also compare with the configurations  with $l=l_{mbo}^-$ at $\omega=\omega_{mso}$ in  Fig.\il(\ref{Fig:Sisecon1})-b
\item[-)${l\in]l_{mbo}^+,l_{mbo}^-[}$ ] See Fig.\il(\ref{Fig:Rip-second})-a. The values $W_{min}^-<0$ and $W_{Max}>0$ decrease  with the fluid angular momentum: there are one closed stable configuration as $W\in]W_{min}^+,W_{min}^-[$,  double closed stable configurations as $W\in]W_{min}^-,0[$, the value $W=0$ corresponds to a centered closed configuration $C^-$ centered  in $r_{min}^-$ and an open outer one, as $W\in]0,W_{Max}[$, there are closed stable  $C^+$ toroidal configurations  with center in $r_{min}^+$, finally at $W_{Max}$ a crossed surface with an outer cusp appears, this configuration is associated with the excretion. These are $\mathbf{I}$-configurations.
\item[-) $l=l_{mbo}^+$] In this case a the maximum of the configuration is located at $r_{mbo}^+$ as such $W_{max}=0$. The situation, sketched in Fig.\il(\ref{Fig:Rip-second})-b is qualitatively similar to Fig.\il(\ref{Fig:Rip-second})-a. This is  a  $\mathbf{I}$-configuration.
\item[-) ${l\in]l_{\Omega}^{Max},l_{mbo}^+}$] It is $W_{Max}<0$. At $W=W_{Max}$ there is a double centered  crossed closed configuration,  Fig.\il(\ref{Fig:Rip-second})-c and this is  a  $\mathbf{I}$-configuration%The situation is similar to
\item[-) $l=l_{\Omega}^{Max}$] It is $r_{min}^-=r_{\Omega}^{Max}$, consequently there is an inner set of closed $C^-$ surfaces centered in $r_{\Omega}^{Max}$, see Fig.\il(\ref{Fig:ira-n-b})-a. In this case it is $W_{min}^-<W_{min}^+<0$, therefore it is a \textbf{III}-configurations. There is  continuum shift between a \textbf{I} and \textbf{III}  configuration, with  a \textbf{II} type with  $W_{min}^-=W_{min}^+<0$. There are no excretion point.
\item[-) ${l\in]l_{mso}^+,l_{\Omega}^{Max}[ }$] This plot is sketched in Fig.\il(\ref{Fig:ira-n-b})-b. This situation is analogue to Fig.\il(\ref{Fig:ira-n-b})-a, there is a couple $C^{\pm}$ of closed \textbf{III}  configurations.
\item[ -) $l=l_{mso}^+$ ] There is a saddle point located in $r_{mso}^+$ This correspond to an  critical configuration, ``outer cusp'', as in  Fig.\il(\ref{Fig:ira-n-b})-c.
\item[-) ${l\in]l_{stat},l_{mso}^+[}$] There is only one minimum in $r_{min}^-$, and correspondingly one set of closed $C^-$  toroidal configurations centred in $r_{min}^-$ Fig.\il(\ref{Fig:ira-n-b}-d).
\end{description}
%%
%\begin{figure}[h]
%%\clearpage
%
\subsubsection{Naked singularity: $\omega=\omega_{c}$}\label{NSparc}
\begin{figure}[h]
%%CPlotoiscom
%\\
\includegraphics[width=.481\textwidth]{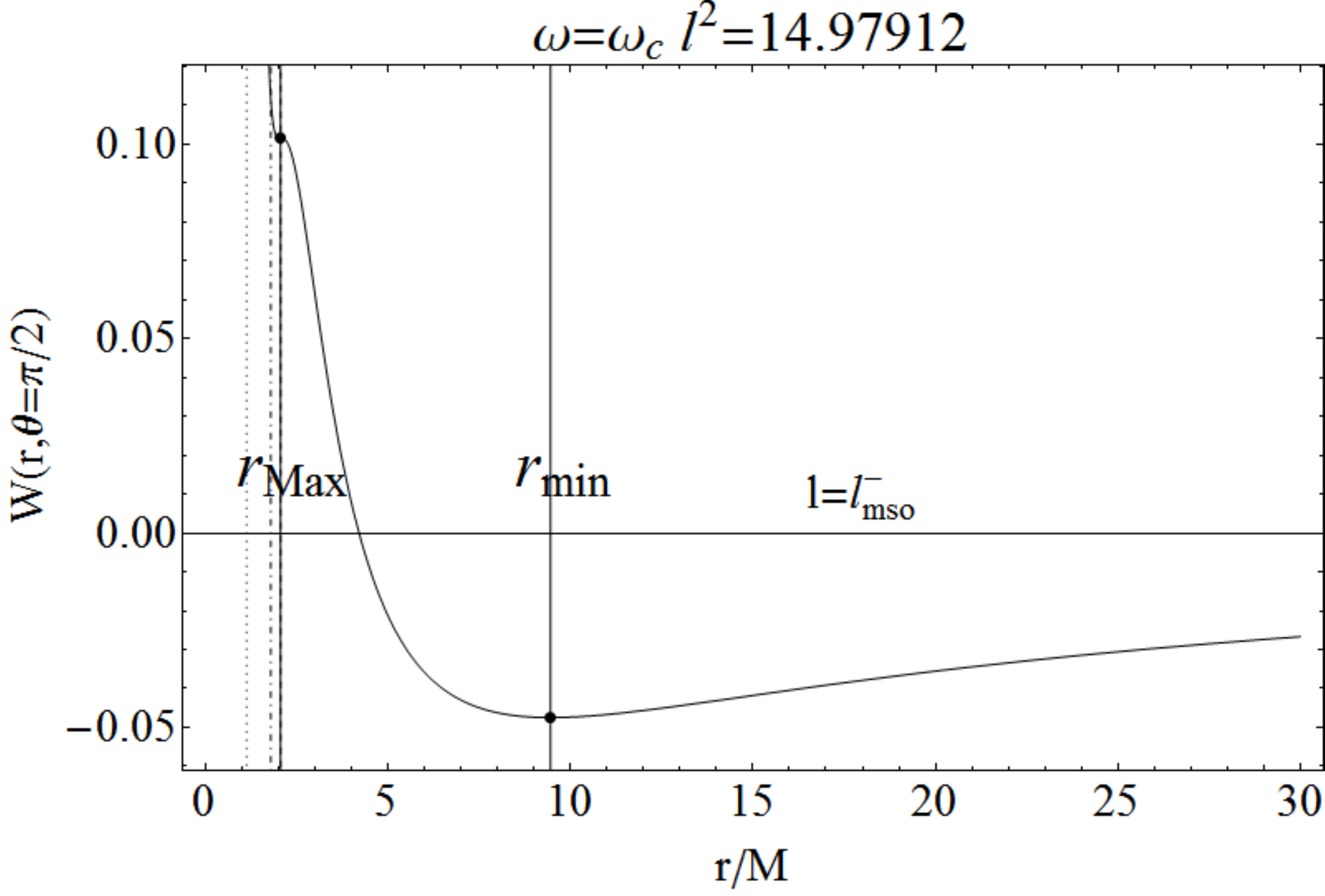}
\includegraphics[width=.31\textwidth]{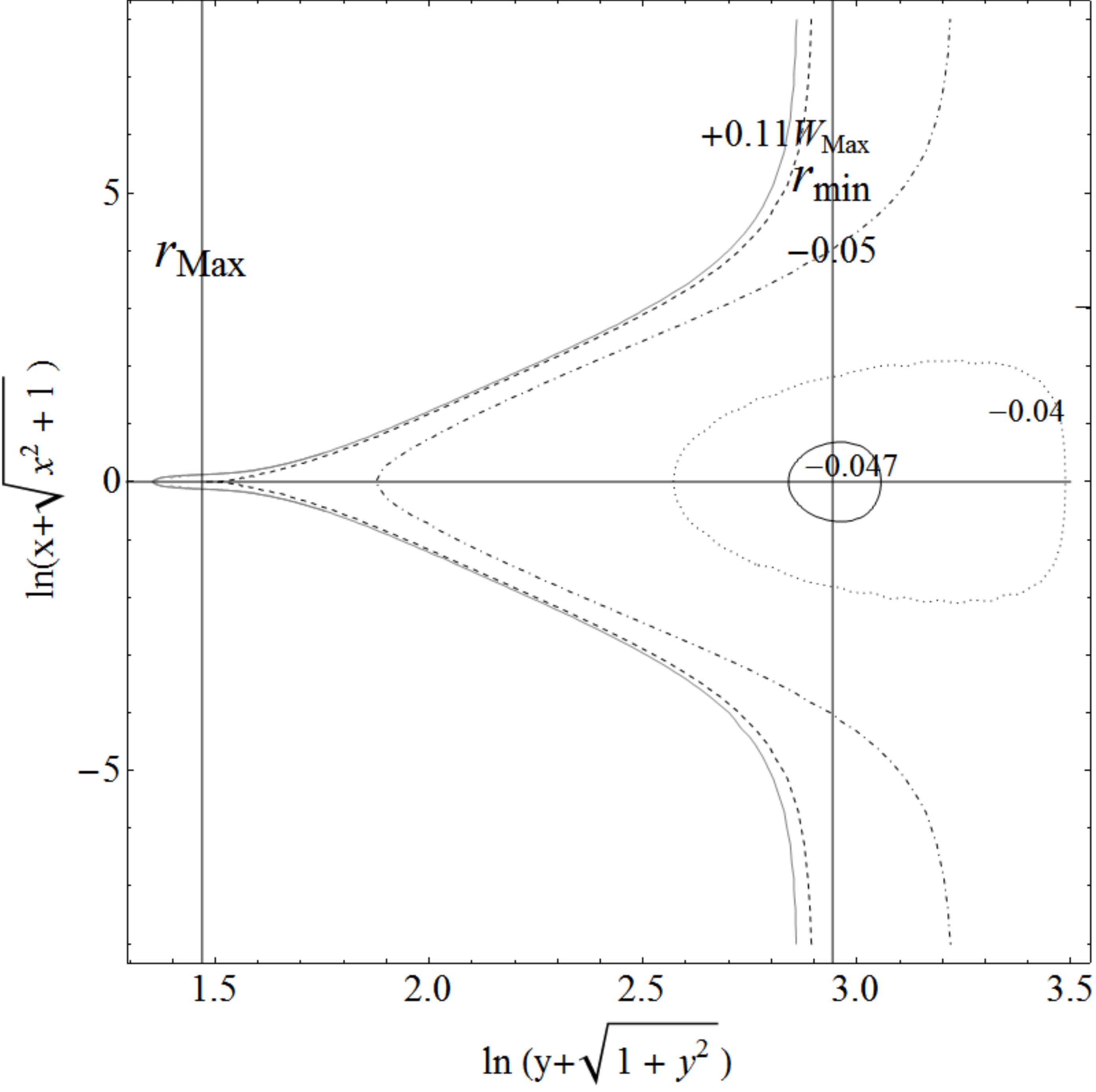}
\\
\includegraphics[width=.481\textwidth]{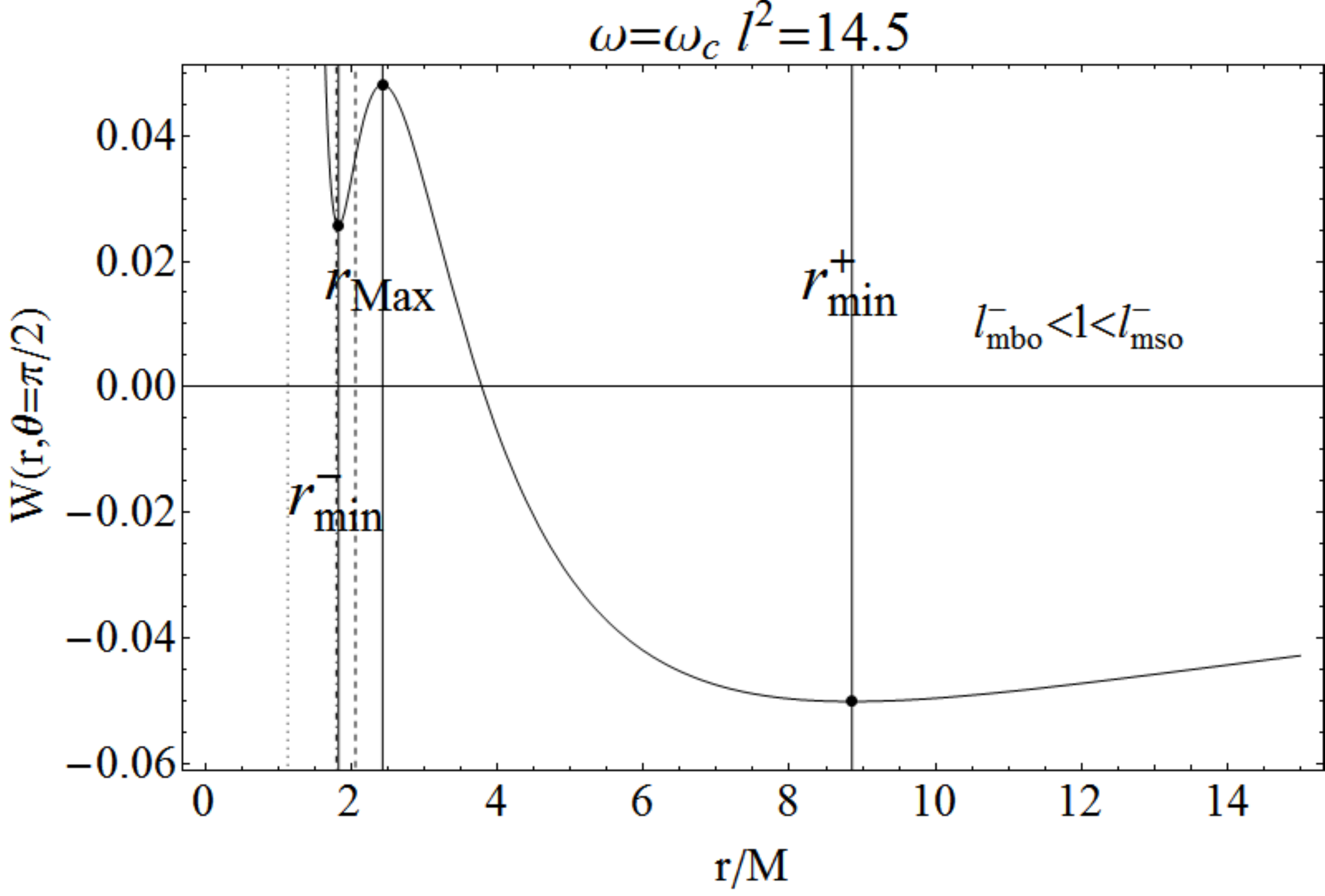}
\includegraphics[width=.41\textwidth]{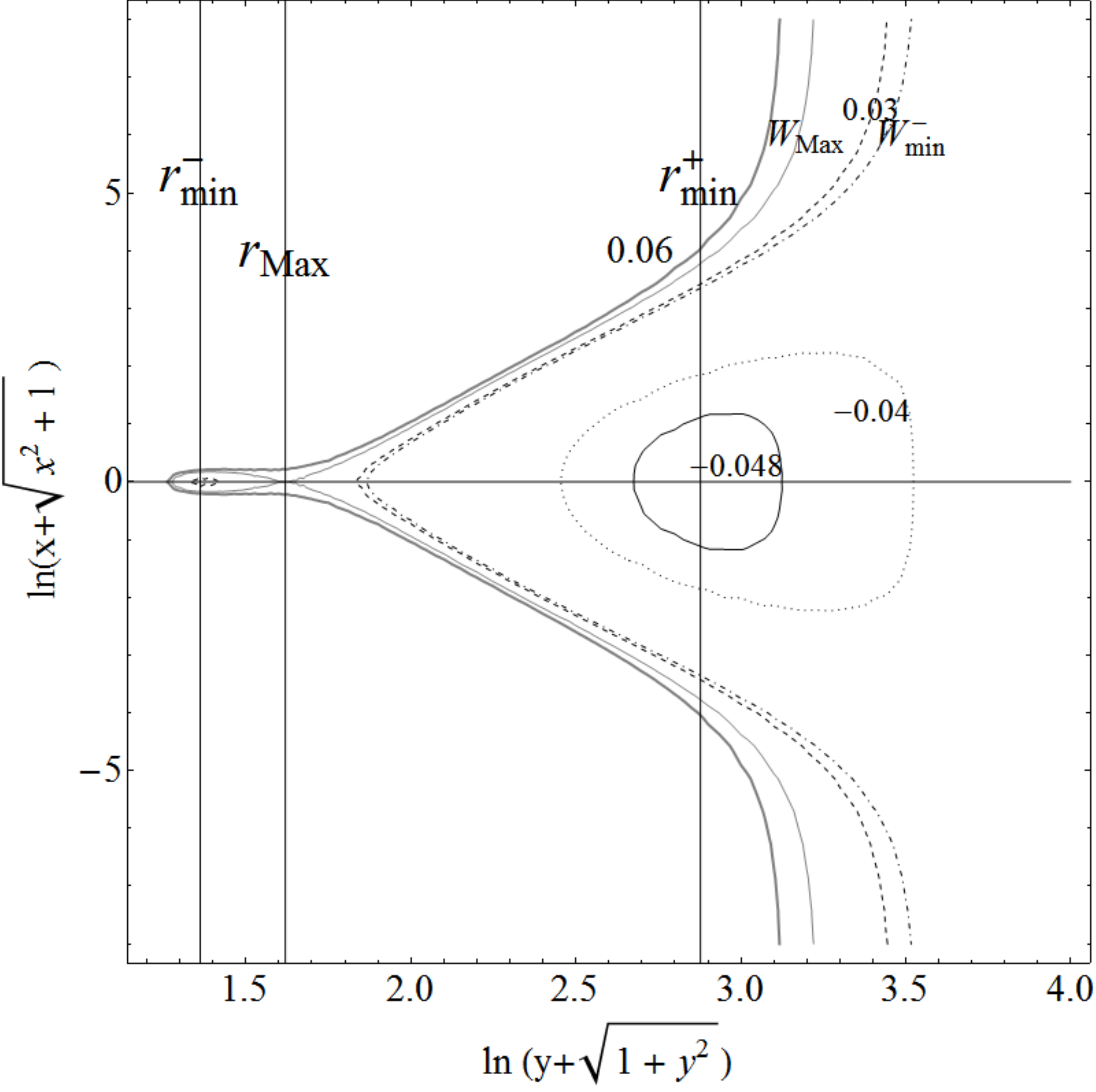}
\\
\includegraphics[width=.481\textwidth]{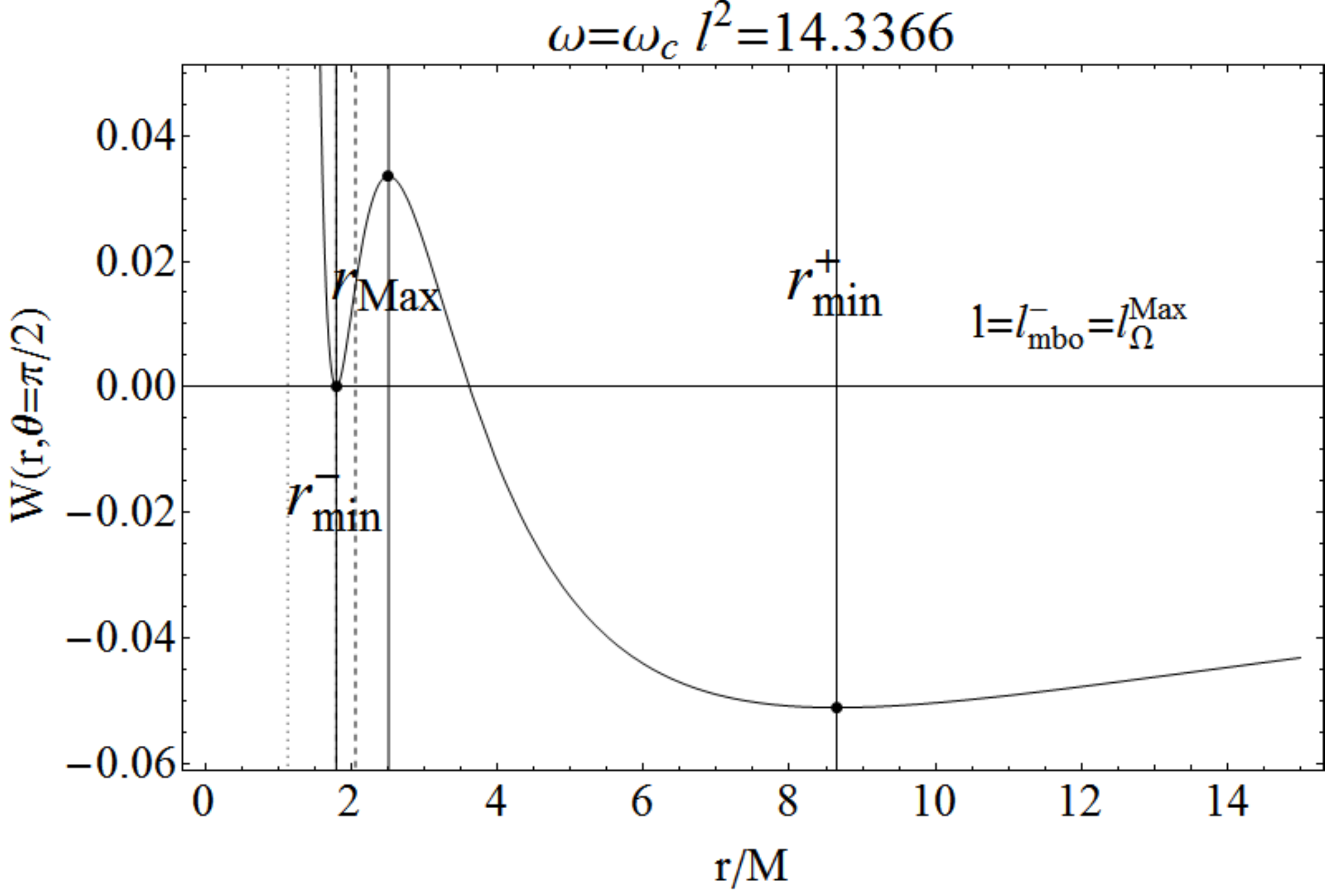}
\includegraphics[width=.481\textwidth]{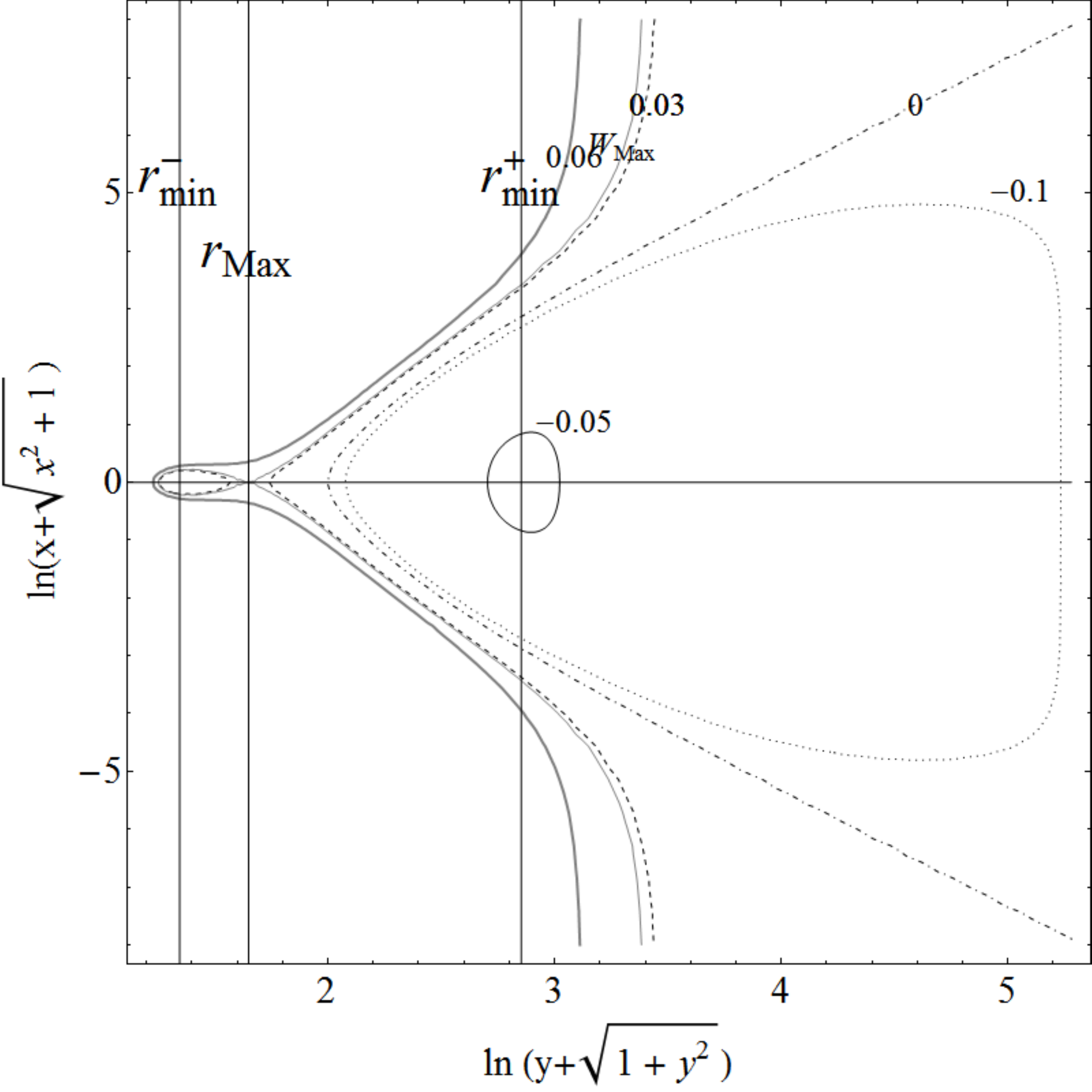}
\caption{ Naked singularity: $\omega M^2=\omega_{c} M^2=0.347506$. Where $\omega M^2\rightarrow \omega$. It  $l_{mso}^->l_{\Omega}^{Max}= l^-_{mbo}> l^+_{mbo}> l_{mso}^+$, and $r_{stat}<r_{mbo}^-=r_{\Omega}^{Max}<r_{mso}^-<r_{mbo}^+<r_{mso}^+$. It is $r/M=\sqrt{x^2+y^2}$ and  $(x,y)$ are Cartesian coordinates.  Vertical lines in right panels set the $r_i\in\mathfrak{R}$ and  the effective potential critical points.}
\label{Fig:namij-n}
\end{figure}

\begin{figure}[h]
%%CPlotoiscomPlotoocl14Plotoocl14
\includegraphics[width=.481\textwidth]{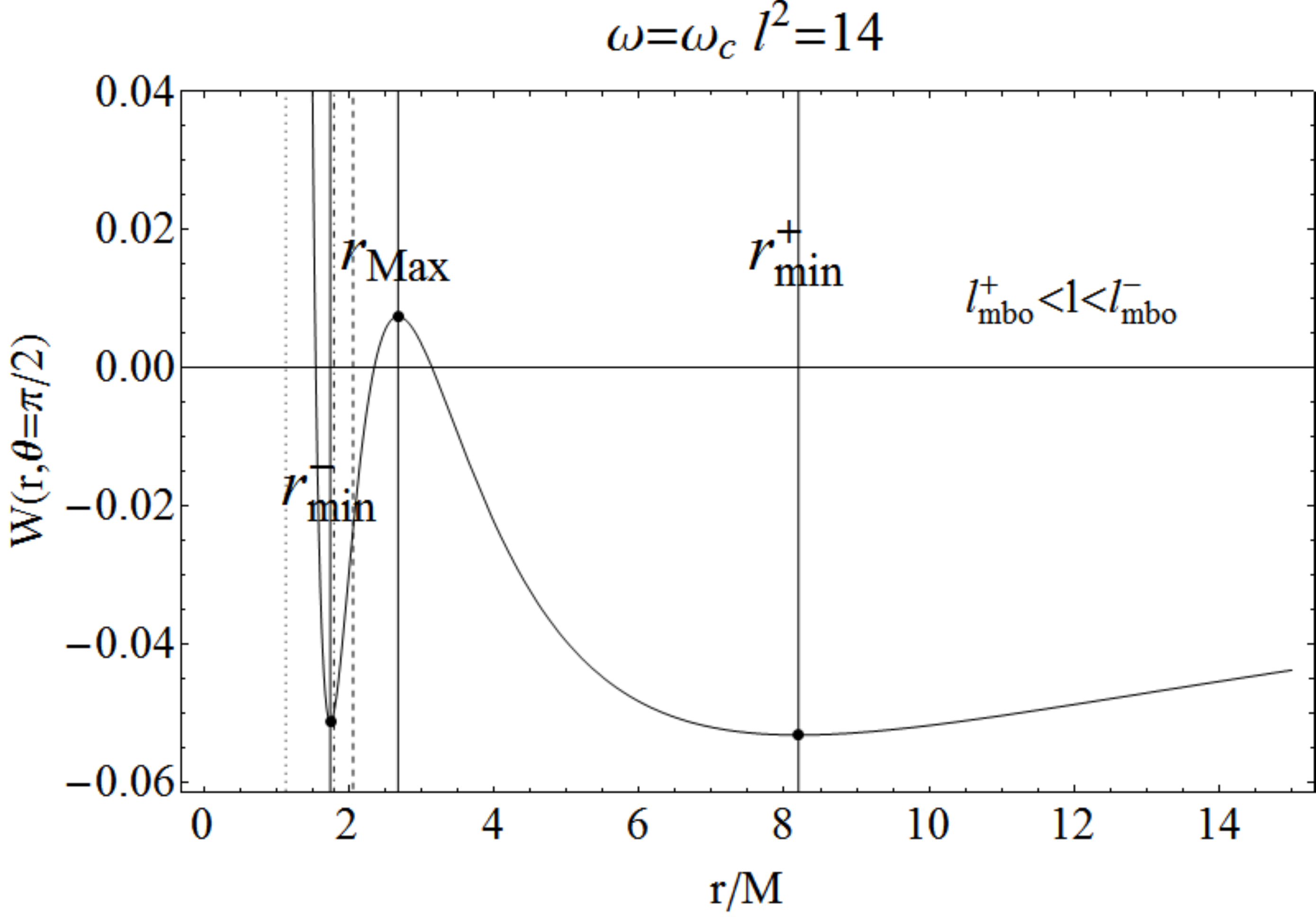}
\includegraphics[width=.31\textwidth]{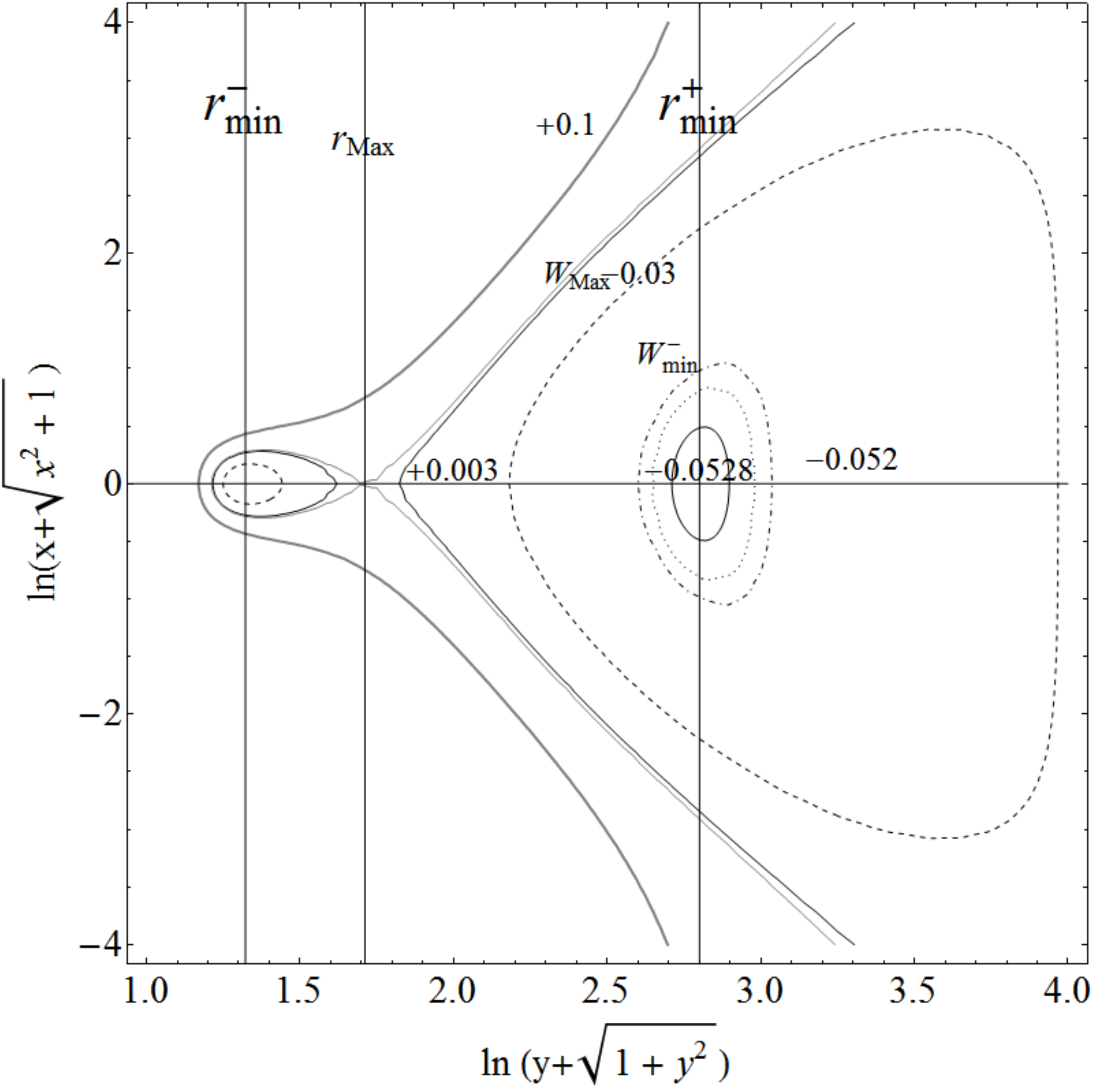}
\\
\includegraphics[width=.481\textwidth]{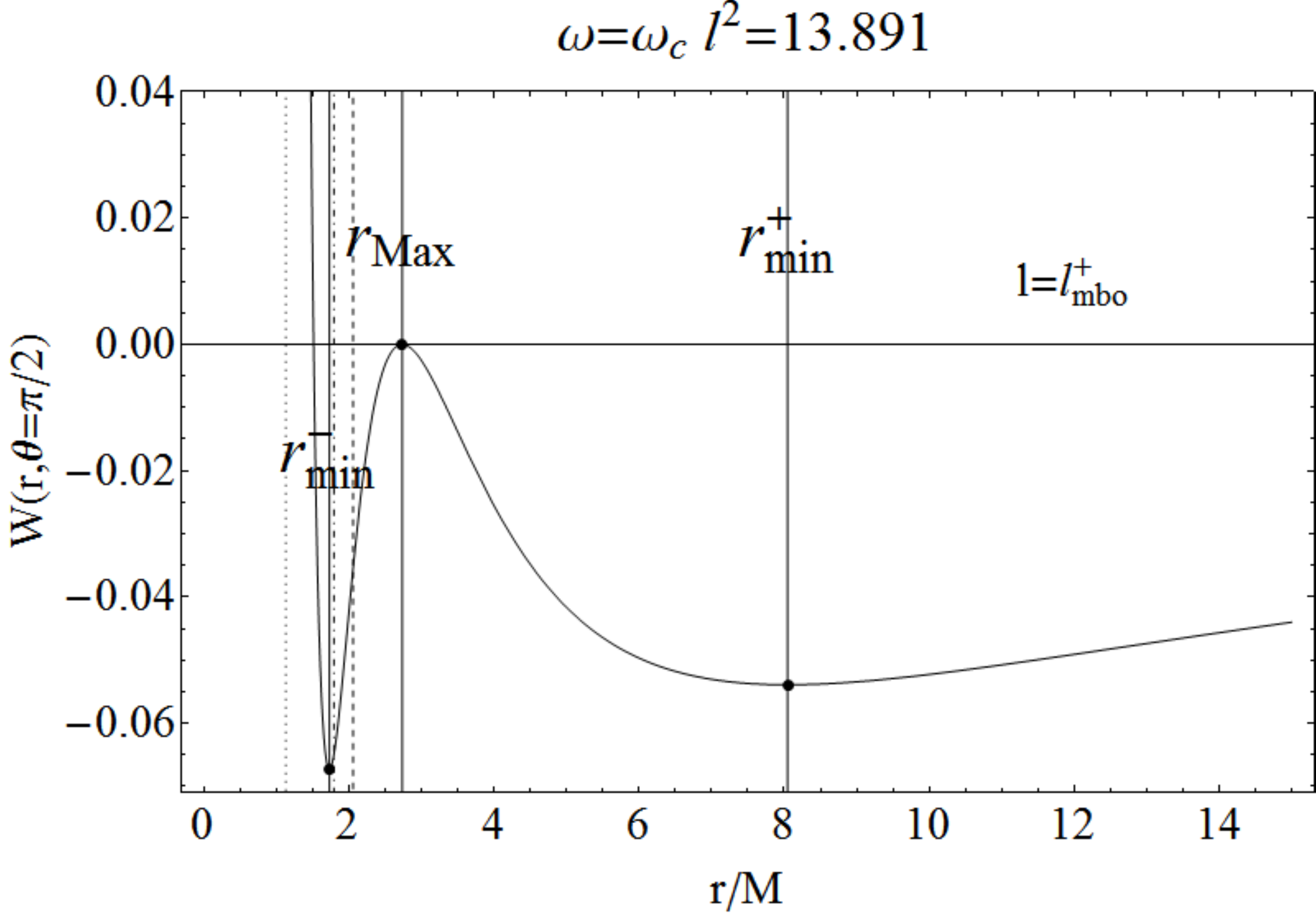}
\includegraphics[width=.31\textwidth]{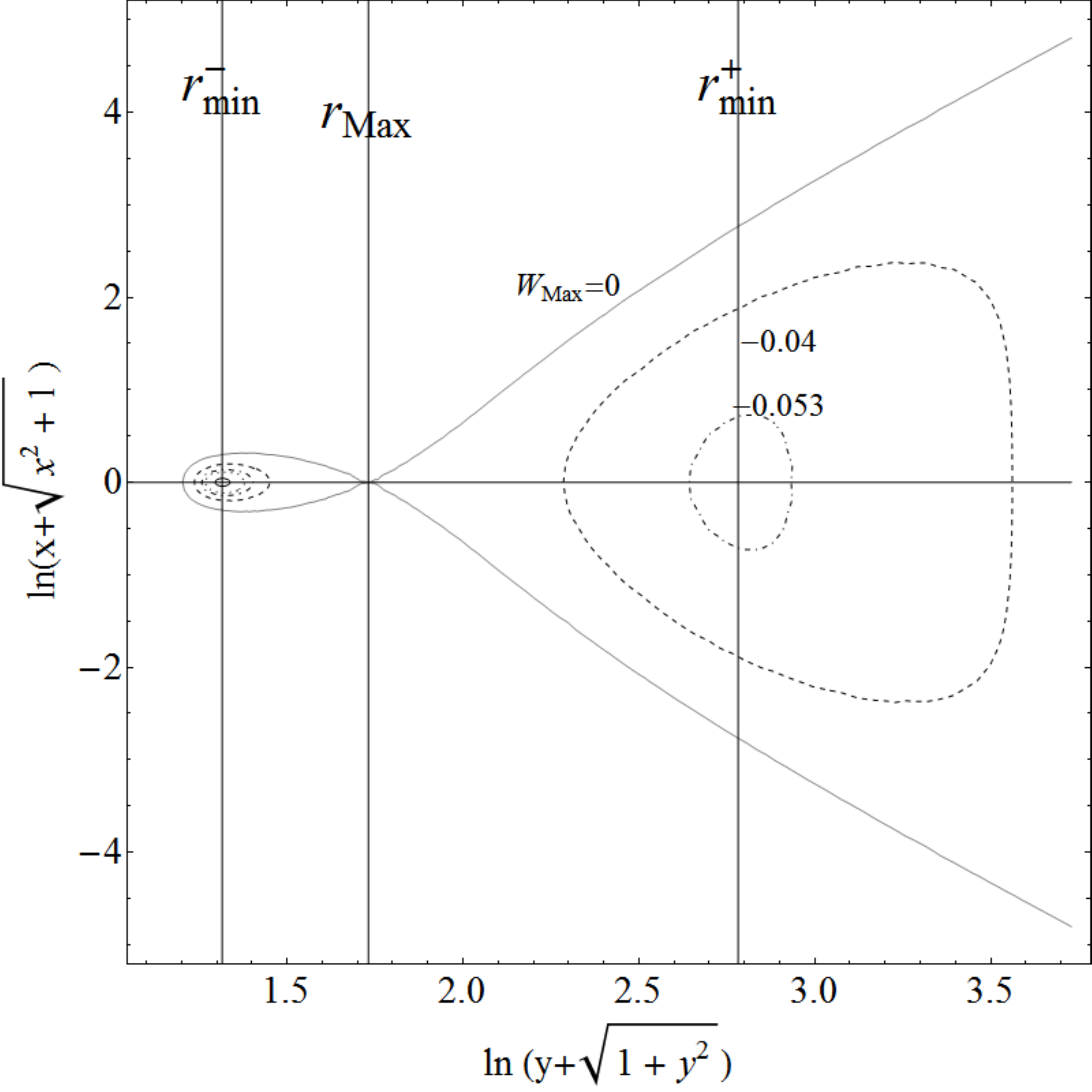}
\\
\includegraphics[width=.481\textwidth]{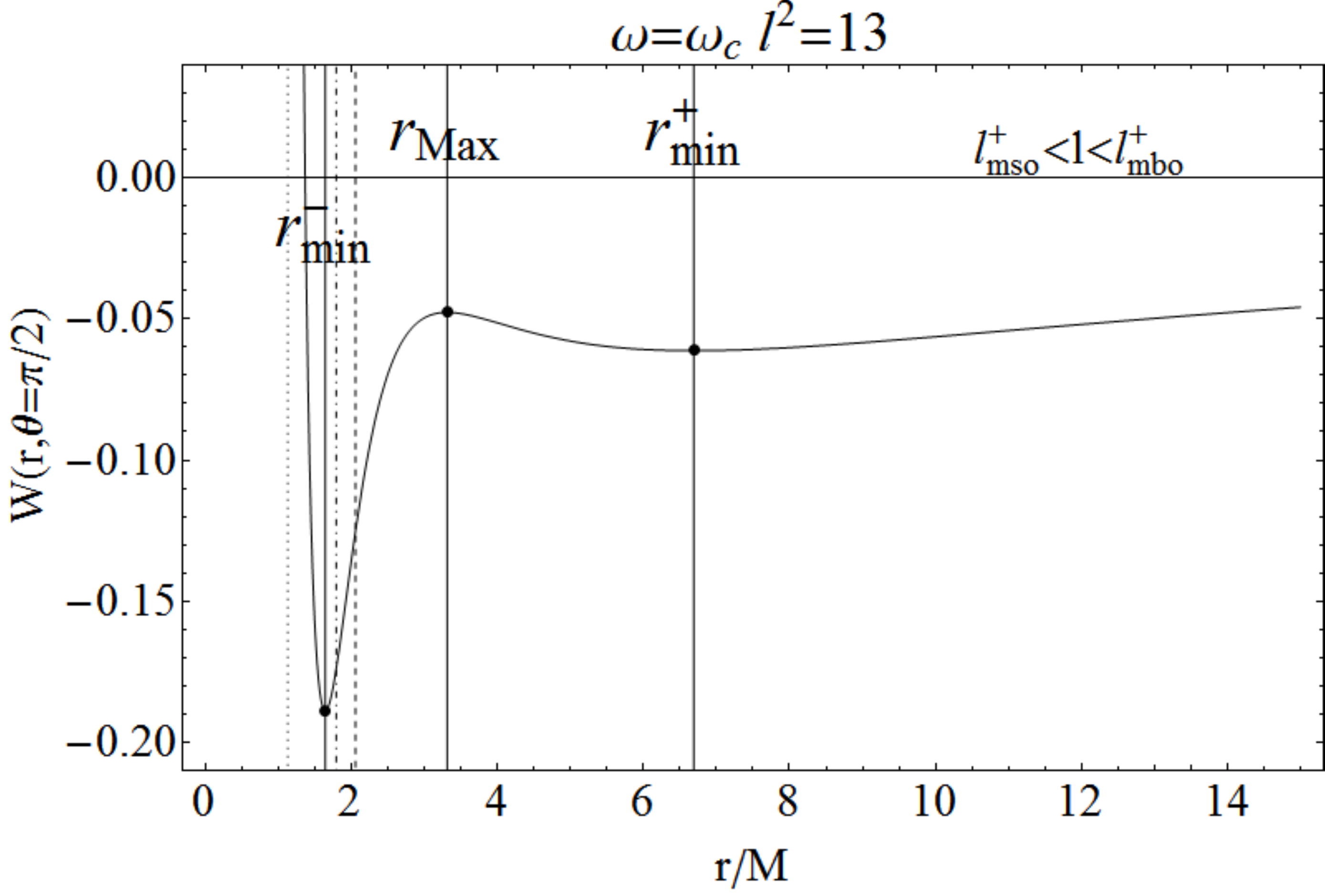}
\includegraphics[width=.41\textwidth]{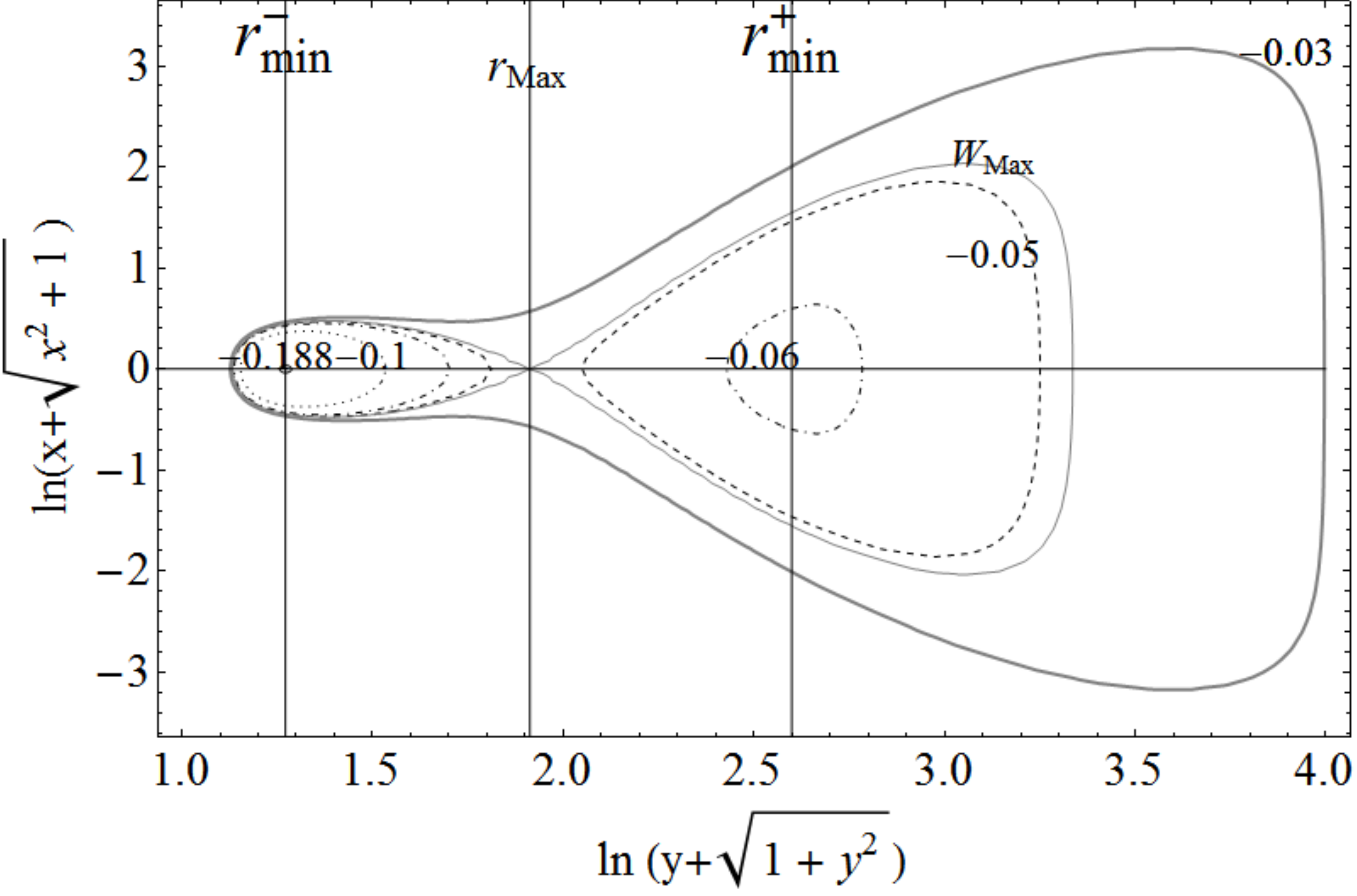}
\\
\includegraphics[width=.481\textwidth]{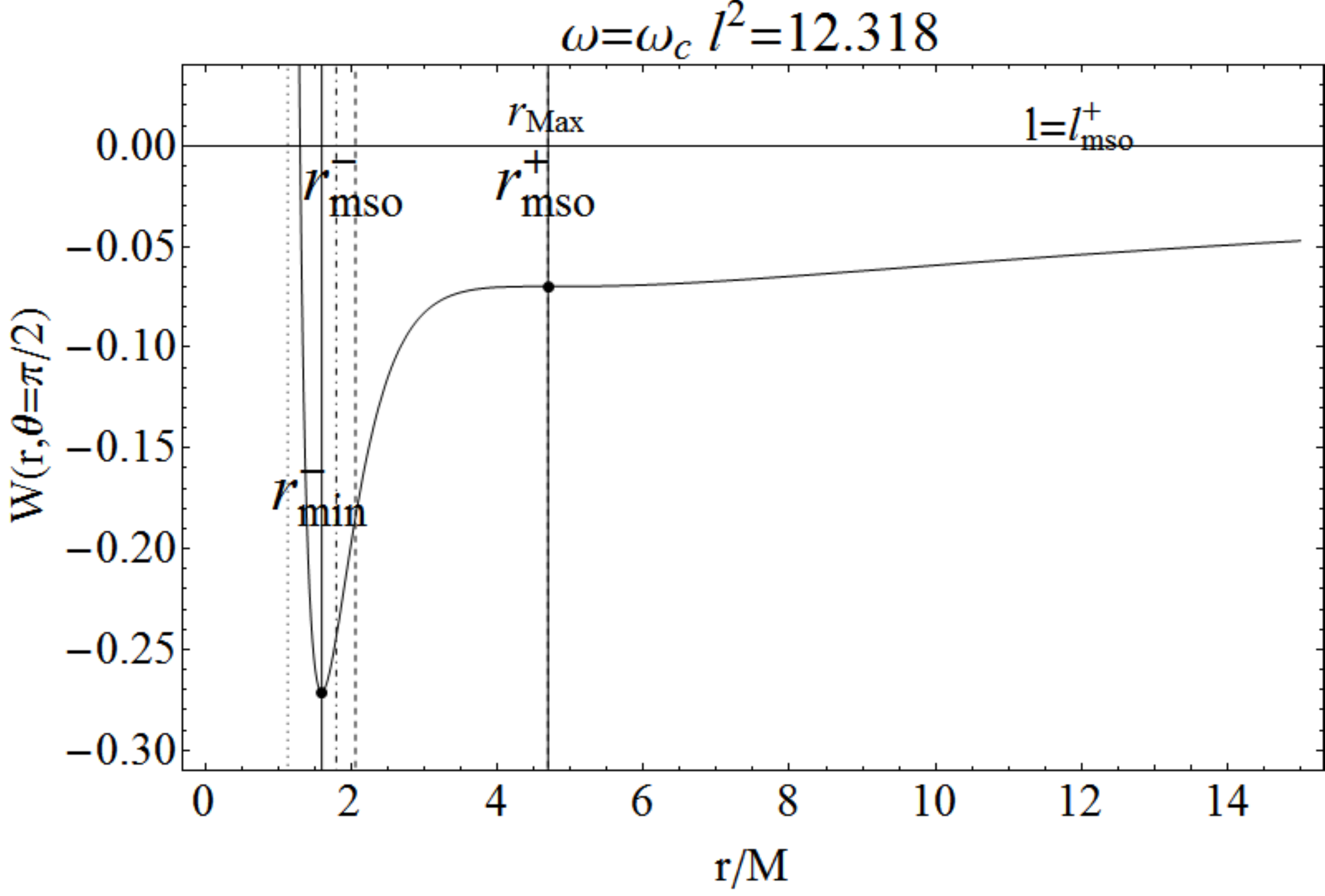}
\includegraphics[width=.481\textwidth]{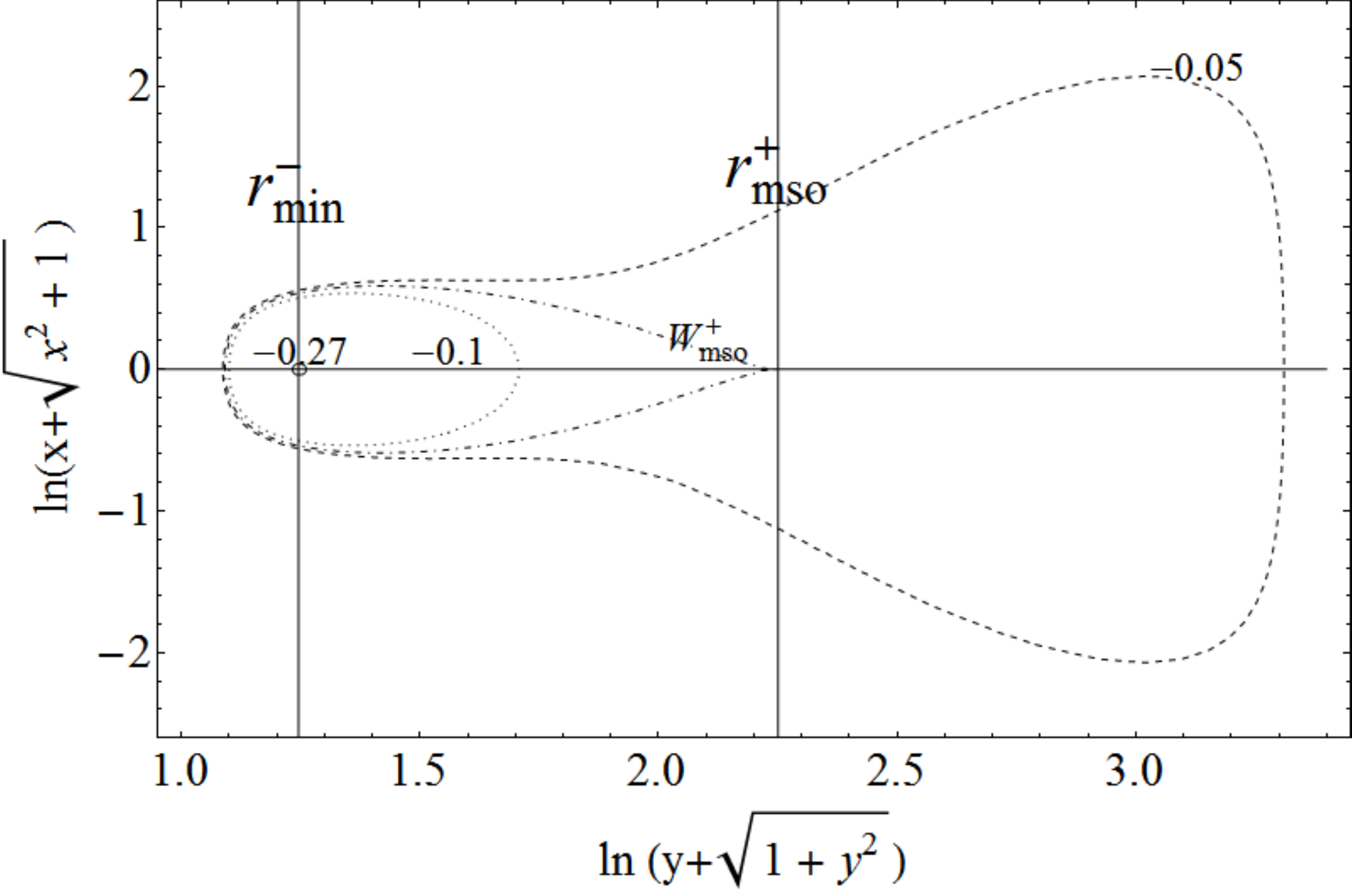}
\caption{ Naked singularity: $\omega M^2=\omega_{c} M^2=0.347506$, with  $\omega M^2\rightarrow \omega$. It  is $l_{mso}^->l_{\Omega}^{Max}= l^-_{mbo}> l^+_{mbo}> l_{mso}^+$, and $r_{stat}<r_{mbo}^-=r_{\Omega}^{Max}<r_{mso}^-<r_{mbo}^+<r_{mso}^+$.  Vertical lines in right panels set the $r_i\in\mathfrak{R}$ and  the effective potential critical points.  It is $r/M=\sqrt{x^2+y^2}$ and  $(x,y)$ are Cartesian coordinates.}
\label{Fig:Os-spe}
\end{figure}

\begin{figure}[h]
%%CPlotoiscomPlotoocl14Plotoocl14
\includegraphics[width=.481\textwidth]{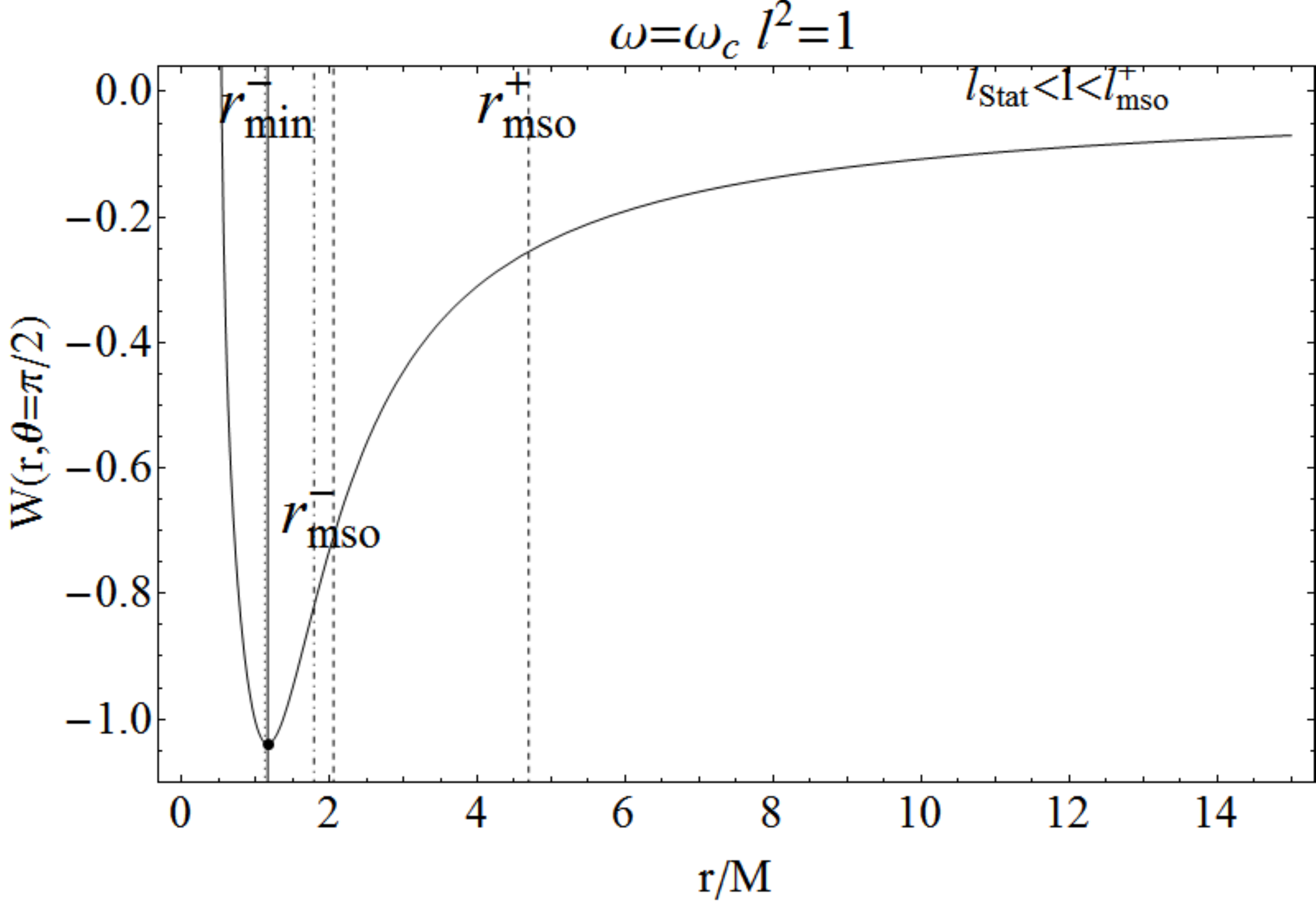}
\includegraphics[width=.481\textwidth]{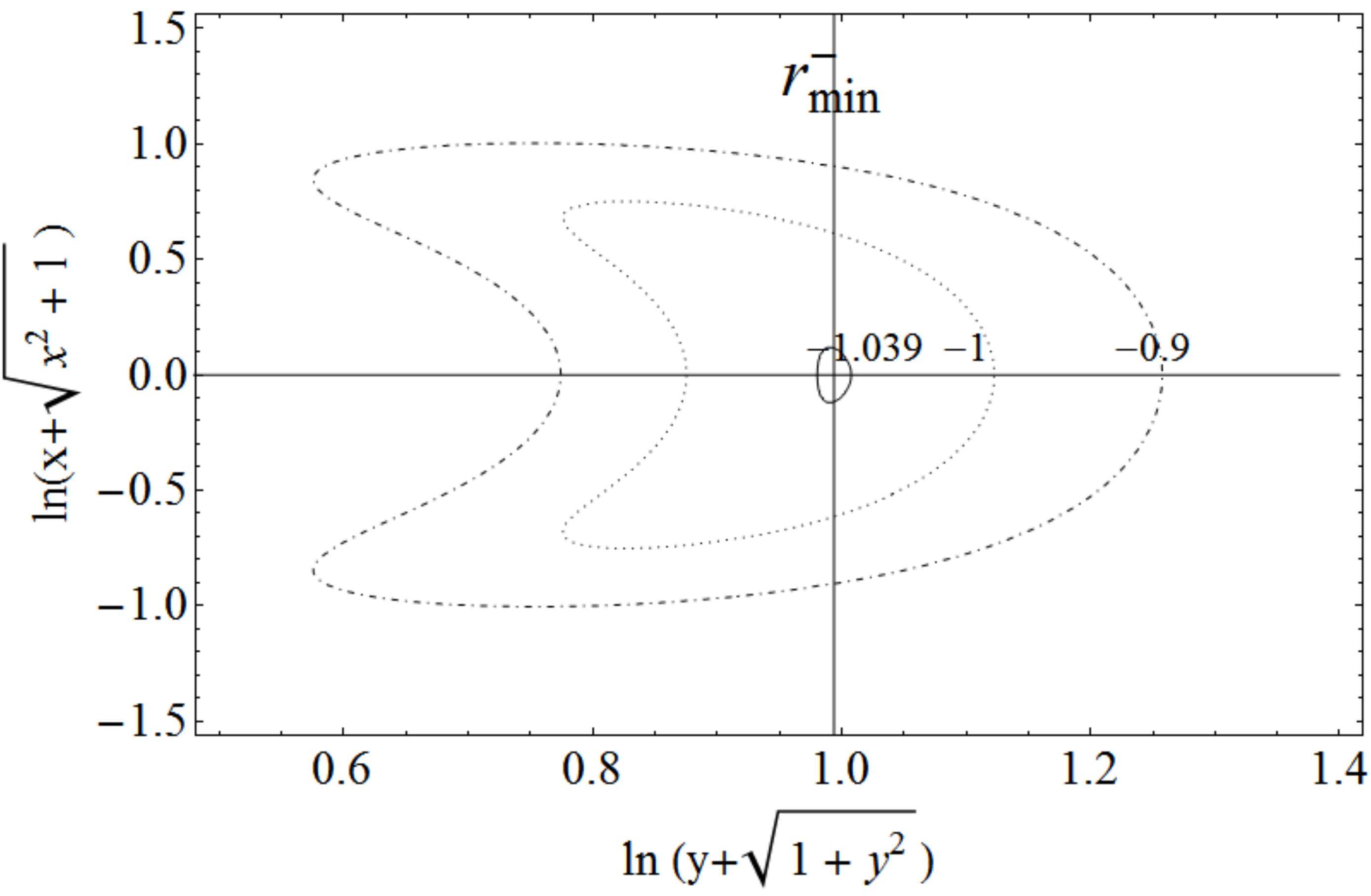}
\caption{ Naked singularity $\omega M^2=\omega_{c} M^2=0.347506$, with  $\omega M^2\rightarrow \omega$. It  is $l_{mso}^->l_{\Omega}^{Max}= l^-_{mbo}> l^+_{mbo}> l_{mso}^+$ and $r_{stat}<r_{mbo}^-=r_{\Omega}^{Max}<r_{mso}^-<r_{mbo}^+<r_{mso}^+$. With $r/M=\sqrt{x^2+y^2}$ and  $(x,y)$ are Cartesian coordinates.  Vertical lines in right panels set the $r_i\in\mathfrak{R}$ and  the effective potential critical points.}
\label{Fig:otheo-ri}
\end{figure}

We focus on the special K-S naked singularity case spacetime  $\omega M^2=\omega_{c} M^2=0.347506$, where    $r_{stat}<r_{mbo}^-=r_{\Omega}^{Max}<r_{mso}^-<r_{mbo}^+<r_{mso}^+$ and  $l_{mso}^->l_{\Omega}^{Max}= l^-_{mbo}> l^+_{mbo}> l_{mso}^+$.
This peculiar source separates
spacetimes where $r_{\Omega}^{Max}<r_{mbo}^-$ (\textbf{RegionIII-a}) from those $r_{\Omega}^{Max}>r_{mbo}^-$  (\textbf{RegionIII-b}).  We detail the situation as follows:
\begin{description}
\item[-) $l>l_{mso}^-$] There only one set of closed configurations with center in $r_{min}$,  it is $W_{min}<0$,  the situation is similar to Fig.\il(\ref{Fig:newfigokey}-a).
\item[-)$l=l_{mso}^-$] There is a family of closed $C^+$ surfaces with center in $r_{min}^+$, and a critical surface with an ``inner cusp'' located at the saddle point of the $W$ potential where: $W_{mso}^->0$, Fig.\il(\ref{Fig:namij-n})-a, this situation is  similar to the configuration at $l=l_{mso}^-$ for Ho\v{r}ava parameter in \textbf{Region III-a} (Fig.\il(\ref{Fig:lun-1})-b).
\item[-)${l\in]l_{mbo}^-,l_{mso}^-}[$] This case is illustrated in Fig.\il(\ref{Fig:namij-n})-b. As it is $W_{min}^-\in]0, W_{Max}[$ and $W_{min}^+<0$, the maximum point $r_{Max}$, gives rise to a critical crossed open surface, with outer cusp, associated to an matter excretion  from the inner configuration to the outer (larger) one. To every  inner closed surface centered in $r_{min}^-$ correspond an outer open surface, see also Fig.\il(\ref{Fig:lun-1})-c.
\item[-)${l=l_{mbo}^-=l_{\Omega}^{Max}}$]
The center, $r_{\Omega}^{Max}$, of the inner set of closed configuration is for  $W_{min}^-=0$, the configurations structure is shown in  Fig.\il(\ref{Fig:namij-n})-c, and can be compared with  Fig.\il(\ref{Fig:namij-n})-a.
\item[-) ${l\in]l_{mbo}^+,l_{mbo}^-[}$] See Fig.\il(\ref{Fig:Os-spe})-a. It is $W_{Max}>0$, $W_{min}^->W_{min}^+<0$ (\textbf{I}-class of configurations). The crossed surface, with an outer cusp is opened and can be associated to an excretion configuration. There exist at fixed $W\in]W_{min}^-,0[$ a double class of closed configurations, the inner centred in $r_{min}^-$ and the  outer in $r_{min}^+$. One class of closed configurations are for $W\in]W_{min}^+, W_{min}^-[$ centred in $r_{min}^+$ (outer disc),  a second  single set of inner toroidal configuration with center in $r_{min}^-$ is for $W\in]0,W_{Max}[$.
\item[-) ${l=l_{mbo}^+}$] The function $W$ has a saddle point in $r_{mbo}^+$ where $W_{mbo}^+=0$. It is $W_{min}^-<W_{min}^+$ (\textbf{III}-class of configurations). This situation sketched in Fig.\il(\ref{Fig:Os-spe})-b, and it is similar to the configuration structure of Fig.\il(\ref{Fig:Rip-second})-\textbf{a} representing the situation in spacetimes in \textbf{Region III-a}, however  in this last case it was $W_{min}^->W_{min}^+$ (\textbf{I-class}). One could say there is shift between a \textbf{I} and \textbf{III}  configuration, with  a \textbf{II} type with  $W_{min}^-=W_{min}^+<0$
\item[-) ${l\in]l_{mso}^+,l_{mbo}^+[}$] This case is shown in  Fig.\il(\ref{Fig:Os-spe})-c. This is a  \textbf{III}-class of configurations, as $W_{Max}<0$ the crossed critical surface is closed and at $W\in]W_{Max},0[$ there is one only closed surface, with two maximum points for the Boyer surfaces with  the centers located  on   the maximum points for the pressure $r_{min}^{\pm}$, see also Figs.\il(\ref{Fig:ira-n-b})-a-b and
Fig.\il(\ref{Fig:Rip-second})-c.
\item[-) $l=l_{mso}^+$]  At $r_{mso}^+$ there is an outer cusp, for the closed configuration with center in $r_{min}^-$.  There is only one closed set surface  from $W\in]W_{min}^-,0[$. We can consider two subclasses those configurations belonging to $W\in]W_{min}^-, W_{mso}^+[$, centerd in $r_{min}^-$ and with outer edge in $r\in]r_{min},r_{mso}^+[$, and the second class for $W\in]W_{mso}^+,0[$. The maximum of pressure for these configurations is still located in $r_{min}^-$, but the outer edge of the  discs are at $r>r_{mso}^+$. Similar configurations where  in Figs.\il(\ref{Fig:ira-n-b})-c,
(\ref{Fig:ban-or})-c
(\ref{Fig:tis-bre})-b.
\item[-) ${l\in]l_{stat},l_{mso}^+[}$] There is one set of closed configurations with center in $r_{min}^-$, see Fig.\il(\ref{Fig:otheo-ri}).
\end{description}
%
%\end{figure}
%%
%%%%%%%%%%%%%%%%%%%%%%%%%%%%%%%%%%%%%%%%%%%%%%%%%%%%%%%%%%%%%%%%%%%%%
%-------------------------------------------------------------------
%%%%%%%%%%%%%%%%%%%%%%%%%%%%%%%%%%%%%%%%%%%%%%%%%%%%%%%%%%%%%%%%%%%%%
%\clearpage
\subsubsection{Region III-b: $\omega\in]\omega_c,\omega_{\gamma}[$}\label{Sec:NSRegionIIIb}
\begin{figure}[h]
%%CPlotoiscom
%\includegraphics[width=.481\textwidth]{Ploto3718.pdf}
%\includegraphics[width=.31\textwidth]{CPloto3718.pdf}
%\\
\includegraphics[width=.481\textwidth]{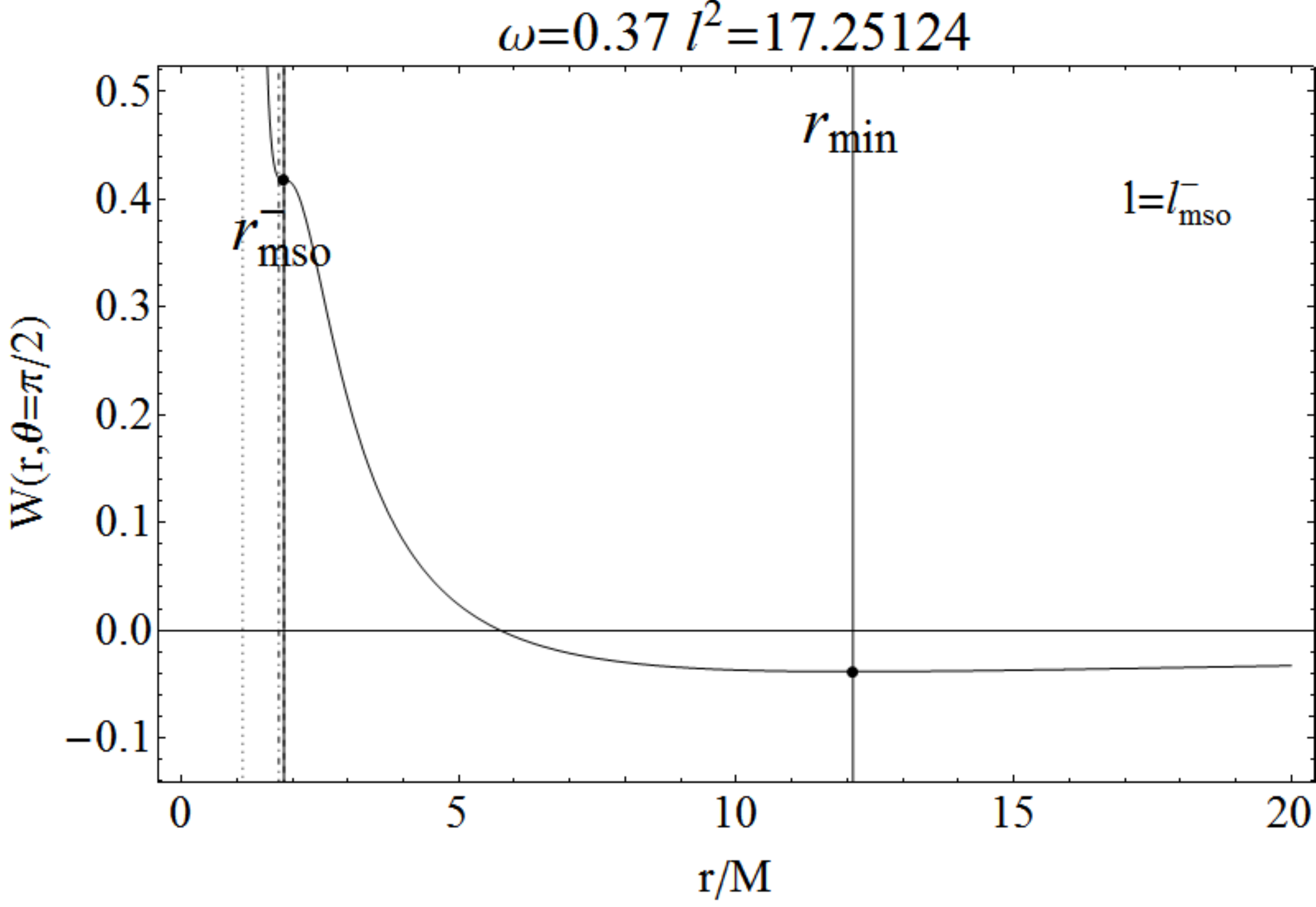}
\includegraphics[width=.31\textwidth]{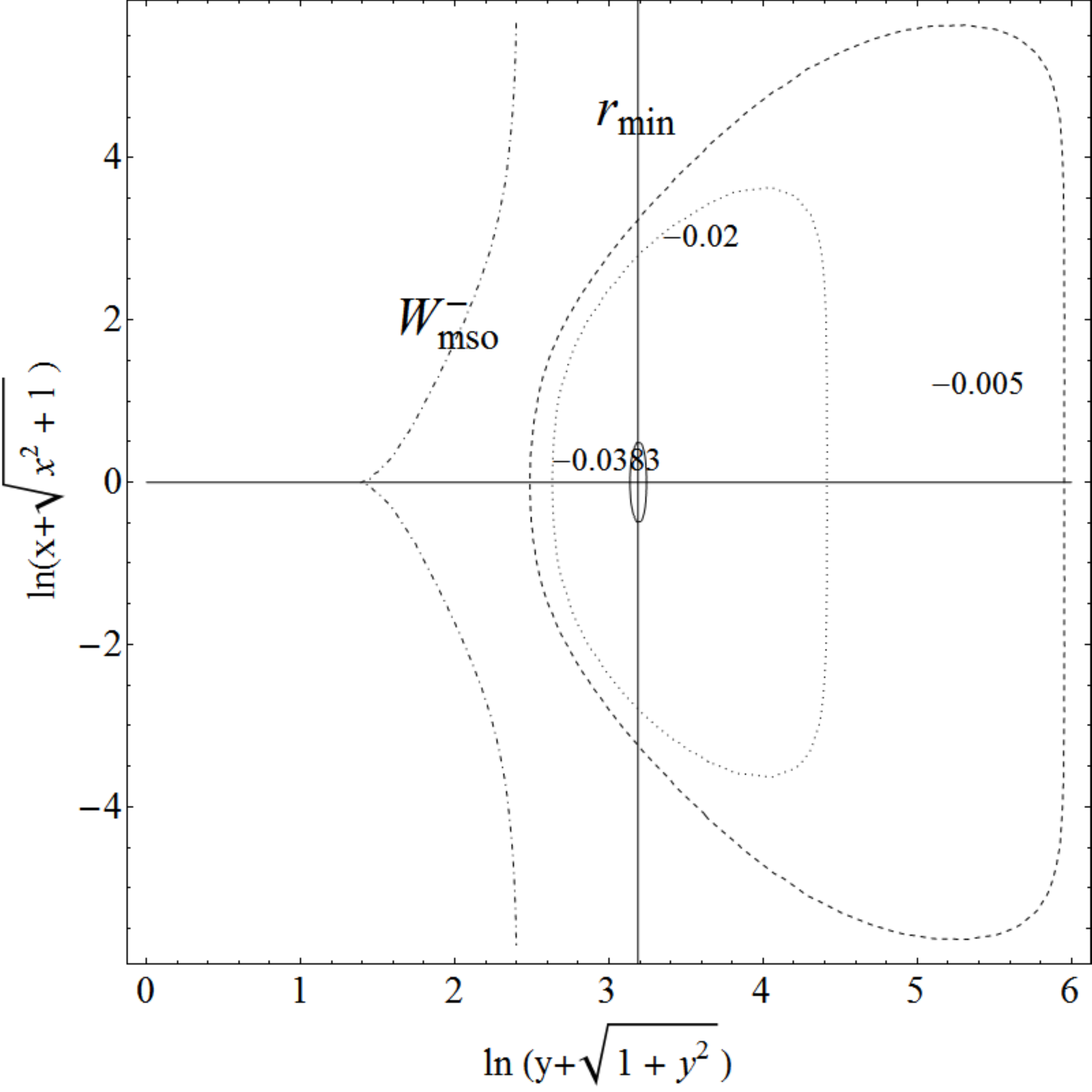}
\includegraphics[width=.481\textwidth]{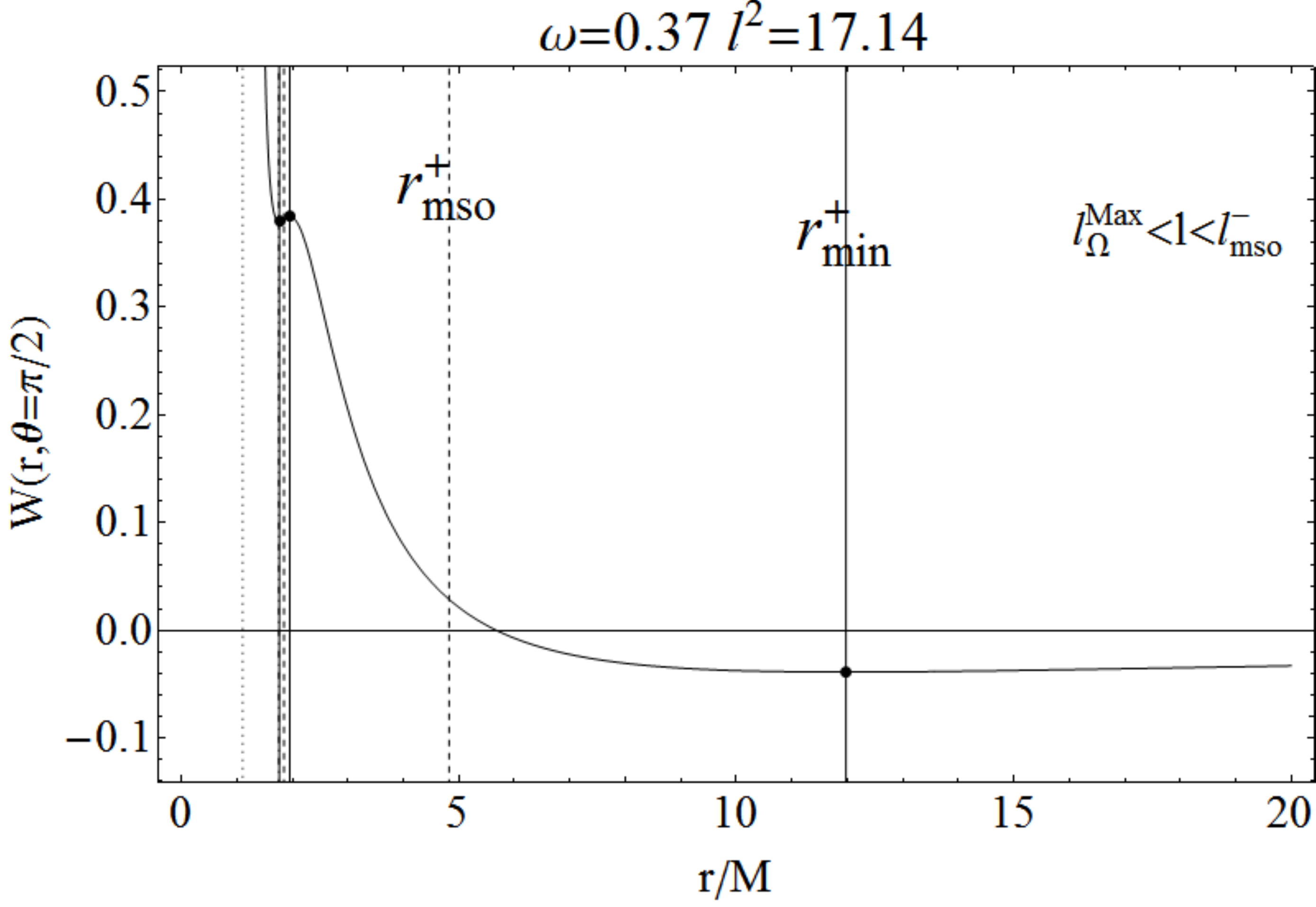}
\includegraphics[width=.41\textwidth]{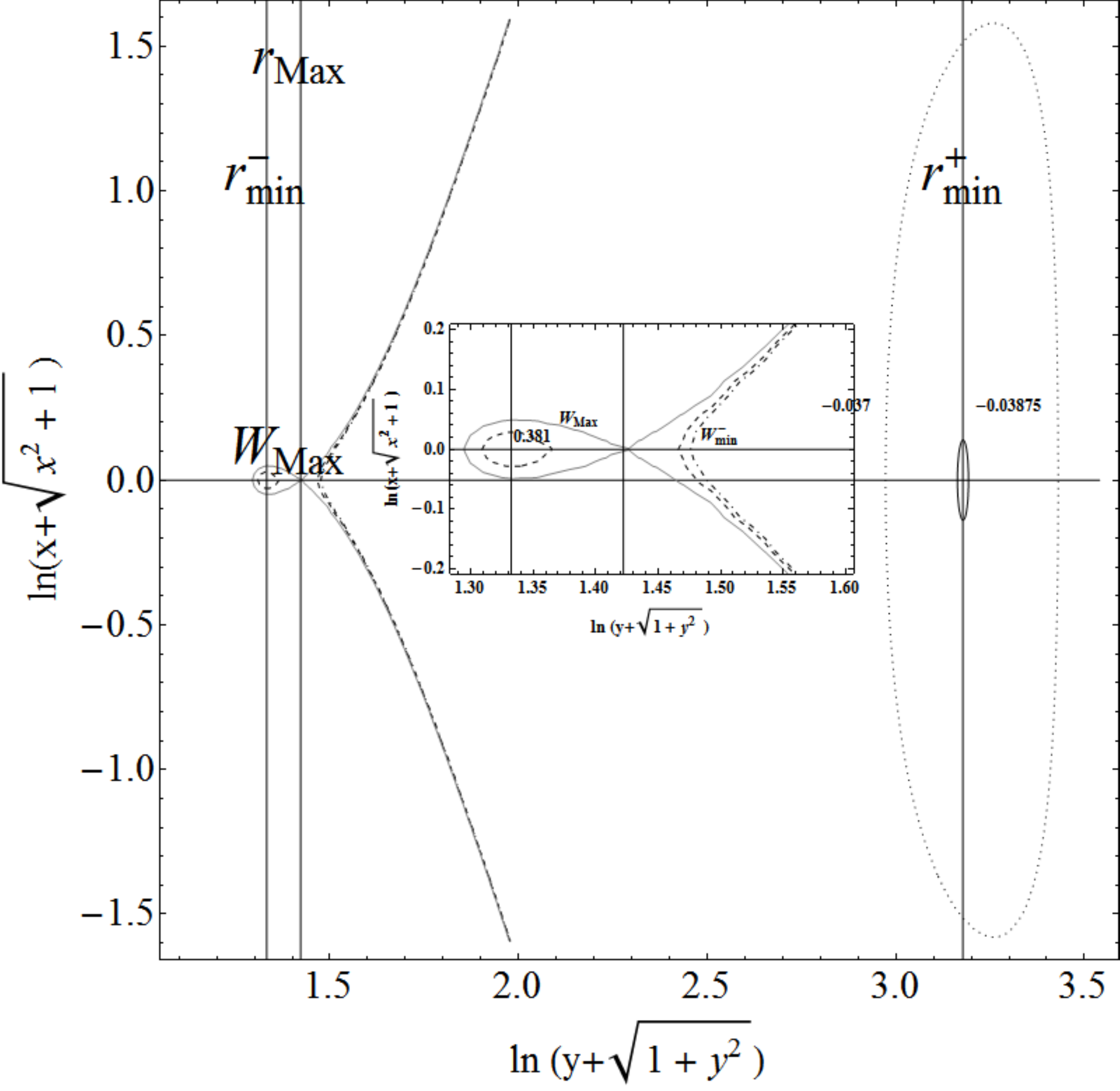}
\\
\includegraphics[width=.481\textwidth]{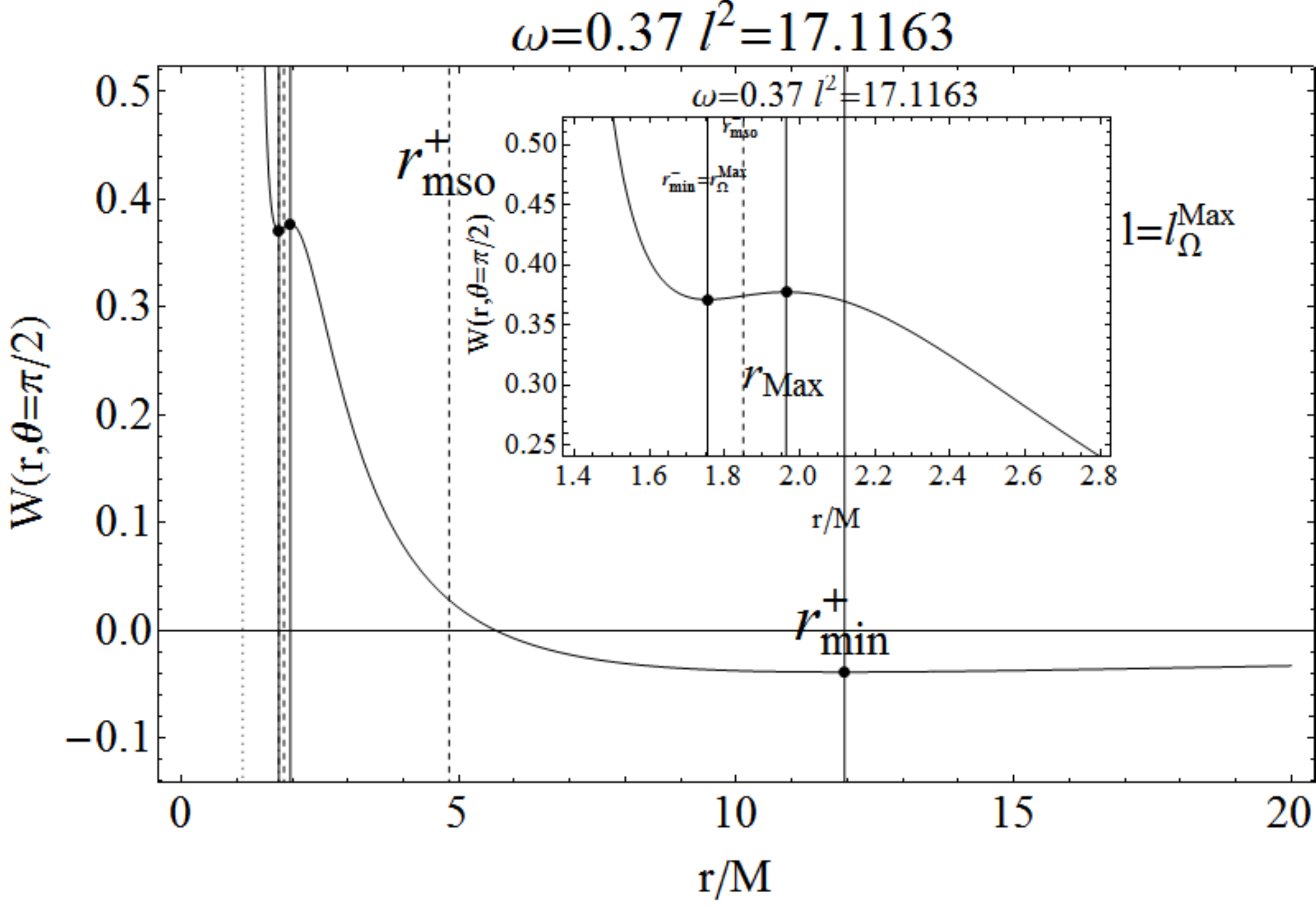}
\includegraphics[width=.481\textwidth]{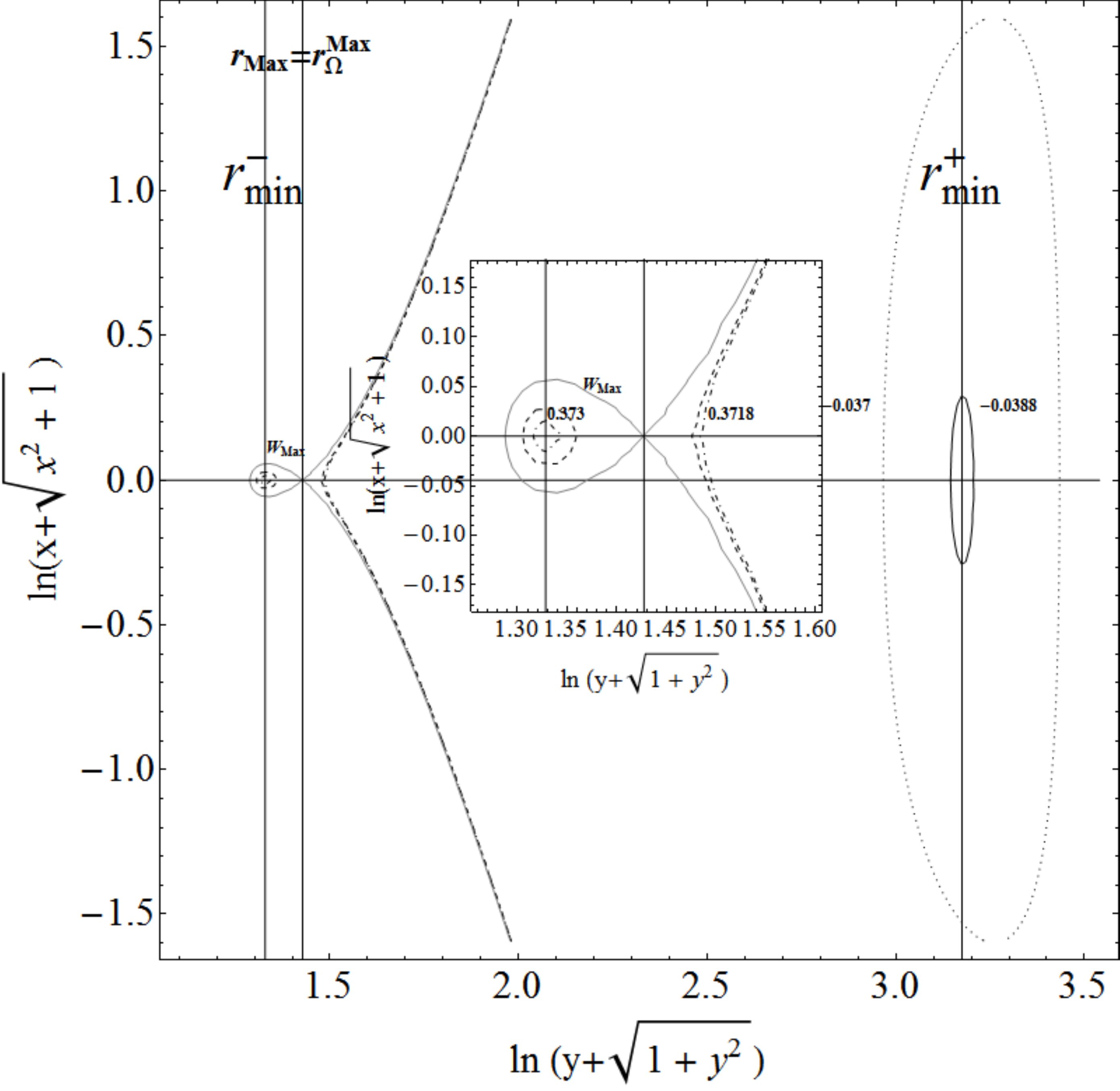}
\caption{Region III-b: $\omega\in]\omega_c,\omega_{\gamma}[$, naked singularity  $\omega M^2=0.37$, where $\omega M^2\rightarrow \omega$. It is  $l_{mso}^->l_{\Omega}^{Max}> l^-_{mbo}> l^+_{mbo}> l_{mso}^+$, and $r_{stat}<r_{mbo}^-<r_{\Omega}^{Max}<r_{mso}^-<r_{mbo}^+<r_{mso}^+$.  Vertical lines in right panels set the $r_i\in\mathfrak{R}$ and  the effective potential critical points. With $r/M=\sqrt{x^2+y^2}$ and  $(x,y)$ are Cartesian coordinates.}
\label{Fig:NBib-for}
\end{figure}

\begin{figure}[h]
%\\
\includegraphics[width=.481\textwidth]{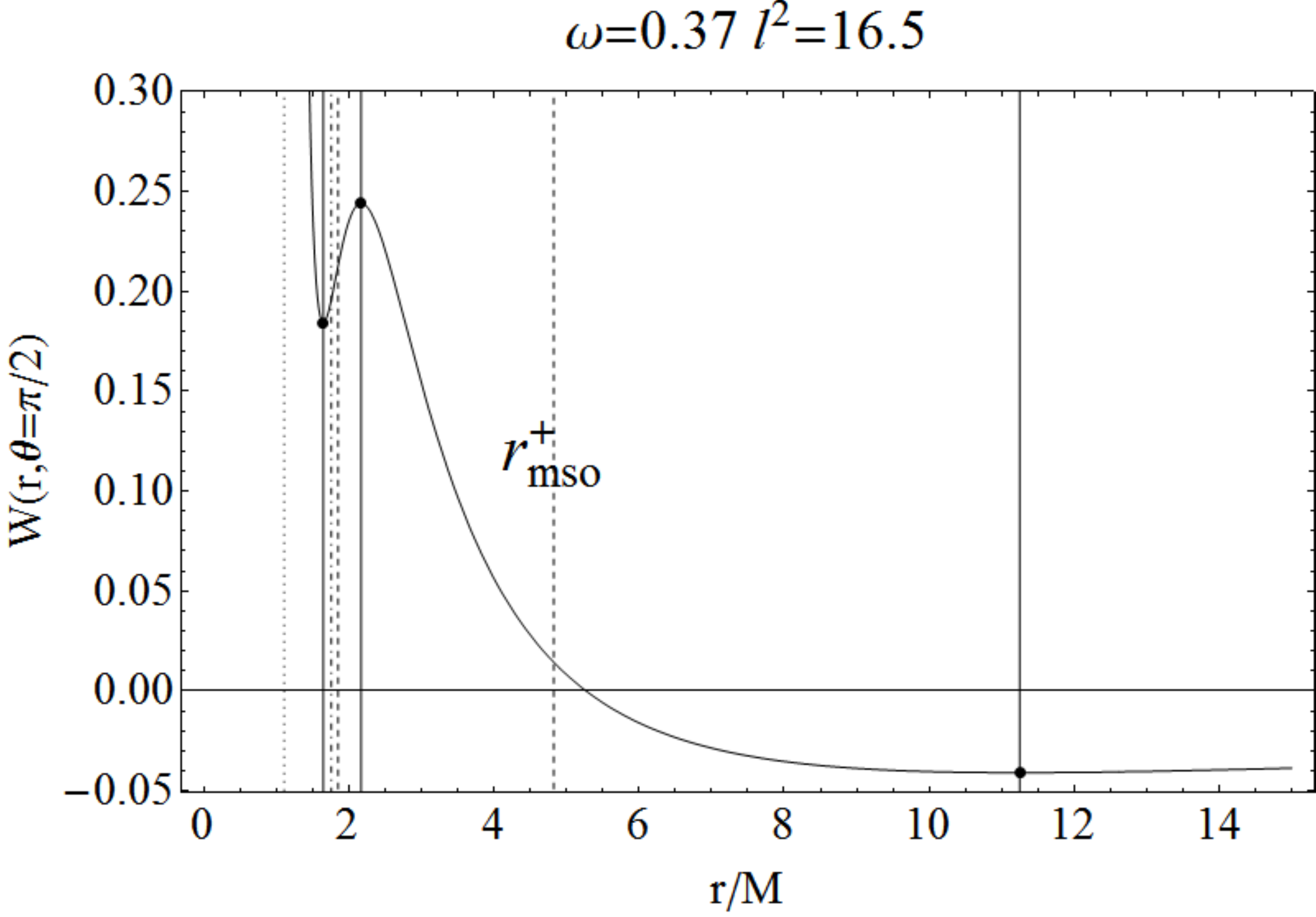}
\includegraphics[width=.451\textwidth]{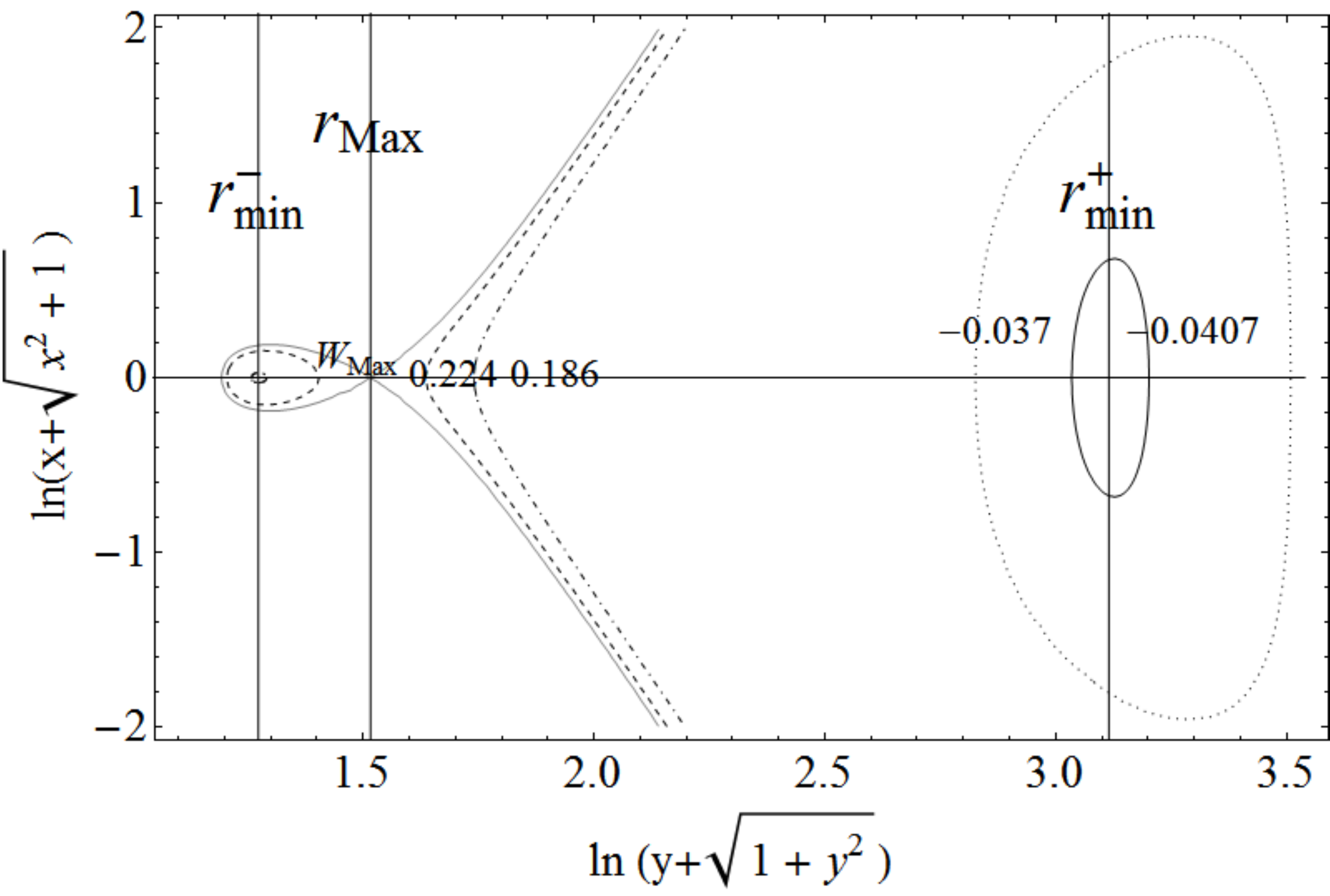}
\\
\includegraphics[width=.481\textwidth]{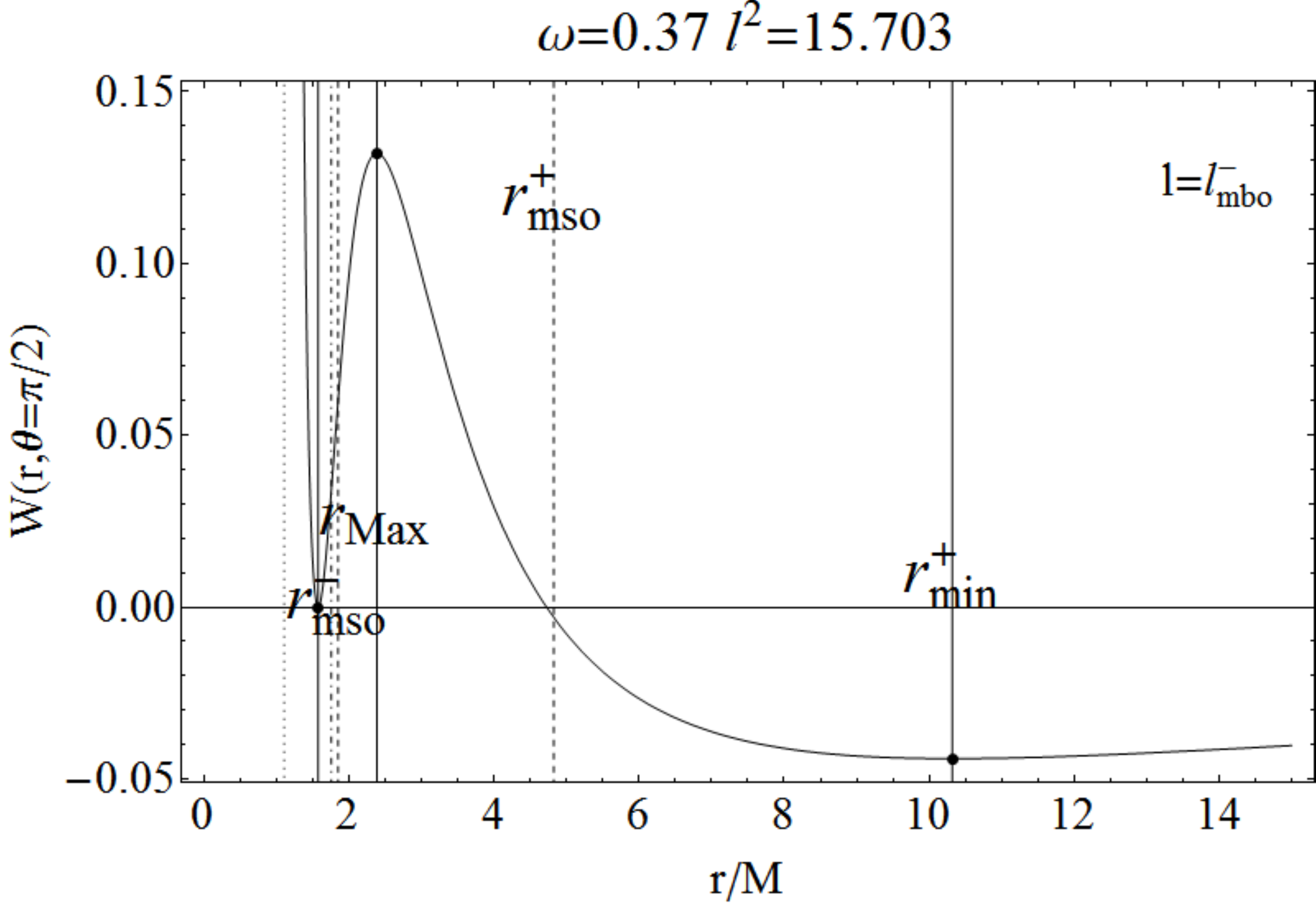}
\includegraphics[width=.451\textwidth]{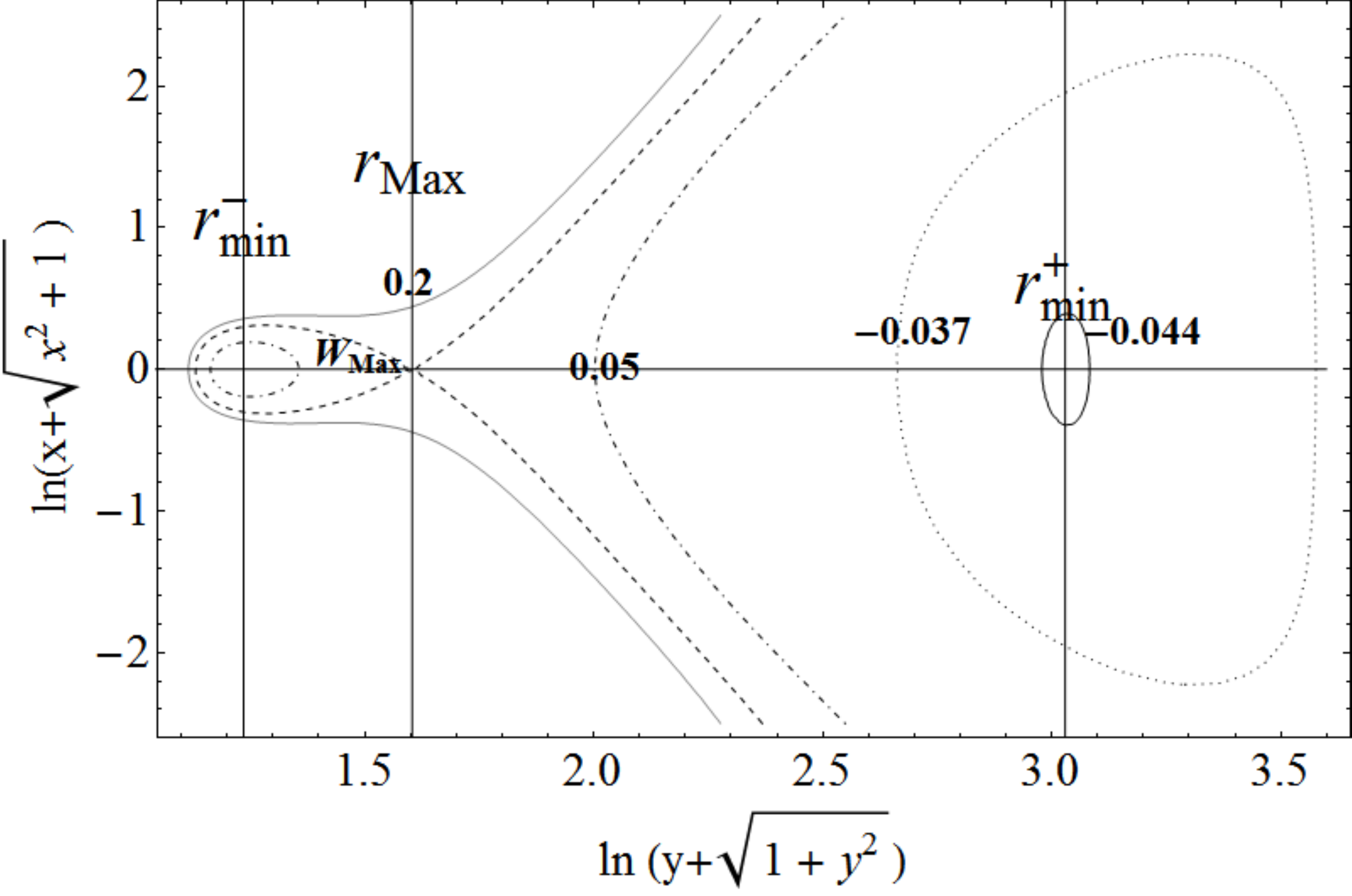}
\\
\includegraphics[width=.481\textwidth]{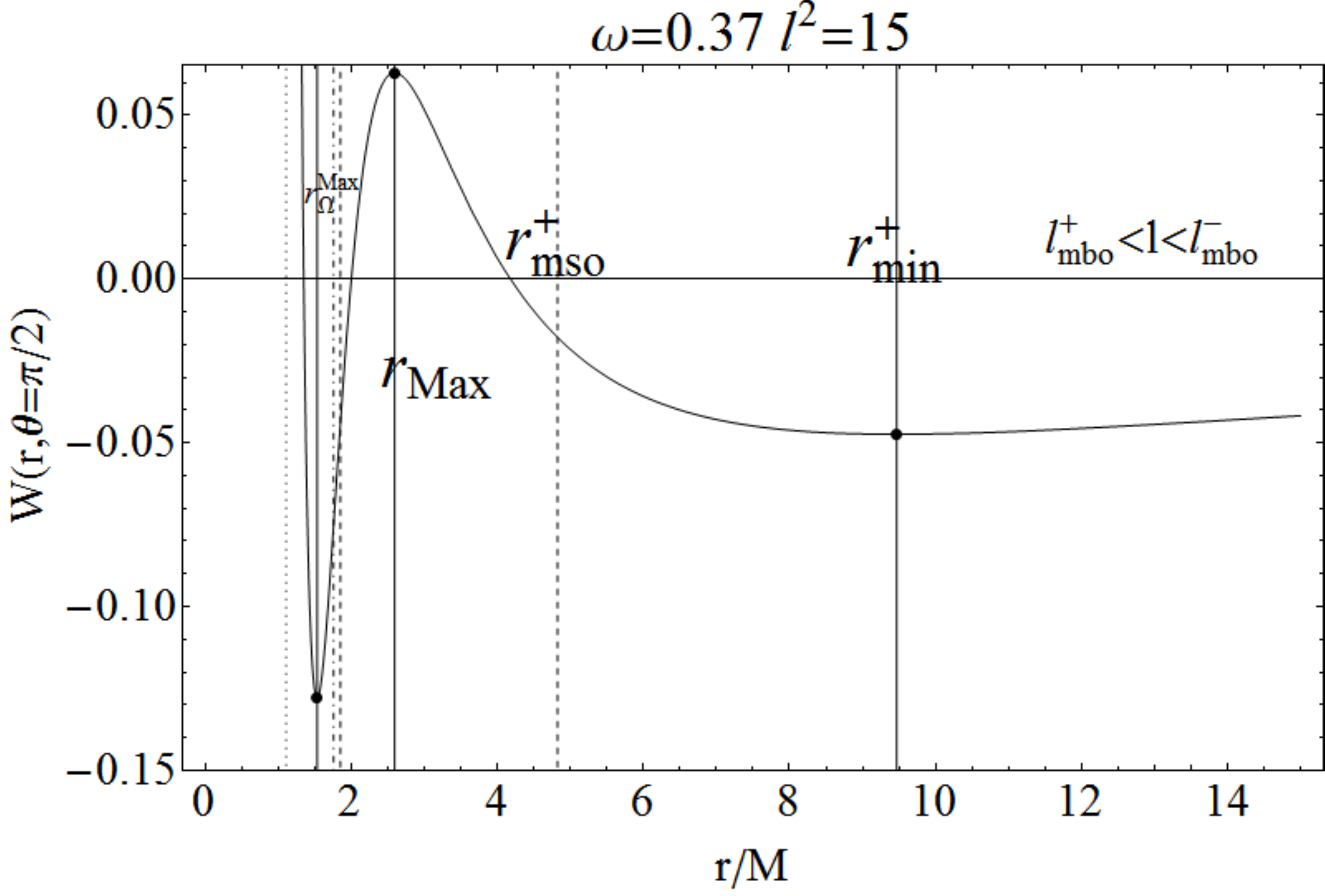}
\includegraphics[width=.33\textwidth]{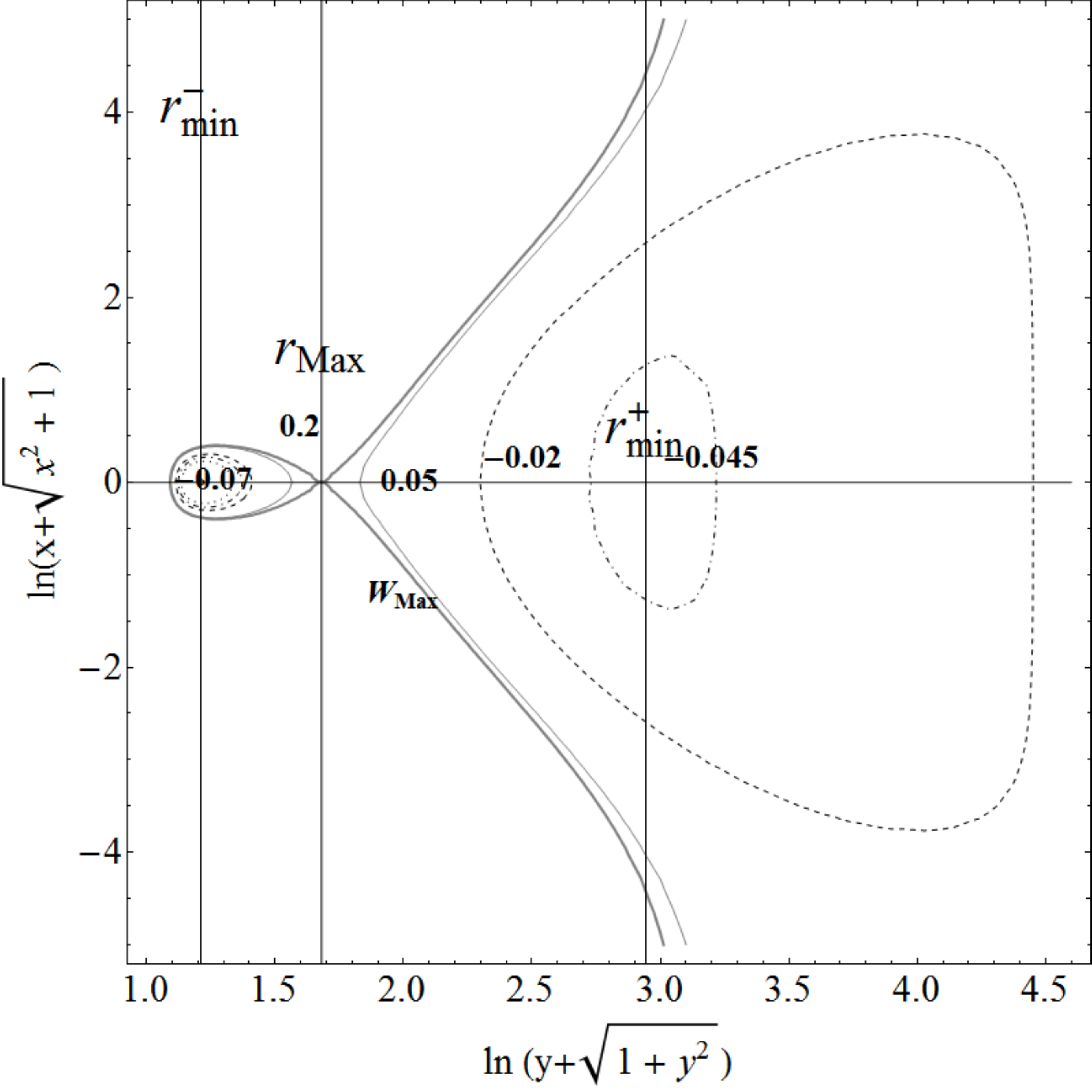}
\\
\includegraphics[width=.481\textwidth]{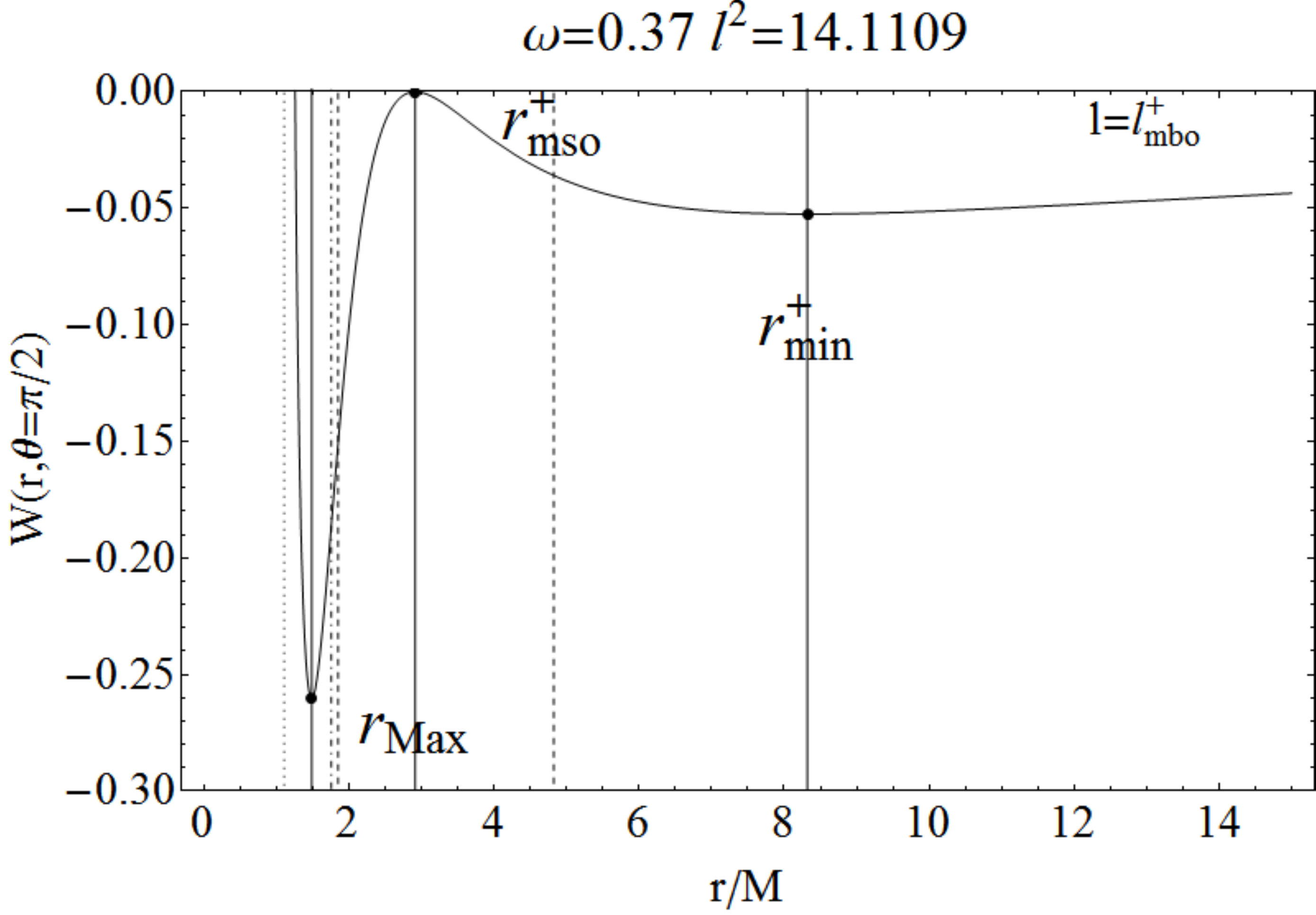}
\includegraphics[width=.41\textwidth]{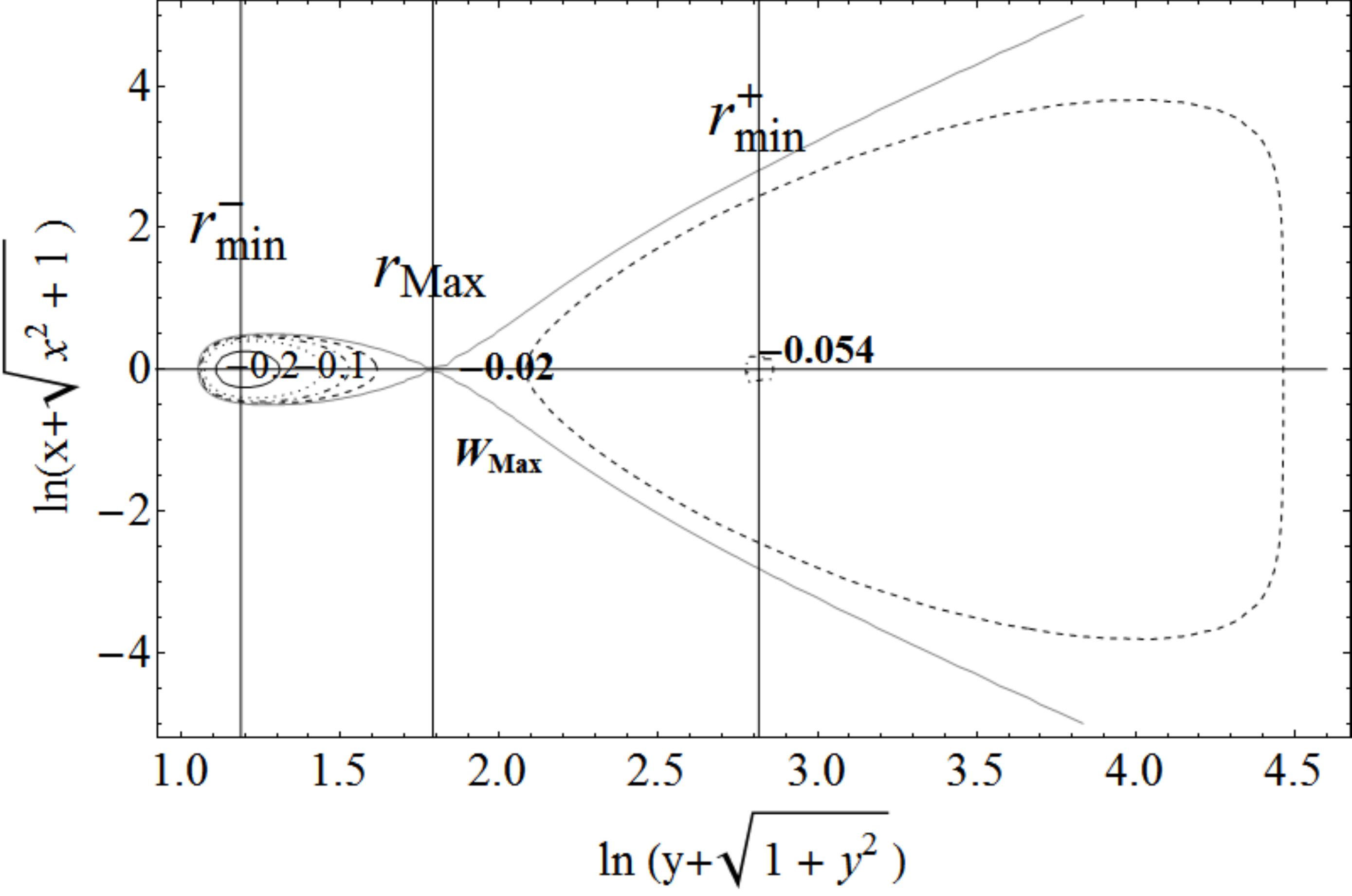}
\caption{Region III-b: $\omega\in]\omega_c,\omega_{\gamma}[$, naked singularity  $\omega M^2=0.37$, where $\omega M^2\rightarrow \omega$, it is $l_{mso}^->l_{\Omega}^{Max}> l^-_{mbo}> l^+_{mbo}> l_{mso}^+$, and $r_{stat}<r_{mbo}^-<r_{\Omega}^{Max}<r_{mso}^-<r_{mbo}^+<r_{mso}^+$. Vertical lines in right panels set the $r_i\in\mathfrak{R}$ and  the effective potential critical points.}
\label{Fig:ceg-his}
\end{figure}

\begin{figure}[h]
%\\
\includegraphics[width=.481\textwidth]{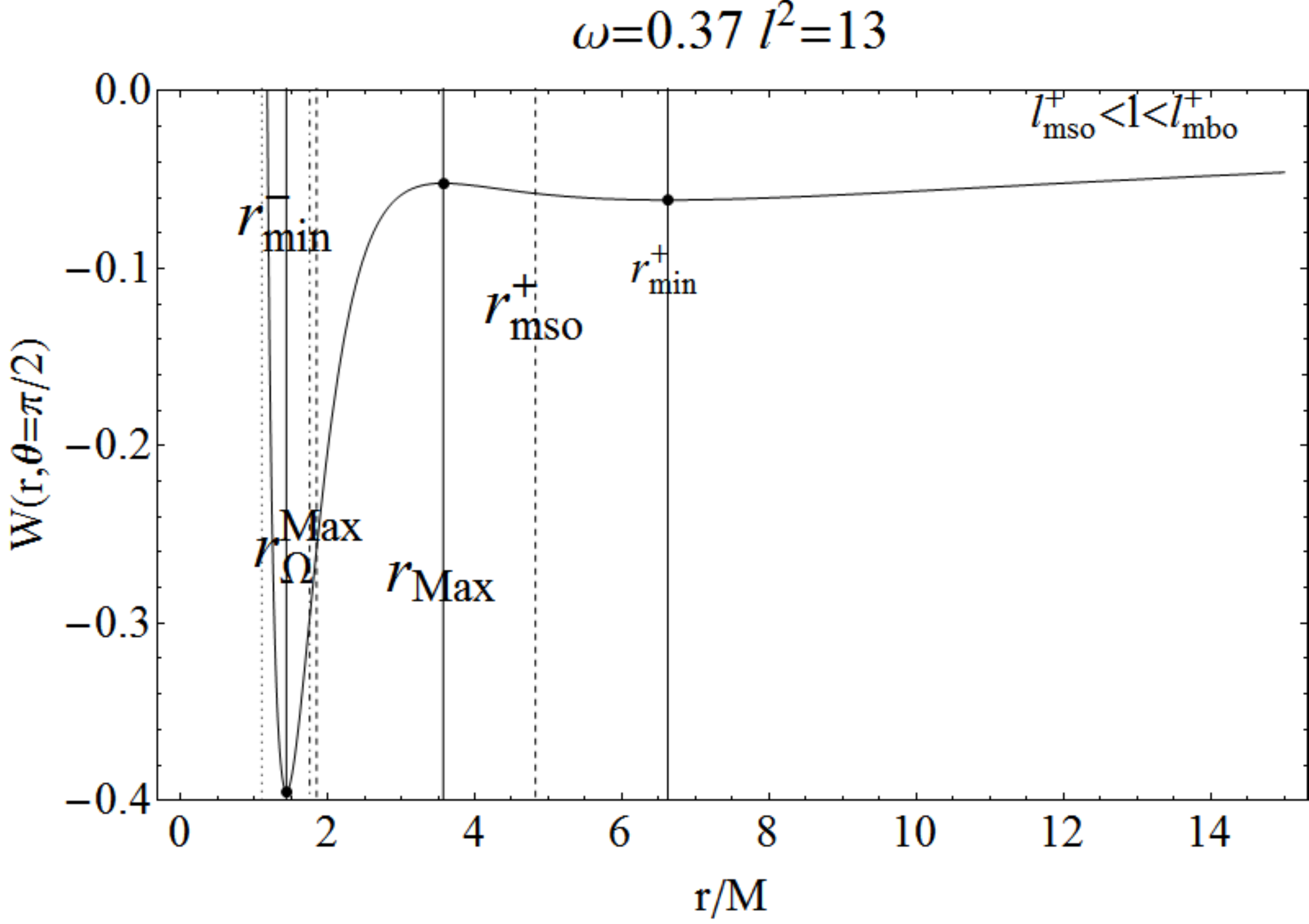}
\includegraphics[width=.451\textwidth]{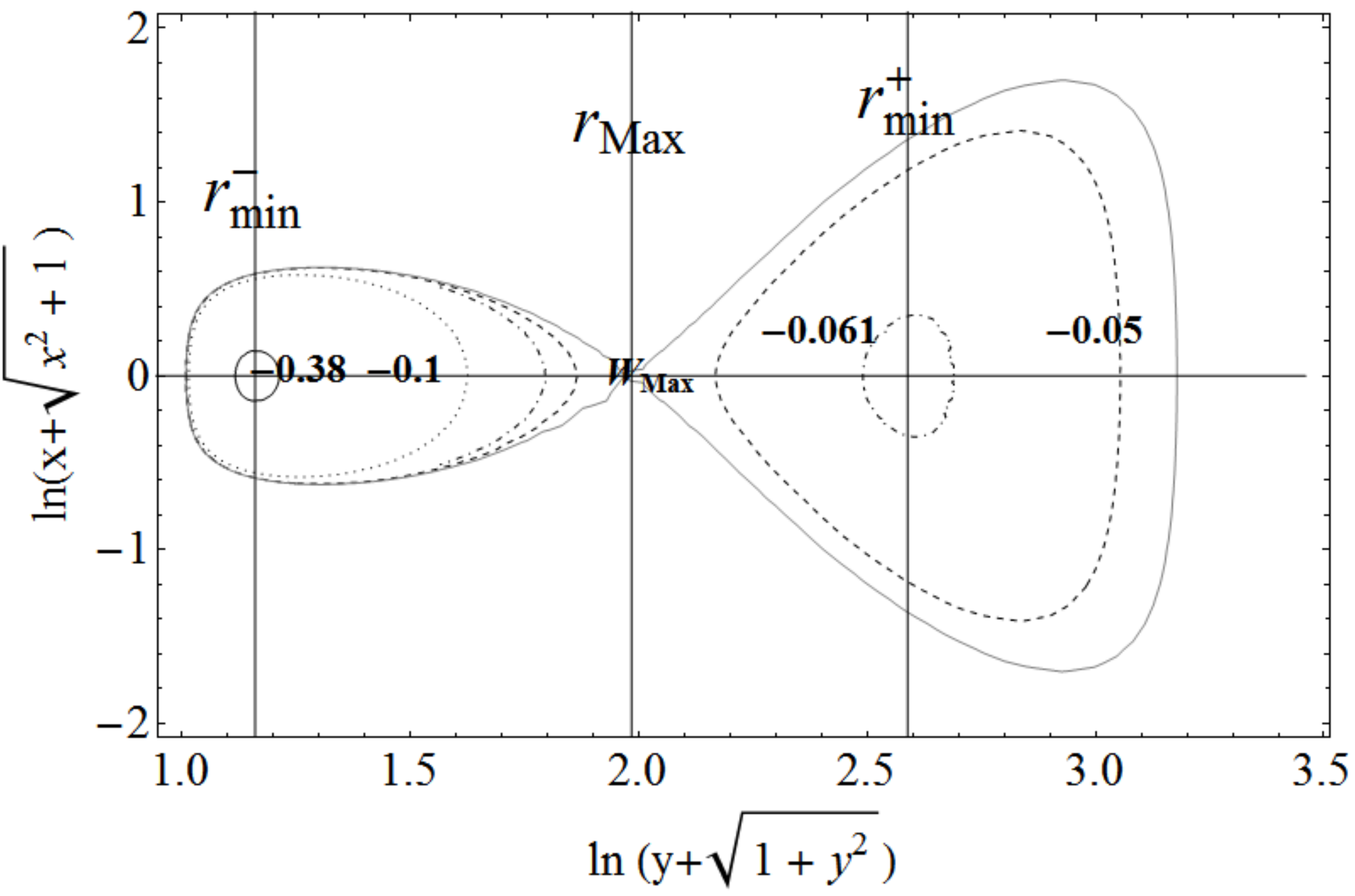}
\\
\includegraphics[width=.481\textwidth]{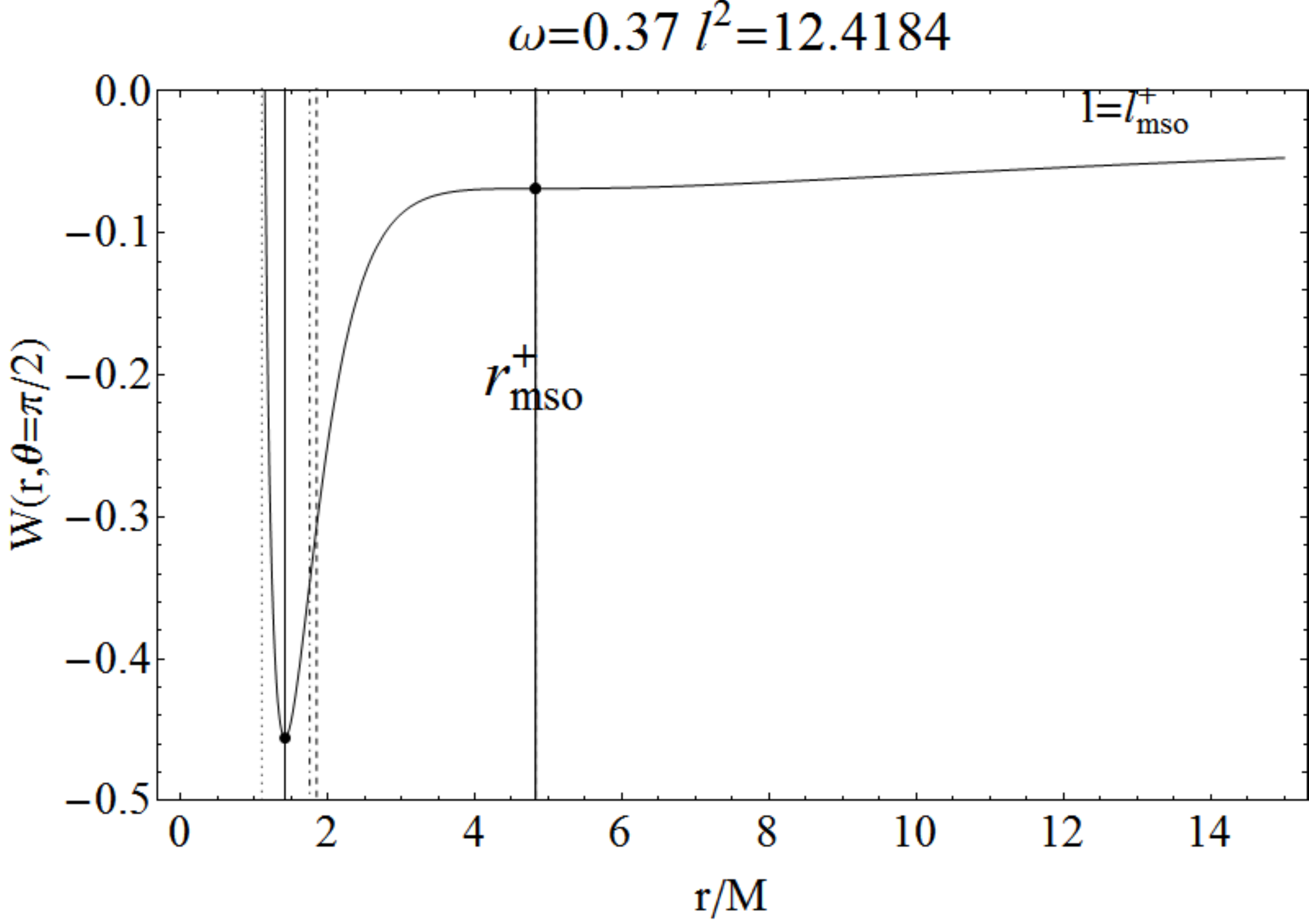}
\includegraphics[width=.451\textwidth]{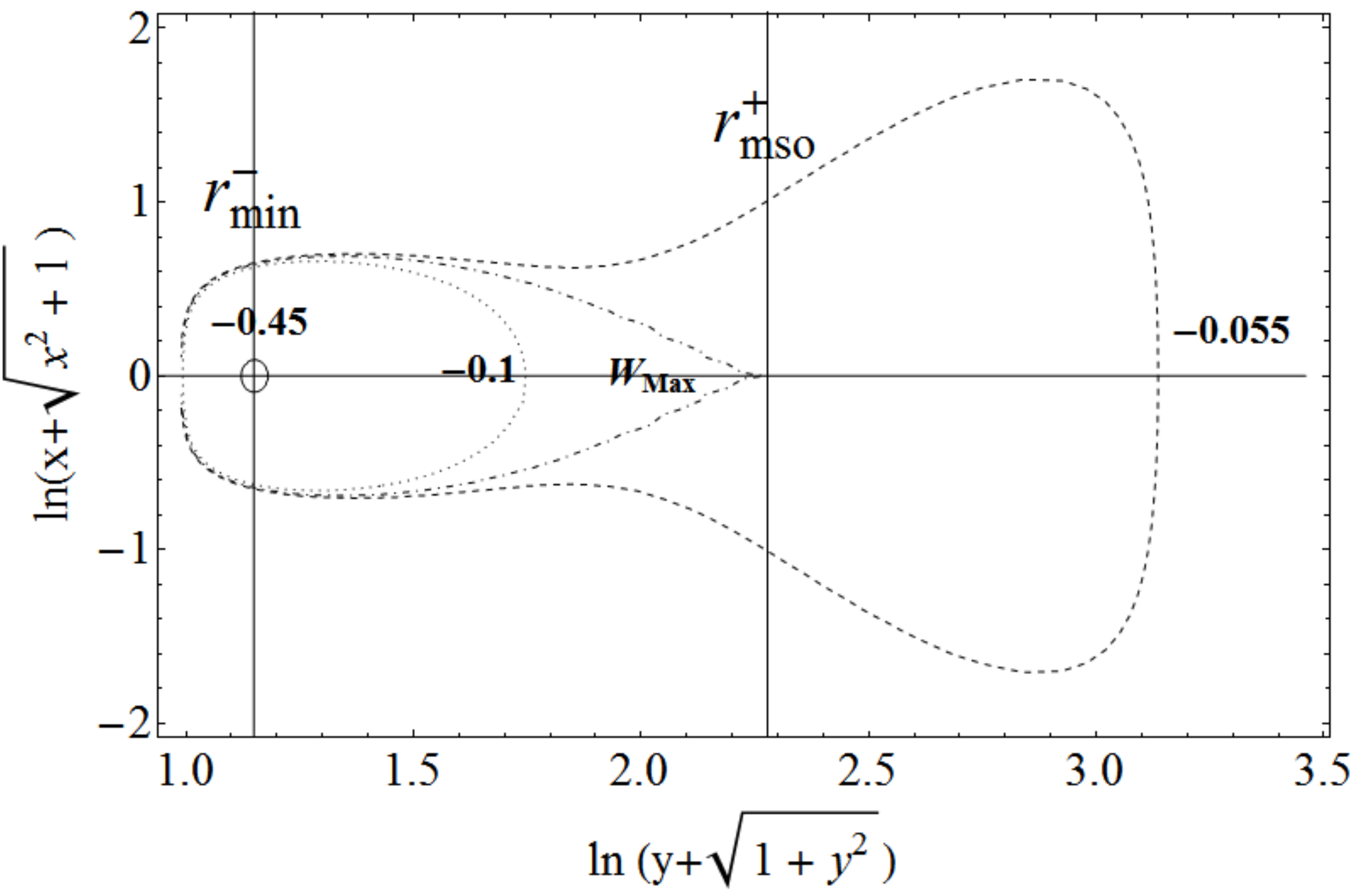}
\\
\includegraphics[width=.481\textwidth]{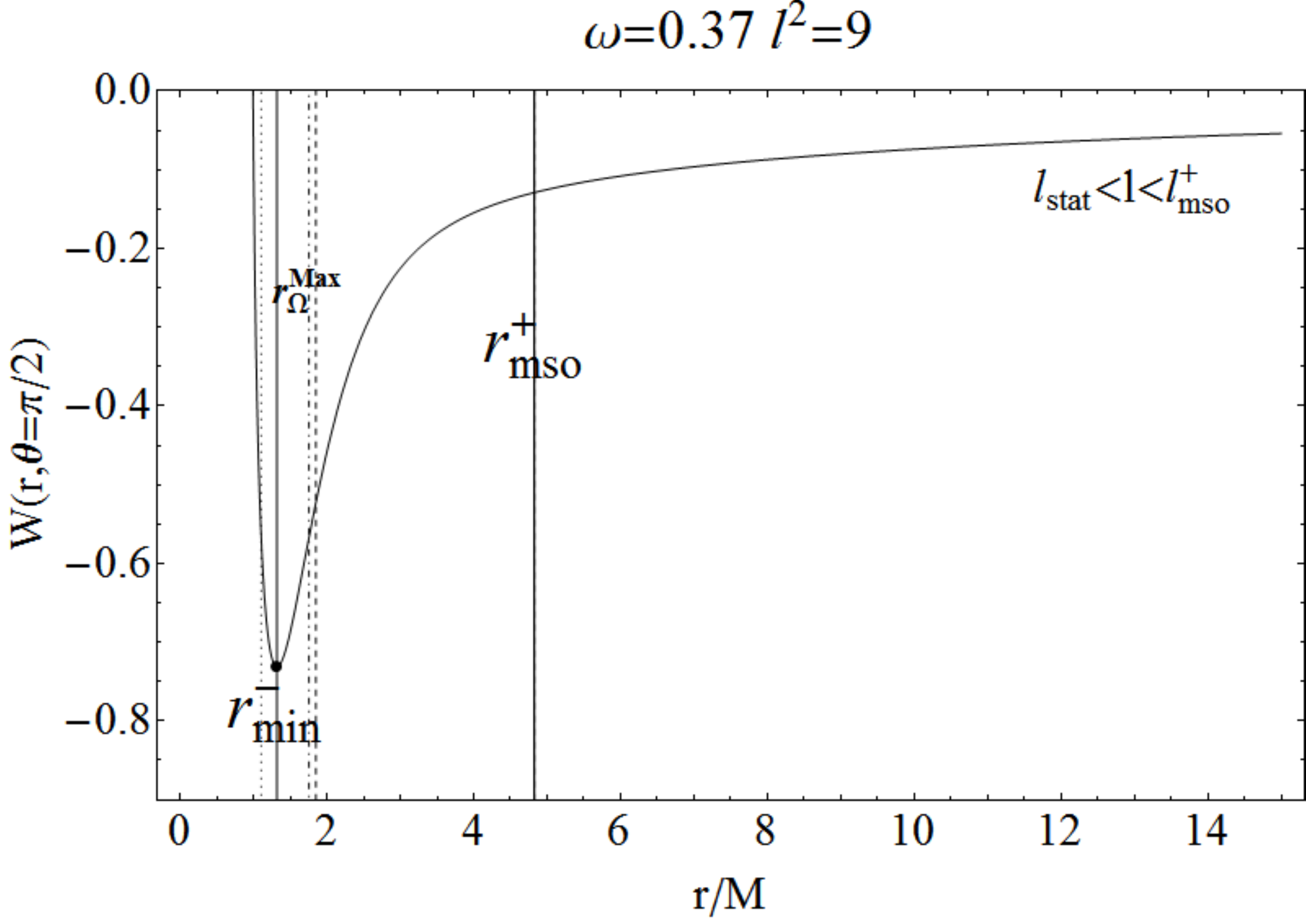}
\includegraphics[width=.451\textwidth]{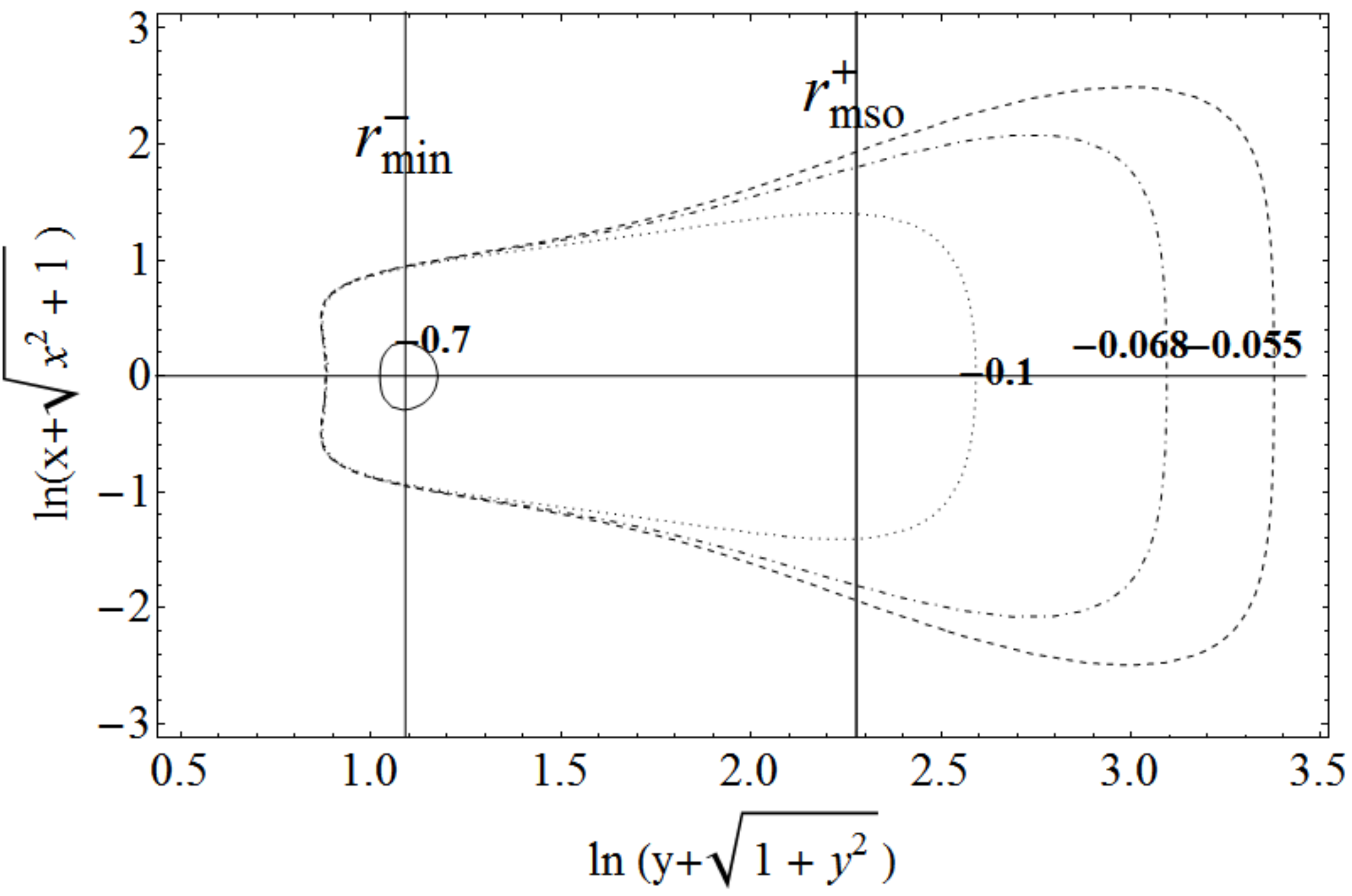}
%\\
\caption{Region III-b: $\omega\in]\omega_c,\omega_{\gamma}[$, naked singularity  $\omega M^2=0.37$, where $\omega M^2\rightarrow \omega$. It is $l_{mso}^->l_{\Omega}^{Max}> l^-_{mbo}> l^+_{mbo}> l_{mso}^+$, and $r_{stat}<r_{mbo}^-<r_{\Omega}^{Max}<r_{mso}^-<r_{mbo}^+<r_{mso}^+$. With $r/M=\sqrt{x^2+y^2}$ and  $(x,y)$ are Cartesian coordinates.  Vertical lines in right panels set the $r_i\in\mathfrak{R}$ and  the effective potential critical points.}
\label{Fig:outo-ty}
\end{figure}

In the naked singularity spacetimes  with Ho\v{r}ava parameter in $\omega\in]\omega_c,\omega_{\gamma}[$, the situation is as follows  $r_{stat}<r_{mbo}^-<r_{\Omega}^{Max}<r_{mso}^-<r_{mbo}^+<r_{mso}^+$ and
 and $l_{mso}^->l_{\Omega}^{Max}> l^-_{mbo}> l^+_{mbo}> l_{mso}^+$. \textbf{Region III-b} is a part of Class II attractors considered in Sec.\il(\ref{Subsec:ClassII}). As in \textbf{Region III-a}, this set of sources are characterized by excretion points.
 We detail the situation as follows:
 \begin{description}
 \item[-)$l>l_{mso}^-$] There are one set of closed configurations with center in $r_{min}: W_{min}<0$, see Fig.\il(\ref{Fig:newfigokey}-a).
 \item[-)$l=l_{mso}^-$] A saddle point of $W$ is at $r_{mso}^-$, as $W_{mso}^->0$  there is a set of closed configuration centered in the minimum $r_{min}$ and an open configuration with inner cusp, the two open branches are aligned on the axis {(this is therefore a descriminant case)}, see Fig.\il(\ref{Fig:NBib-for}).
\item[-) ${l\in]l_{\Omega}^{Max}, l_{mso}^-[}$] There are two minima for the $W$ function:  $W_{min}^->0$ and $W_{min}^+<0$. Consequently a doubled two configuration  $C^{\pm}$ of the \textbf{I}-type is possible, at different $K$  centered  respectively in $r_{min}^-$ the inner $C^-$ one  and $r_{min}^+$ the outer $C^+$  one. The maximum corresponds to an open crossed surface, with an outer cusp. This case is discussed in Fig.\il(\ref{Fig:NBib-for})-b, see also Fig.\il(\ref{Fig:namij-n})-a.
\item[-) $l=l_{\Omega}^{Max}$] The center of the inner disc $C^-$ is located in $r_{\Omega}^{Max}$, where $W_{\Omega}^{Max}>0$.  This case is illustrated in
\il(\ref{Fig:NBib-for})-c and corresponds to a \textbf{I}-type configuration.
\item[-)${l\in]l_{mbo}^-,l_{\Omega}^{Max}[}$]  There is an excretion open crossed \textbf{I} configuration, Fig.\il(\ref{Fig:ceg-his})-a.
\item[-) $l=l_{mbo}^-$] The center of the inner  \textbf{I}  disc is at $W_{mbo}^-$. This situation is similar to  the case $l\in]l_{\Omega}^{Max},l_{mbo}^-[$ and it is sketched in Fig.\il(\ref{Fig:ceg-his})-b.
\item[-) ${l\in]l_{mbo}^+,l_{mbo}^-[}$] This case is shown in Fig.\il(\ref{Fig:ceg-his})-c. This is a \textbf{III} type configuration. Two configurations $C^{\pm}$ are therefore possible at equal $W\in]W_{min}^+, 0[$. Excretion towards the exterior is possible. In general there is a  sequence of \textbf{I-II-III} configurations.
    \item[-) $l=l_{mbo}^+$] Respect to the case  ${l\in]l_{mbo}^+,l_{mbo}^-[}$ the maximum point is at $r_{mbo}^+$, where $W_{mbo}^+=0$, there is then a crossed open \textbf{III} configuration, Fig.\il(\ref{Fig:ceg-his})-d
\item[-) ${l\in]l_{mso}^+,l_{mbo}^+[}$]  There is a crossed \textbf{III}  closed surface Fig.\il(\ref{Fig:outo-ty})-a.
\item[-) $l=l_{mso}^+$]
There is and outer ``cusp'' in $r_{mso}^+$  Fig.\il(\ref{Fig:outo-ty})-b.
\item[-)${l\in]l_{stat},l_{mso}^+[}$] Only closed toroidal surfaces with center in $r_{min}$ are possible, Fig.\il(\ref{Fig:outo-ty})-c.
 \end{description}
\subsubsection{Naked singularity: $\omega=\omega_{\gamma}$}\label{Sec:NSpar4}
\begin{figure}[h]
%%CPlotoiscom
\includegraphics[width=.481\textwidth]{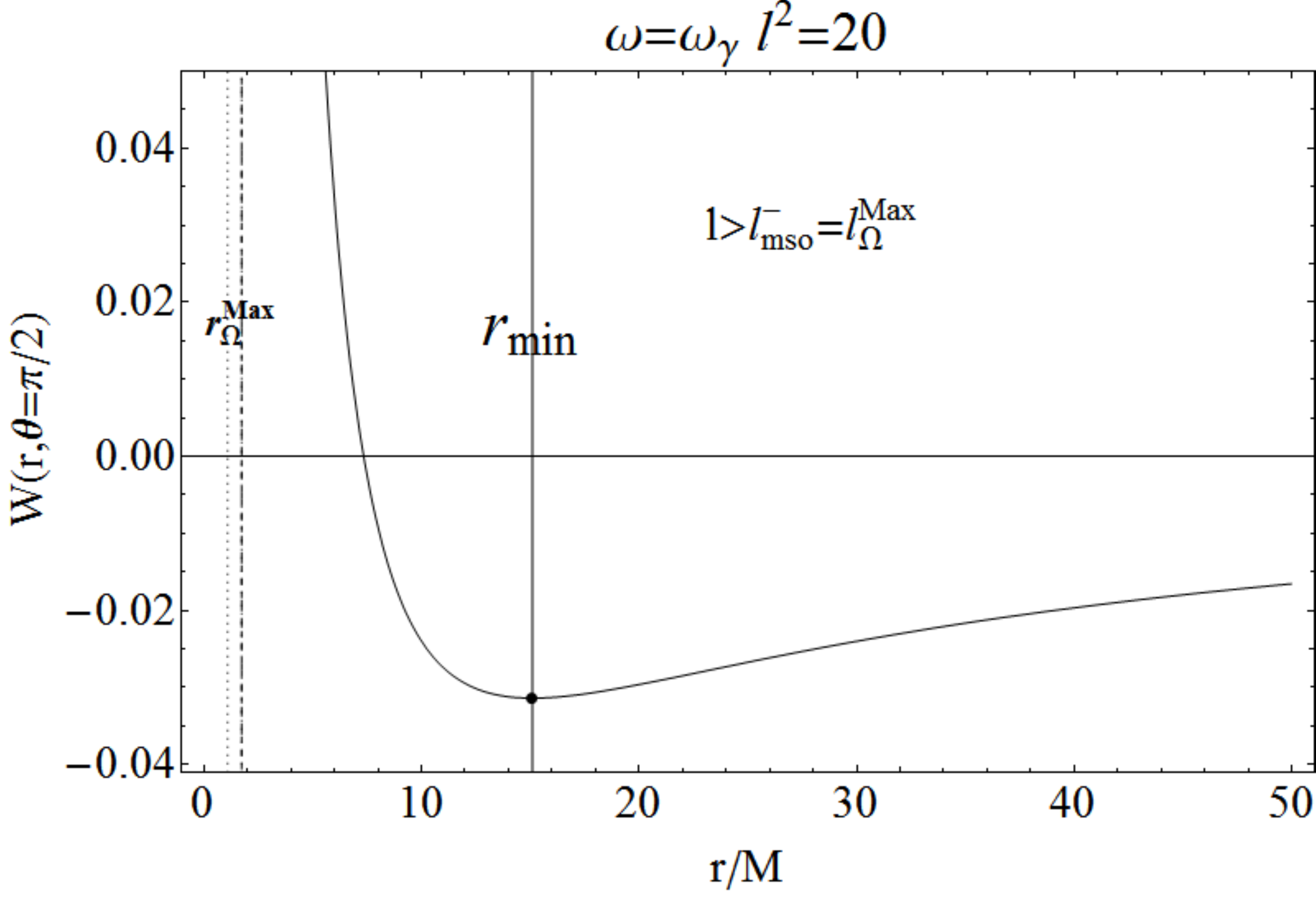}
\includegraphics[width=.31\textwidth]{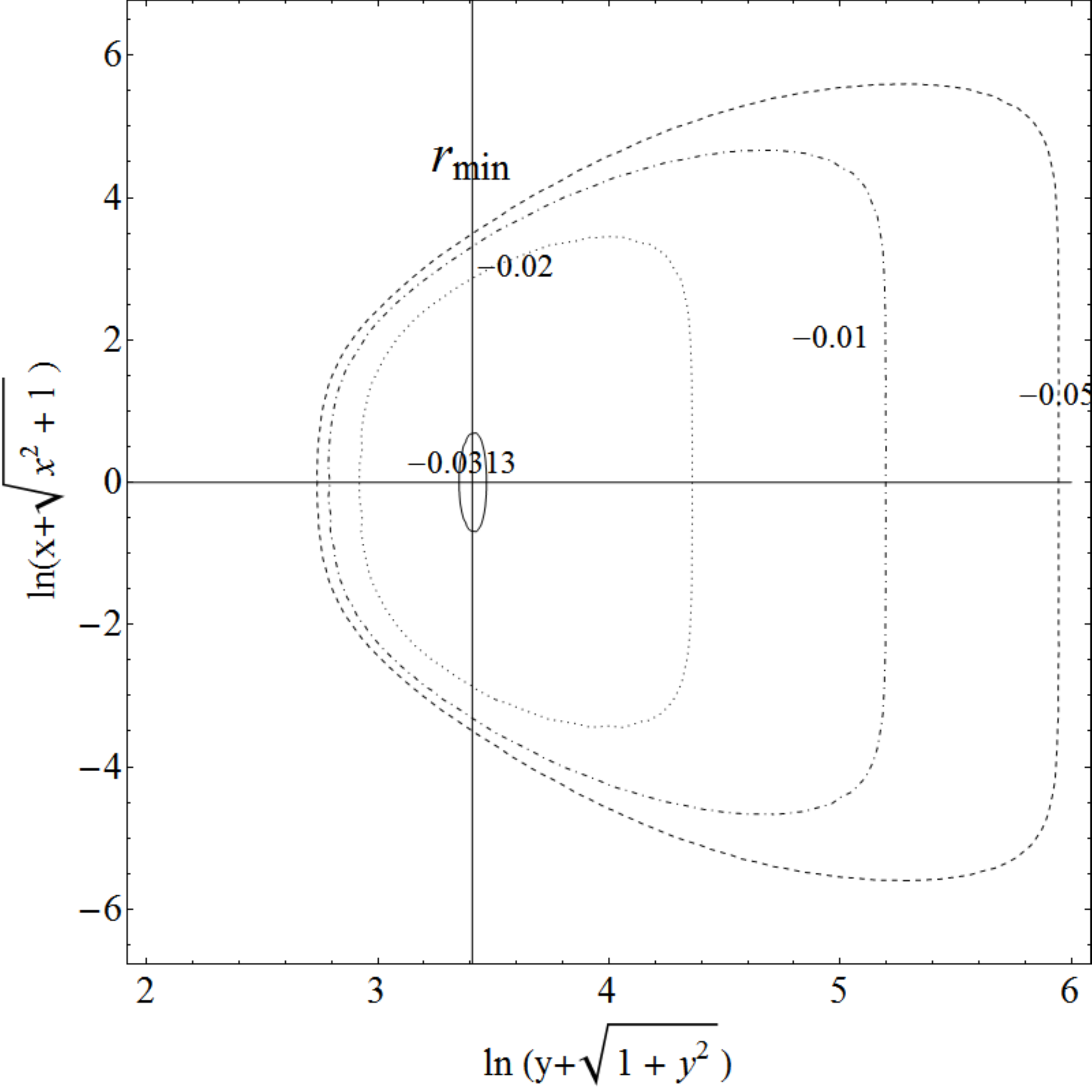}
%\\
\\
\includegraphics[width=.481\textwidth]{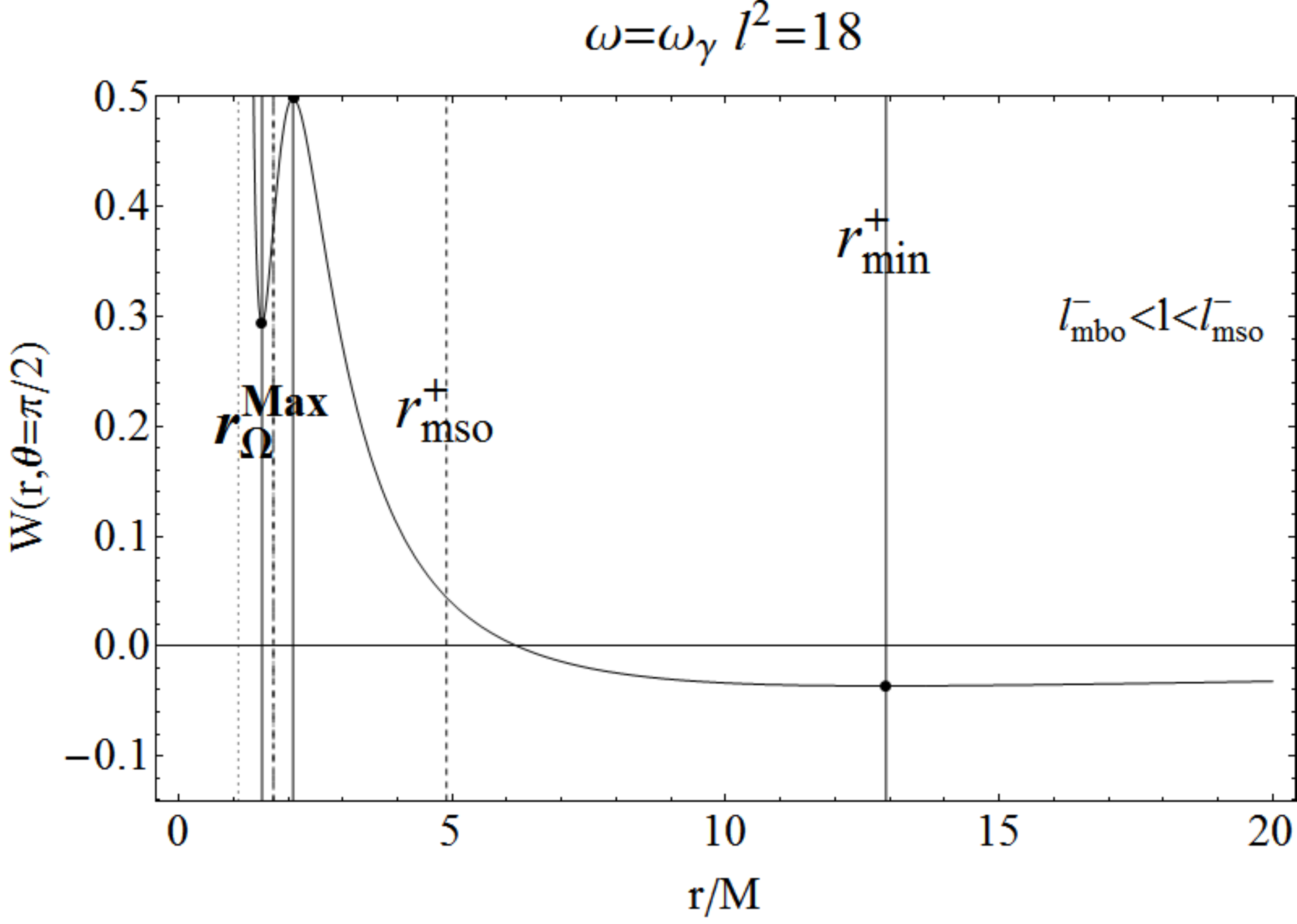}
\includegraphics[width=.31\textwidth]{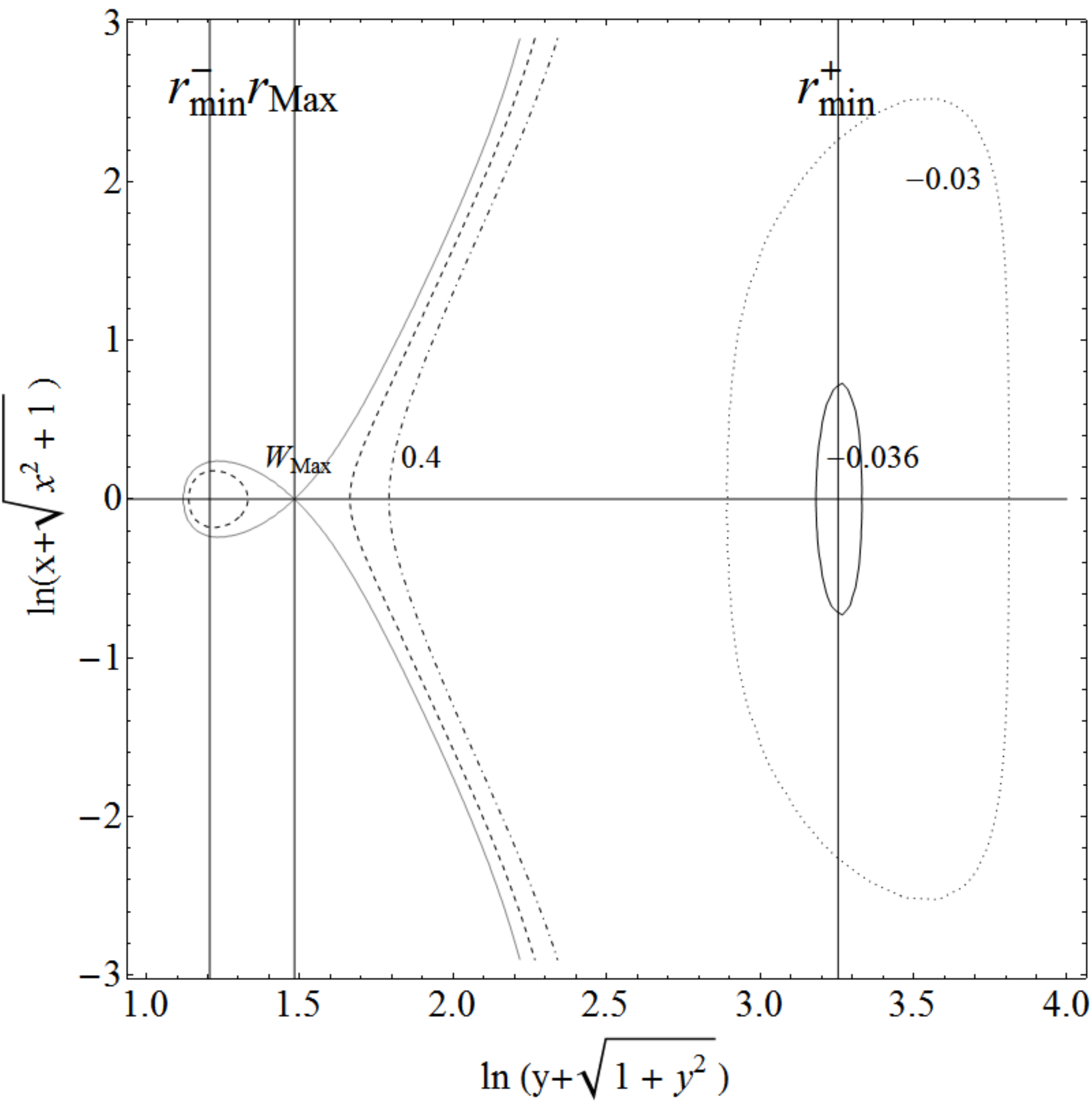}
\caption{Naked singularity: $\omega=\omega_{\gamma}$. Where $\omega M^2\rightarrow \omega$. It is $l_{mso}^-=l_{\Omega}^{Max}> l^-_{mbo}> l^+_{mbo}> l_{mso}^+$, and $r_{stat}<r_{mbo}^-<r_{\Omega}^{Max}=r_{\gamma}^{\pm}=r_{mso}^-<r_{mbo}^+<r_{mso}^+$. With $r/M=\sqrt{x^2+y^2}$ and  $(x,y)$ are Cartesian coordinates. Vertical lines in right panels set the $r_i\in\mathfrak{R}$ and  the effective potential critical points.}
\label{Fig:cioc-e1}
\end{figure}

\begin{figure}[h]
%%CPlotoiscom
\includegraphics[width=.481\textwidth]{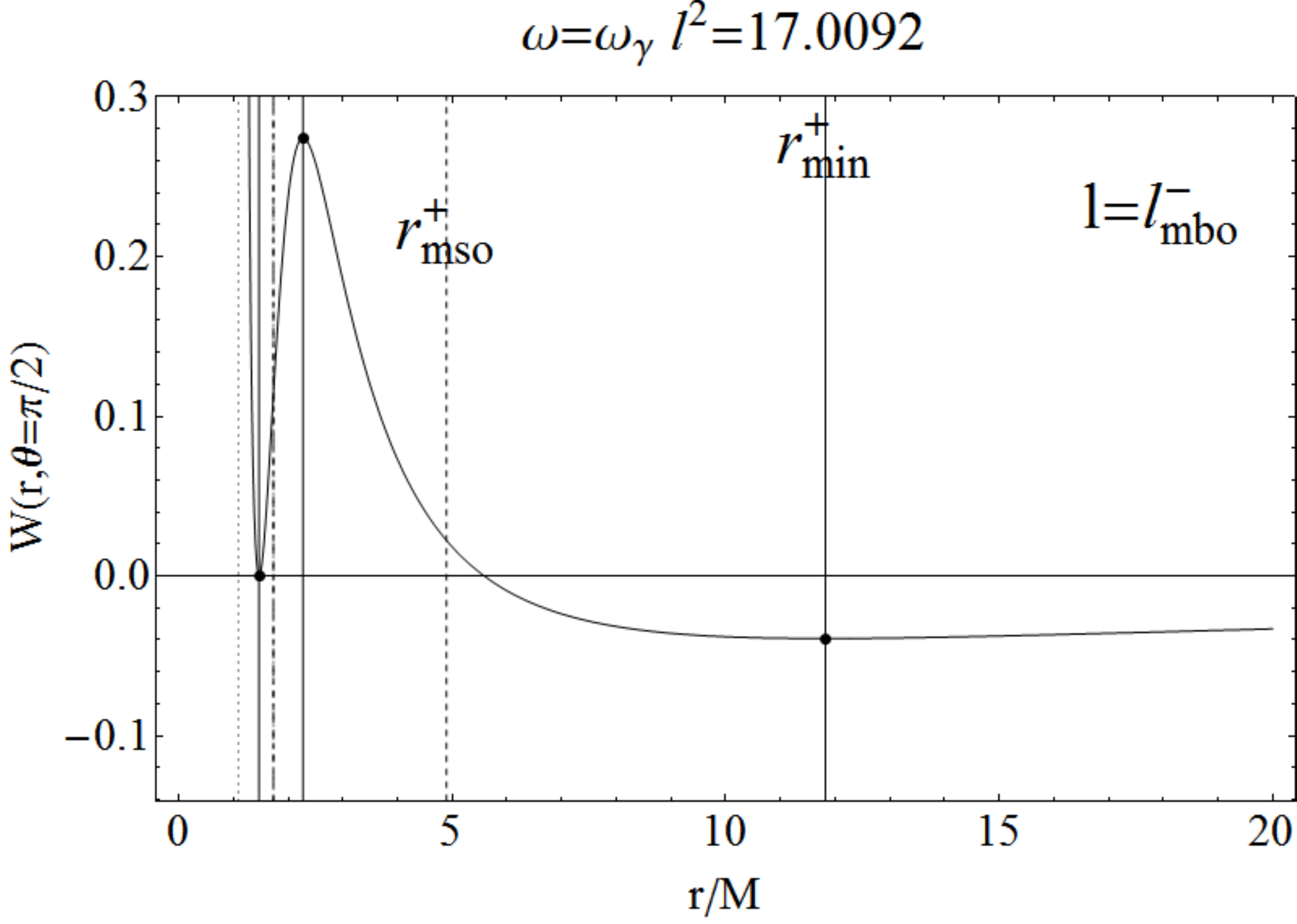}
\includegraphics[width=.31\textwidth]{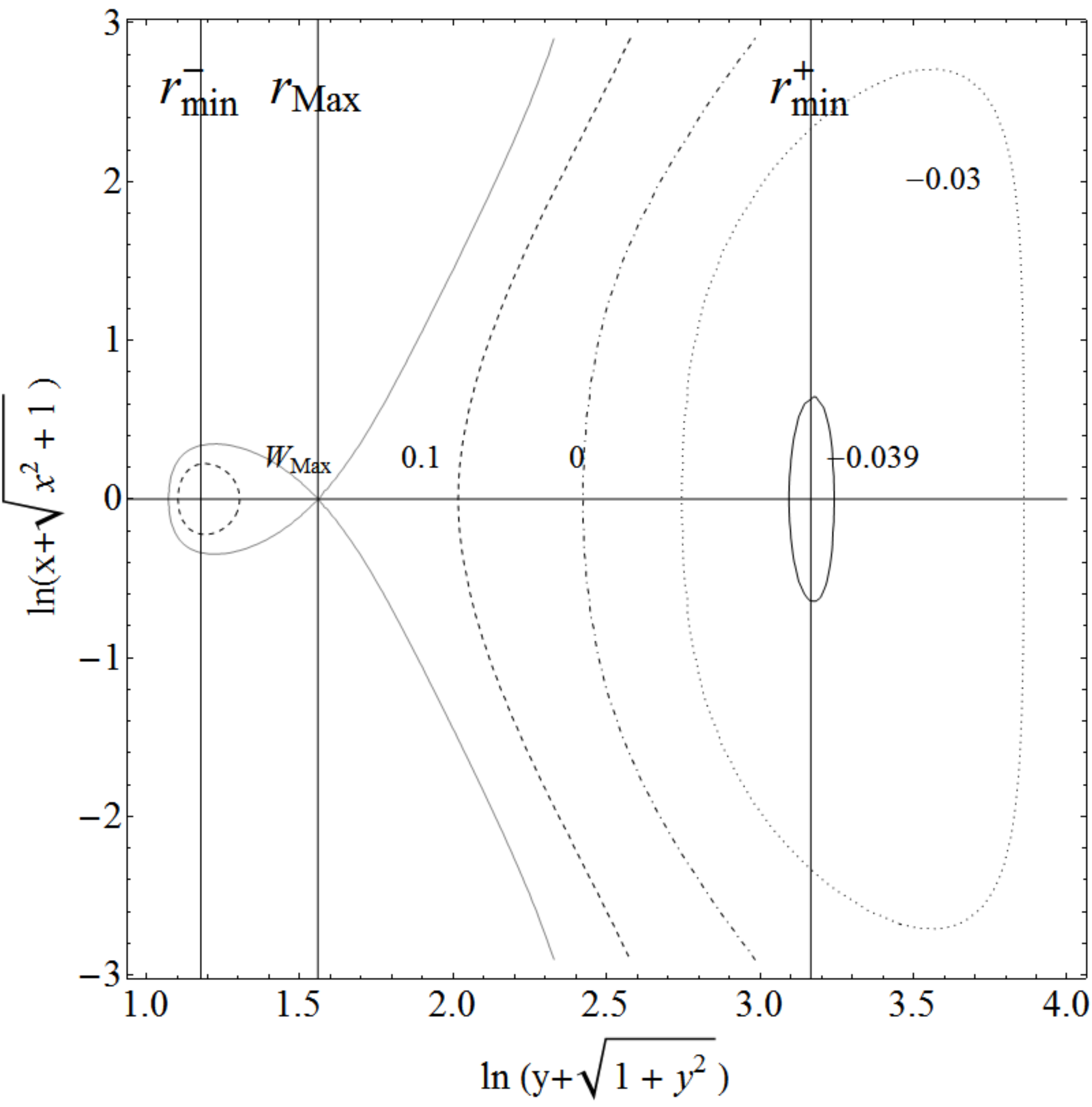}
\\
\includegraphics[width=.481\textwidth]{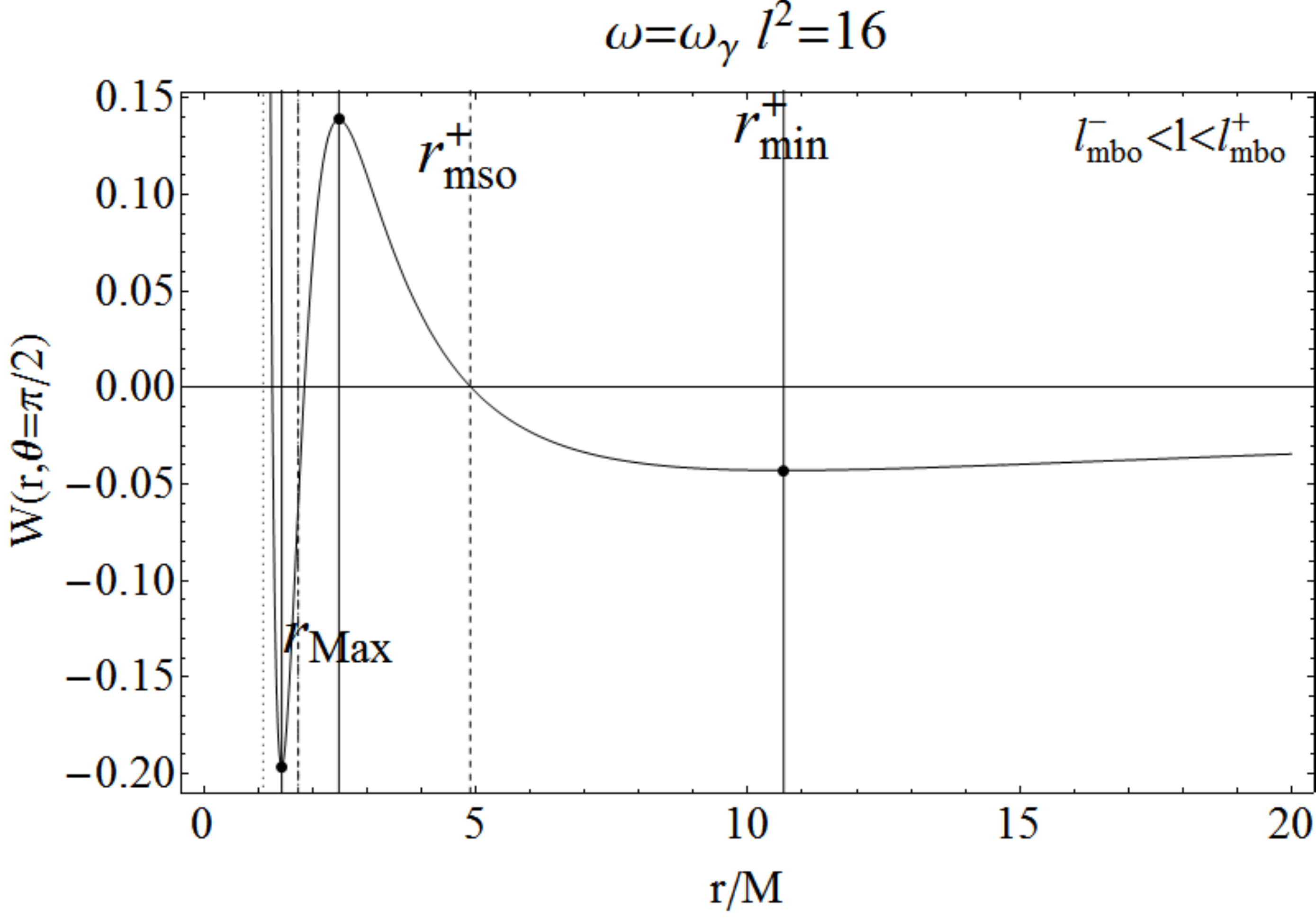}
\includegraphics[width=.31\textwidth]{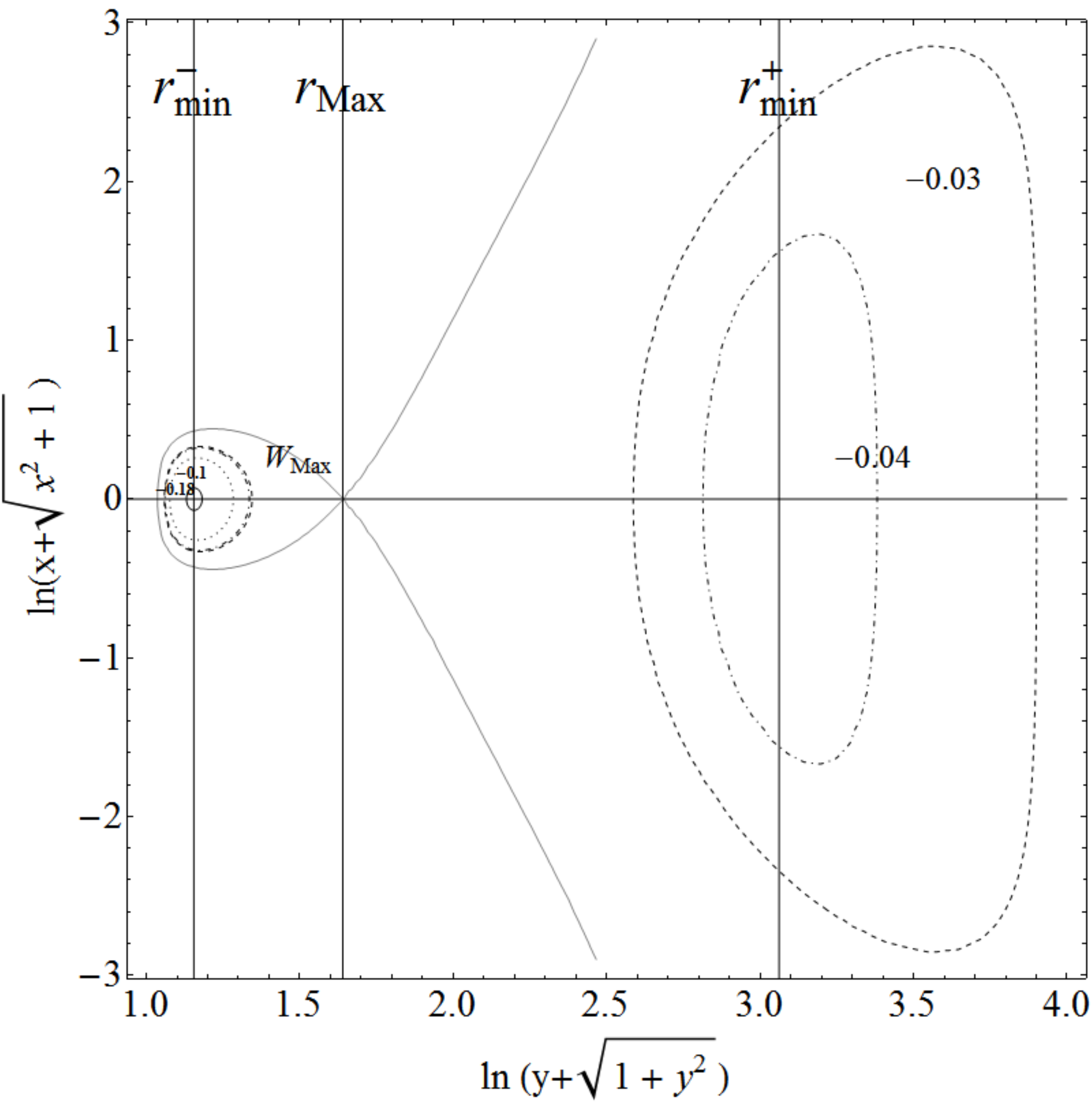}
\\
\includegraphics[width=.481\textwidth]{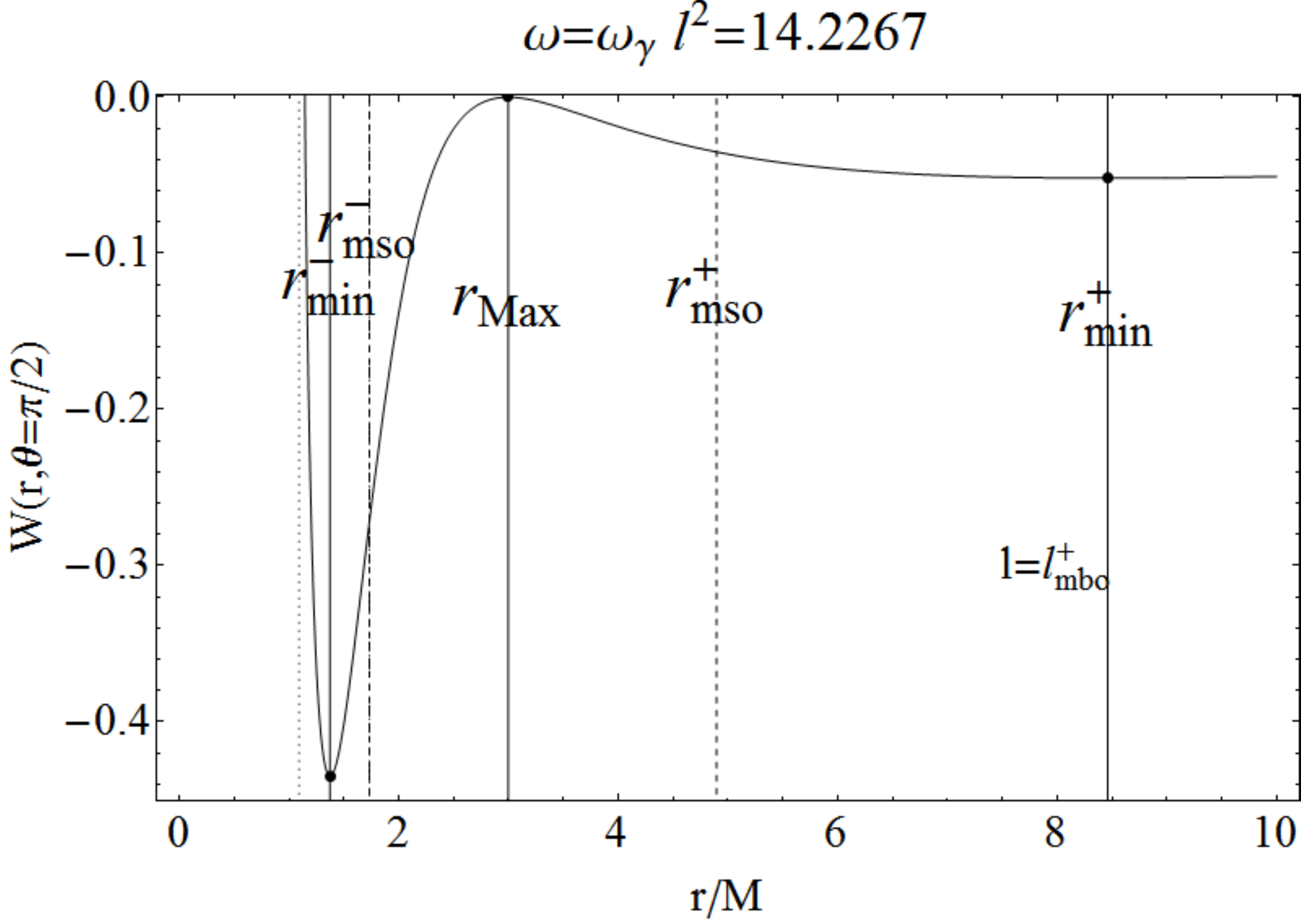}
\includegraphics[width=.31\textwidth]{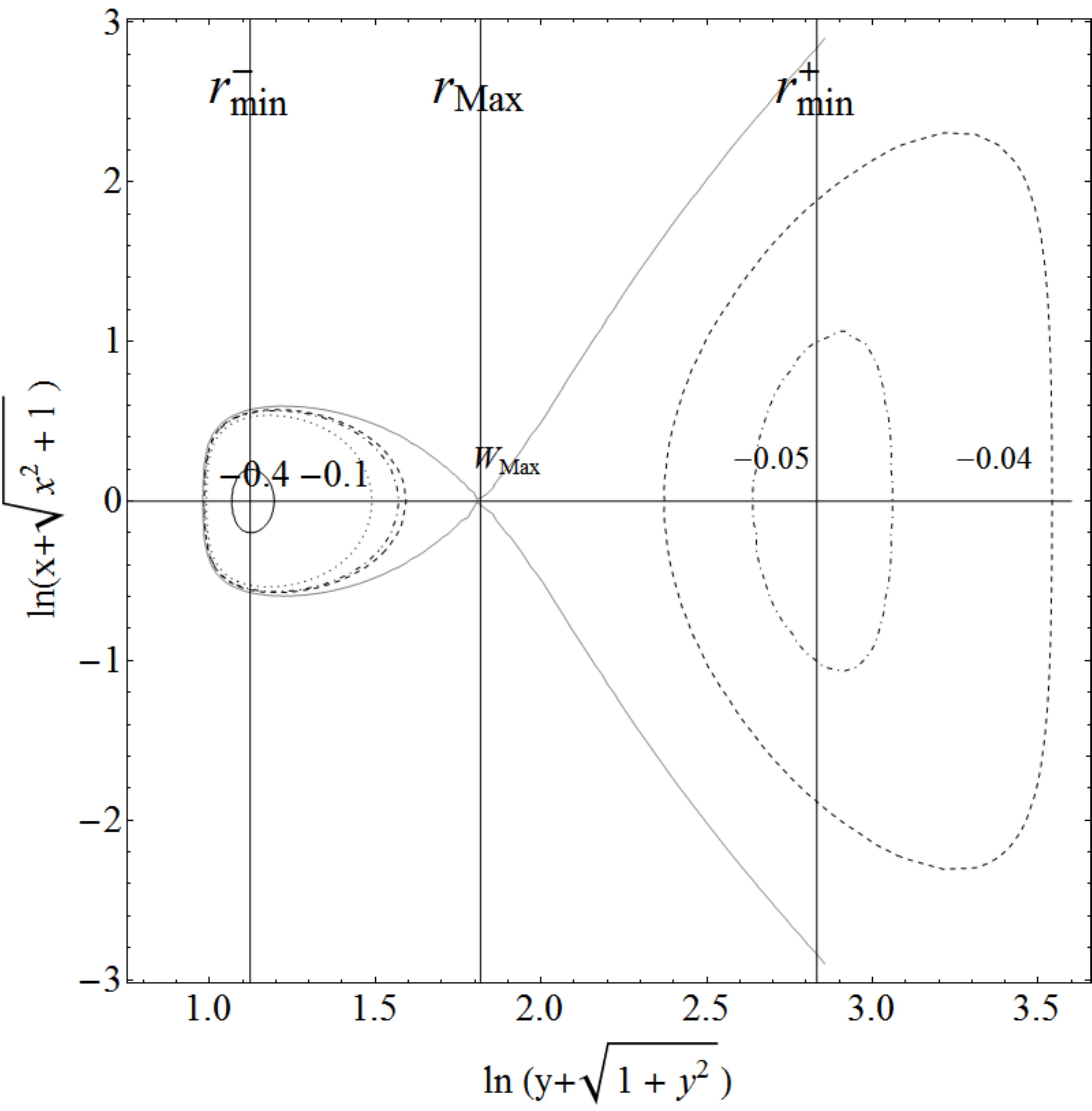}
\\
\includegraphics[width=.481\textwidth]{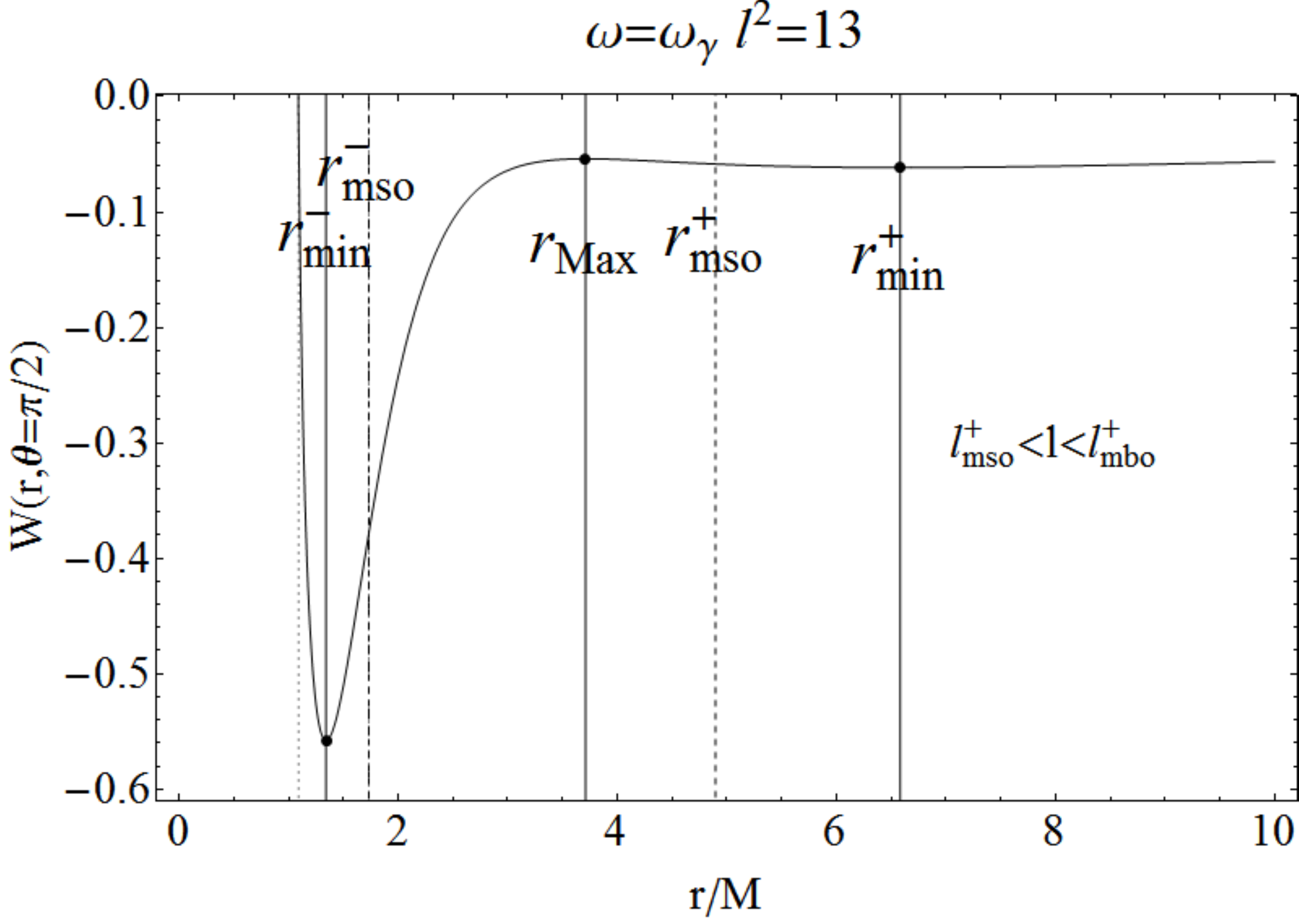}
\includegraphics[width=.31\textwidth]{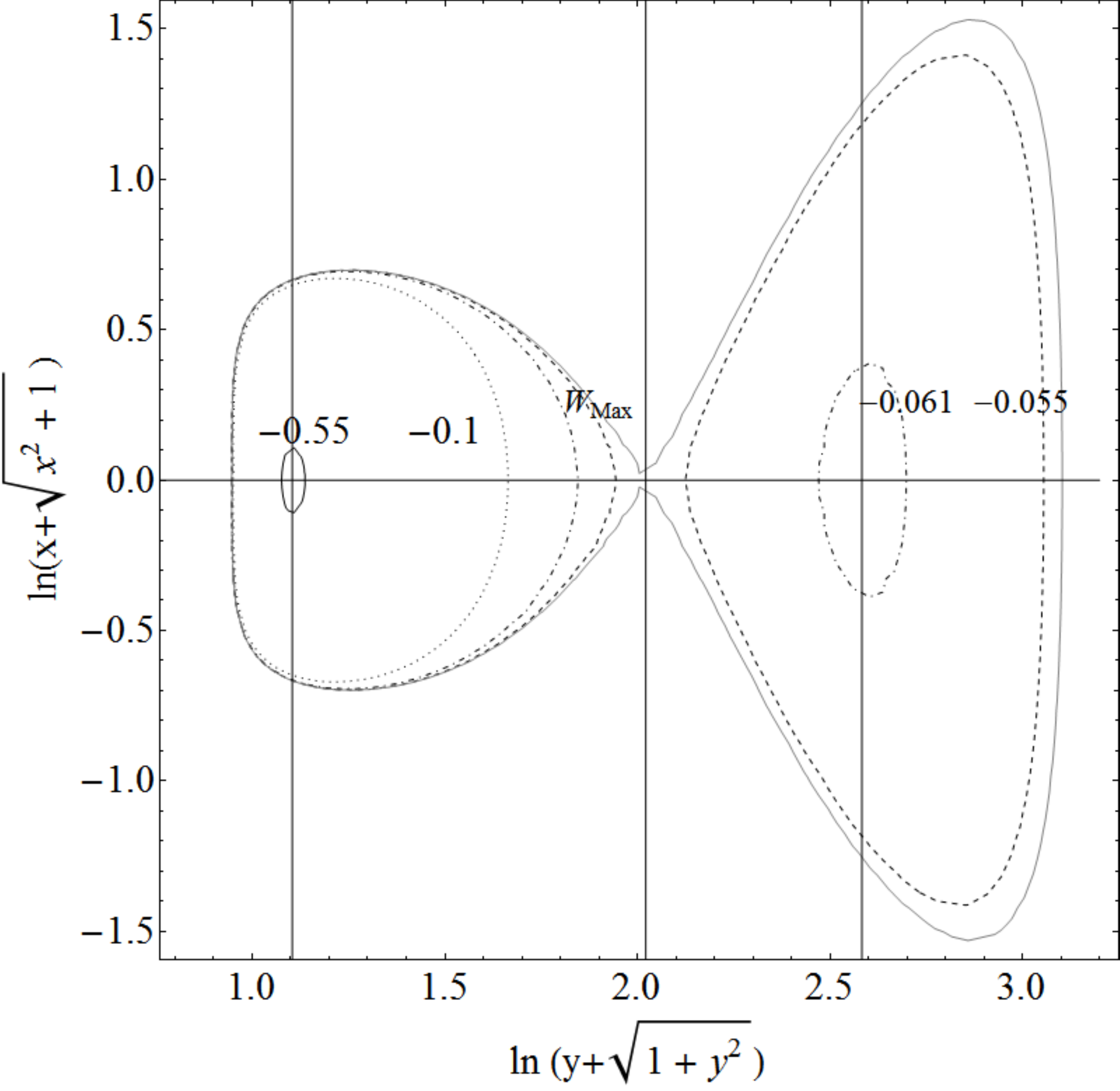}
\caption{Naked singularity $\omega=\omega_{\gamma}$. Where $\omega M^2\rightarrow \omega$.  Vertical lines in right panels set the $r_i\in\mathfrak{R}$ and  the effective potential critical points. It is $l_{mso}^-=l_{\Omega}^{Max}> l^-_{mbo}> l^+_{mbo}> l_{mso}^+$, and $r_{stat}<r_{mbo}^-<r_{\Omega}^{Max}=r_{\gamma}^{\pm}=r_{mso}^-<r_{mbo}^+<r_{mso}^+$.  With $r/M=\sqrt{x^2+y^2}$ and  $(x,y)$ are Cartesian coordinates.}
\label{Fig:Oh-De1}
\end{figure}

\begin{figure}[h]
%%CPlotoiscom
\includegraphics[width=.481\textwidth]{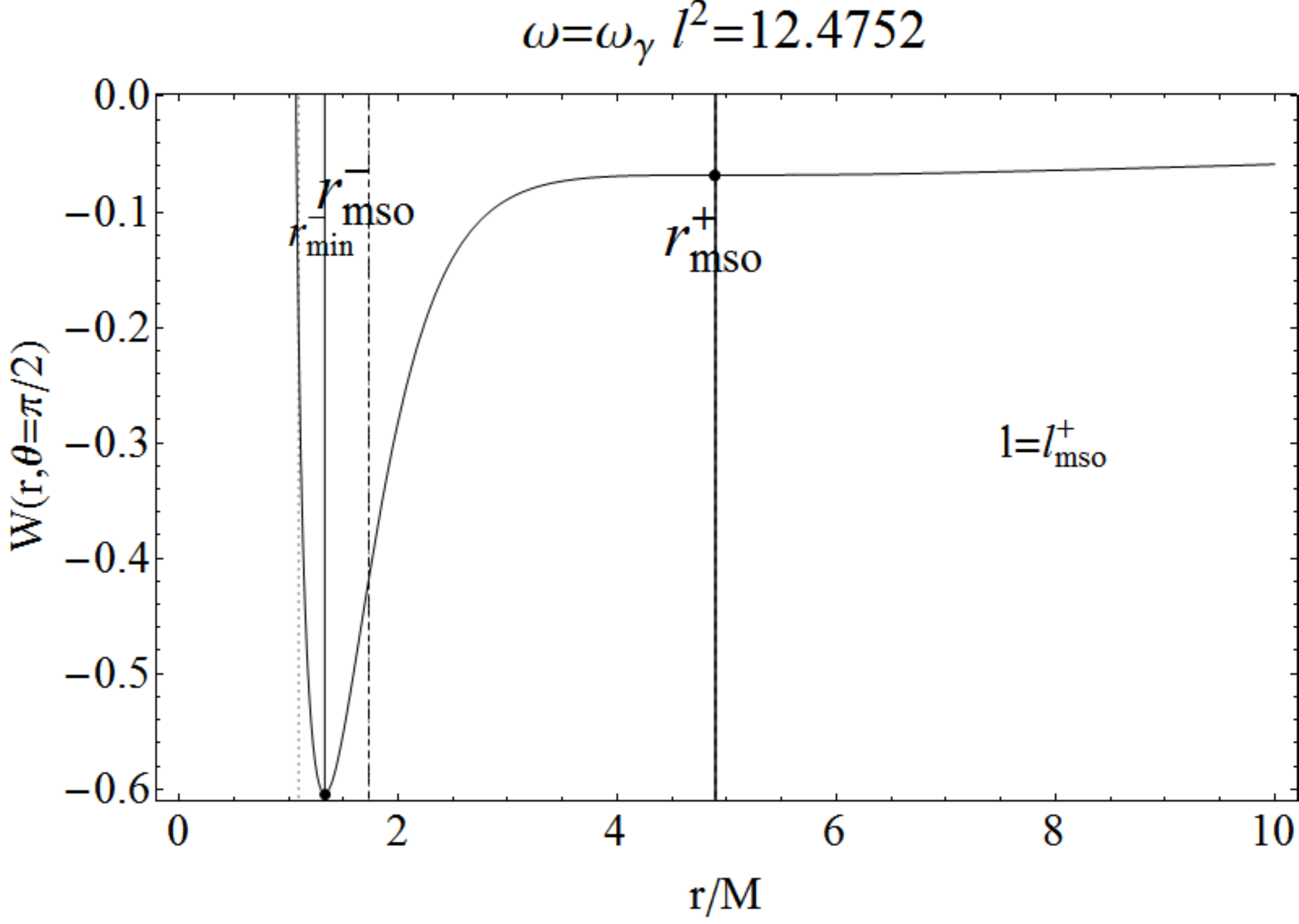}
\includegraphics[width=.31\textwidth]{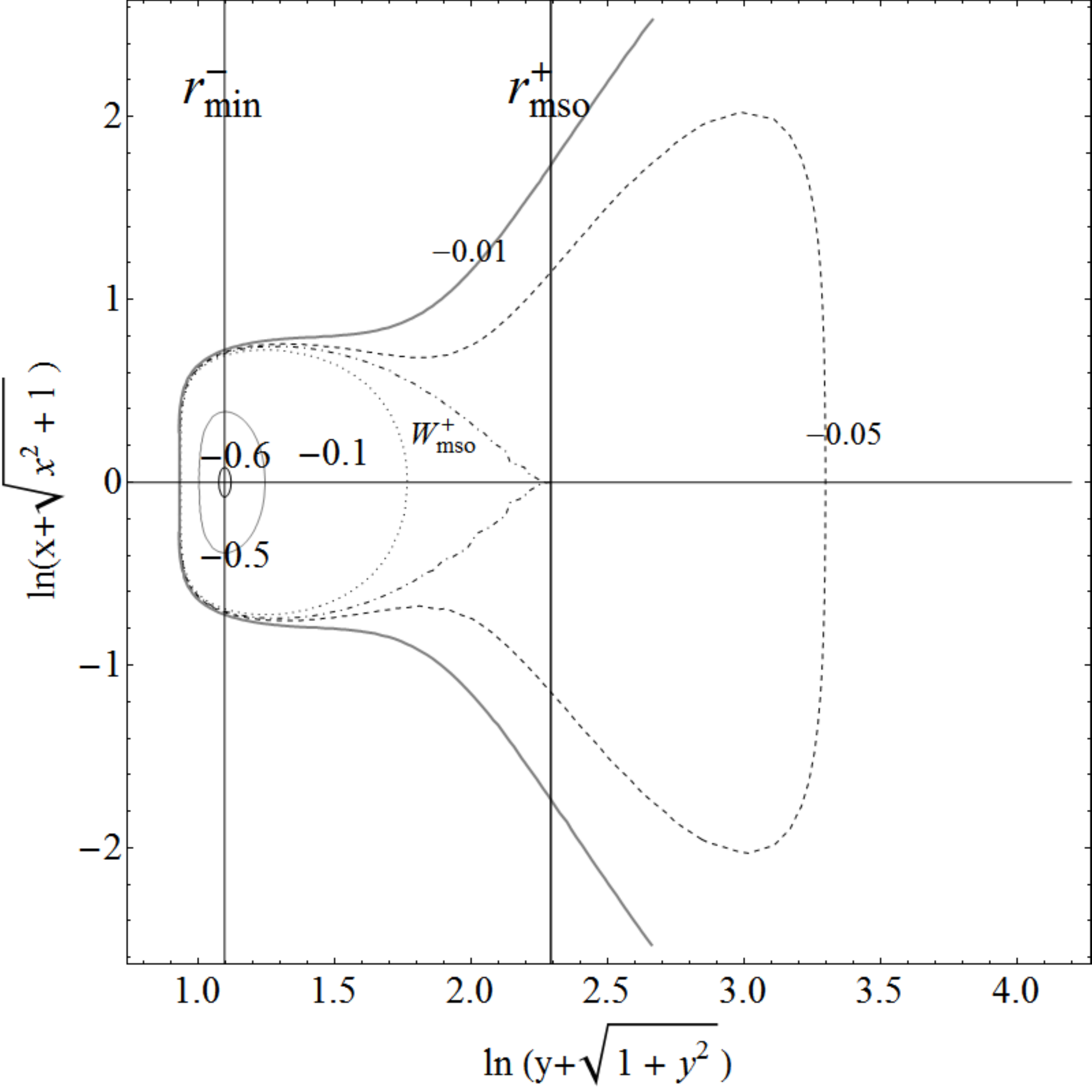}
\\
\includegraphics[width=.481\textwidth]{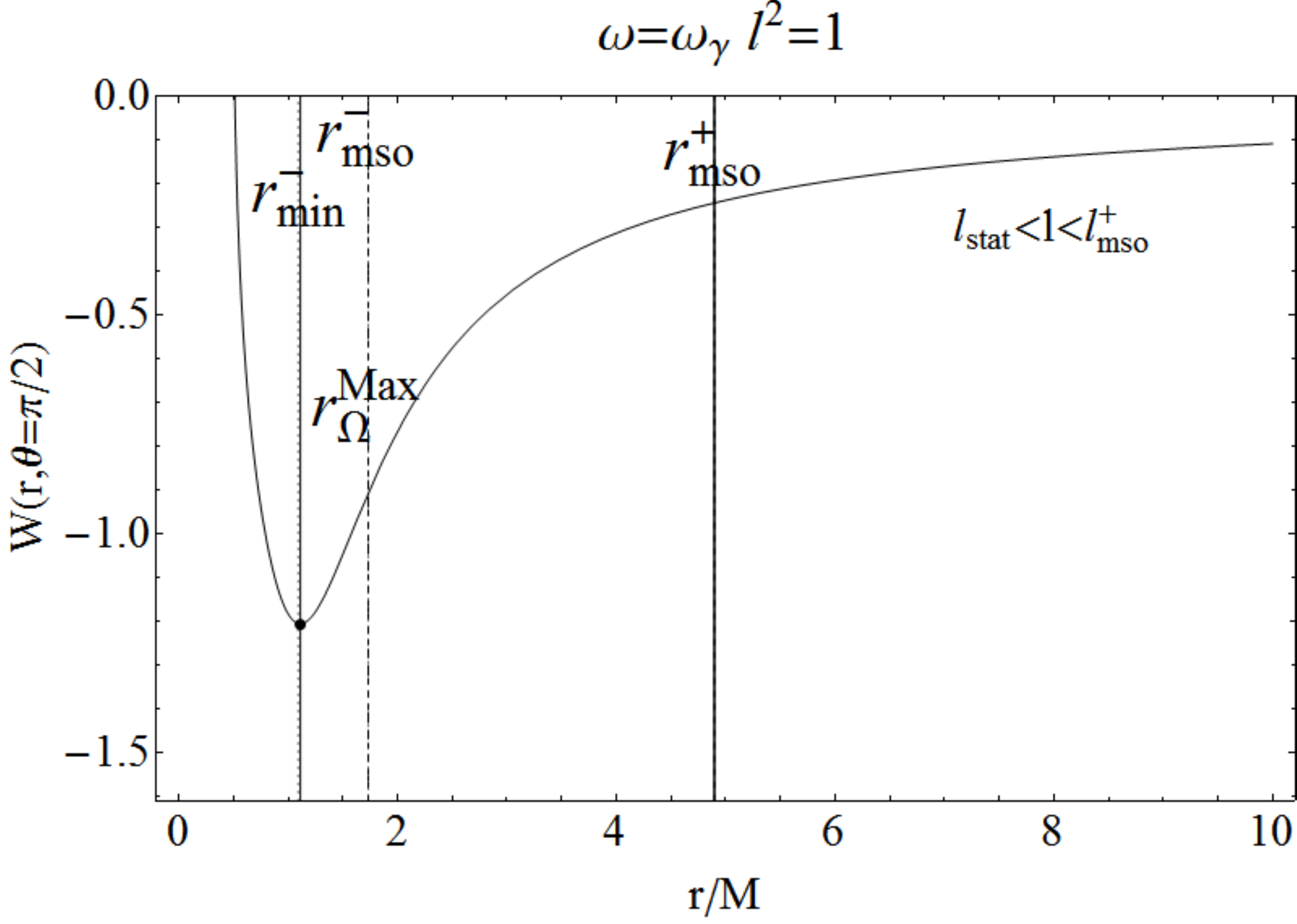}
\includegraphics[width=.31\textwidth]{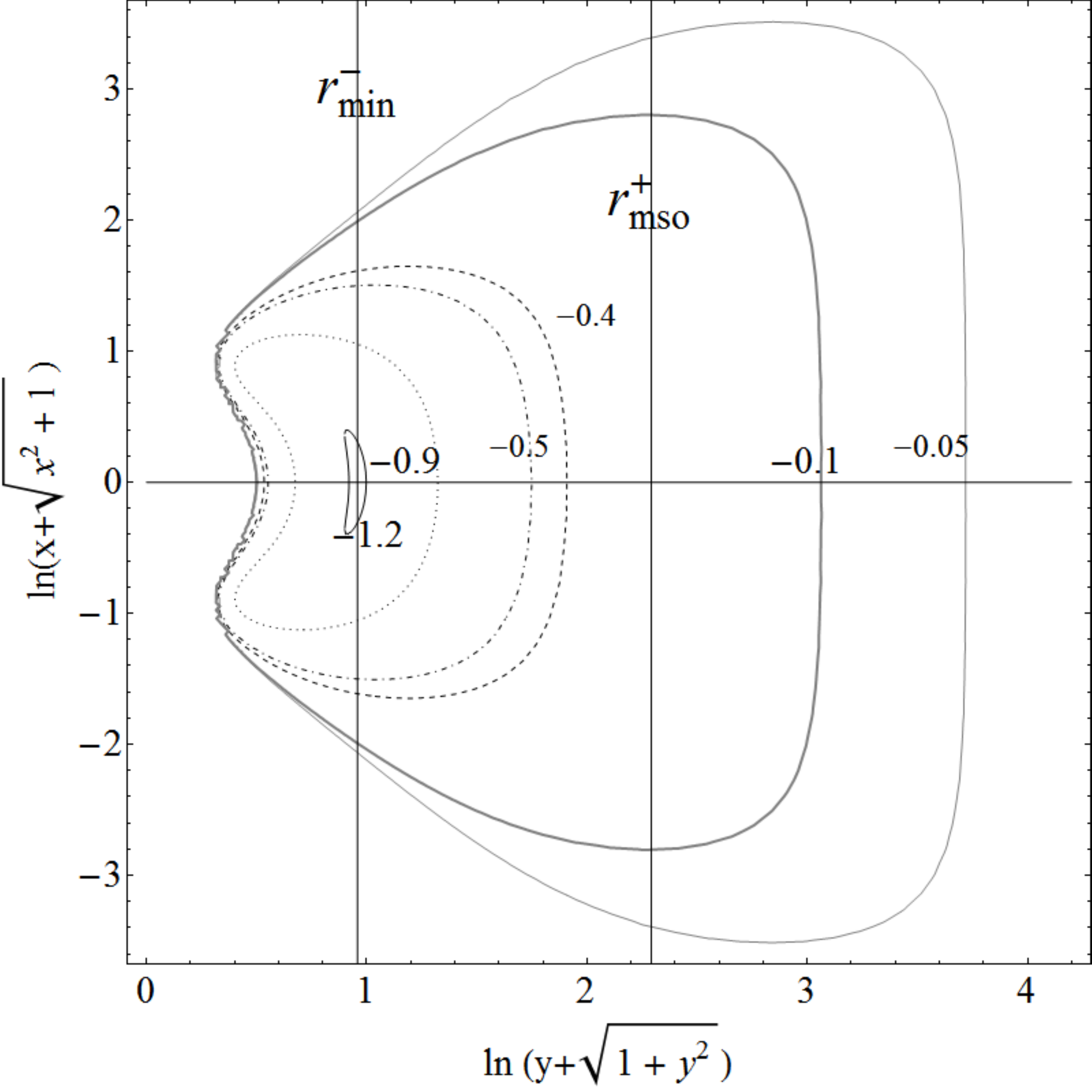}
%\\
\caption{Naked singularity $\omega=\omega_{\gamma}$. Where $\omega M^2\rightarrow \omega$. It is $l_{mso}^-=l_{\Omega}^{Max}> l^-_{mbo}> l^+_{mbo}> l_{mso}^+$, and $r_{stat}<r_{mbo}^-<r_{\Omega}^{Max}=r_{\gamma}^{\pm}=r_{mso}^-<r_{mbo}^+<r_{mso}^+$. Vertical lines in right panels set the $r_i\in\mathfrak{R}$ and  the effective potential critical points.  With $r/M=\sqrt{x^2+y^2}$ and  $(x,y)$ are Cartesian coordinates.}
\label{Fig:tar-De1}
\end{figure}

In this section we analyse the properties of fluid tori in special the K-S-spacetime  with $\omega=\omega_{\gamma}$ where
$r_{stat}<r_{mbo}^-<r_{\Omega}^{Max}=r_{\gamma}^{\pm}=r_{mso}^-<r_{mbo}^+<r_{mso}^+$ and $l_{mso}^-=l_{\Omega}^{Max}> l^-_{mbo}> l^+_{mbo}> l_{mso}^+$.
This is a discriminant case splitting the K-S naked singularity spacetimes containing two photon circular orbits $\omega\in]\omega_{\gamma},1/2[$, from the sources in the \textbf{Region I-II-IIIa-IIIb} with no photon orbits.
\begin{description}
\item[-)$l\geq l_{mso}^-=l_{\Omega}^{Max}$] There are only closed configuration, see Fig.\il(\ref{Fig:cioc-e1})-(a)
\item[-) ${l\in[l_{mbo}^-,l_{\Omega}^{Max}[}$] These cases are discussed in Fig.\il(\ref{Fig:cioc-e1})-b and Fig.\il(\ref{Fig:Oh-De1})-a. There is a closed $C^+$ configuration centered in $r_{min}^+$ and an inner $C^-$ one with center in $r_{min}^-$ with an open excretion point. This is a \textbf{I} type configuration.
\item[-)${l\in]l_{mbo}^+,l_{mbo}^-]}$] This case is illustrated in Figs.\il(\ref{Fig:Oh-De1})-(b, c), these are \textbf{III}-class configurations.  There is  a shift between a \textbf{I} and \textbf{III}  configuration, where a \textbf{II} type with  $W_{min}^-=W_{min}^+<0$ appears.
\item[-) ${l\in]l_{mso}^+,l_{mbo}^+[}$] There is a double closed $C^{\pm}$ \textbf{III} configurations eventually closing into a single  closed  crossed configuration Fig.\il(\ref{Fig:Oh-De1})-d.
\item[-) $l=l_{mso}^+$] An outer cusp is located in  $r_{mso}^+$, there is only one class of closed $C^-$ surfaces see Figs.\il(\ref{Fig:tar-De1})-a and an excretion configuration.
\item[-) ${l\in]l_{stat},l_{mso}^+[}$] Only closed toroidal surfaces are possible, Fig.\il(\ref{Fig:tar-De1})-b.
\end{description}
%
%%%%%%%%%%%%%%%%%%%%%%%%%%%%%%%%%%%%%%%%%%%%%%%%%%%%%%%%
%
%\clearpage
\subsubsection{{Region IV}: $\omega\in]\omega_{\gamma},0.5[$}\label{Sec:NSregionIV}
\begin{figure}[h]
%%CPlotoiscomPlol219
\includegraphics[width=.481\textwidth]{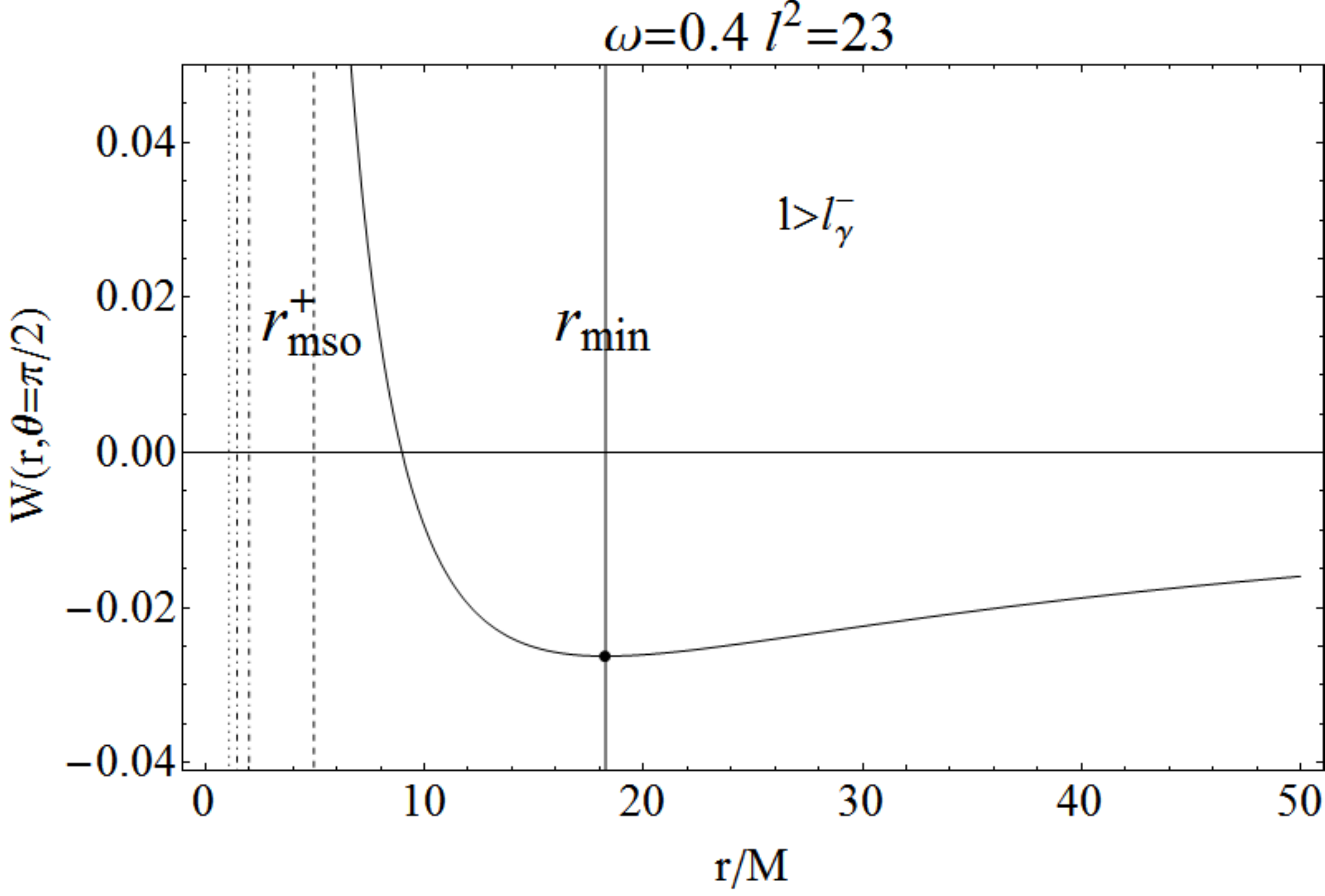}
\includegraphics[width=.31\textwidth]{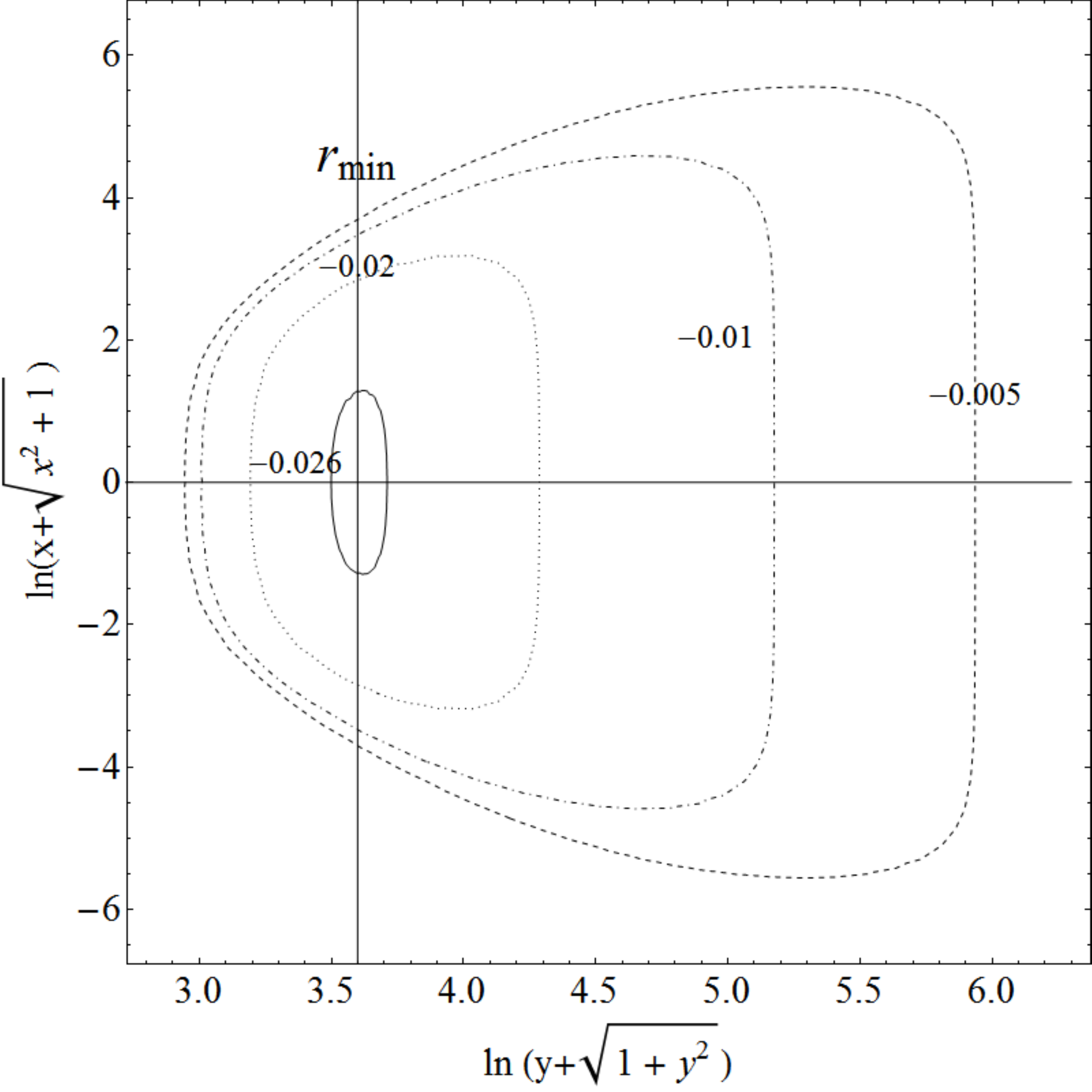}
%\\
\\
\includegraphics[width=.481\textwidth]{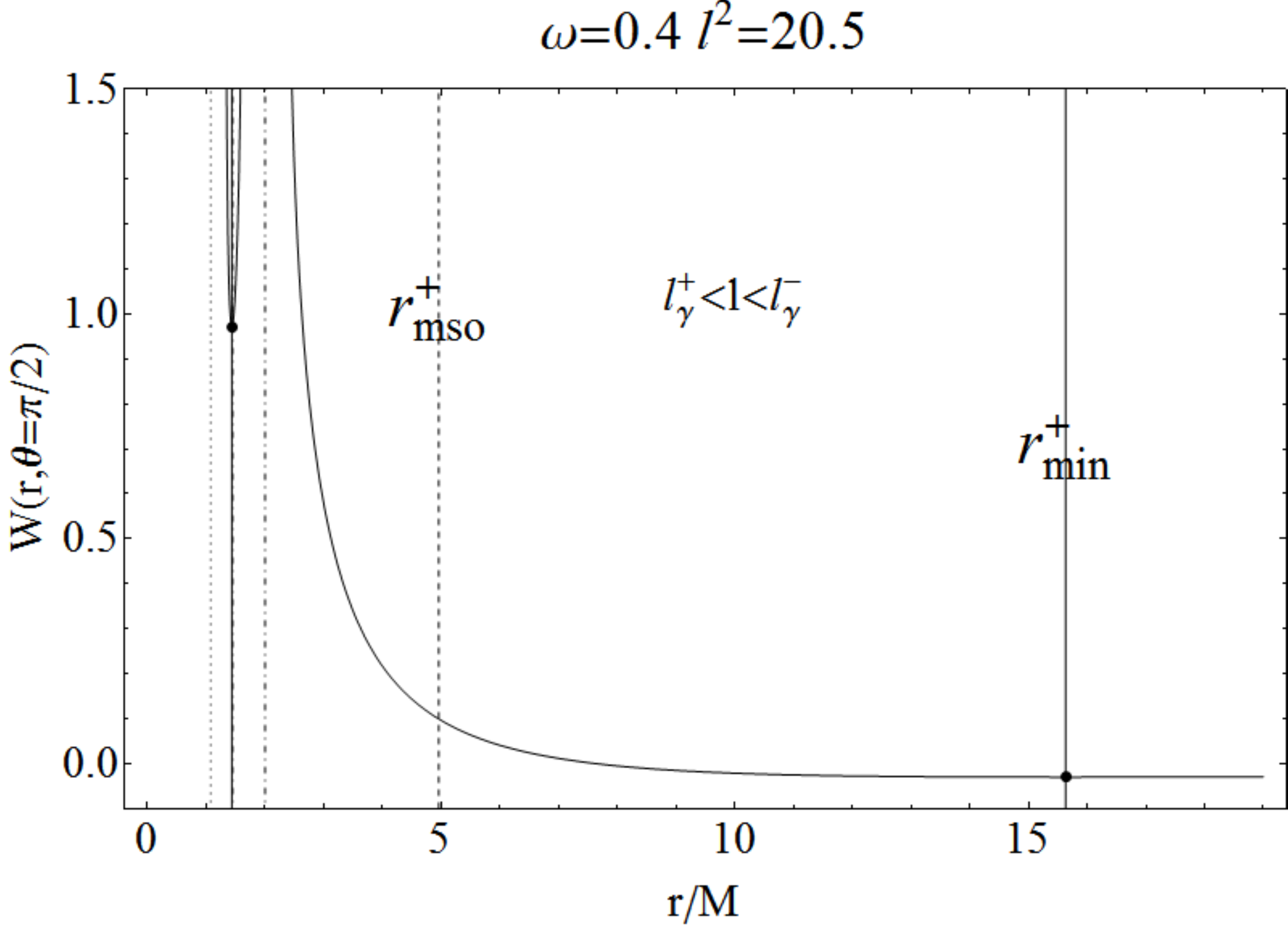}
\includegraphics[width=.31\textwidth]{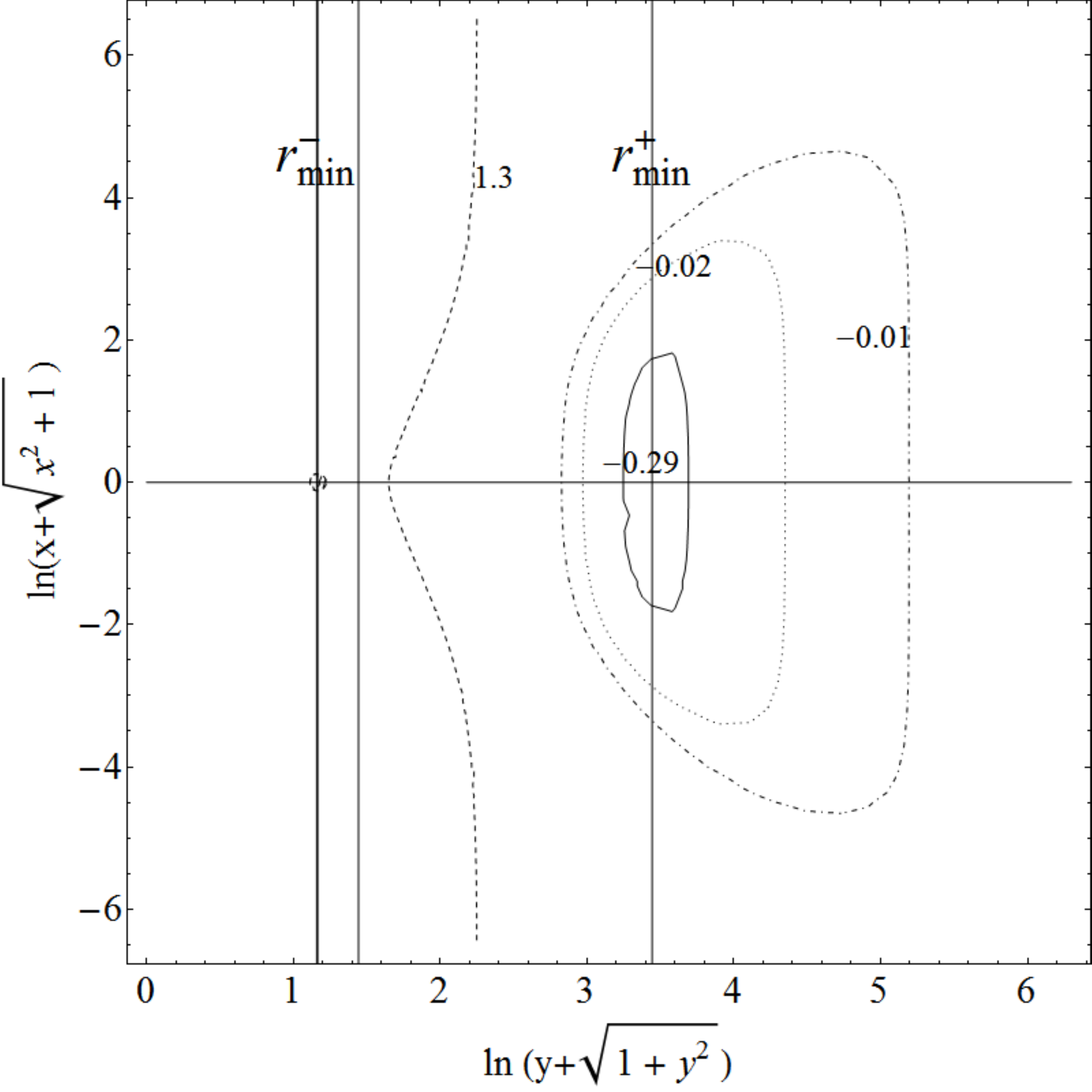}
%\\
\caption{Region IV: $\omega\in]\omega_{\gamma},0.5[$. Where $\omega M^2\rightarrow \omega$. It is $\omega M^2=0.4$ with  $l^-_{\gamma}>l^+_{\gamma}> l^-_{mbo}> l^+_{mbo}> l_{mso}^+$, and $r_{stat}<r_{mbo}^-<r_{\gamma}^-<r_{\gamma}^+<r_{mbo}^+<r_{mso}^+$.  With $r/M=\sqrt{x^2+y^2}$ and  $(x,y)$ are Cartesian coordinates. Left panels: dotted-dashed lines are the radii $r_{\gamma}^-<r_{\gamma}^+$, dotted line is $r_{stat}$.  Vertical lines in right panels set the $r_i\in\mathfrak{R}$ and  the effective potential critical points.}
\label{Fig:pens-ma}
\end{figure}

\begin{figure}[h]
%%CPlotoiscom
\includegraphics[width=.481\textwidth]{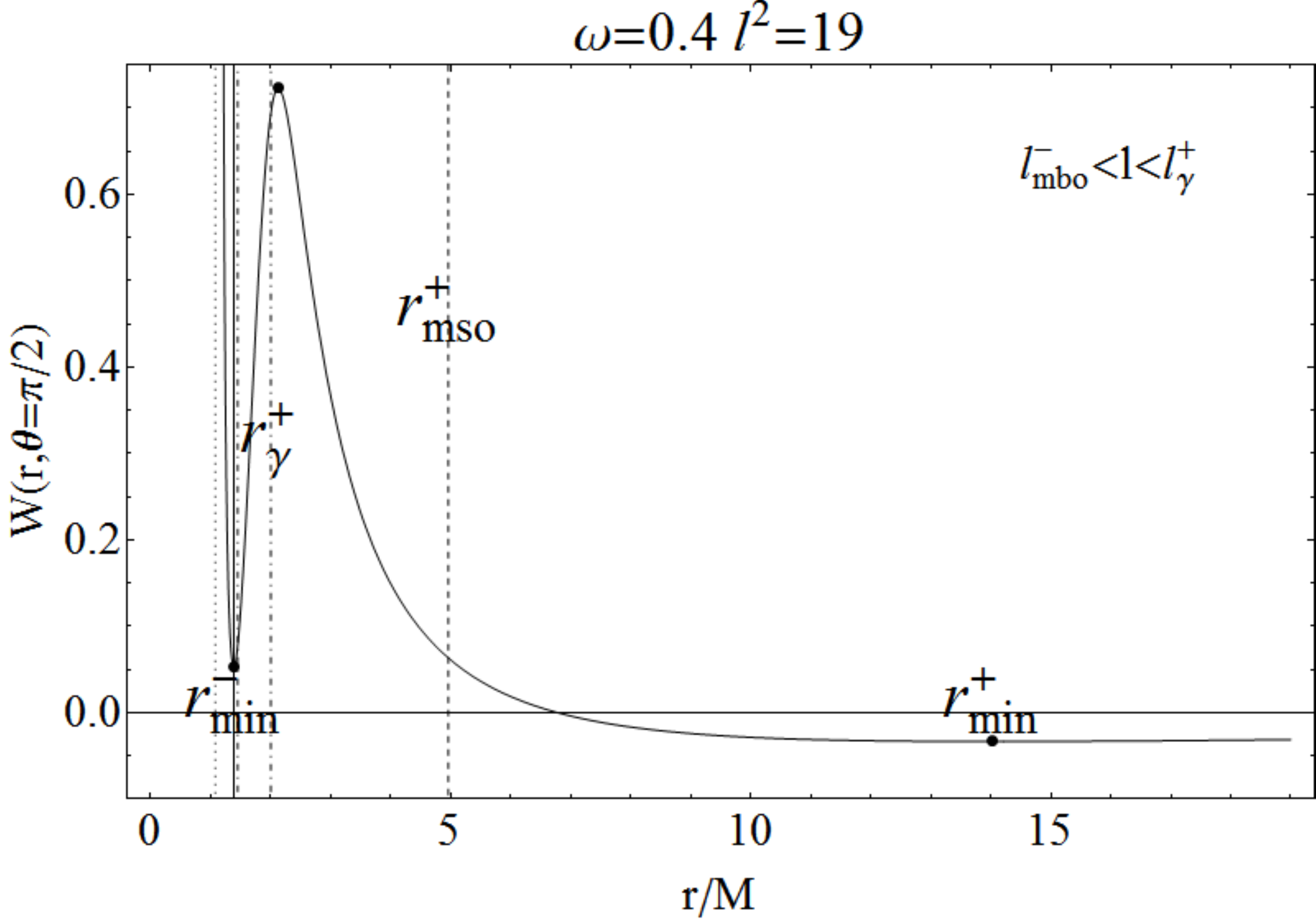}
\includegraphics[width=.31\textwidth]{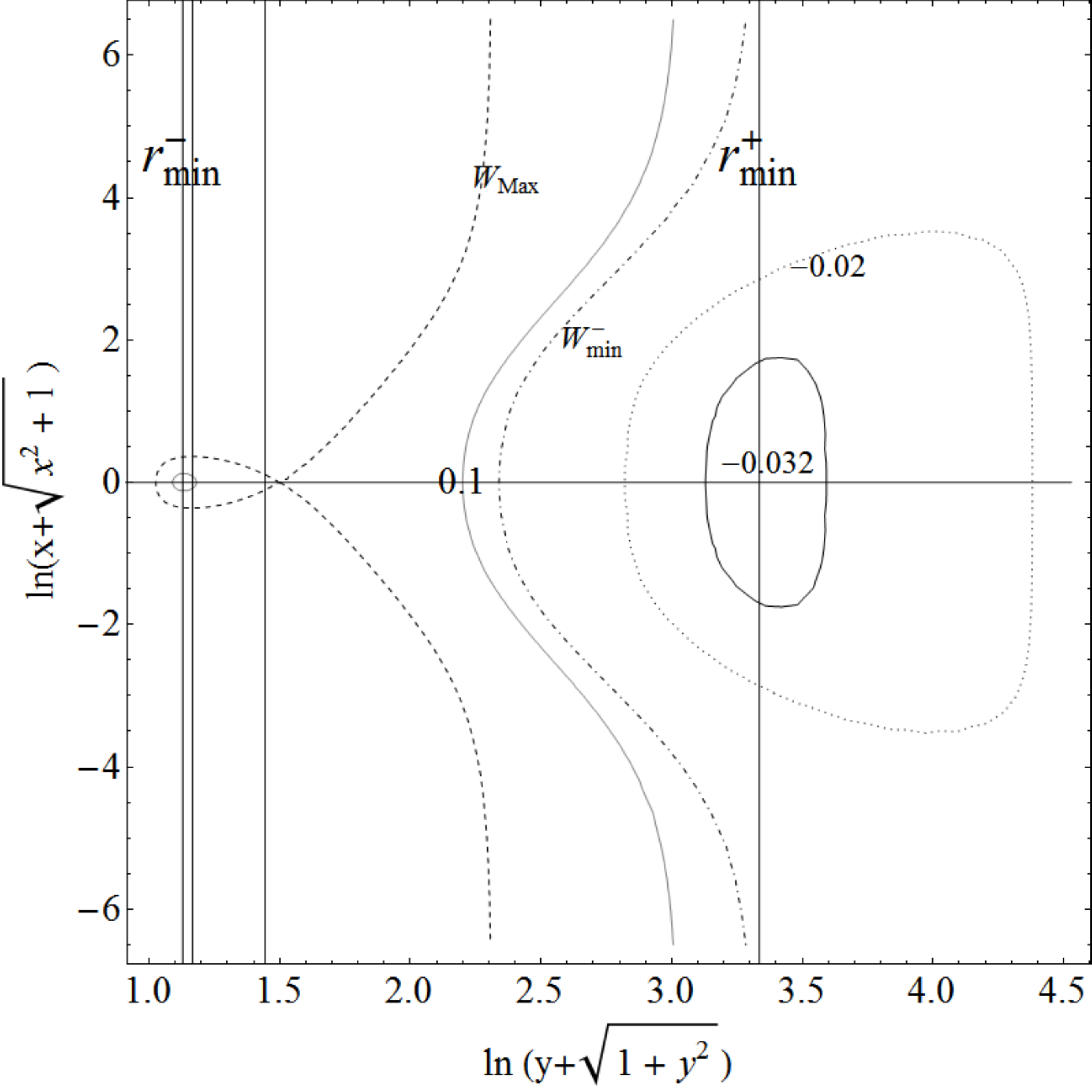}
\\
\includegraphics[width=.481\textwidth]{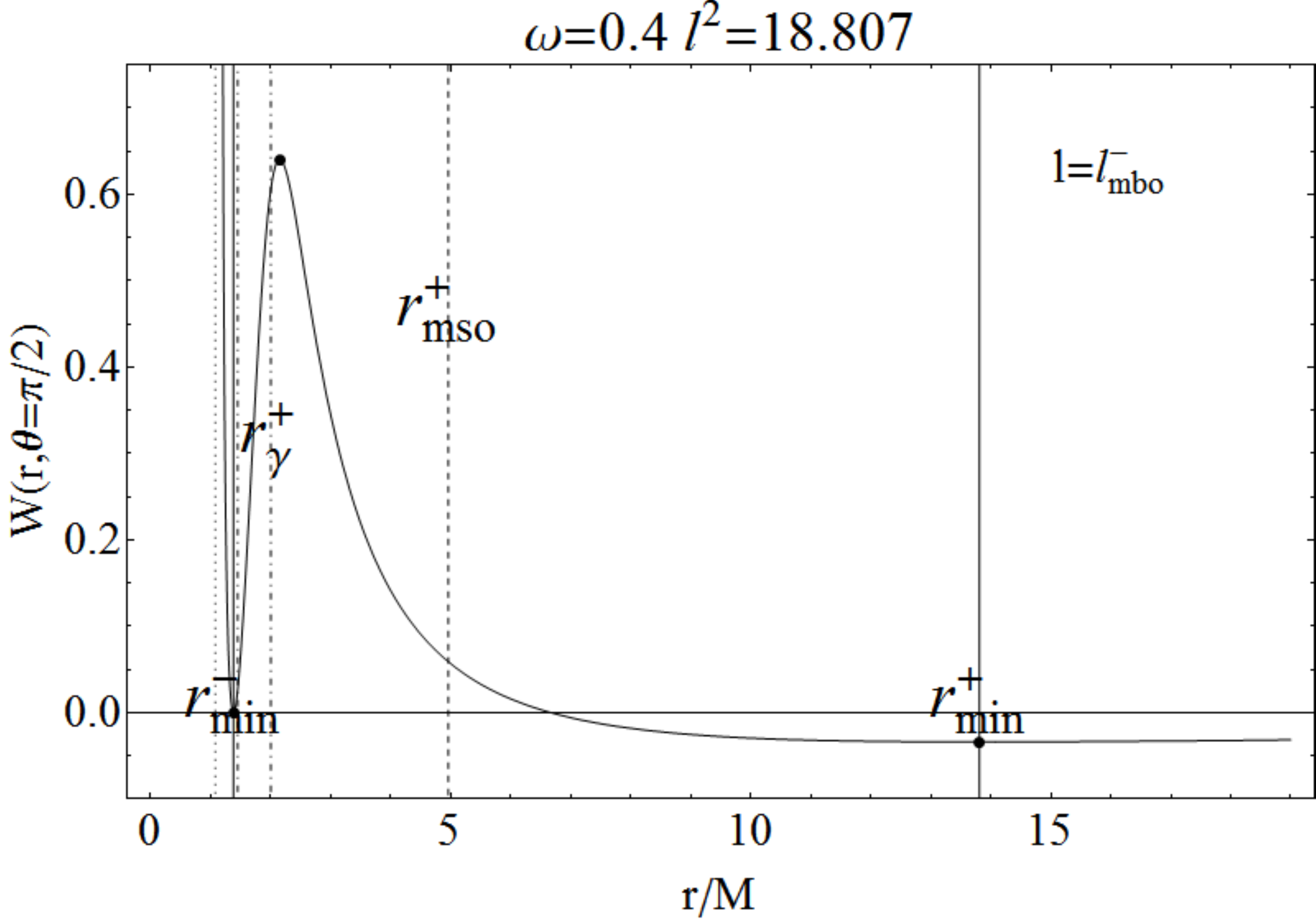}
\includegraphics[width=.31\textwidth]{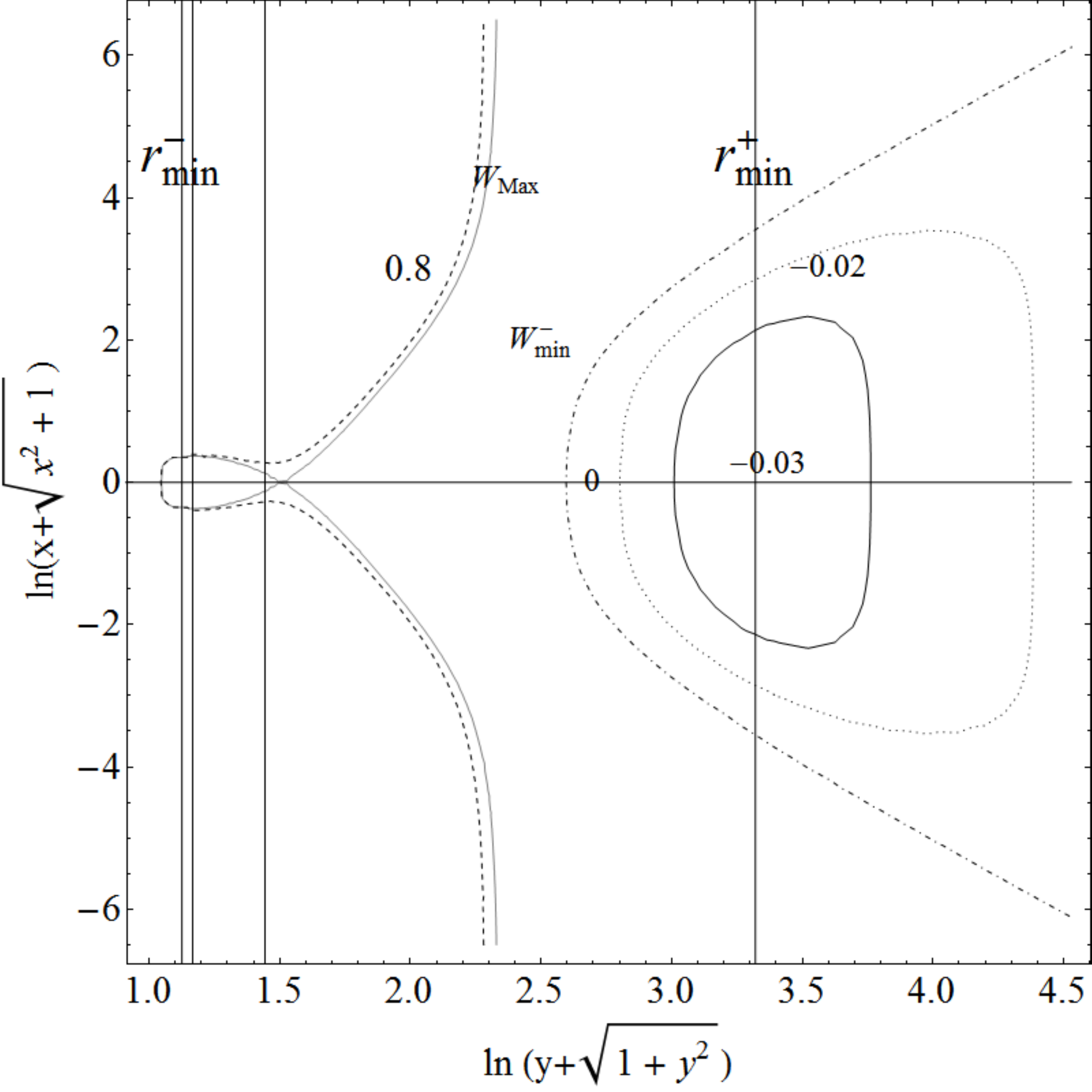}
\\
\includegraphics[width=.481\textwidth]{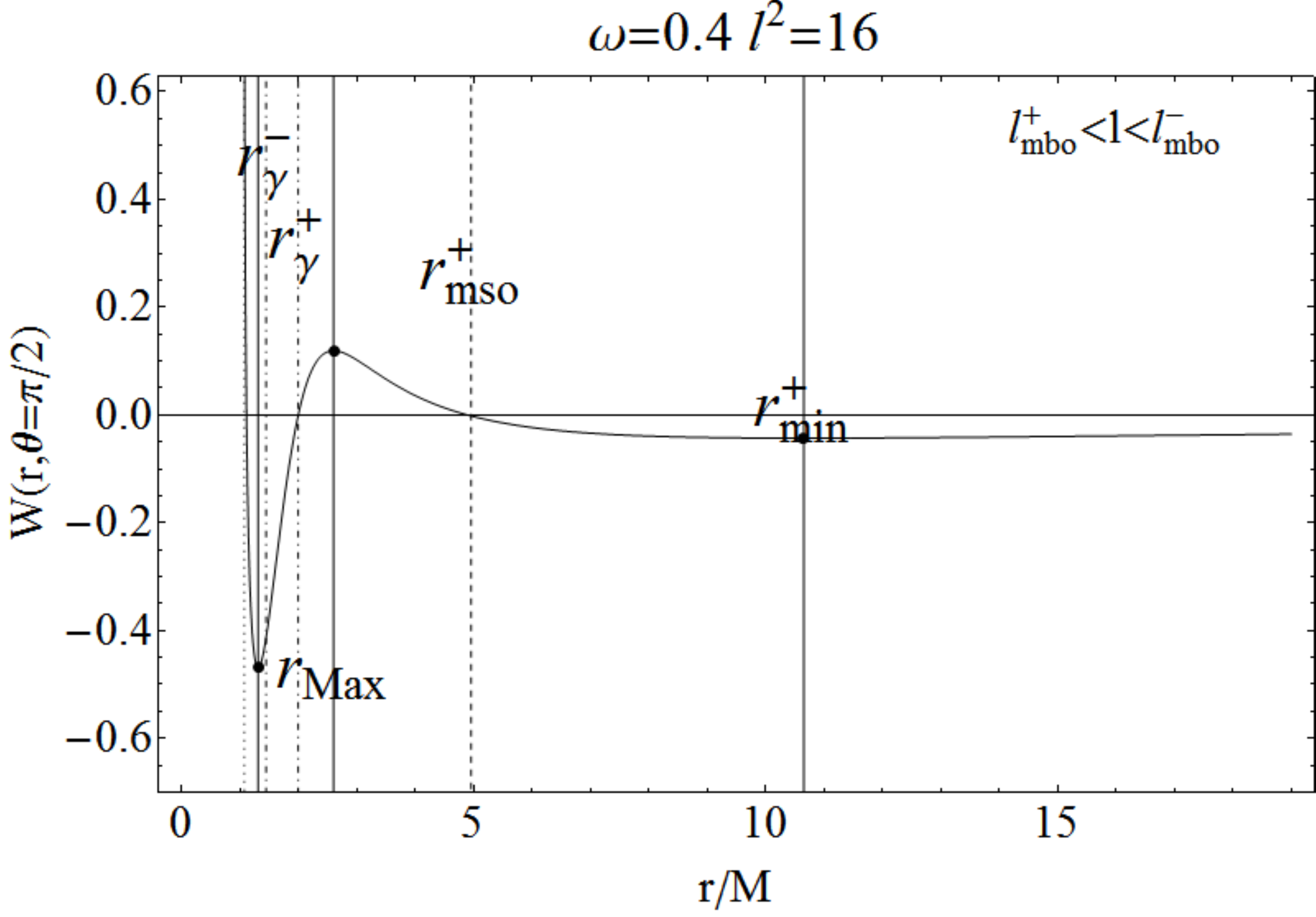}
\includegraphics[width=.31\textwidth]{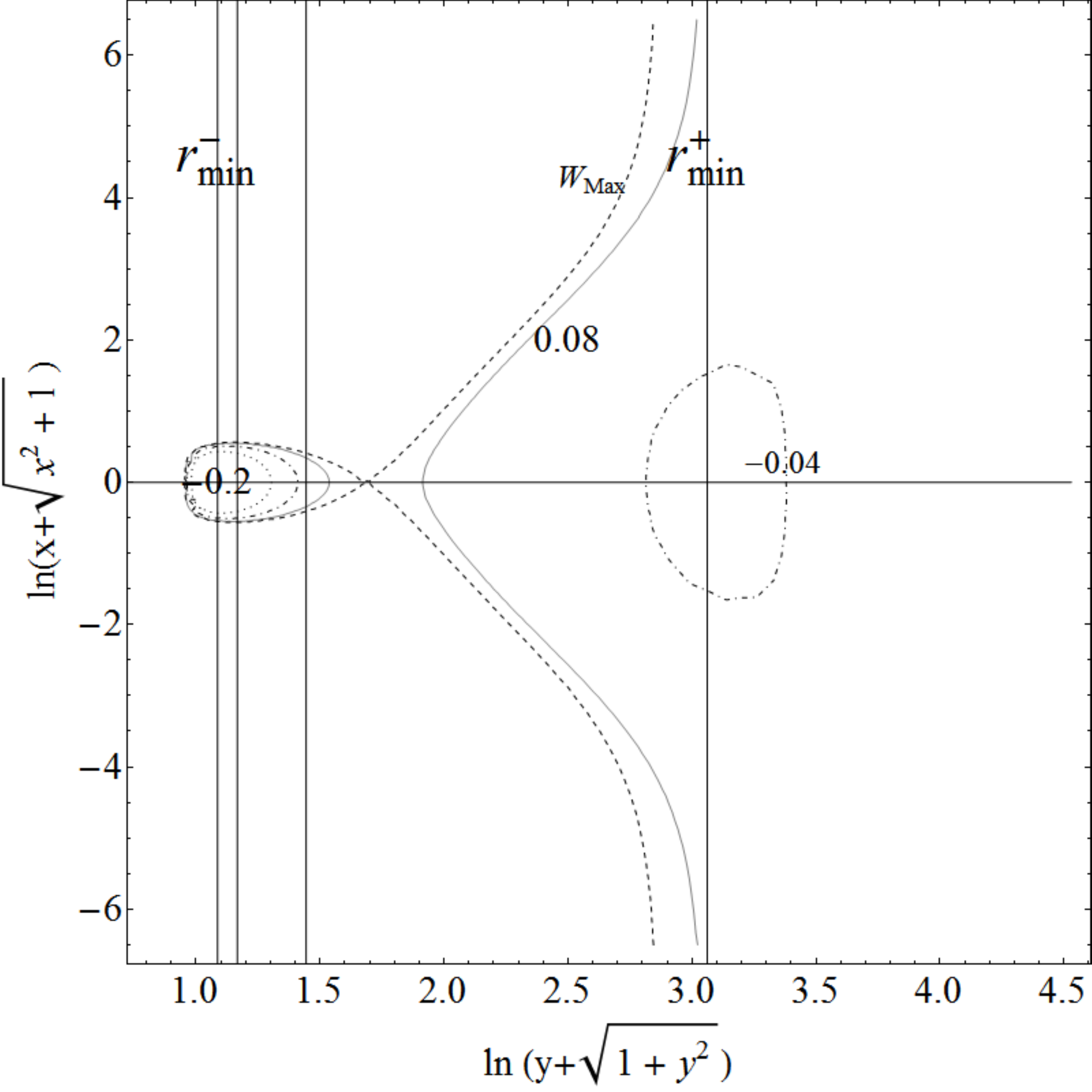}
\\
\includegraphics[width=.481\textwidth]{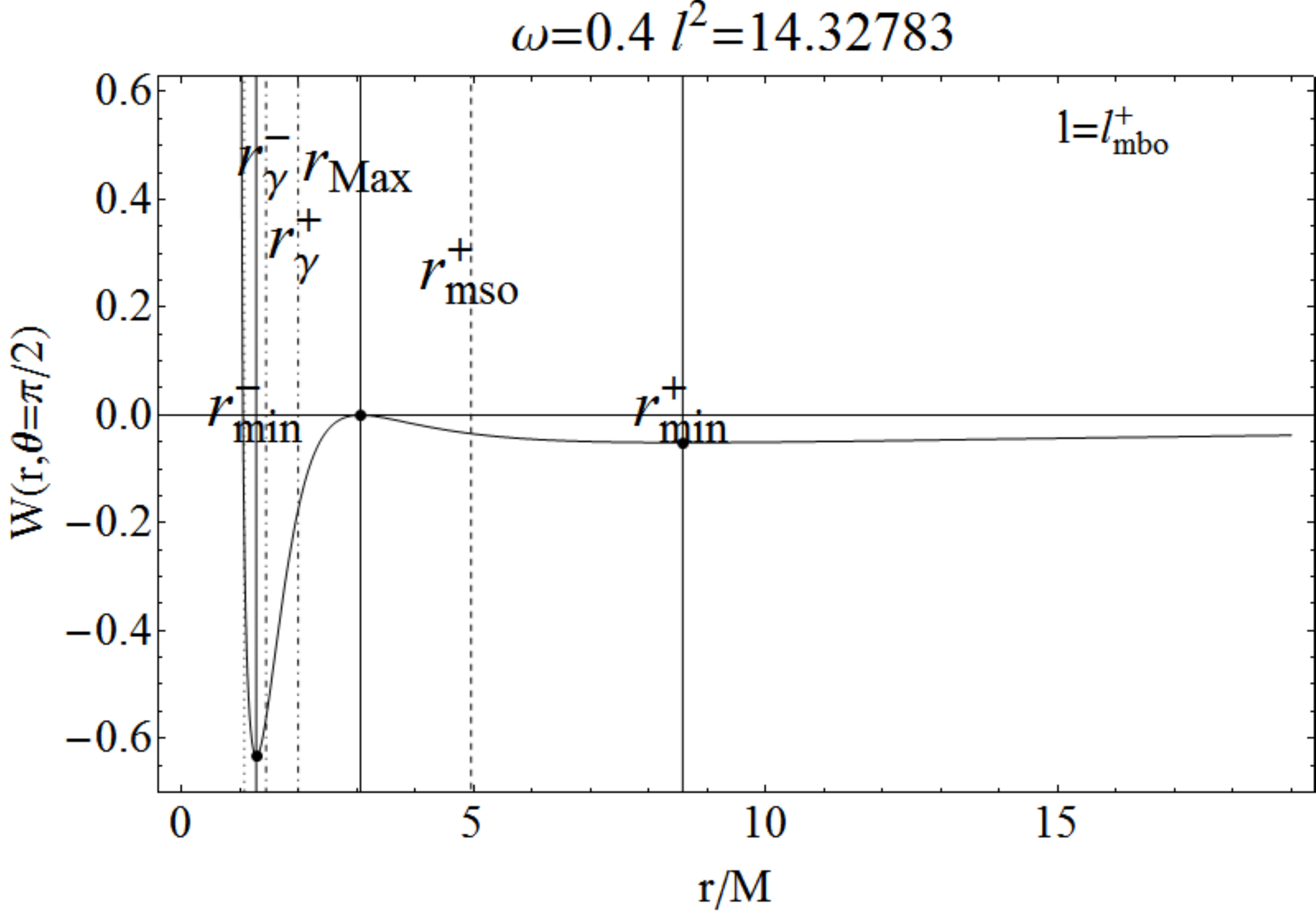}
\includegraphics[width=.41\textwidth]{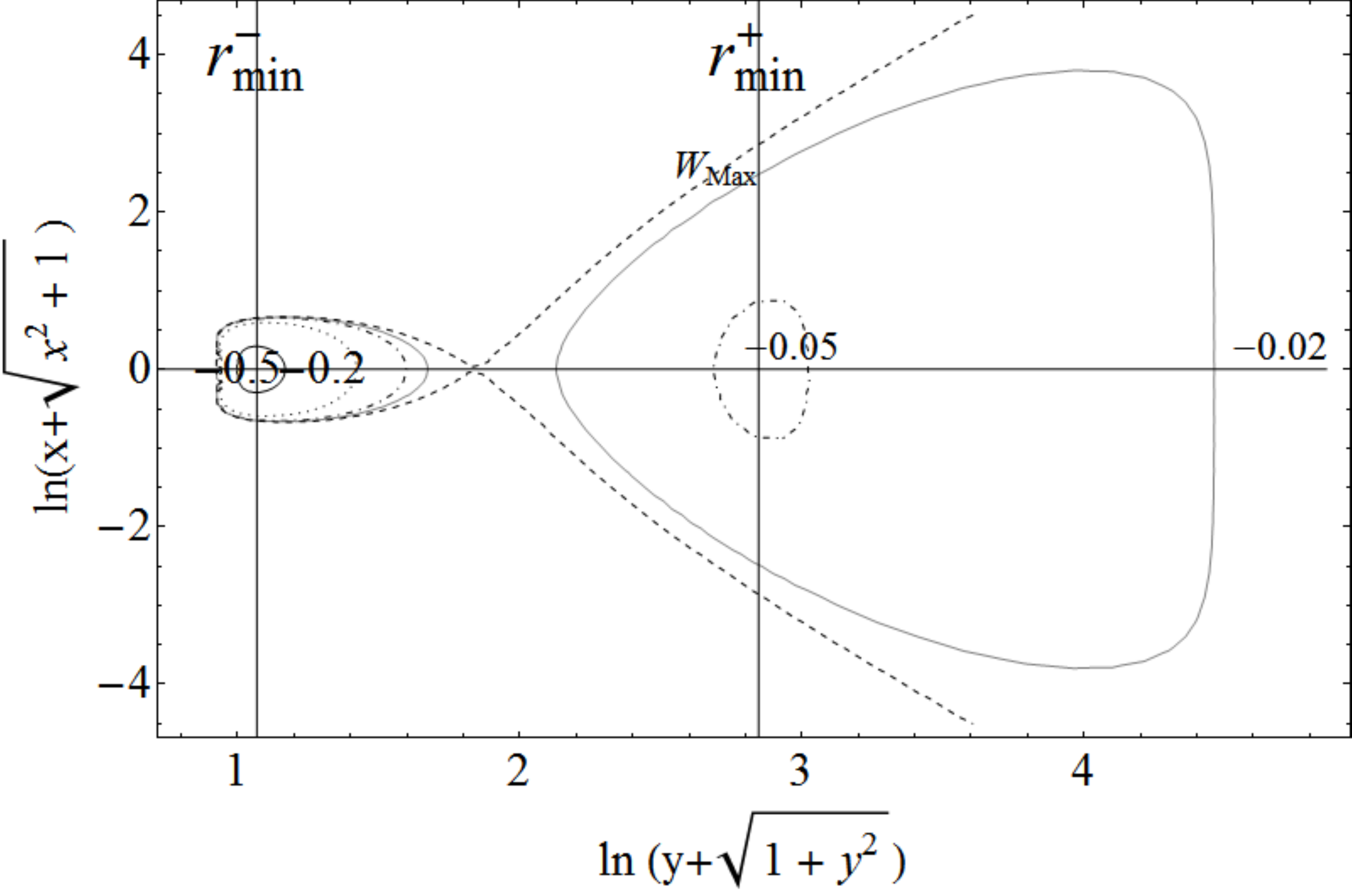}
\caption{Region IV: $\omega\in]\omega_{\gamma},0.5[$. It is $\omega M^2=0.4$ with $l^-_{\gamma}>l^+_{\gamma}> l^-_{mbo}> l^+_{mbo}> l_{mso}^+$, and $r_{stat}<r_{mbo}^-<r_{\gamma}^-<r_{\gamma}^+<r_{mbo}^+<r_{mso}^+$. Where $\omega M^2\rightarrow \omega$.  With $r/M=\sqrt{x^2+y^2}$ and  $(x,y)$ are Cartesian coordinates.   Vertical lines in right panels set the $r_i\in\mathfrak{R}$ and  the effective potential critical points. Left panels: dotted-dashed lines are the radii $r_{\gamma}^-<r_{\gamma}^+$, dotted line is $r_{stat}$.}
\label{Fig:c-tak}
\end{figure}

\begin{figure}[h]
%%CPlotoiscom
\includegraphics[width=.481\textwidth]{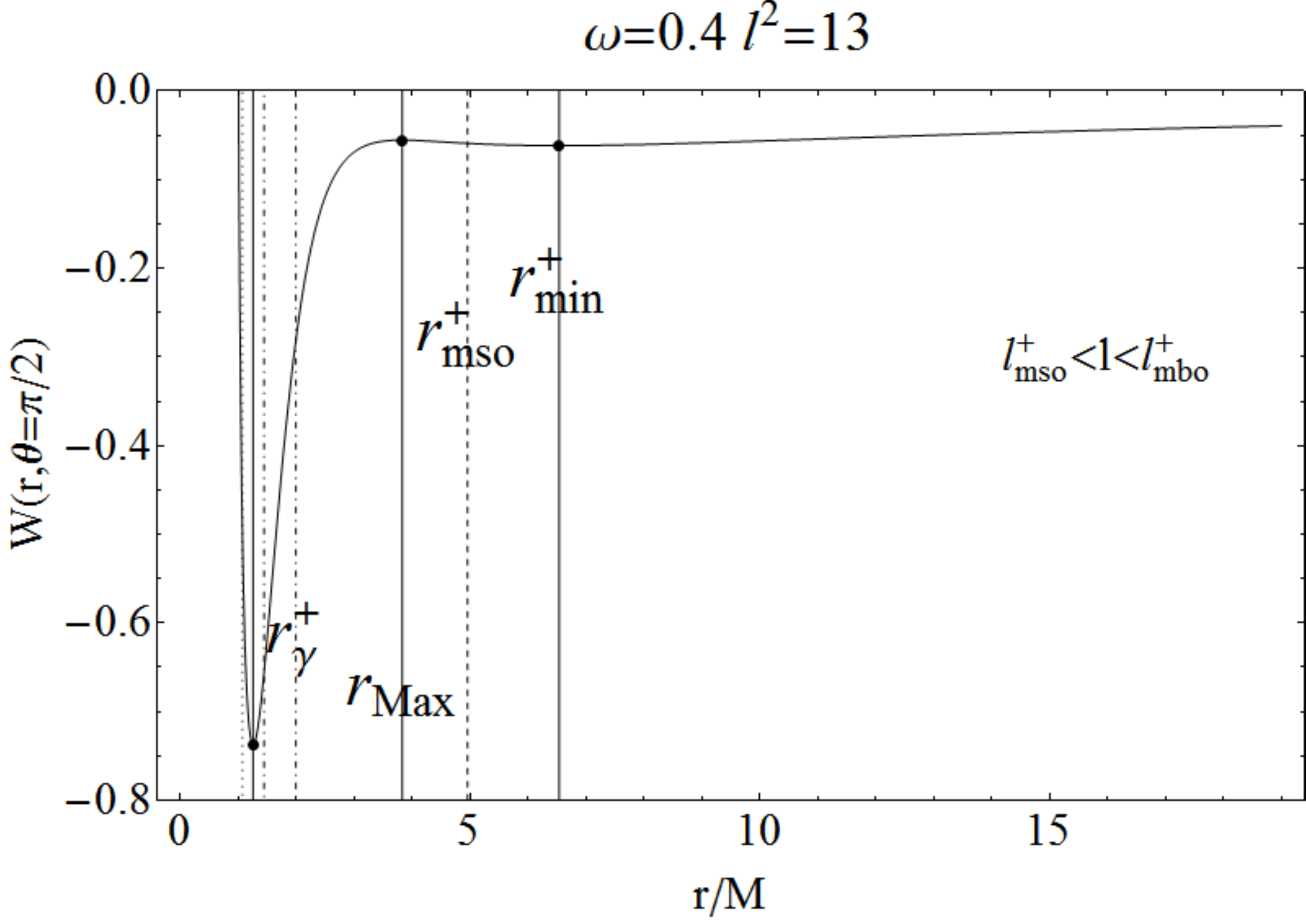}
\includegraphics[width=.481\textwidth]{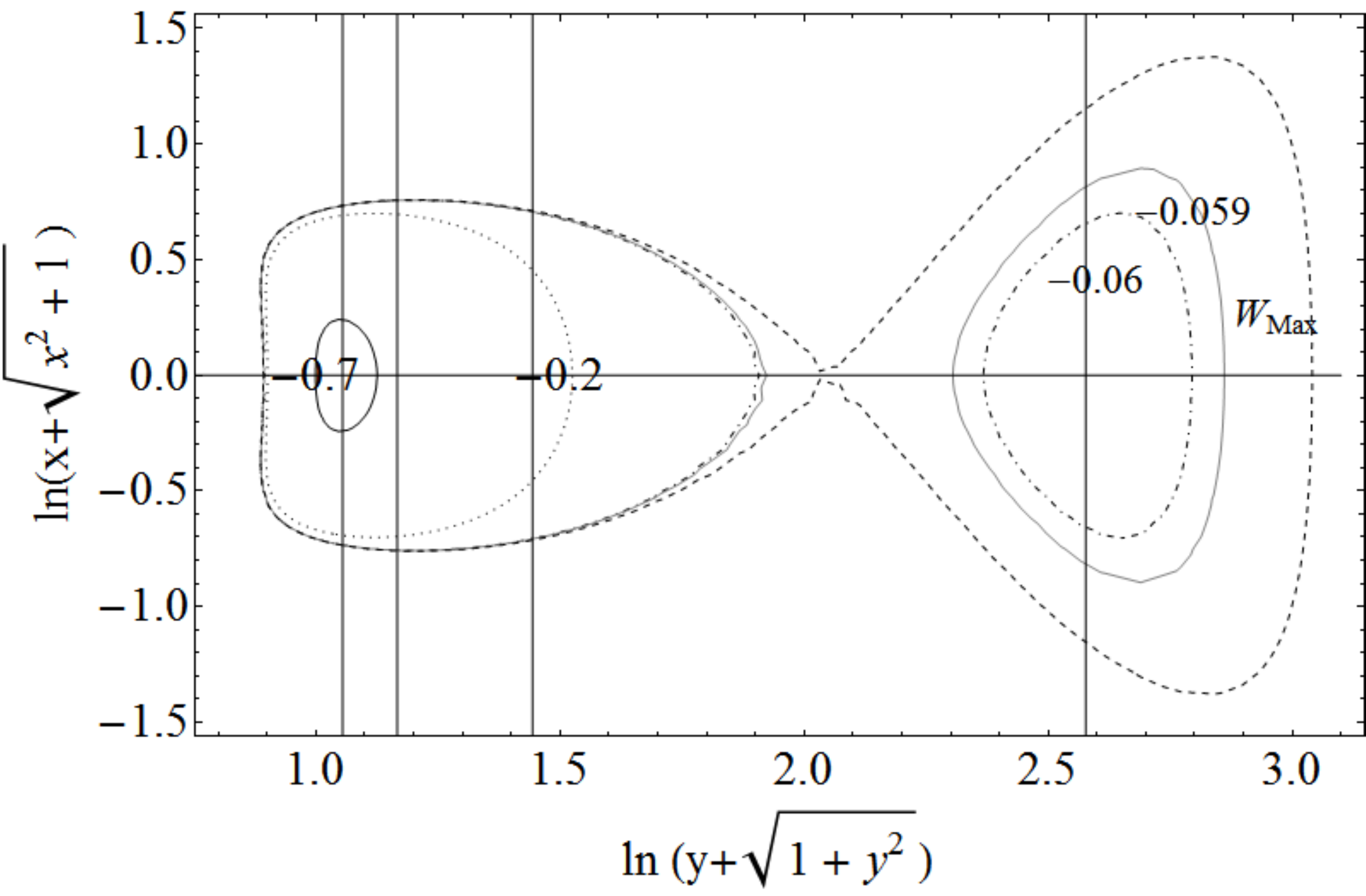}
\\
\includegraphics[width=.481\textwidth]{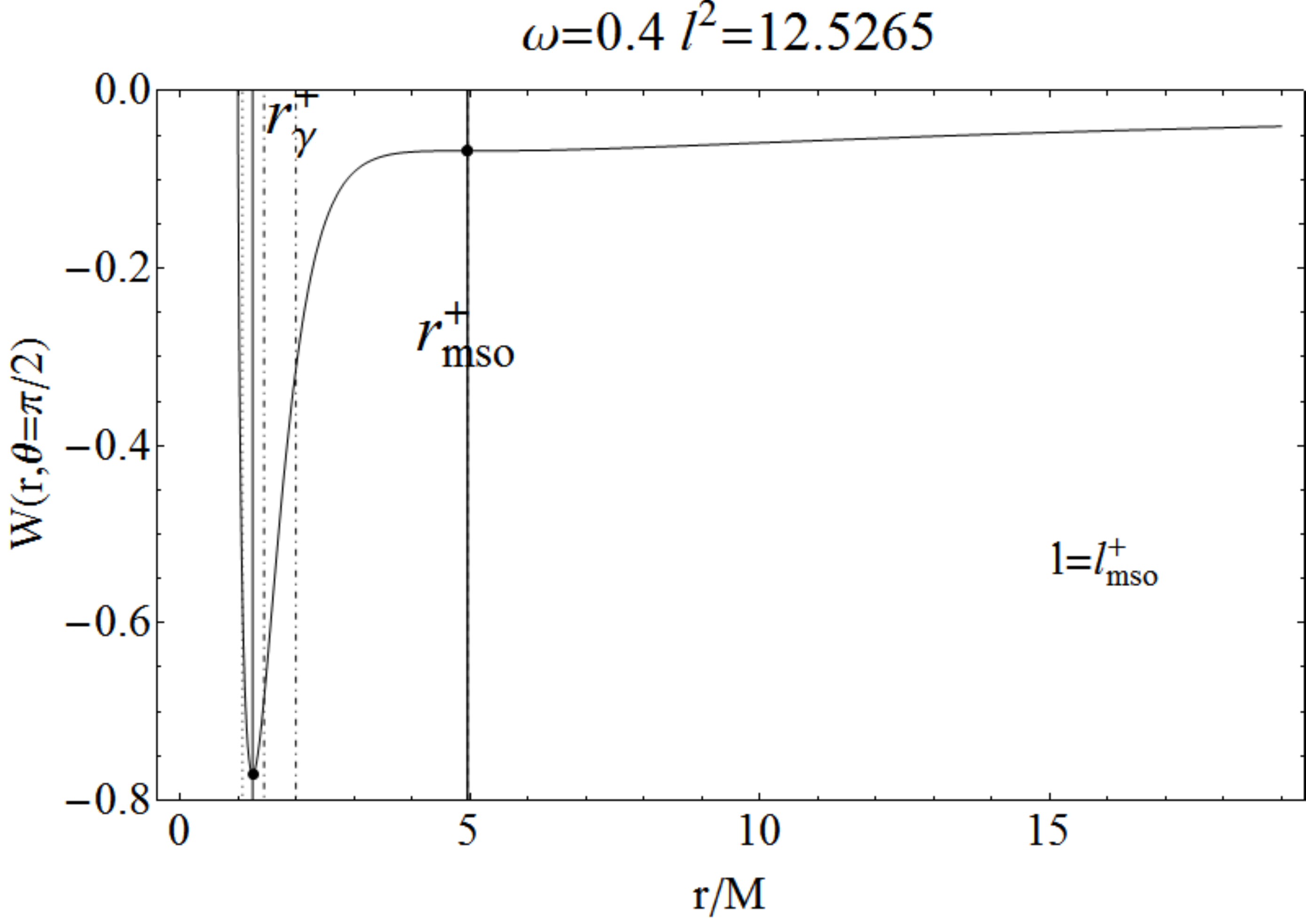}
\includegraphics[width=.481\textwidth]{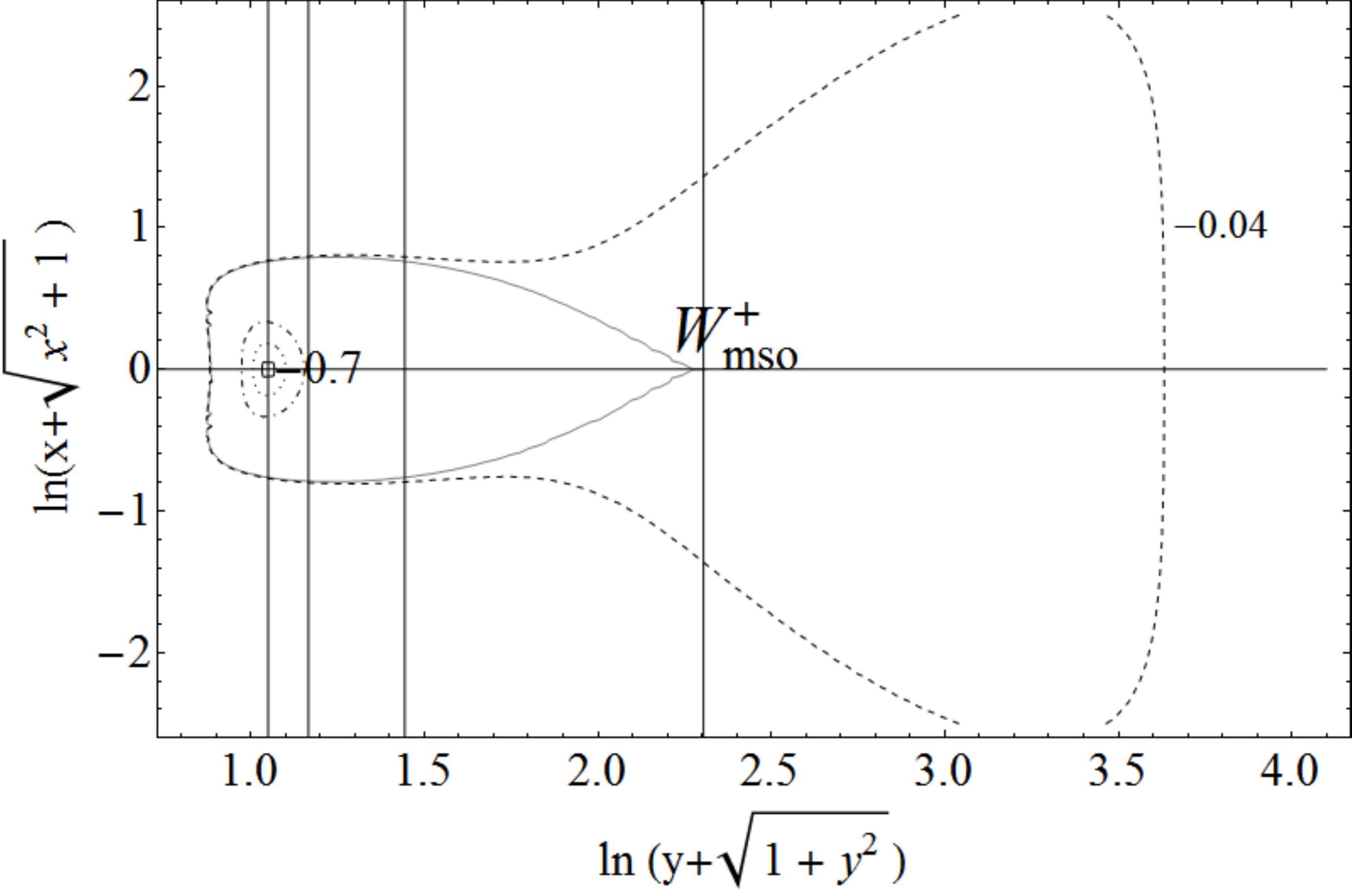}
\\
\includegraphics[width=.481\textwidth]{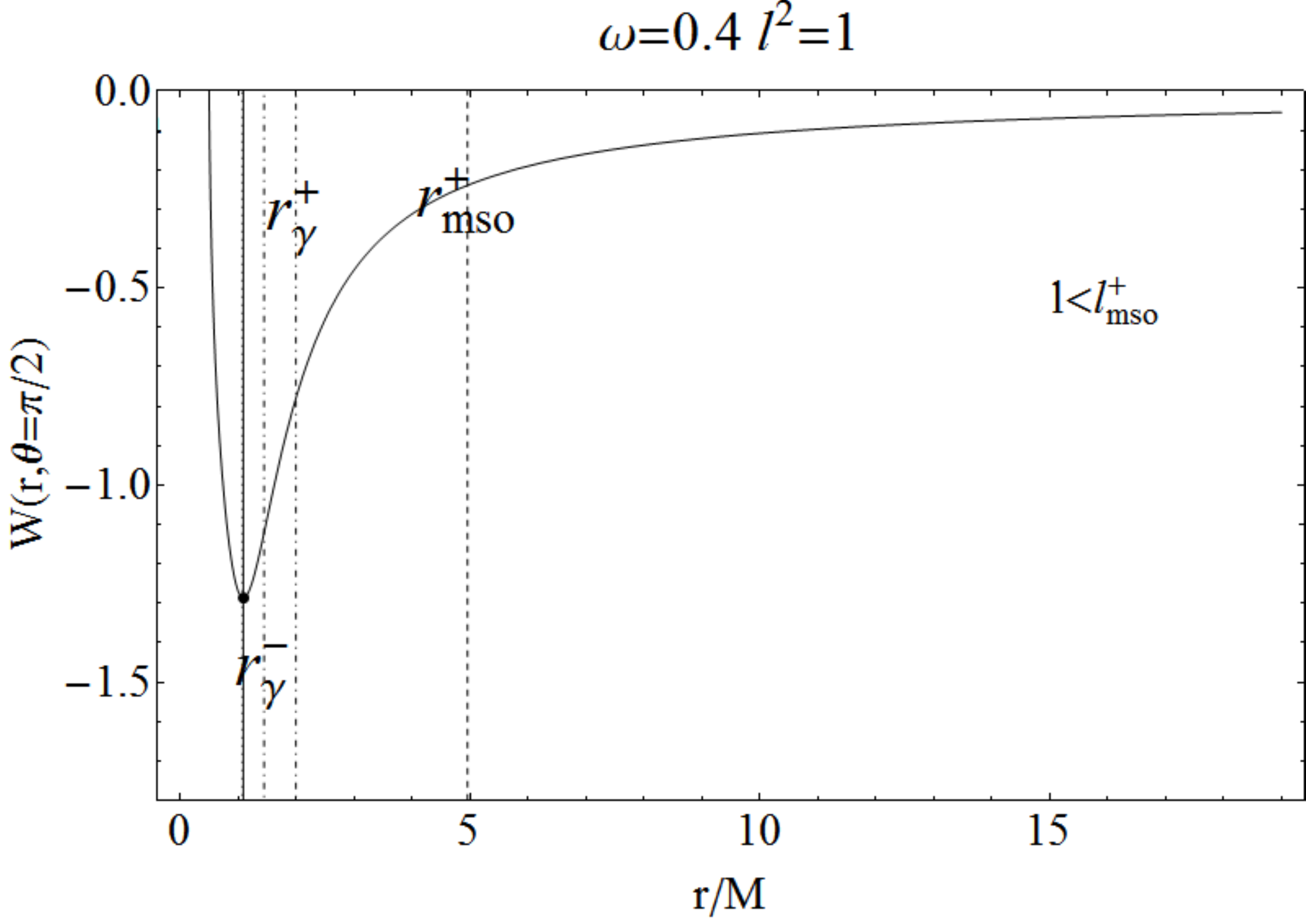}
\includegraphics[width=.481\textwidth]{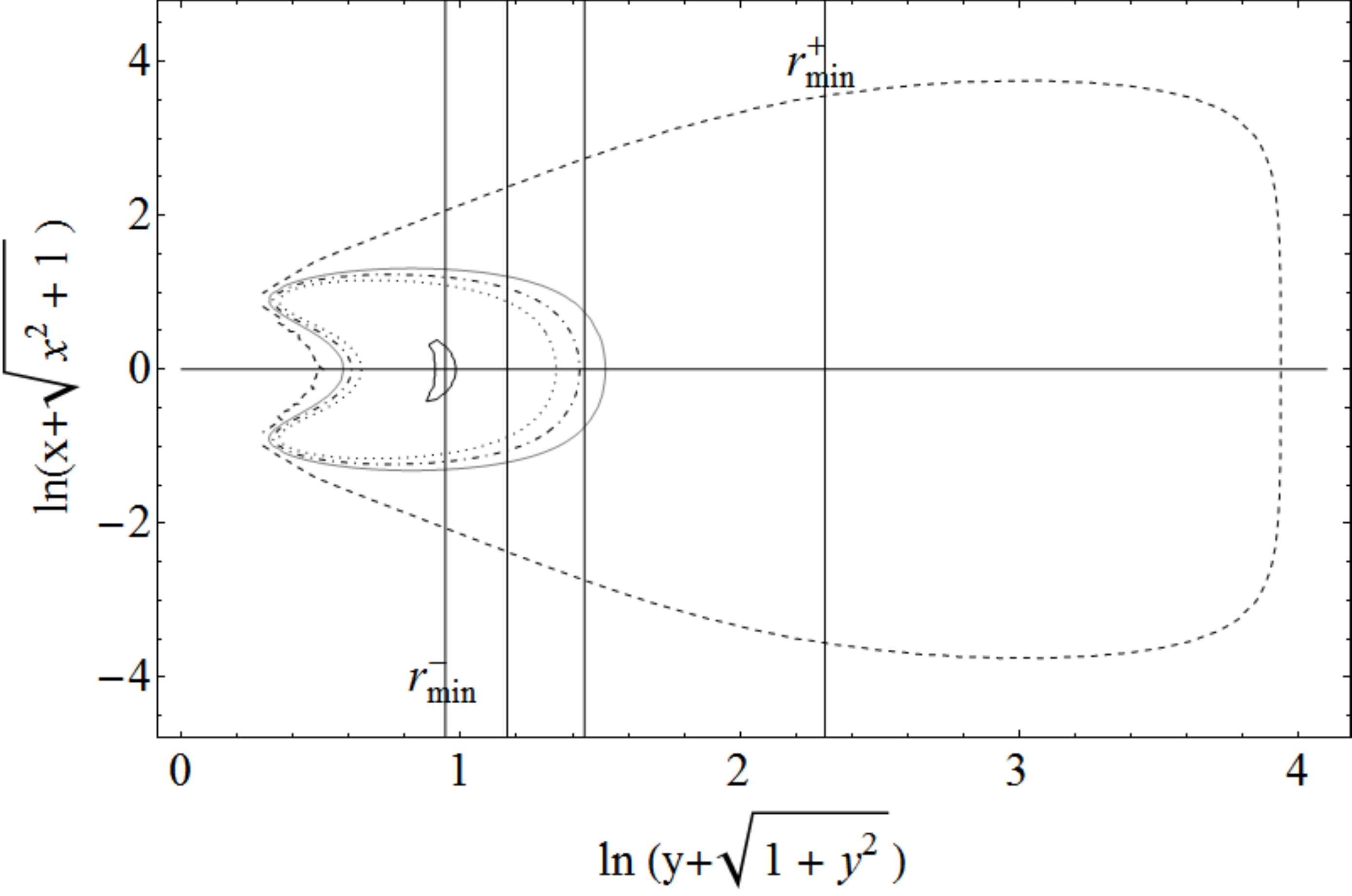}
%\\
\caption{Region IV: $\omega\in]\omega_{\gamma},0.5[$. It is $\omega M^2=0.4$ with   $\omega M^2\rightarrow \omega$ and  $l^-_{\gamma}>l^+_{\gamma}> l^-_{mbo}> l^+_{mbo}> l_{mso}^+$, and $r_{stat}<r_{mbo}^-<r_{\gamma}^-<r_{\gamma}^+<r_{mbo}^+<r_{mso}^+$.  With $r/M=\sqrt{x^2+y^2}$ and  $(x,y)$ are Cartesian coordinates.  Left panels: dotted-dashed lines are the radii $r_{\gamma}^-<r_{\gamma}^+$, dotted line is $r_{stat}$.  Vertical lines in right panels set the $r_i\in\mathfrak{R}$ and  the effective potential critical points.}
\label{Fig:Abe-1}
\end{figure}

Kehagias-Sfetsos  naked singularity spacetimes in \textbf{Region IV} contain two photon circular orbits, the
inner stable one  $r_{\gamma}^-$, the outer unstable one at $r_{\gamma}^+$, and the  stable circular orbit $r_{mso}^+$, indeed it is
   $r_{stat}<r_{mbo}^-<r_{\gamma}^-<r_{\gamma}^+<r_{mbo}^+<r_{mso}^+$ and $l^-_{\gamma}>l^+_{\gamma}> l^-_{mbo}> l^+_{mbo}> l_{mso}^+$.
This Region of attractors correspond to the Class III  detailed in
Sec.\il(\ref{Sub:ClassIII})). In the bend $\Delta_{\gamma}^{\pm}\equiv]r_{\gamma}^-,r_{\gamma}^+[$, no circular geodesics
are allowed i.e. there is no critical point  of the effective potential and therefore no disc center or accretion/excretion point or cusp can be located in $\Delta_{\gamma}^{\pm}$. Then there is no critical point for the angular frequency  $\Omega$.
\begin{description}
\item[-) $l\geq l_{\gamma}^-$] This situation is sketched in  Fig.\il(\ref{Fig:pens-ma})-(a): there is one class of closed configurations.
\item[-) ${l\in]l_{\gamma}^+,l_{\gamma}^-]}$] There is a minimum in $r_{min}^+$ with $W_{min}^+<0$, center of a set of closed $C^+$ toroidal surfaces. The effective potential has a further critical point in $r_{min}^-<r_{\gamma}^-$   with  $W_{min}^->0$. Correspondingly there is a sequence of inner (very narrow) $C^-$ discs centered in $r_{min}^-$. However, two sets of closed configurations, are separated by means of the open surface correspondent to the second solution for $W={\rm const}:\; W\in[W_{min}^-,+\infty[$ slightly ``cusped'' from the interior towards the exterior  solution, and aligned along the axis,  see Fig.\il(\ref{Fig:pens-ma})-(b). This is a \textbf{I} type configuration.
\item[-) ${l\in]l_{mbo}^-,l_{\gamma}^+[}$] This situation is illustrated in Fig.\il(\ref{Fig:c-tak})-a. As $W_{min}^->0>W_{min}^+$ (\textbf{I}-class)  there is a $r_{Max}>r_{\gamma}^+$, and  a crossed open surface, with exterior cusp (excretion configuration) and  with branches aligned to the axis, separated from the  inner closed discs (with $W>0$) and the disc center in $r_{min}^+$.
\item[-)$l=l_{mbo}^-$] This case is shown in Fig.\il(\ref{Fig:c-tak})-b, the minimum is located in $r_{mbo}^-$  with $W_{min}^-=0$, the situation is  similar to  configurations at ${l\in]l_{mbo}^+,l_{\gamma}^+[}$ in  Fig.\il(\ref{Fig:c-tak})-a.
\item[-) ${l\in]l_{mbo}^+, l_{mbo}^-[}$]  In Fig.\il(\ref{Fig:c-tak})-c  it is shown a  \textbf{III} class configuration, there are two sets $C^{\pm}$ of closed configurations  separated by an open crossed excretion surface.
\item[-) $l=l_{mbo}^+$] The  maximum is at $r_{mbo}^+$ and this correspond to an open crossed excretion surface, see Fig.\il(\ref{Fig:c-tak})-d.
\item[-) ${l\in]l_{mbo}^-,l_{mso}^+[}$] The critical points $W_{crit}<0$,  there are closed surfaces ranged by the closed crossed one as in Fig.\il(\ref{Fig:Abe-1})-a.
\item[-) $l=l_{mso}^+$] There is an outer cusp for $W_{mso}^+$, see Fig.\il(\ref{Fig:Abe-1})-b.
\item[-)$l<l_{mso}^+$] There are closed crossed surfaces centered in $r_{min}^+$, Fig.\il(\ref{Fig:Abe-1})-c.
\end{description}
\clearpage
\subsection{Extreme case: $\omega M^2=\omega_h M^2=0.5$}\label{Sec:K-S-ExBH}
\begin{figure}[h]
%%CPlotoiscom
\includegraphics[width=.481\textwidth]{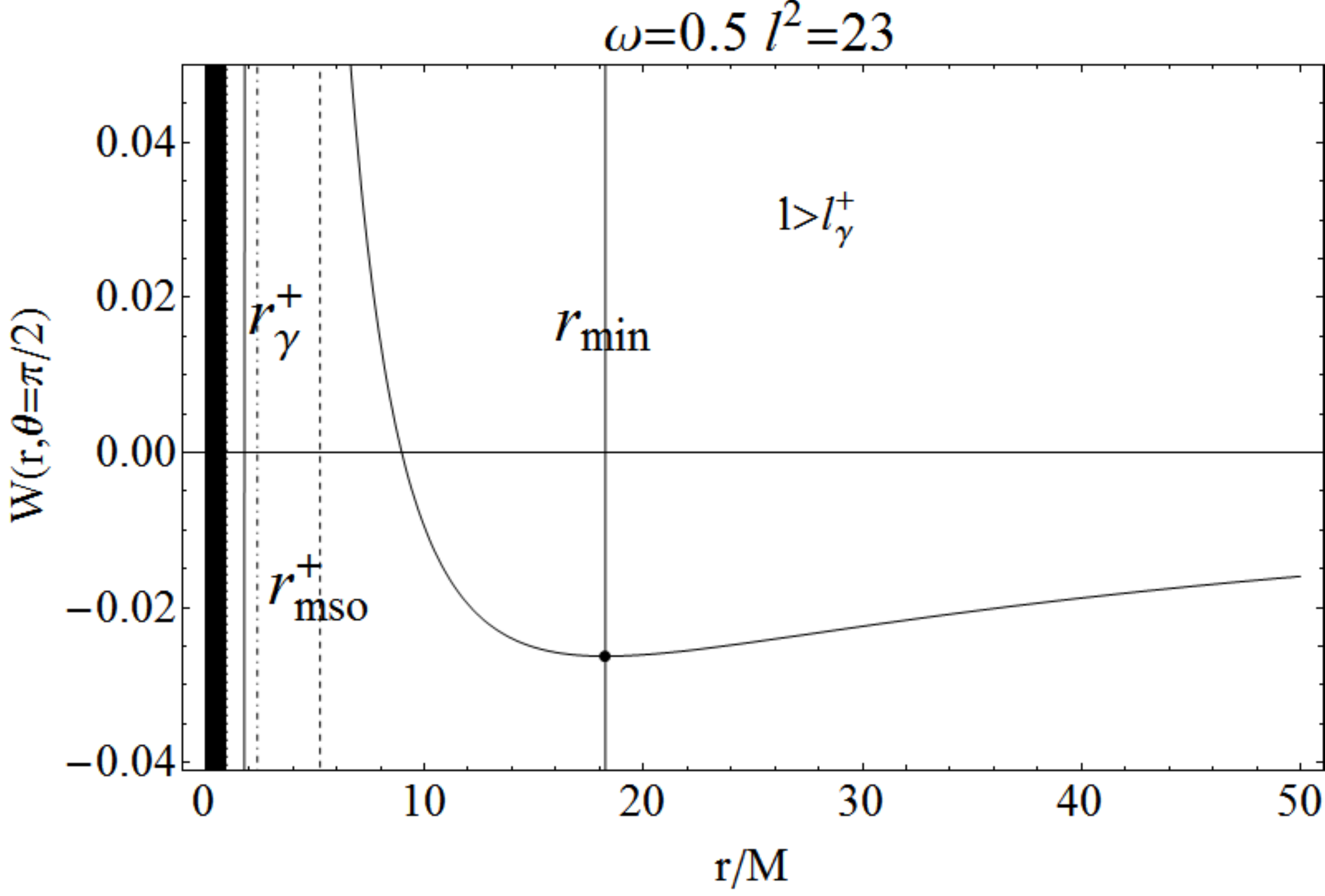}
\includegraphics[width=.31\textwidth]{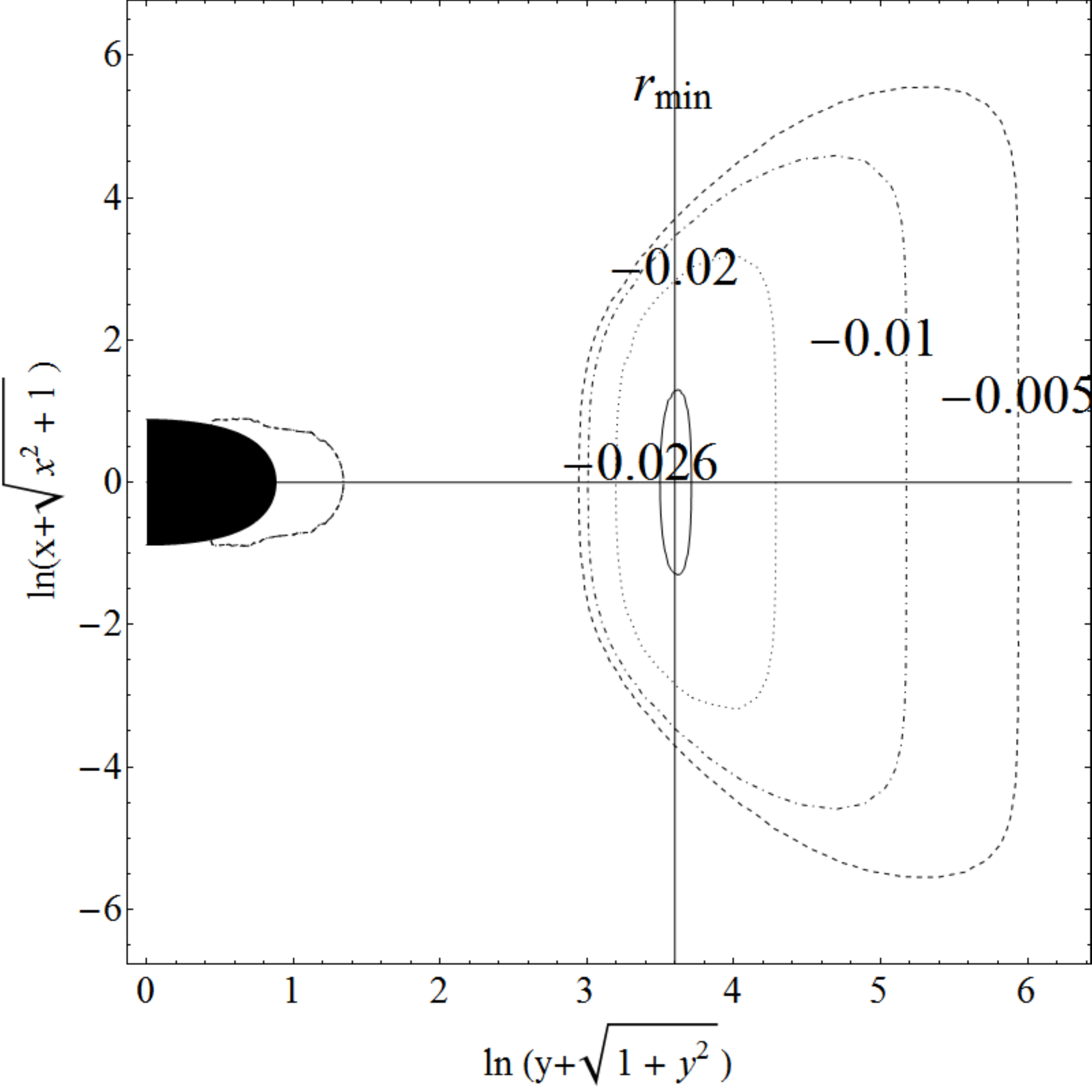}
%\\
\\
\includegraphics[width=.481\textwidth]{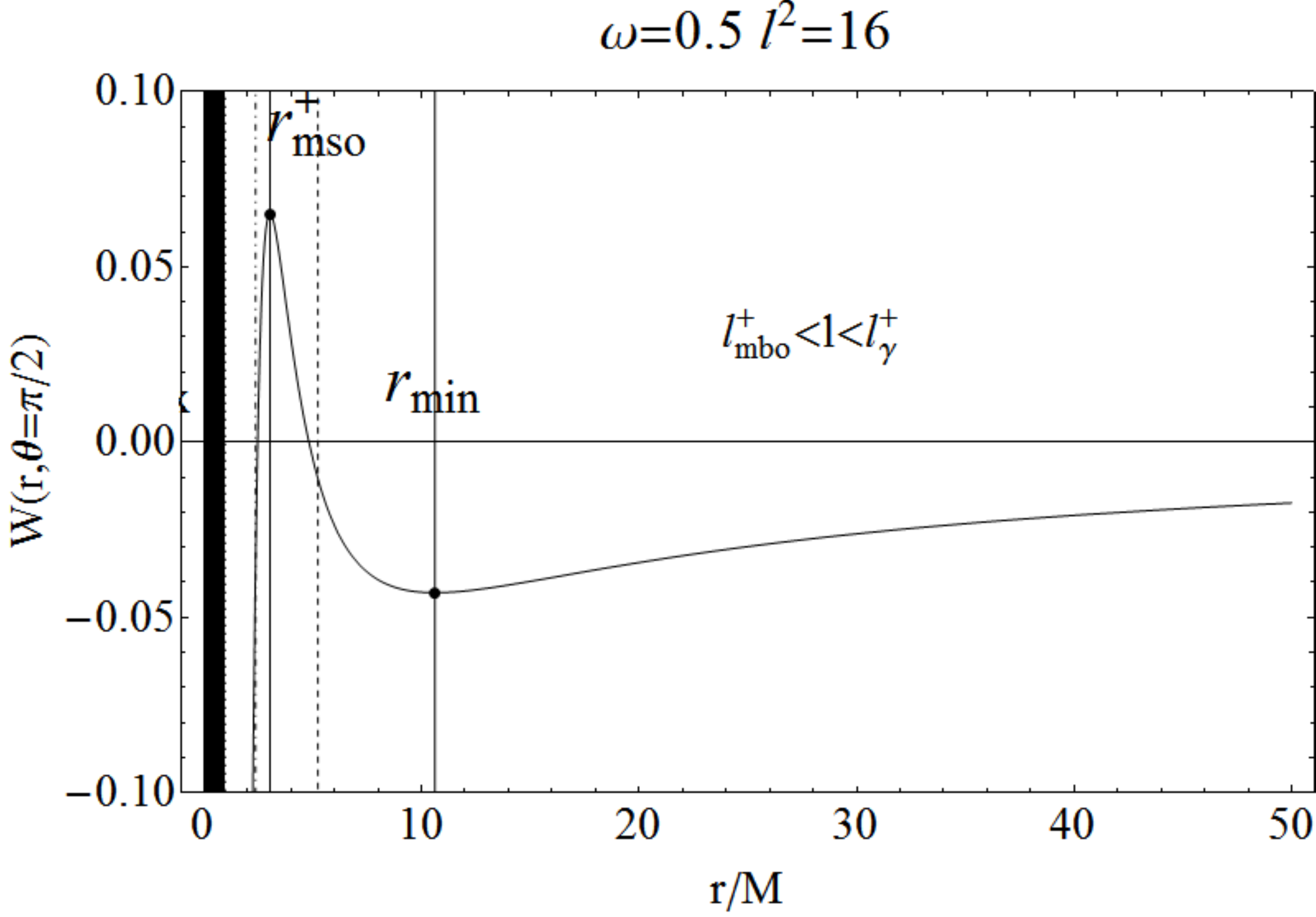}
\includegraphics[width=.31\textwidth]{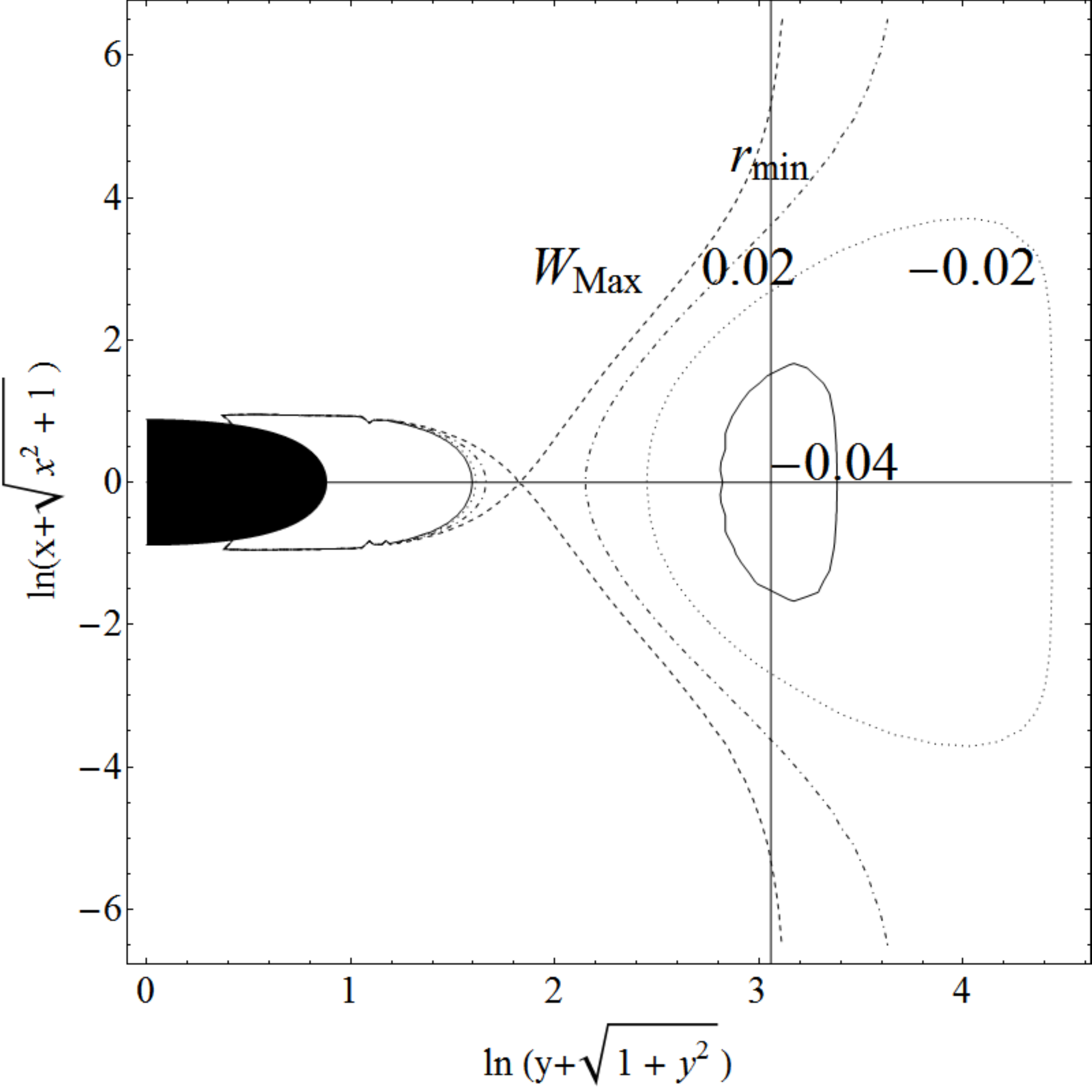}
\\
\includegraphics[width=.481\textwidth]{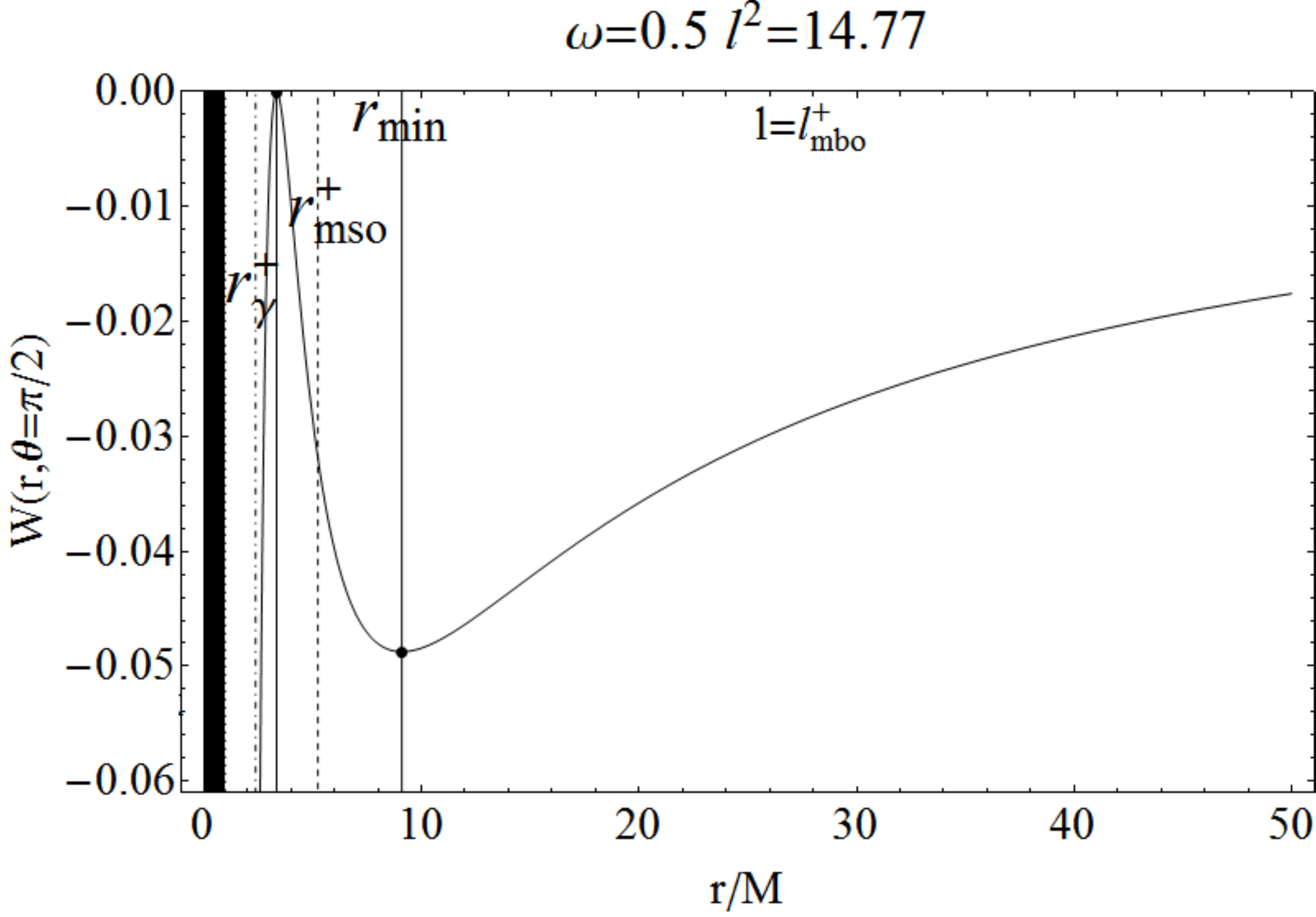}
\includegraphics[width=.31\textwidth]{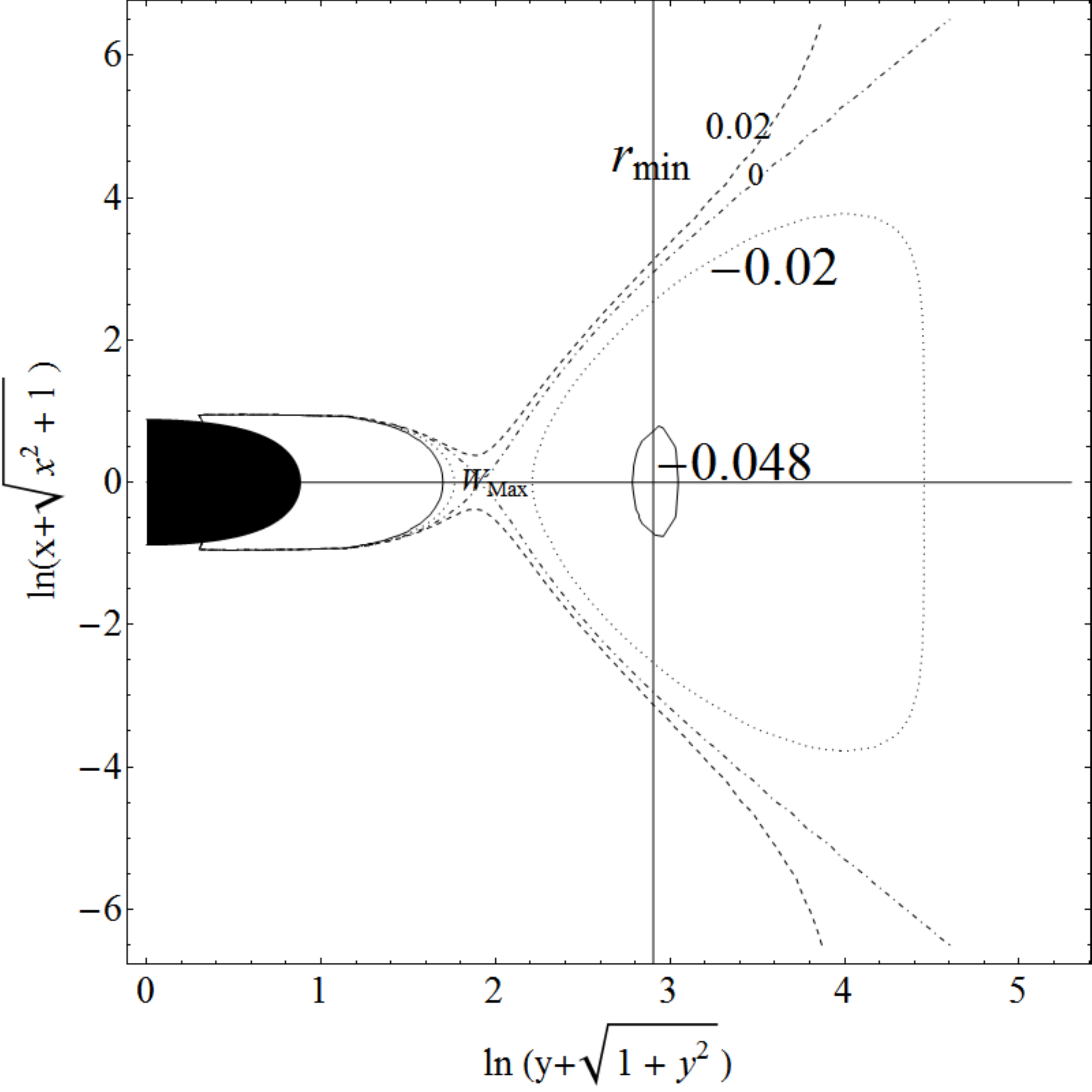}
\caption{Extreme black hole case: $\omega M^2=0.5$. It is $r_{\pm}=M$, $r_{\pm}<r_{\gamma}^+<r_{mbo}^+<r_{mso}^+$ and $l^+_{\gamma}>  l^+_{mbo}> l_{mso}^+$. Black region is $r<r_+$. Where $\omega M^2\rightarrow \omega$.  With $r/M=\sqrt{x^2+y^2}$ and  $(x,y)$ are Cartesian coordinates.  Vertical lines in right panels set the $r_i\in\mathfrak{R}$ and  the effective potential critical points. Black region is $r<r_+$. dotted-dashed line in the left panels is $r_{\gamma}^{\pm}$.}
\label{Fig:mel-eBH}
\end{figure}

\begin{figure}[h]
%%CPlotoiscom
\includegraphics[width=.481\textwidth]{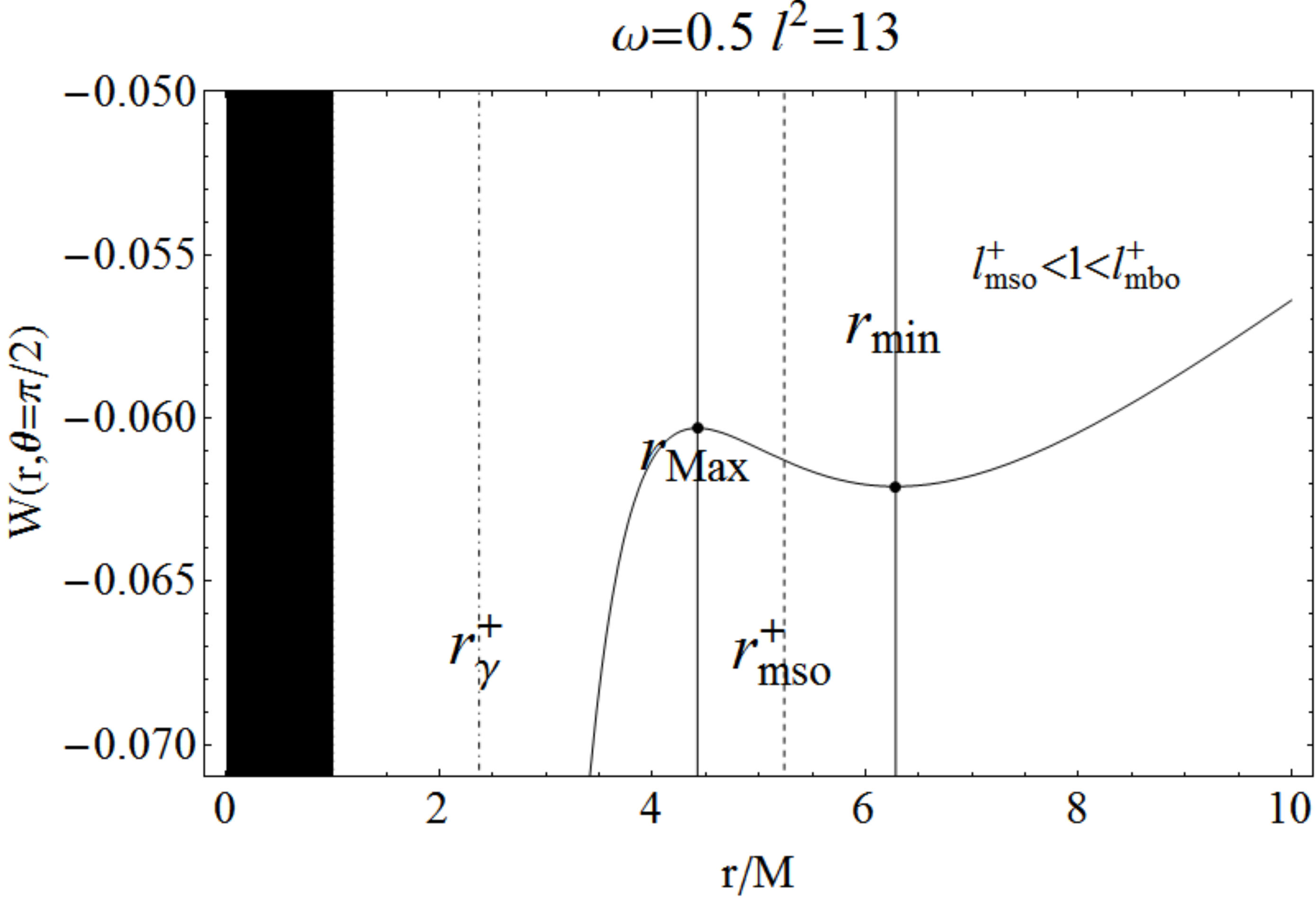}
\includegraphics[width=.31\textwidth]{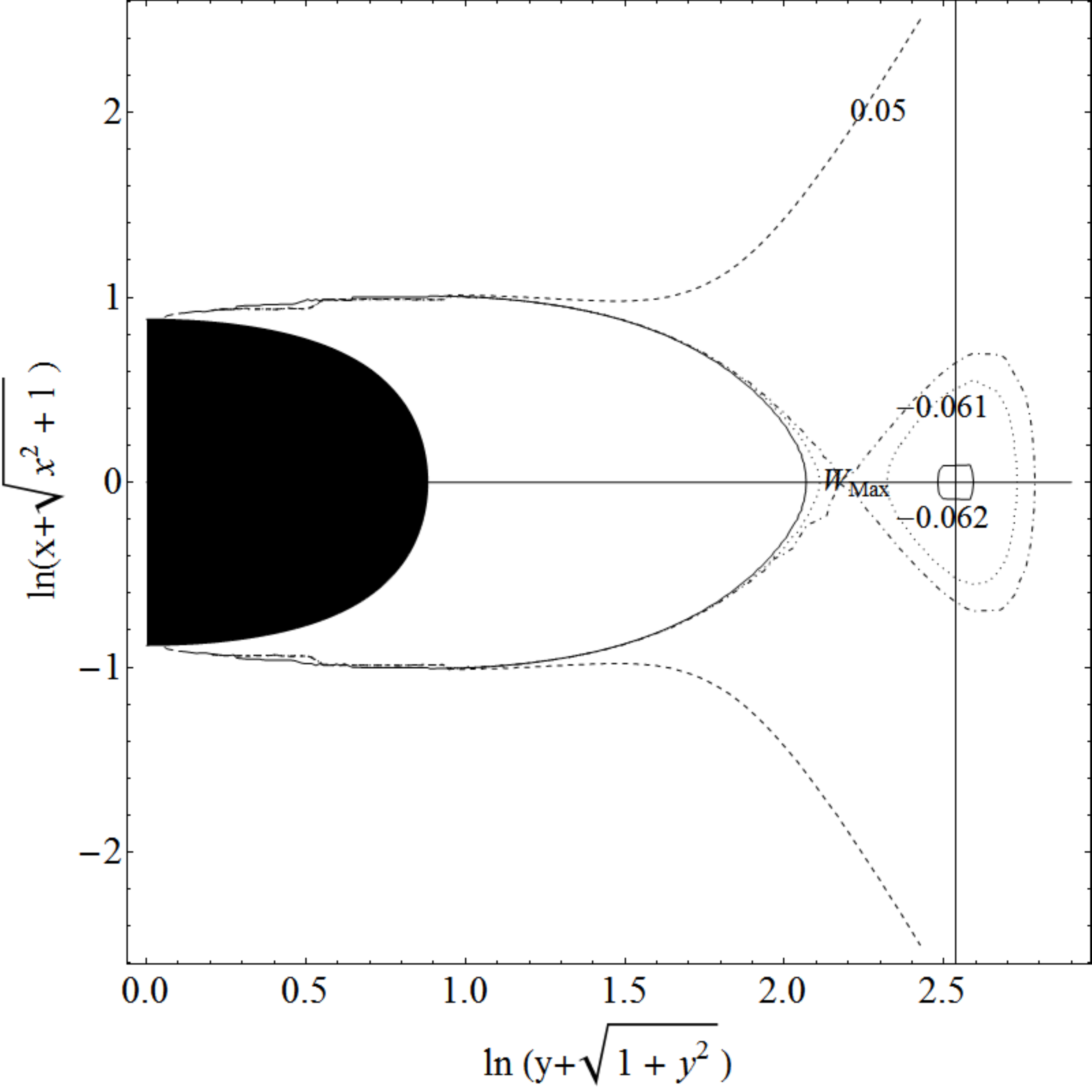}
\\
\includegraphics[width=.481\textwidth]{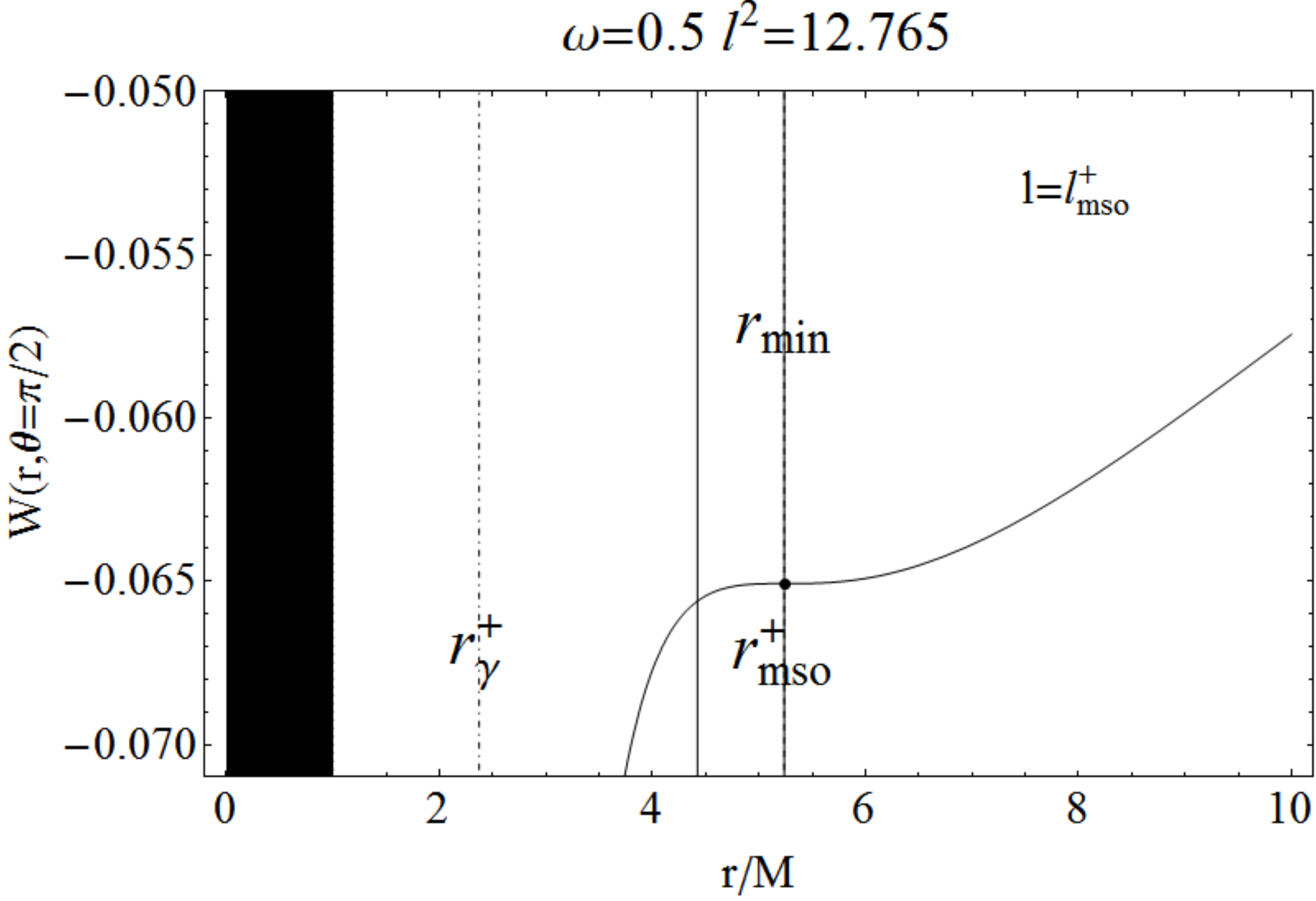}
\includegraphics[width=.31\textwidth]{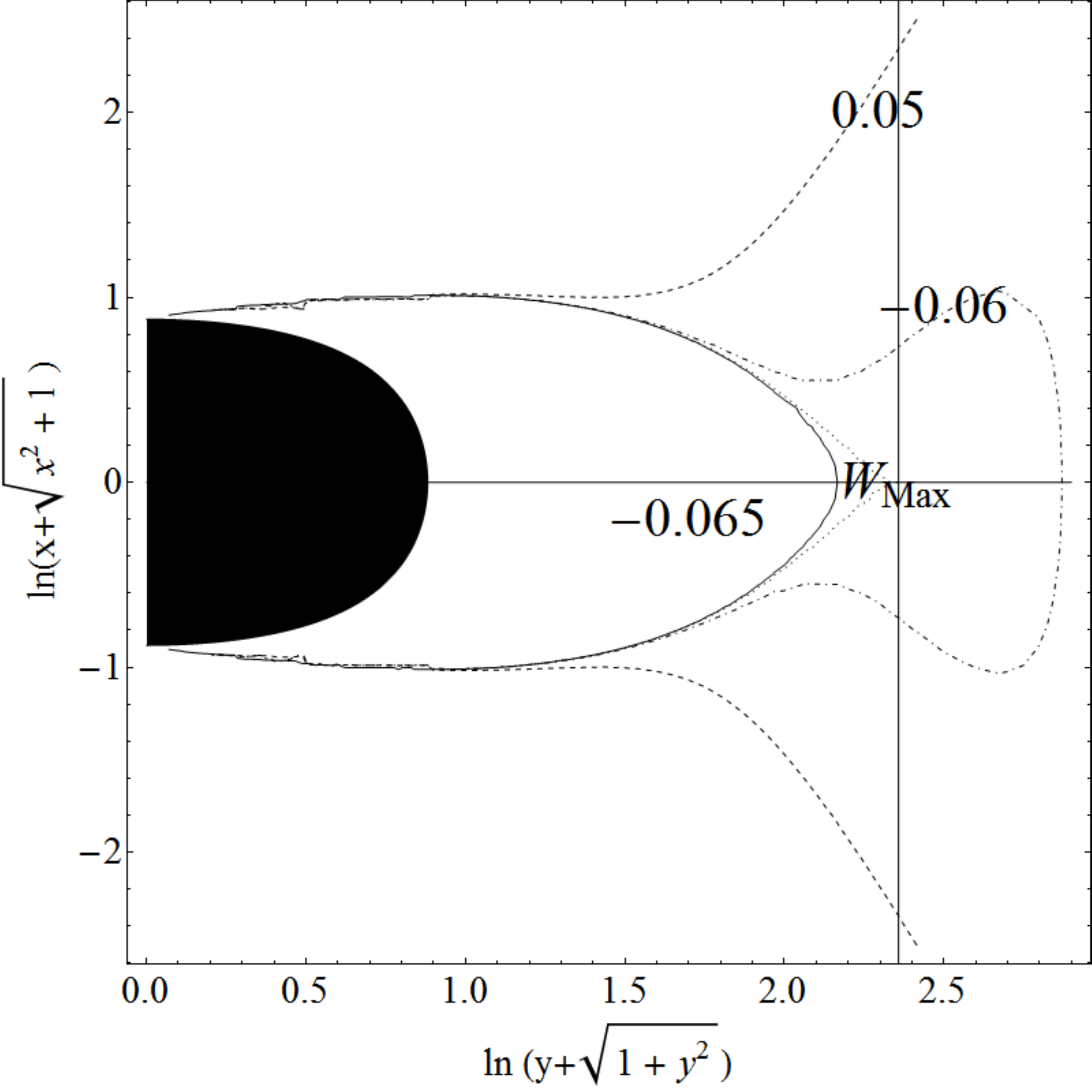}
\\
\includegraphics[width=.481\textwidth]{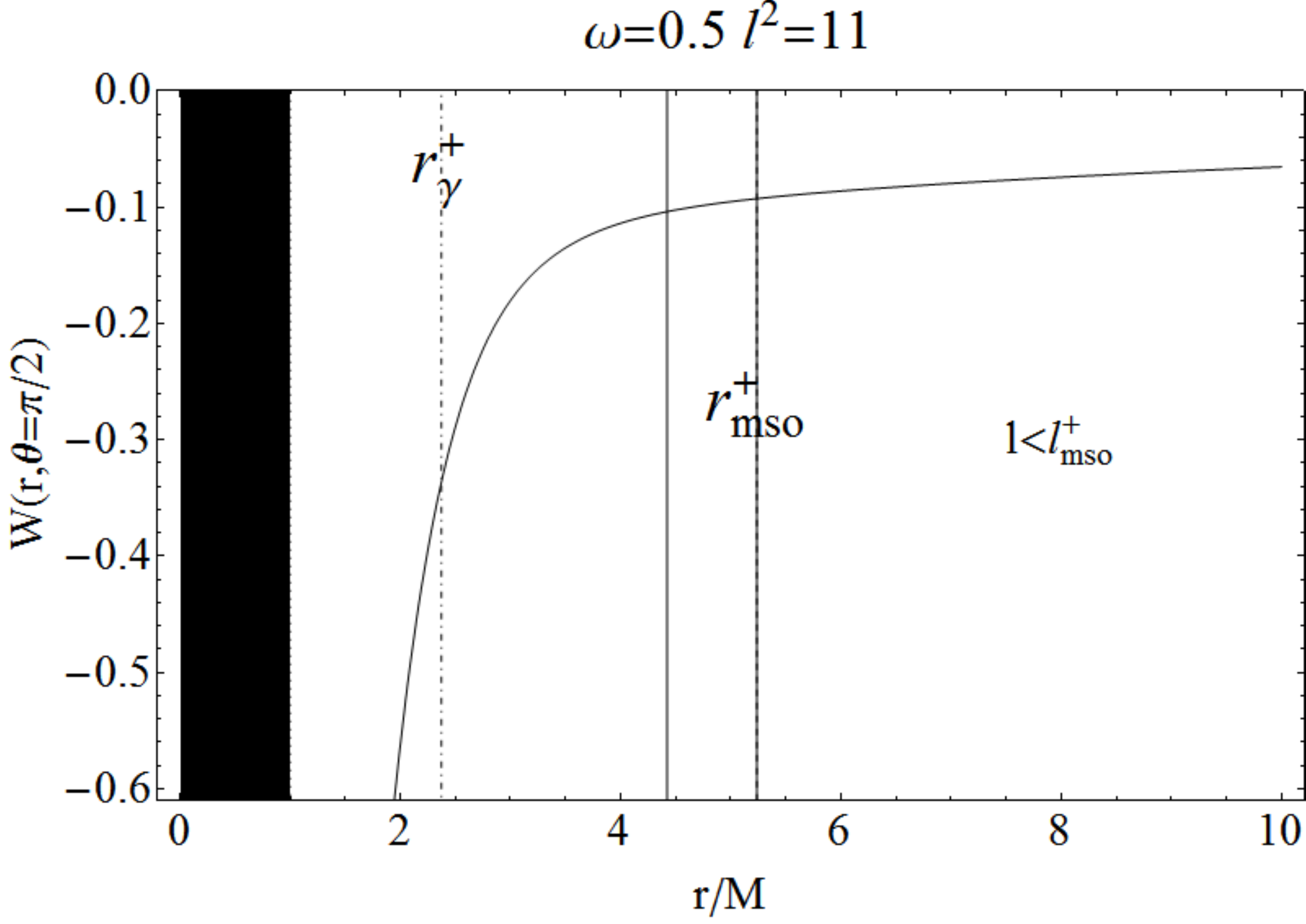}
\includegraphics[width=.31\textwidth]{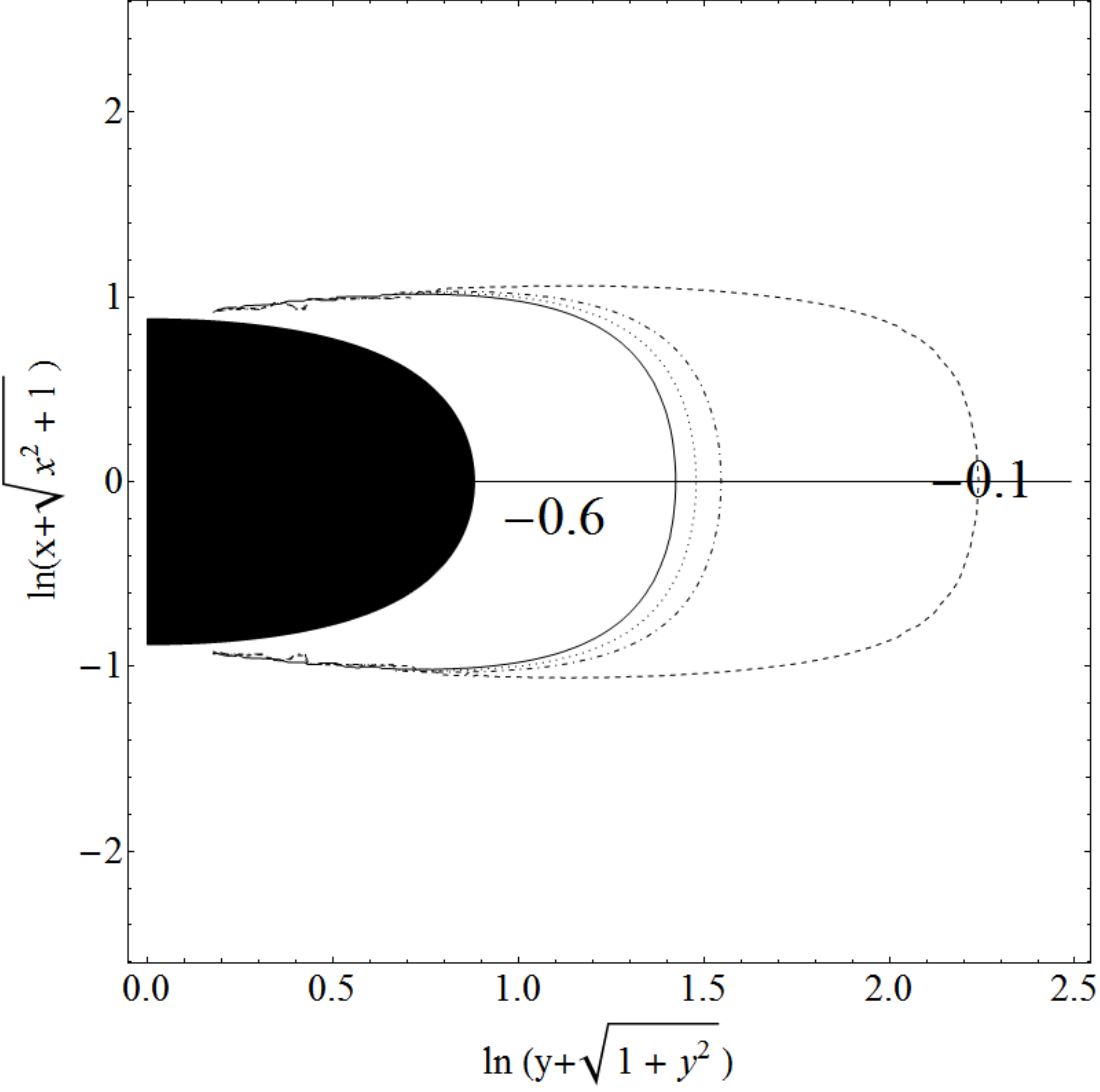}
%\\
\caption{Extreme black hole case: $\omega M^2=0.5$. It is $r_{\pm}=M$ with $r_{\pm}<r_{\gamma}^+<r_{mbo}^+<r_{mso}^+$ and $l^+_{\gamma}>  l^+_{mbo}> l_{mso}^+$. With $r/M=\sqrt{x^2+y^2}$ and  $(x,y)$ are Cartesian coordinates. Black region is $r<r_+$. Where $\omega M^2\rightarrow \omega$.  Vertical lines in right panels set the $r_i\in\mathfrak{R}$ and  the effective potential critical points.}
\label{Fig:hom-x}
\end{figure}

In this section we consider the extreme Kehagias-Sfetsos BH spacetime: $\omega M^2=0.5$. It is  $r_{\pm}=M$, $r_{\pm}<r_{\gamma}^+<r_{mbo}^+<r_{mso}^+$ and $l^+_{\gamma}>  l^+_{mbo}> l_{mso}^+$.  The circular orbits arrangement has been described in Sec.\il(\ref{Sub: ClassIV}) There is only one photon orbit located in $r_{\gamma}^+$, that is the inner limit for circular orbits or the cusp configurations.
\begin{description}
\item[-) $l\geq l_{\gamma}^+$] There is a set of closed configuration and correspondingly
a  set of inner ones, embracing  the black hole, Figs.\il(\ref{Fig:mel-eBH})-(a).
\item[-)${l\in]l_{mbo}^+,l_{\gamma}^+[}$] There are open configurations with  a  cross point see Figs.\il(\ref{Fig:mel-eBH})-c and a closed set of outer ones.
\item[-) $l=l_{mbo}^+$] This is a critical case where the maximum  $W_{Max}=0$ is located in $r_{mbo}^+$, Figs.\il(\ref{Fig:mel-eBH})-d.
\item[-) ${l\in]l_{mso}^+, l_{mbo}^+[}$] There is one maximum of the potential $W_{Max}<0$ and a minimum $W_{min}<0$, as such the inner and outer configurations are connected by the crossed closed surface, Fig.\il(\ref{Fig:hom-x})-a
\item[-)$l=l_{mso}^+$]
These is a saddle point of $W$ located in $r_{mso}^+$ consequently there is an outer cusped surface for the inner set of surfaces see   Fig.\il(\ref{Fig:hom-x})-b.
\item[-) $l<l_{mso}^+$] There are only inner surfaces as in Fig.\il(\ref{Fig:hom-x})-c.
\end{description}
%\clearpage
\subsection{The Black hole case $\omega>0.5$}\label{Sec:K-S-BH}
\begin{figure}[h]
%%CPlotoiscom
\includegraphics[width=.481\textwidth]{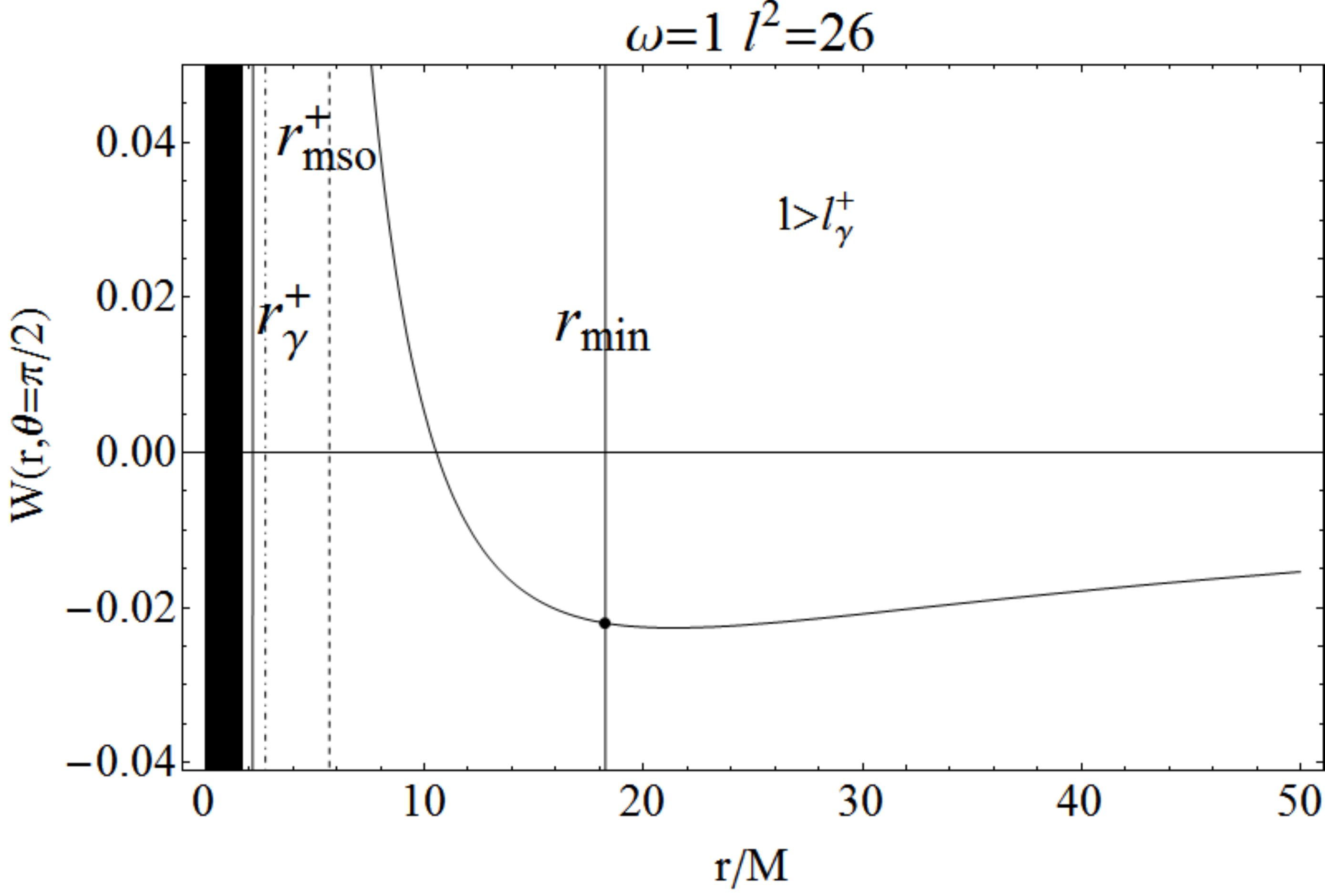}
\includegraphics[width=.31\textwidth]{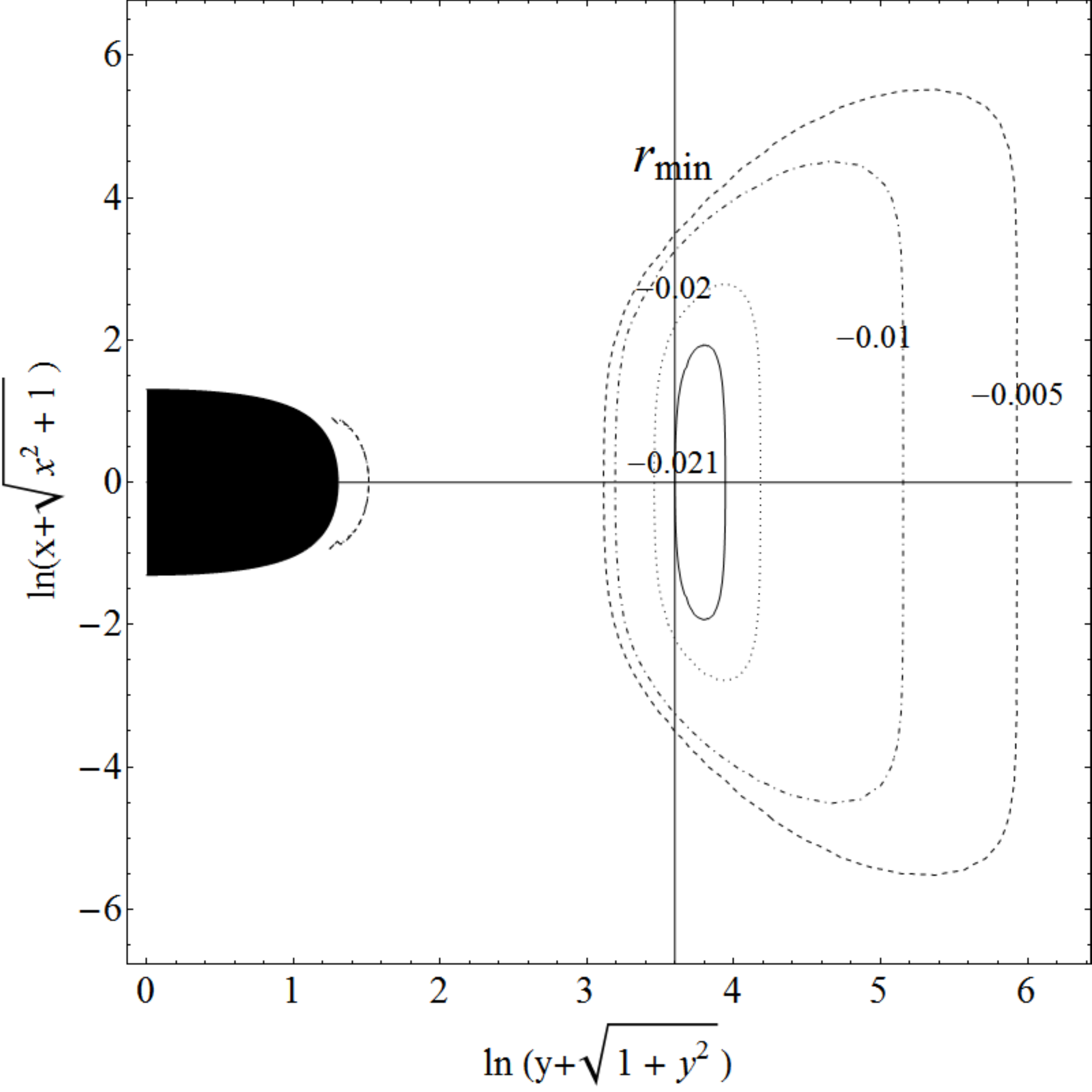}
%\\
\\
\includegraphics[width=.481\textwidth]{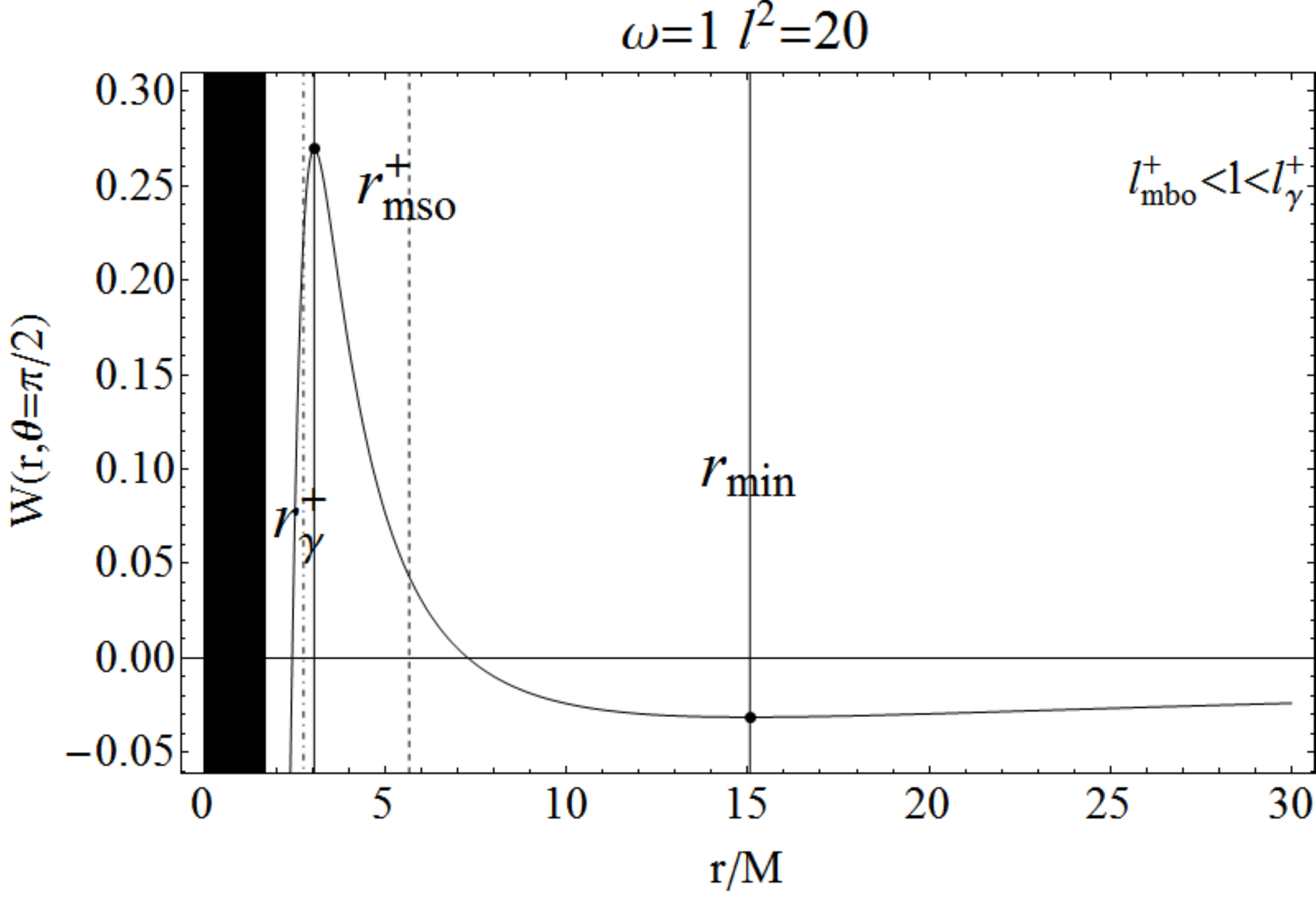}
\includegraphics[width=.31\textwidth]{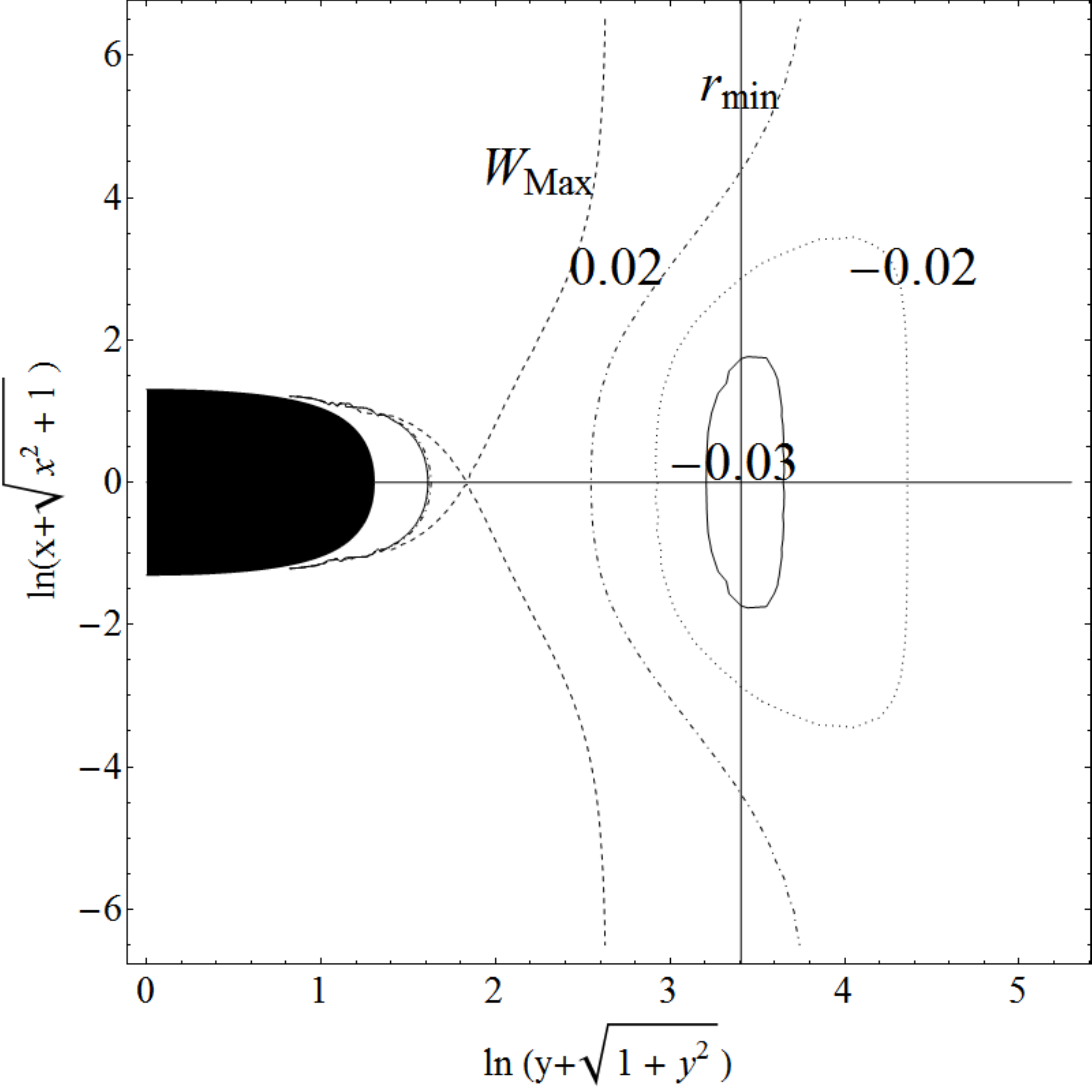}
\\
\includegraphics[width=.481\textwidth]{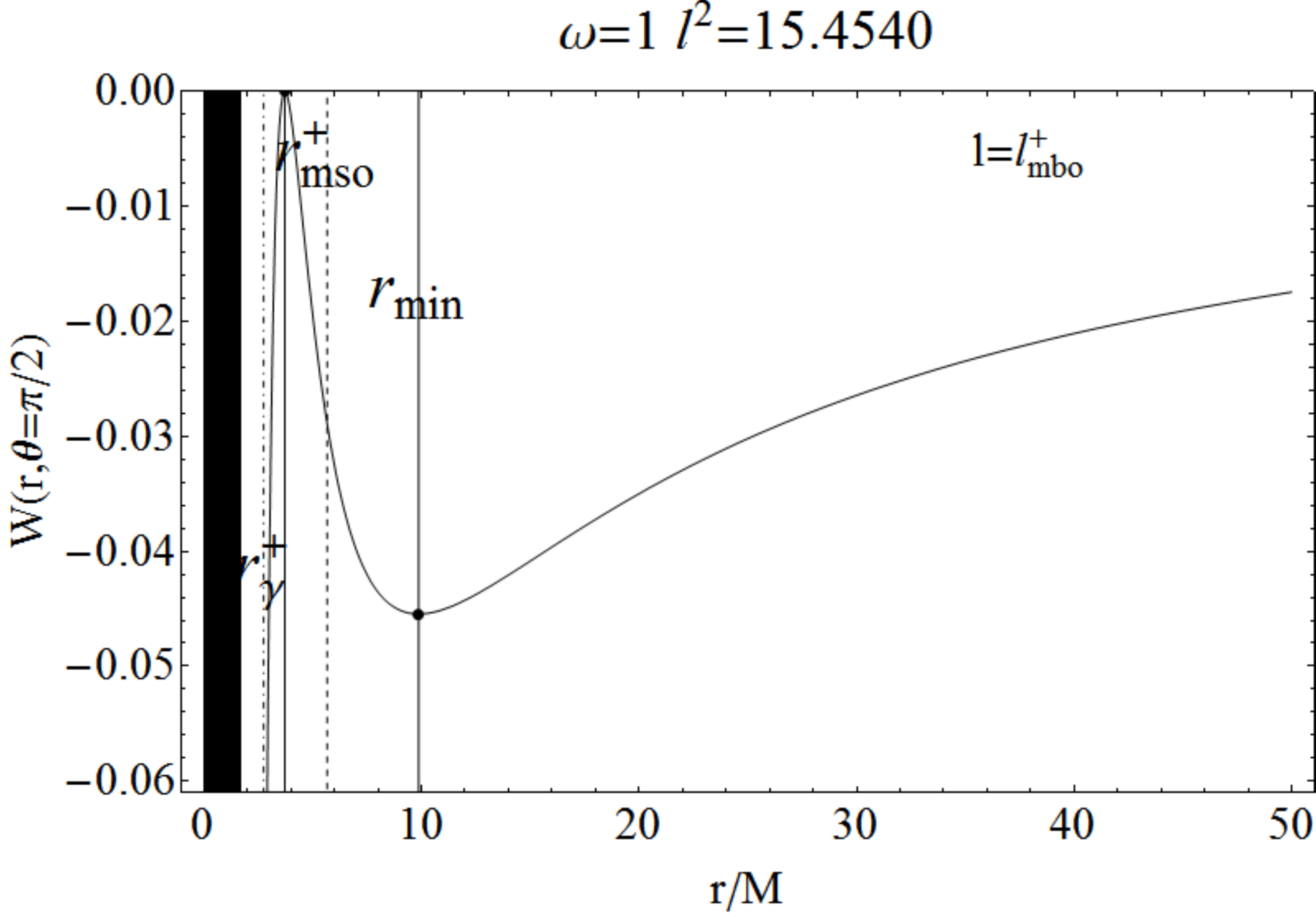}
\includegraphics[width=.31\textwidth]{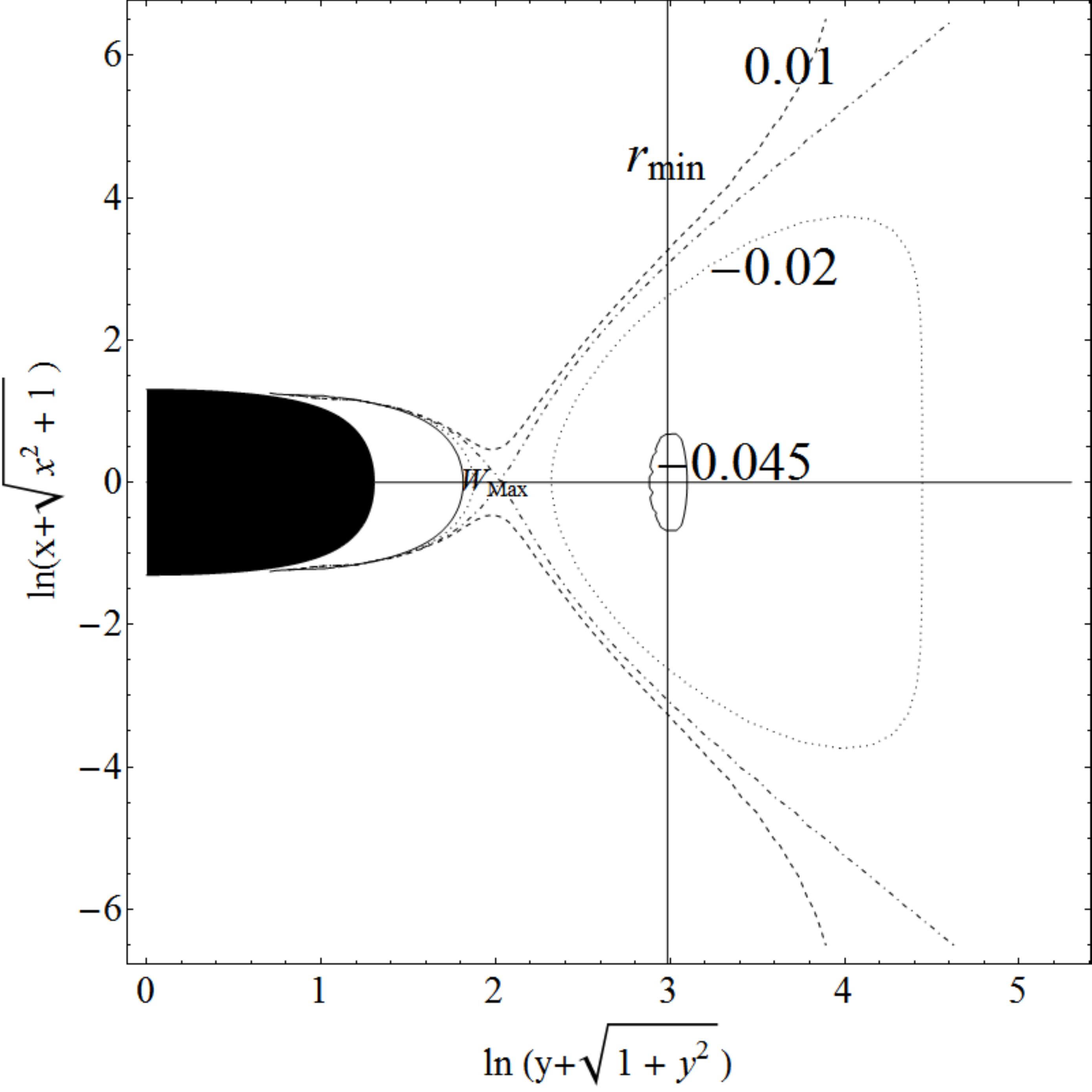}
\caption{Black hole  case.  Vertical lines in right panels set the $r_i\in\mathfrak{R}$ and  the effective potential critical points. It is $\omega M^2=1$ with $r_{+}=(1+1/\sqrt{2})M$ with $r_{+}<r_{\gamma}^+<r_{mbo}^+<r_{mso}^+$ and $l^+_{\gamma}>  l^+_{mbo}> l_{mso}^+$.  With $r/M=\sqrt{x^2+y^2}$ and  $(x,y)$ are Cartesian coordinates. Black region is $r<r_+$. It is $\omega M^2\rightarrow \omega$.}
\label{Fig:xcole-c}
\end{figure}

\begin{figure}[h]
%%CPlotoiscom
\includegraphics[width=.481\textwidth]{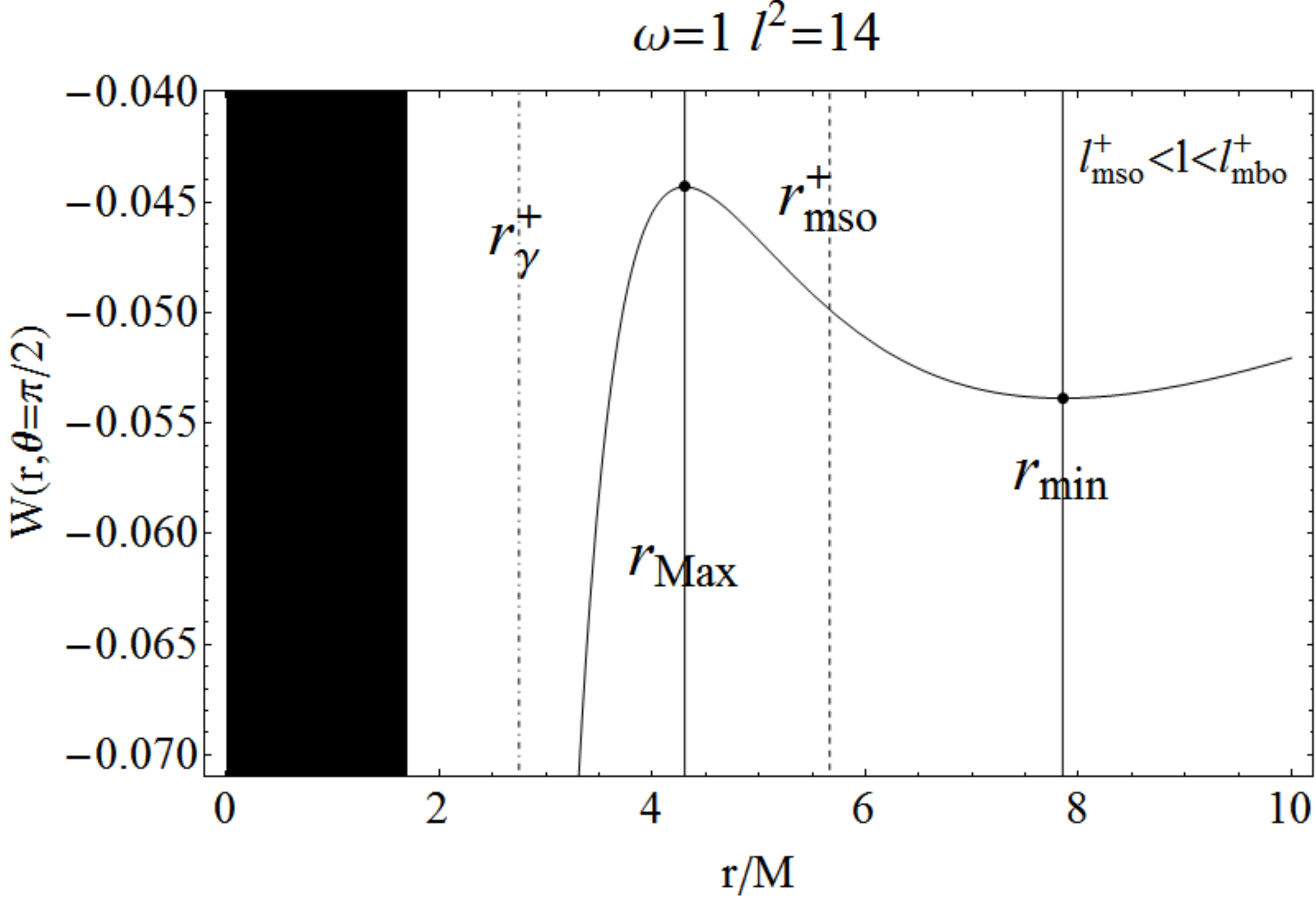}
\includegraphics[width=.31\textwidth]{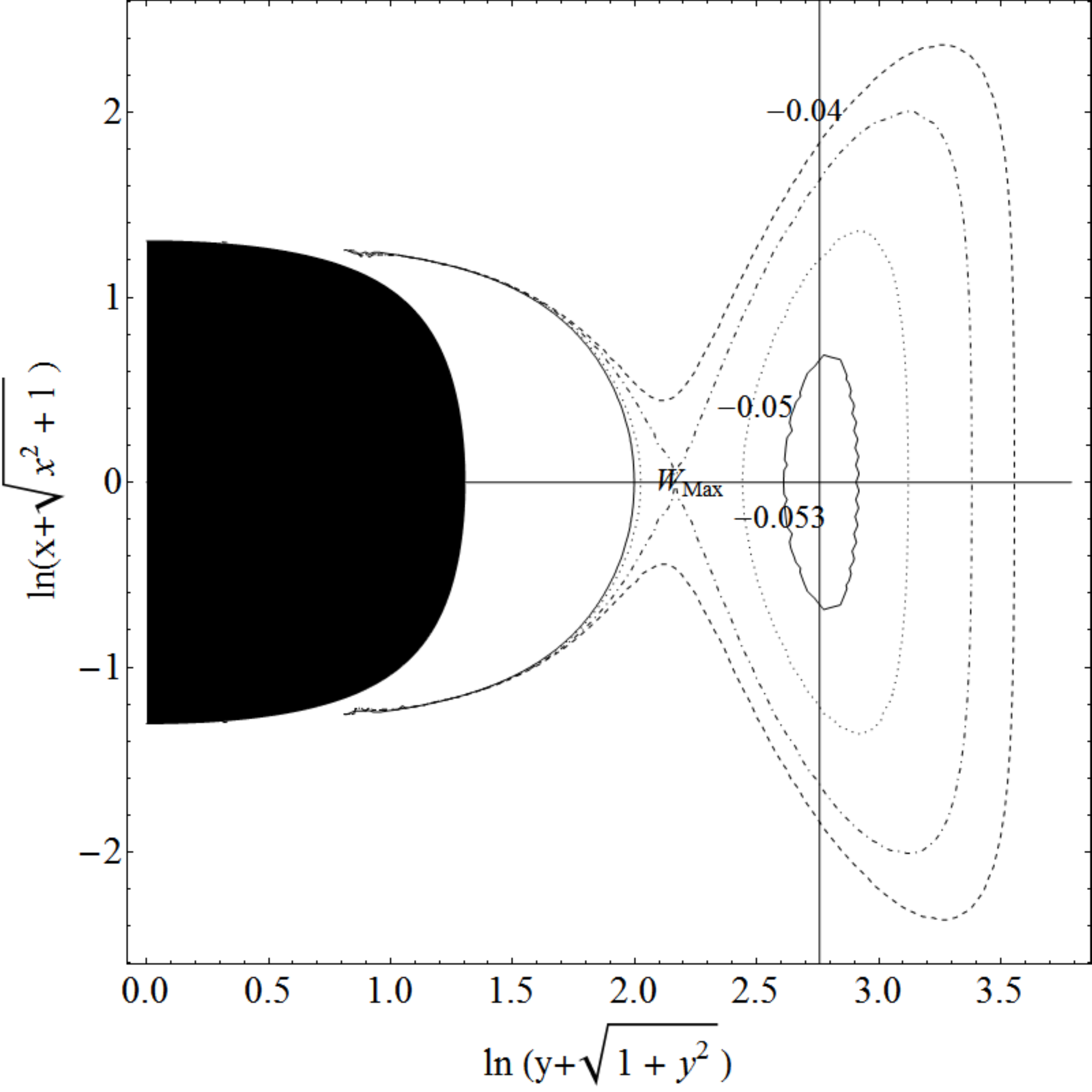}
\\
\includegraphics[width=.481\textwidth]{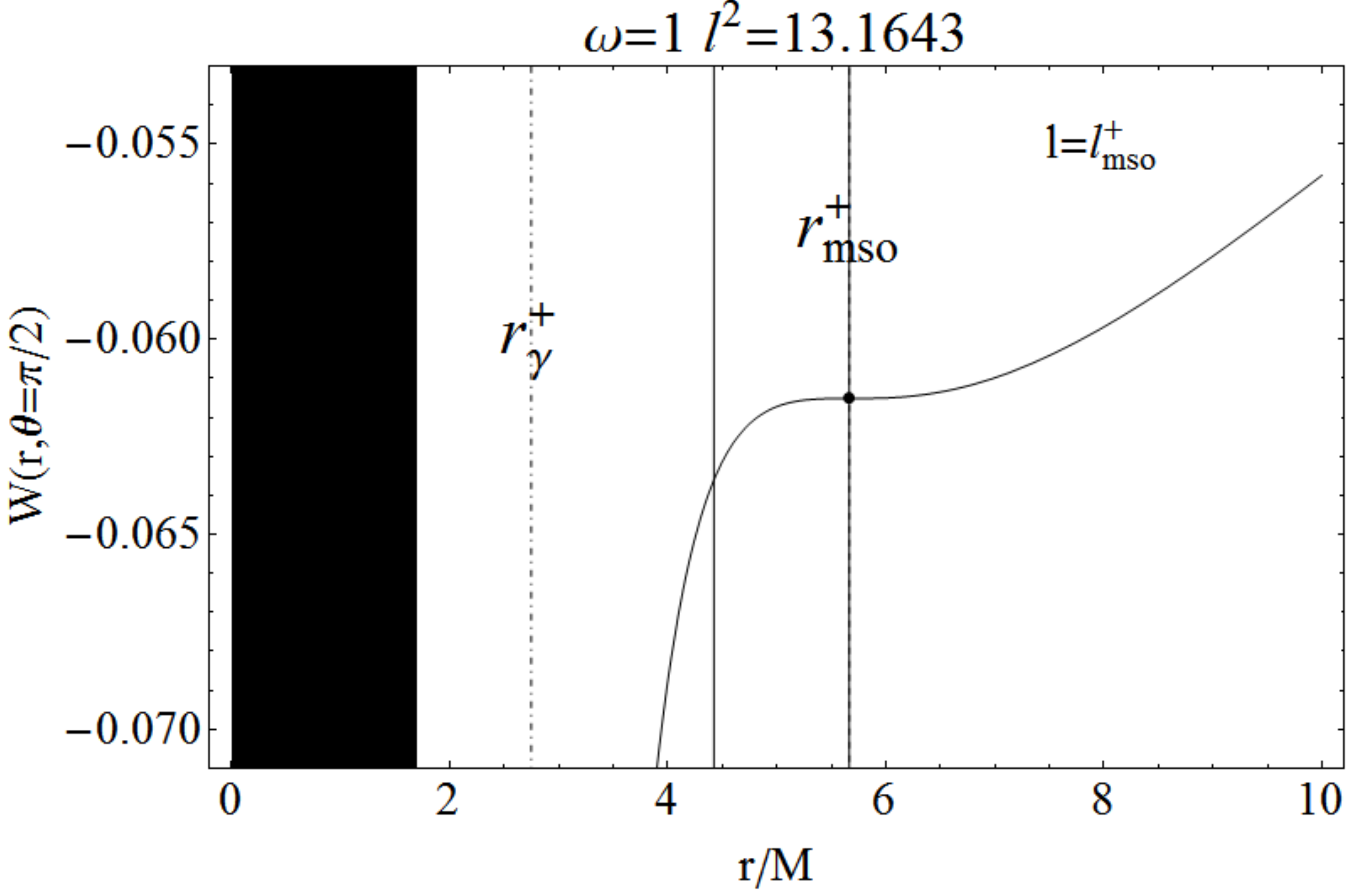}
\includegraphics[width=.31\textwidth]{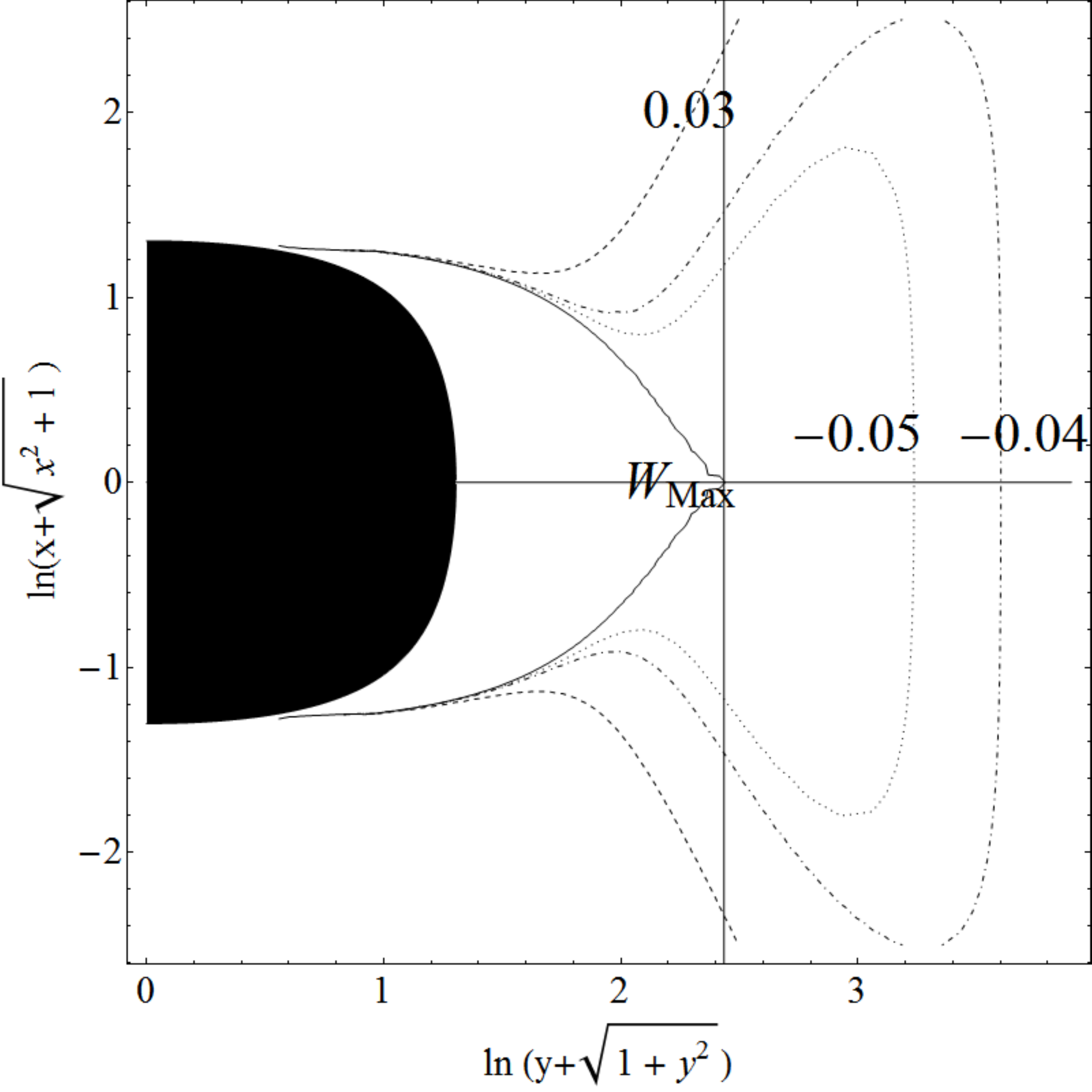}
\\
\includegraphics[width=.481\textwidth]{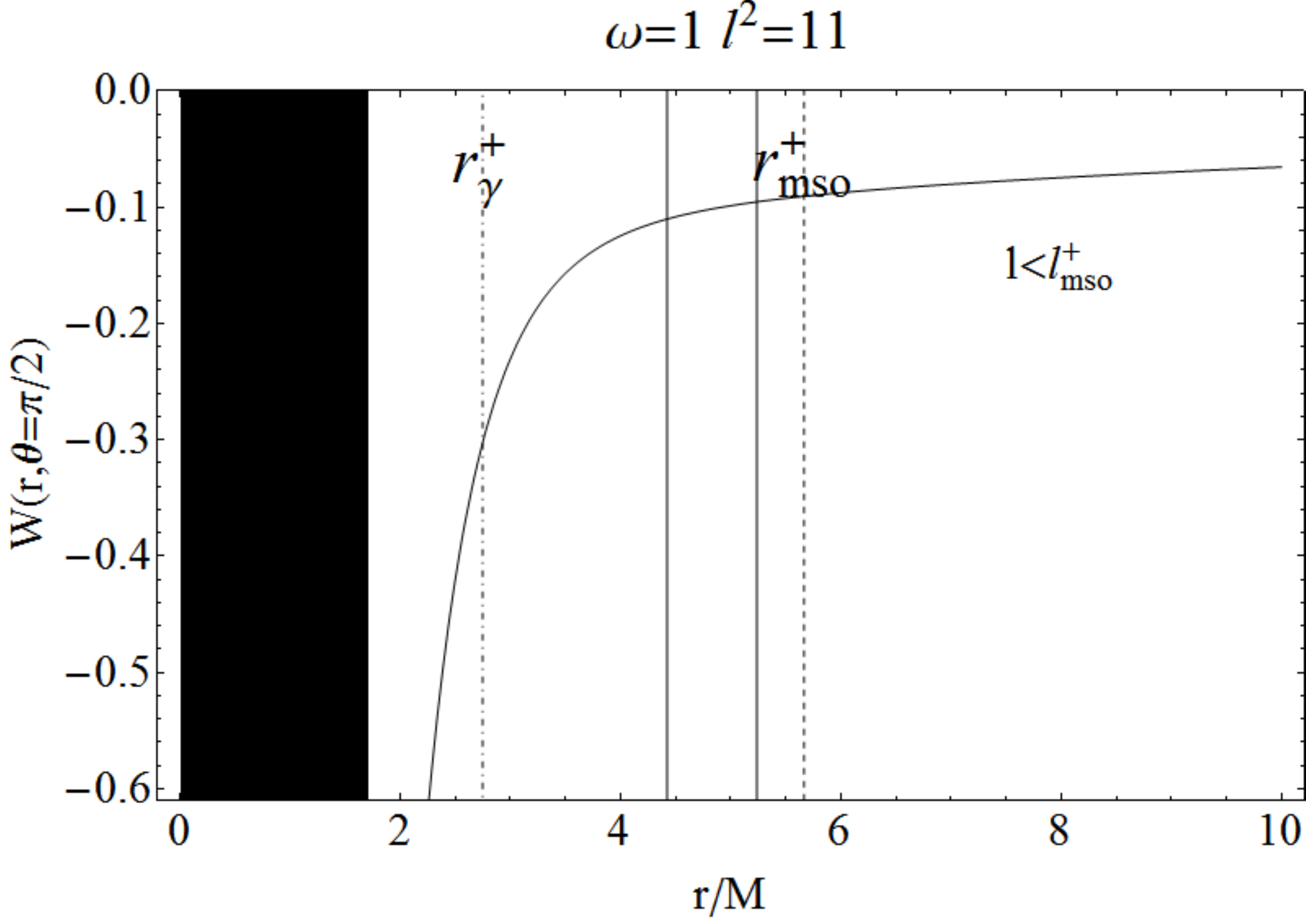}
\includegraphics[width=.31\textwidth]{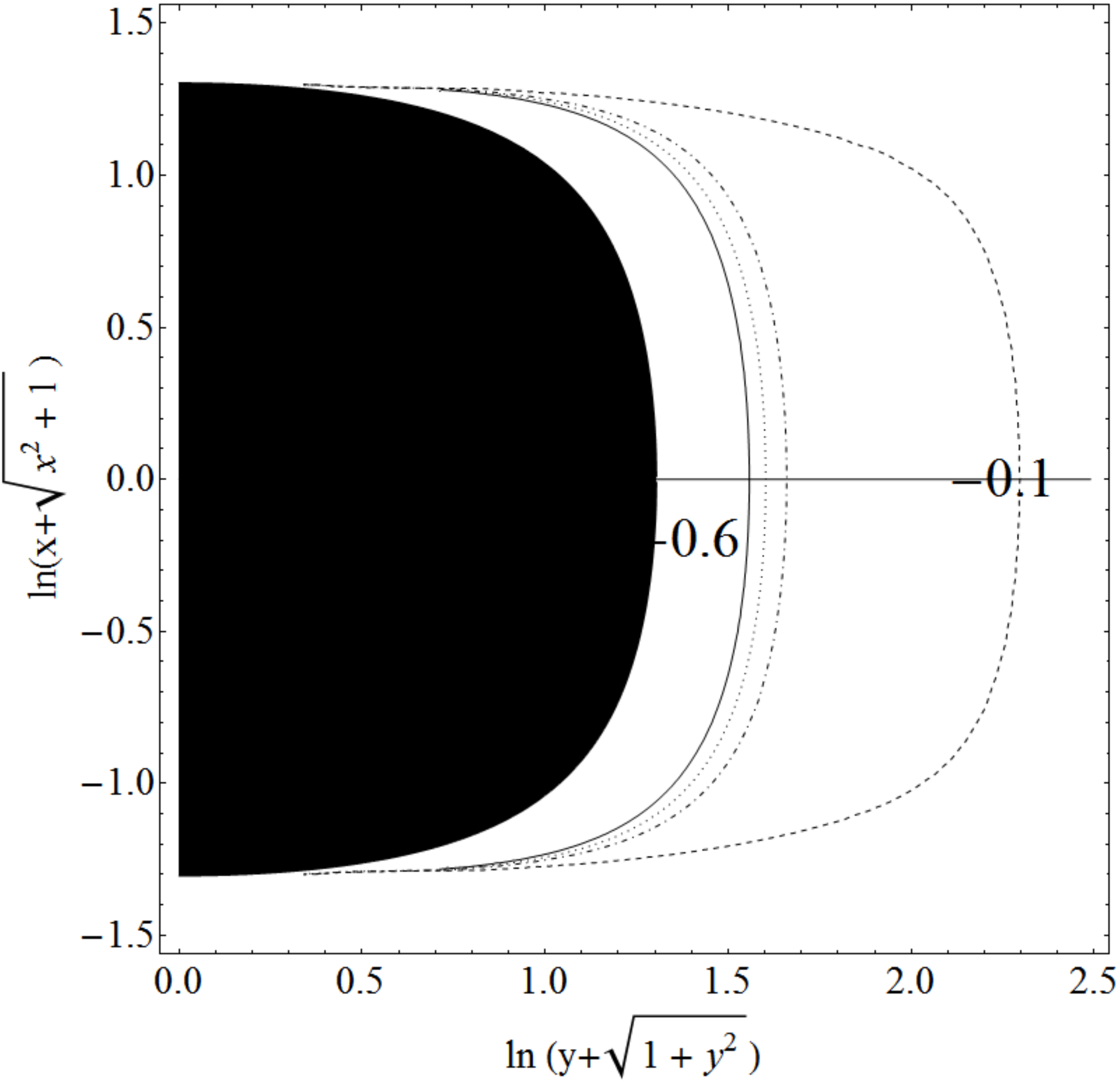}
%\\
\caption{Black hole  case. It is $\omega M^2=1$ with  $r_{+}=(1+1/\sqrt{2})M$, $r_{+}<r_{\gamma}^+<r_{mbo}^+<r_{mso}^+$ and $l^+_{\gamma}>  l^+_{mbo}> l_{mso}^+$. With $r/M=\sqrt{x^2+y^2}$ and  $(x,y)$ are Cartesian coordinates. Black region is $r<r_+$. It is $\omega M^2\rightarrow \omega$.  Vertical lines in right panels set the $r_i\in\mathfrak{R}$ and  the effective potential critical points.}
\label{Fig:Ta-topo-Miln}
\end{figure}

We consider the perfect fluid orbiting The Kehagias-Sfetsos   black hole spacetimes in the region $r>r_+$.
The properties of circular motion  were discussed in Sec.\il(\ref{Sub: ClassIV}). The structure of the Boyer surfaces is similar to the extreme-BH case analysed in Sec.\il(\ref{Sec:K-S-ExBH}), and it can be summarized as follows:
\begin{description}
\item[-) $l\geq l_{\gamma}^+$] This case is illustrated in Figs.\il(\ref{Fig:xcole-c})-(a), there is a set of outer closed surfaces  and an inner one. No cusped, critical, surfaces are possible.
\item[-) $l{\in[l_{mbo}^+,l_{\gamma}^+[}$] There is a set of outer closed surfaces and the inner ones, a crossed open one is for $W_{Max}$, Figs.\il(\ref{Fig:xcole-c})-(b, c)
\item[-) ${l\in]l_{mso}^+,l_{mbo}^+]}$] There is a set of outer closed surfaces and the  inner ones, a  crossed one is for $W=W_{Max}$, Fig.\il(\ref{Fig:Ta-topo-Miln})-a.
\item[-) $l=l_{mso}^+$]
The outer cusp appears for the inner closed  surfaces, Fig.\il(\ref{Fig:Ta-topo-Miln})-b.
\item[-)$l<l_{mso}^+$]
There is no critical point for the effective potential and the configurations are as in  Fig.\il(\ref{Fig:Ta-topo-Miln})-c.
\end{description}
\section{Discussion and  Conclusions}\label{Sec:concl}
We analyzed the structure and shape of perfect fluid toroidal equilibrium tori orbiting the Kehagias-Sfetsos (K-S) sources.
We considered the tori for whole range of the dimensionless parameter $\omega M^2 < 1/2$ correspondent to naked singularity spacetimes and the values $\omega M^2\geq 1/2$ for the black hole solutions.
From the characterization  of the toroidal configurations we then create  a classification of K-S spacetimes having peculiar and distinctive features with respect to the matter dynamics.
According with the properties of the orbiting
fluids  we are able to distinguish  four classes of naked singularity sources correspondent to  four regions of values of the quantum parameter $\omega$.
The extreme BH-case $(\omega M^2=1/2)$ is addressed in Sec.\il(\ref{Sec:K-S-ExBH}) and the geometries defined  with the limiting  values of the quantum parameter in $C_i$ with respect to the stability and morphology  of   the configurations were explored in details.

The first class of naked singularities, identified as {\textbf{Region I} $\omega\in]0,\omega_{mso}[$, is studied
Sec.\il(\ref{Sec:NS-RegionI}): only stable circular orbits are allowed.
Geometries belonging to
\textbf{Region II} with $\omega\in]\omega_{mso},\omega_{mbo}[$, as detailed in Sec.\il(\ref{Sec:NSRegionII}),
are characterized by
two marginally stable circular orbits
 is more articulated as contains two regions of orbital stability  separated by an instability region.
\textbf{Region III} is split in  two sets:
{\textbf{Region III-a} with $\omega\in]\omega_{mbo},\omega_c[$}, considered in Sec.\il(\ref{Sec:NSRegionIII}), presents two marginally bounded orbits and the situation is as follows:
$r_{stat}<r_{\Omega}^{Max}<r_{mbo}^-<r_{mso}^-<r_{mbo}^+<r_{mso}^+$, the second region {\textbf{Region III-b}, as investigated in Sec.\il(\ref{Sec:NSRegionIIIb}),
with Ho\v{r}ava parameter in $\omega\in]\omega_c,\omega_{\gamma}[$, is defined by the radii  $r_{stat}<r_{mbo}^-<r_{\Omega}^{Max}<r_{mso}^-<r_{mbo}^+<r_{mso}^+$ and finally
\textbf{{{Region IV}} } where $\omega\in]\omega_{\gamma},0.5[$, and  considered in Sec.\il(\ref{Sec:NSregionIV}),
contains two photon circular orbits, the
inner stable one  $r_{\gamma}^-$, the outer unstable one at $r_{\gamma}^+$, and the  marginally stable circular orbit $r_{mso}^+$, where  it is
   $r_{stat}<r_{mbo}^-<r_{\gamma}^-<r_{\gamma}^+<r_{mbo}^+<r_{mso}^+$.
In the bend $\Delta_{\gamma}^{\pm}\equiv]r_{\gamma}^-,r_{\gamma}^+[$, no circular geodesics
are allowed.
The situation in the { Black hole case $\omega>0.5$}, detailed in Sec.\il(\ref{Sec:K-S-BH})
 in the region $r>r_+$ is regulated by the existence of a  photon orbit $r_{\gamma}>r_+$, a marginally bounded orbit $r_{mbo}>r_{\gamma}$ and  a marginally stable orbit $r_{mso}>r_{mbo}$.

The different properties of  toroidal configurations in K-S spacetimes are such  to catch signatures of some K-S naked singularity spacetimes that enable to clearly distinguish them from the black hole backgrounds due to the appearance of accretion phenomena.
Thus we can summarize the main differences between the fluid configurations in
K-S naked singularity and black hole  spacetimes
as follows.

For the spacetimes  in the \textbf{Regions II-III-IV}  there are two
types of particle orbits  and fluid configurations   to be considered, associated to  the   outer  and inner circular orbits  $(\pm)$, the farest and closest to the attractor respectively.
This distinction  makes the matter dynamics in  these geometries   very similar to the
case of corotating and  counterrotating particles  orbiting in the Kerr geometry or electrically neutral particles   in the
case of  charged  spherically symmetric attractors, but  in the K-S cases, as for the electrically neutral particles orbiting a  Reissner-Nordstr\"om spacetime, this peculiar feature is due exclusively to the
geometric properties of naked singularity spacetimes when the Ho\v{r}ava parameter is $\omega\in]\omega_{mbo},0.5[$. A more general discussion on the  analogies between the Kehagias-Sfetsos   and the Reissner-Nordstr\"om solutions can be found in \cite{Vie-etal:2014:PHYSR4,Janke:2010zn}.
%\item[b)]

Associated to the two sets of orbits  in naked singularity of \textbf{Region  IV } there is a couple of
 photon  orbits,  the inner stable one and  the outer unstable one, defining a halo where no cusp or disc center can be located. In the  black hole case
case only one limiting light-orbit   exists.
As well in naked singularity spacetimes   of \textbf{Regions  II-IV} there are a couple of stable orbits,  and a couple of  last bounded orbits in  \textbf{Regions III-IV}.
The
properties  of test particle orbits in the  BH and NS attractors   are detailed and discussed in
 \cite{Vie-etal:2014:PHYSR4,Stuchlik:2014yaa}.
%\item[c)]

In the  spacetimes of \textbf{Regions I-III}  the relativistic angular velocity  $\Omega$ has a critical point (as function of $r/M$) located in $r_{\Omega}^{Max}$. The exact location of this  orbit with respect to the  radii in the set  $\mathfrak{R}$ and in particular $r_{mbo}^-$, as $r_{mbo}^-(\omega_{c})=r_{\Omega}^{Max}(\omega_c)$ makes convenient to distinguish naked singularity  sources in \textbf{Region III} in those with $\omega\in]\omega_{mbo},\omega_c[$ (\textbf{Region IIIa}) and those with $\omega\in]\omega_c,\omega_{\gamma}[$ (\textbf{Region IIIb}). These cases are discussion Sec.\il(\ref{Sec:NSRegionIII}) and Sec.\il(\ref{Sec:NSRegionIIIb}) respectively, the main difference being in the   location of $W_{crit}>0$, and  leading therefore to the open or closed critical configurations.

%\item[d)]
In all the   naked singularity spacetimes there is a ``static limit'' where $l_{stat}=0$. Generally stable orbits can be located in $r>r_{stat}$.  The region  $r>r_{stat}$  includes  in geometries of  \textbf{Regions II-IV}  a disconnected, unstable  orbital range in \textbf{Regions II-IIIb}, and   forbidden   halo in \textbf{Regions IV}. The existence  of  a  static limit could be  interpreted as a repulsive gravity  effect due to the  naked singularity. We should note however that $r_{stat}$ decreases with $\omega$,  and as a consequence of this fact as $\omega$ approaches the value for the extreme black hole  case, the regions where (stable) matter configuration can be located closes the singularity, even if  there cannot be a proper accretion point.  This peculiar structure of the stability properties of naked singularity spacetime can be compared with analogue situation for other exact solutions of Einstein equations, for example in the  Reissner Nordstr\"om case  \cite{Pugliese:2010ps,Pugliese:2013zma}, or the  cosmological solutions of  Kerr-de Sitter or    Reissner Nordstr\"om de Sitter-case \cite{Kuc-Sla-Stu:2011:JCAP,Sla-Stu:2005:CLAQG}. However at $r>r_{stat}$ there is always a bend of stable (closed) configurations.
%    \item[e)]

The closed  configurations in the black hole geometries  are associated to an inner matter surface embracing the black hole, see for example Fig.\il(\ref{Fig:xcole-c}). In the naked singularity spacetimes in general there are exterior open surfaces  (as $W_{Max}\geq0$) associated to the closed inner ones, whose distance is regulated by the gap between the critical points of the function $W$, see for example Fig.\il(\ref{Fig:namij-n}). The open equipotential surfaces one the other hand  could be relevant for the description of the outflow of matter in jets or winds.
%\item[f)]

Because of the  presence of two minimum points for the fluid effective potential in the naked singularities of  \textbf{Regions II-IV} it is  possible to identify  three types  of configurations correspondent to the following three cases:  \textbf{I} $W_{min}^->W_{min}^+$,  \textbf{II} $W_{min}^-=W_{min}^+$ and finally  \textbf{III} $W_{min}^-<W_{min}^+$. Their properties are discussed in Sec.\il(\ref{Sec:NSRegionII}).
The alternation of these configurations characterizes and distinguishes the different cases from the \textbf{Region II} for classes of fluids with specific angular momenta.  However when   $W_{min}^-> 0$,  as it is  $W_{min}^+ <0$, the configuration is always of the \textbf{I}-kind
and a typical profile is shown in  Fig.\il(\ref{Fig:outo-ty})-a, in this case it is $W_{Max}>0$, corresponding to crossed configurations opened towards the  outer regions, and  excretion configurations.
These discs are in naked singularity spacetimes of  \textbf{Regions III-IV} where radii $r_{mbo}^{\pm}$ are defined.
Then it could be  $W_{min}^{-}<0$, and it can be as well $W_{Max}>0$, see for example Fig.\il(\ref{Fig:c-tak})-c, or $W_{Max}=0$ Fig.\il(\ref{Fig:c-tak})-d or finally $W_{Max}>0$  as in Fig.\il(\ref{Fig:Abe-1})-a. However the sign of the maximum values only determines the open or closed topology of the unstable configurations. Focusing on the minimum values:
it can be \textbf{I}-kind $W_{min}^->W_{min}^+$, see for example  Figs.\il(\ref{Fig:Rip-second}) or $W_{min}^-<W_{min}^+$, see for example  Figs.\il(\ref{Fig:ira-n-b}). The second, limiting case, is \textbf{II}-$W_{min}^-=W_{Max}^+$ satisfies the conditions: $W(r_{min}^-,l)=W({r_{min}^+,l})$, $\partial_r W(r_{min}^-,l)=\partial_r W(r_{min}^+,l)=0$ and $\partial^{2}_r W(r_{min}^-,l)>0$, $\partial^{2}_r W(r_{min}^+,l)>0$. %and we should consider the radius $r_{mso}^{\pm}$ and $r_{mbo}^{\pm}$.
%\\
%\item[g)]
In geometries of   \textbf{Regions II-IV},  configurations can lead  to  excretion of matter towards the outer space but   not accretion, see also \cite{Sla-Stu:2005:CLAQG} and \cite{Kuc-Sla-Stu:2011:JCAP} for similar cases.
 Correlated to this,
there is an inner cusp of the potential, in $r_{mso}^-$ for $l=l_{mso}^-$,  and an outer  cusp in $r_{mso}^+$ with $l=l_{mso}^+$.
Consequently there is  couple of stable (or crossed) closed
configurations  at equal $l$ and $K$, generating an inner  and outer discs
(corresponding  to  two minima for the Keplerian disc effective potential).
The two sets of  closed configurations can be separated by an open crossed configuration with an outer
 cusp (as only excretion is possible  not accretion), when $W_{Max}\geq0$, or  by a closed crossed surface  with two centers  into the maximums points of the pressure when $W_{Max}<0$.  In this binary system  can occur then a feeding of matter from one  configuration (the inner smaller one) to the companion (the outer larger  one). It is possible for $W\in]W_{Max},0[$ one closed configuration with two maxima of the thickness for the two
extrema of the pressure $p_{Max}$ for the outer    and inner configurations respectively. These double ``centers'' are clearly related to the three  classes  of  configurations introduced in \textbf{Region III}.

We can conclude that character of the toroidal equilibrium configurations of perfect fluid could give  clear signatures of the presence of the Kehagias-Sfetsos naked singularities representing a simple solution of the laws of the Horava quantum gravity.

%\end{description}
 \section*{Acknowledgements}
The authors acknowledge the Institutional support of the Faculty of Philosophy and Science of the Silesian University at Opava.  ZS and JS  acknowledge the Albert Einstein Centre for gravitation and astrophysics supported by the Czech Science Foundation Grant No. 14-37086G.

%%%%%%%%%%%%%%%%%%%%%%%%%%%%%%%%%%%%%%%%%%%%%%%%%%%%%%%%%%%%%%%%%%%%%%%%%%%%%%%%%%%%%%%%%%%%%%%%%%%%%%%
\end{document}